\pdfoutput=1
\newcommand*{\ATLASLATEXPATH}{latex/}
\documentclass[cernpreprint,texlive=2016,txfonts,UKenglish]{\ATLASLATEXPATH atlasdoc}
\usepackage[subfigure,biblatex=true]{\ATLASLATEXPATH atlaspackage}
\usepackage{\ATLASLATEXPATH atlasbiblatex}
\usepackage{\ATLASLATEXPATH atlasphysics}

\graphicspath{{logos/}{figures/}}

\usepackage{WZInclusive-defs}

\AtlasTitle{Precision measurement and interpretation of inclusive
$W^+$, $W^-$ and $Z/\gamma^*$ production cross sections with the ATLAS detector}

\author{The ATLAS Collaboration}

\AtlasRefCode{STDM-2012-20}

\PreprintIdNumber{CERN-EP-2016-272}

\AtlasJournal{EPJC}
\AtlasJournalRef{Eur. Phys. J. C 77 (2017) 367}
\AtlasDOI{10.1140/epjc/s10052-017-4911-9}

\AtlasAbstract{High-precision measurements by the ATLAS Collaboration
are presented of inclusive \Wpluslnu, \Wminuslnu\ and \Zgll\
($\ell=e,\mu$) Drell--Yan production cross sections 
at the LHC.
The data were collected in proton--proton collisions at $\sqrt{s} =
7\tev$ with an integrated luminosity of \lumi. Differential $W^+$ and $W^-$
cross sections are measured in a lepton pseudorapidity range 
$|\etal|<2.5$. Differential \Zg\ cross sections are measured as a
function of the absolute dilepton rapidity, for $|\yll| < 3.6$, for three
intervals of dilepton mass, \mll, extending from $46$ to
$150\gev$. The integrated and differential electron- and muon-channel
cross sections are combined and compared to theoretical predictions
using recent  sets of 
parton distribution functions. The data, together with the final
inclusive $e^{\pm}p$ scattering cross-section data from H1 and ZEUS,
are interpreted in a next-to-next-to-leading-order QCD analysis, and a new
set of parton distribution functions, \epWZ16, is obtained. The ratio of
strange-to-light sea-quark densities in the proton is determined 
more accurately than in previous determinations based on collider data only, and is established to be close to unity
in the sensitivity range of the data. A new
measurement of the CKM matrix element \Vcs\ is also provided.}

\hypersetup{pdftitle={ATLAS document},pdfauthor={The ATLAS Collaboration}}

\newcommand{\sigfidZmu}{\ensuremath{  477.8 \pm 0.4\,\mathrm{(stat)}  \pm 2.0\,\mathrm{(syst)}  \pm 8.6\,\mathrm{(lumi)}  }}
\newcommand{\sigfidWmuplus}{\ensuremath{  2839 \pm 1\,\mathrm{(stat)}    \pm 17\,\mathrm{(syst)}  \pm 51\,\mathrm{(lumi)}  }}
\newcommand{\sigfidWmuminus}{\ensuremath{   1901 \pm 1\,\mathrm{(stat)}  \pm 11\,\mathrm{(syst)}  \pm 34\,\mathrm{(lumi)}  }}

\renewcommand{\arraystretch}{1.3}

\begin{document}

\maketitle

\tableofcontents

\clearpage

\section{Introduction}
\label{sec:intro}
The precise measurement of  inclusive $W^+$, $W^-$ and $Z/\gamma^*$
production in $pp$ scattering at the LHC constitutes a sensitive test
of perturbative Quantum Chromodynamics (QCD).  The rapidity
dependence of boson production in the Drell--Yan process provides
constraints on the parton distribution functions (PDFs) of the proton,
as the boson rapidity is strongly correlated with the proton momentum
fractions $x_1$, $x_2$ carried by the partons participating in the
hard scattering subprocess. The weak and electromagnetic components of the
neutral current (NC) process, \Zgll, combined with the weak charged
current (CC) reactions, \Wpluslnu\ and \Wminuslnu, probe the quark flavours of the proton in a way that complements the information from deep inelastic
lepton--hadron scattering (DIS).

The previous differential $W,\,Z$ cross-section measurement of
ATLAS~\cite{Aad:2011dm} at a centre-of-mass energy of $\sqrt{s}=7\TeV$
was based on a data sample taken in 2010 with an integrated luminosity
of $36\ipb$, determined with an uncertainty of $3.5\%$. The
precision of that measurement -- not including the luminosity
uncertainty -- reached about $2$--$3\%$. The new $W^{\pm},~Z$
cross-section measurement  presented here uses the data taken
at $\sqrt{s}=7\TeV$ by ATLAS in 2011. This data sample has a hundred
times more integrated luminosity, \lumi, measured with an improved
precision of $1.8\%$~\cite{Aad:2013ucp}.
A deeper understanding of detector performance and refined
analysis techniques are crucial to reach a measurement precision at
the sub-percent level, apart from the luminosity uncertainty.

Compared to the previous analysis~\cite{Aad:2011dm}, in this article the NC
measurement range is extended to values of dilepton mass, \mll\ ,
significantly below and above the $Z$ peak, covering the range $46 <
\mll < 150\GeV$.  ATLAS NC data have also been presented at
even lower \cite{Aad:2014qja} ($12 < \mll < 66\GeV$) and higher
dilepton masses \cite{Aad:2013iua, Aad:2016zzw} ($116 < \mll <
1500\GeV$). 
Precise NC measurements at $\sqrt{s}=8\TeV$ over a range
of dilepton masses of $12<\mll<150\gev$ focused on boson transverse
momentum distributions have been provided in
Ref.~\cite{Aad:2015auj}. 
Recently, first integrated cross-section results on
inclusive $W^{\pm}$ and $Z$ production at $\sqrt{s}=13\TeV$ were
published by ATLAS~\cite{Aad:2016naf}.

Weak boson cross-section measurements at forward rapidity
were presented by 
LHCb~\cite{Aaij:2012vn, Aaij:2012mda, Aaij:2014wba, Aaij:2015gna,
Aaij:2015vua, Aaij:2015zlq, Aaij:2016qqz, Aaij:2016mgv} in the muon and electron
channels. The CMS Collaboration has measured 
NC cross sections as a function of boson mass
and rapidity~\cite{Chatrchyan:2013tia, CMS:2014jea},
of boson transverse momentum and rapidity~\cite{Khachatryan:2015oaa}, as well as
differential $W^\pm$ charge asymmetries~\cite{Chatrchyan:2012xt,
Chatrchyan:2013mza, Khachatryan:2016pev}, and integrated $W$ and $Z$
cross sections~\cite{CMS:2011aa, Chatrchyan:2014mua}.

The precision of the present measurement of the $W^\pm$ and
$Z/\gamma^*$ cross sections exceeds that of the previous related
measurements. The
analysis is performed in both the electron channels, \Wpmenu\ and
\Zgee, and the muon channels, \Wpmmunu\ and \Zgmumu, in a common
fiducial phase space. These measurements provide a new sensitive test
of electron--muon universality in the weak interaction
sector. The electron and muon data are combined, accounting for all
correlations of systematic uncertainties.

Cross-section calculations of the Drell--Yan process are available at
up to next-to-next-to-leading order in the strong coupling constant
$\alphas$ (NNLO QCD) and up to next-to-leading order for electroweak
effects (NLO electroweak). 
The NNLO QCD predictions are calculated 
with kinematic requirements applied to 
match the detector acceptance using
the DYNNLO~\cite{Catani:2007vq,Catani:2009sm} and 
FEWZ~\cite{Gavin:2010az, Gavin:2012sy, Li:2012wn} programs.  The NLO
electroweak corrections are an important ingredient at this level of
precision and can be evaluated with FEWZ for the NC processes 
and with the SANC
programs~\cite{Andonov:2004hi} for both NC and CC processes.
The measured integrated and
differential cross sections are compared to calculations using various
recent PDF sets:  ABM12~\cite{Alekhin:2013nda},
CT14~\cite{Dulat:2015mca}, HERAPDF2.0~\cite{Abramowicz:2015mha},
JR14~\cite{Jimenez-Delgado:2014twa},
MMHT14~\cite{Harland-Lang:2014zoa}, and NNPDF3.0~\cite{Ball:2014uwa}.
A quantitative analysis within a profiling
procedure~\cite{Paukkunen:2014zia,Camarda:2015zba} is presented to
test the compatibility of the new $W,~Z$ cross-section data with
theoretical predictions using these PDF sets, and to illustrate the impact
of the data on PDF determinations.

The previous ATLAS $W,~Z$ cross-section
measurement~\cite{Aad:2011dm} and its QCD interpretation~\cite{Aad:2012sb} suggested that the
light quark sea ($u,~d,~s$) is flavour symmetric, i.e. the
ratio of the strange-to-anti-down 
quark densities, $r_s=(s+\bar{s})/2\bar{d}$,
was found to be close to unity at $x \simeq 0.023$ within an
experimental uncertainty of about $20\%$. 
This is re-examined here in a new
QCD fit analysis using the present ATLAS measurement together with the
final, combined  NC and CC DIS
cross-section data
from the H1 and ZEUS experiments at the HERA
collider~\cite{Abramowicz:2015mha}. The analysis provides a new NNLO
PDF set, \epWZ16, superseding the \epWZ12 set~\cite{Aad:2012sb}.
It also allows the magnitude of the CKM
matrix element \Vcs\ to be determined, 
without assuming unitarity of the CKM matrix, with a precision comparable to the
determinations from charm hadron decays~\cite{Agashe:2014kda}.

The paper is organized as follows. \SSec~\ref{sec:detsim} presents the
detector, data and simulated event samples and cross-section as well as 
kinematic definitions. The measurements, of both
the $W^{\pm}$ and the $Z/\gamma^*$ reactions, are performed
independently for the electron and muon decay channels as  described in
Sections~\ref{sec:elana} and \ref{sec:muana}. 
The cross-section results are presented
in \Sec~\ref{sec:crossres}, which contains the analysis method, a test
of electron--muon universality, and  a description of the
procedure for, and results of, combining the electron and the muon
data. In \Sec~\ref{sec:comnnlo} 
the integrated and  differential cross sections are compared
with theoretical calculations using recent NNLO PDF sets. 
Measurements are also presented
of the $W^{\pm}$ charge asymmetry and various other
cross-section ratios. This section concludes with the results of the
PDF profiling analysis.  Finally, \Sec~\ref{sec:qcdana} presents an
NNLO QCD fit analysis of the present ATLAS data and the 
final HERA NC and
CC DIS cross-section data, resulting in an improved determination of the
strange-quark distribution in the proton and a measurement of
\Vcs.  A summary of the paper is presented in
\Sec~\ref{sec:summary}.

\section{Detector, simulation and definitions}
\label{sec:detsim}

\subsection{Detector and data samples}
\label{sec:detector}
The ATLAS detector~\cite{Aad:2008zzm} comprises a superconducting
solenoid surrounding the inner detector (ID) and a large
superconducting toroid magnet system with muon
detectors enclosing the calorimeters.  The
ID system is immersed in a $2$\,T axial magnetic field and provides
tracking information for charged particles in a pseudorapidity range
matched by the precision measurements of the electromagnetic
calorimeter. The inner silicon pixel and strip tracking detectors
cover the pseudorapidity range $|\eta|< 2.5$.\footnote{ATLAS uses a
right-handed coordinate system with its origin at the nominal
interaction point (IP) in the centre of the detector and the $z$-axis
along the beam pipe. The $x$-axis points from the IP to the centre of
the LHC ring, and the $y$-axis points upward. Cylindrical coordinates
$(r,\phi)$ are used in the transverse plane, $\phi$ being the
azimuthal angle around the $z$-axis. The pseudorapidity is defined in
terms of the polar angle $\theta$ as $\eta=-\ln\tan(\theta/2)$. The
distance in $\eta$--$\phi$ space between two objects is defined as
$\Delta R = \sqrt{(\Delta\eta)^2 + (\Delta\phi)^2}$. The rapidity 
is defined as $y=\frac{1}{2}\ln\frac{E+p_z}{E-p_z}$.} The transition
radiation tracker, surrounding the silicon detectors, contributes to the
tracking and electron identification for $|\eta| < 2.0$.

The liquid argon (LAr) electromagnetic (EM) calorimeter is divided
into one barrel ($|\eta| < 1.475$) and two end-cap components ($1.375
< |\eta| < 3.2$). It uses lead absorbers and has
an accordion geometry to ensure a fast and
uniform response and fine segmentation for optimal reconstruction and
identification of electrons and photons.  The hadronic steel/scintillator-tile
calorimeter consists of a barrel covering the region $|\eta| <
1.0$, and two extended barrels in the range $0.8 < |\eta| < 1.7$.  The
copper/LAr hadronic end-cap calorimeter ($1.5<|\eta|<3.2$) is located behind
the electromagnetic end-cap calorimeter. The forward calorimeter (FCAL)
covers the range $3.2< |\eta| < 4.9$ and also uses LAr as the active
material and copper or tungsten absorbers for the EM and hadronic sections, respectively.

The muon spectrometer (MS) is based on three large superconducting
toroids with coils arranged in an eight-fold symmetry around the
calorimeters, covering a range of $|\eta|<2.7$.  Over most of the
$\eta$ range, precision measurements of the track coordinates in the
principal bending direction of the magnetic field are provided by
monitored drift tubes. At large pseudorapidities ($2.0 < |\eta| <
2.7$), cathode strip chambers with higher granularity are used in the
layer closest to the IP.  The muon trigger detectors consist of
resistive plate chambers in the barrel ($|\eta|<1.05$) and thin gap
chambers in the end-cap regions ($1.05 < |\eta| < 2.4$), with a small
overlap around $|\eta| \simeq 1.05$.

In 2011, the ATLAS detector had a three-level trigger system consisting of
Level-1 (L1), Level-2 (L2) and the Event Filter (EF). The L1
trigger rate was approximately 75\,kHz. The L2 and EF triggers
reduced the event rate to approximately $300$\,Hz before data transfer to
mass storage.

The data for this analysis were collected by the ATLAS Collaboration
during 2011, the final year of operation at $\sqrt{s}=7\TeV$.  The
analysis uses a total luminosity of $\lumi$ with an estimated uncertainty of
$\dlumi\%$~\cite{Aad:2013ucp}, where the main components of the
apparatus were operational. Data and simulated event
samples were processed with common reconstruction software.

\subsection{Simulated event samples}
\label{sec:simulation}
Simulated and reconstructed Monte Carlo (MC) samples are used to model
the properties of signals and background processes and to calculate acceptance
and efficiency corrections for the extraction of cross
sections. Dedicated efficiency and calibration studies with data are
used to derive correction factors to account for the small differences
between experiment and simulation, as is subsequently described. 

The main signal event samples for \Wpmlnu\ and \Zgll\ production are
generated using the
\Powheg~\cite{Nason:2004rx,Frixione:2007vw,Alioli:2008gx,Alioli:2010xd}
event generator, with the simulation of parton showers, hadronization
and underlying events provided by \Pythia~\cite{pythia}. Systematic
uncertainties in the measurements due to imperfect modelling of the
signals are estimated with alternative event samples generated with \Powheg\
interfaced instead to the \Herwig~\cite{Corcella:2000bw} and
\Jimmy~\cite{Butterworth:1996zw} programs (referred to later as the
\Powheg+\Herwig\ sample) as well as \Mcatnlo~\cite{mcatnlo}, also
interfaced to the \Herwig\ and \Jimmy\ programs (referred to
later as the \Mcatnlo+\Herwig\ sample). For the \Mcatnlo\ and \Powheg\
matrix element calculations the \pdfCteq\ PDF~\cite{Lai:2010vv} set is
used, whereas showering is performed with \pdfCteql\ 
\cite{Pumplin:2002vw}. Samples of \Wtau\ and \Ztau\ events are
generated with the \Alpgen\ generator~\cite{Mangano:2002ea} interfaced
to \Herwig\ and \Jimmy\ and using the \pdfCteql\ PDF set, and also \Powheg\
interfaced to \Pythiaeight~\cite{pythia8}.

All simulated samples of \Wpmlnu\ and \Zgll\ production are normalized to the
NNLO cross sections calculated by the FEWZ program with the MSTW2008 NNLO
PDF set~\cite{Martin:2009iq}. When employing these samples for
background subtraction, an uncertainty in the total cross section
of 5\% is assigned to
account for any uncertainties arising from the PDFs as well as
factorization-scale and renormalization-scale uncertainties. As the simulated
transverse momentum spectrum of the \Wpm\ and \Zg\ bosons does not
describe the one observed in data well, all samples are reweighted by
default to the \Powheg+\Pythiaeight\ AZNLO prediction~\cite{Aad:2014xaa},
which describes the \Zll\ data well at low and
medium dilepton transverse momentum $p_{\mathrm{T},\ell\ell} < 50\gev$.

Top-quark pair (\ttbar) and single top-quark production are simulated with
\Mcatnlo\ interfaced to \Herwig\ and \Jimmy. The $\ttbar$
cross section is calculated at a top quark mass of $172.5\gev$ at NNLO
in QCD including resummation of next-to-next-to-leading logarithmic
soft-gluon terms (NNLL) with
top++2.0~\cite{Cacciari:2011hy,Baernreuther:2012ws,Czakon:2012zr,Czakon:2012pz,Czakon:2013goa,Czakon:2011xx}.
The total theoretical uncertainty of the \ttbar\ production cross section is
calculated using the PDF4LHC prescription~\cite{Botje:2011sn} using
the MSTW2008 NNLO~\cite{Martin:2009iq}, CT10 NNLO~\cite{Gao:2013xoa}
and NNPDF2.3 5f FFN~\cite{Ball:2012cx} PDF sets and adding in
quadrature the scale and $\alphas$ uncertainties. The single-top-quark
cross sections are calculated at approximate NNLO+NNLL
accuracy~\cite{Aad:2012ux, Kidonakis:2011wy, Kidonakis:2010ux,
Kidonakis:2010tc}.

Inclusive production of dibosons $WW, WZ$ and $ZZ$ is simulated with
\Herwig. The samples are normalized to their respective NLO QCD
cross sections~\cite{Campbell:1999ah} with 6\% uncertainty.

While most studies of the multijet background are performed using
control samples from data, some studies in the muon channels are
carried out with \Pythia\ samples, where inclusive, heavy-flavour dijet
production ($c\bar{c}$ and $b\bar{b}$) is simulated and the samples
are filtered for high-\pt muons from  charm or bottom hadron
decays.

All generators are interfaced to \Photos~\cite{Golonka:2005pn} to
simulate the effect of final-state QED radiation (QED FSR). The decays of $\tau$
leptons in \Herwig\ and \Pythia\ samples are handled by
\Tauola~\cite{Jadach:1990mz}. The passage of particles through the
ATLAS detector is modelled~\cite{Aad:2010ah}
using GEANT4~\cite{geant4}. The effect of
multiple $pp$ interactions per bunch crossing (``pile-up'') is modelled
by overlaying the hard-scattering event with additional simulated
inelastic collision events following the distribution observed in the
data with about 9 simultaneous inelastic interactions on average.
These events are simulated using \Pythia\ with the AMBT2
tune~\cite{ATL-PHYS-PUB-2011-008}. While the simulation of pile-up
events reproduces
 the observed width of the luminous region along
the beam direction, a reweighting is applied to match the
longitudinal distribution of the hard-scatter vertex to that observed
in the data. This is needed to accurately control 
acceptance and detector effects, which depend on the details of the
detector geometry.

\subsection{Cross-section definition and fiducial regions}
\label{sec:fidureg}
The measurements reported here correspond to inclusive Drell--Yan cross sections
with a direct decay of the intermediate boson, \Zgll\ or \Wln, where
$\ell=e$ or $\mu$. Other processes that may lead to a pair of leptons,
$\ell\ell$ or $\ell\nu$, in the final state are subtracted as
background. These are \ttbar\ pair and single top-quark production,
cascade decays $\Ztau \to \ell^+\ell^- X$ and $\Wtau\to \ell\nu X$,
photon-induced lepton-pair production $\gamma\gamma \to \ell\ell$, and
gauge boson pair production, with both boson masses exceeding
$20\gev$.
Experimental contaminations of signals through other channels, such as
\Zgll\ contributing as background to $W^{\pm}$ or the small, opposite-sign
$W^{\mp}$ fraction in the $W^{\pm}$ selections, are corrected for
as well.

Each channel of the measurement covers somewhat different regions of phase
space.  
For electrons this corresponds to a restriction to
$|\etal|<2.47$ for central electrons, and further the exclusion of the regions
$1.37<|\etal|<1.52$ and $3.16<|\etal|<3.35$. For muons the acceptance
is restricted to $|\etal|<2.4$.

The combined $e-\mu$ cross sections are
reported in common fiducial regions close to the initial experimental
selections so as to involve only minimal extrapolations. The kinematic requirements applied for
the cross-section measurements are as follows:
\begin{eqnarray*}
  \mbox{Central}\ \Zgll : && \ptl > 20\gev,\ |\etal|<2.5,\ 46 < \mll <
  150\gev\\
  \mbox{Forward}\ \Zgll : && \ptl > 20\gev,\ \mbox{one lepton}\ |\etal|<2.5,\ \mbox{other lepton}\  2.5<|\etal|<4.9,\\
  && \ 66 < m_{\ell\ell} < 150\gev\\
  \Wpmlnu\ : && \ptl > 25\gev,\ |\etal|<2.5,\ \ptnu>25\gev,\ \mt > 40\gev \,.
\end{eqnarray*}
Here the charged-lepton transverse momentum and pseudorapidity are
denoted by \ptl\ and \etal, respectively. The transverse momentum of
the neutrino is given by \ptnu\ and the $W$-boson transverse mass is
calculated as $\mt^2 =
2\,\ptl\,\ptnu\,[1-\cos(\Delta\phi_{\ell,\nu})]$, where
$\Delta\phi_{\ell,\nu}$ is the  azimuthal angle between
the charged lepton and the neutrino directions. The lepton kinematics used in the
definition of the cross sections corresponds to the Born level for QED
final-state radiation effects.
These fiducial regions differ slightly from those used in
Ref.~\cite{Aad:2011dm} such that the corresponding cross-section results
cannot be compared directly.

The integrated charged-current fiducial cross sections are presented
separately for $W^+$, $W^-$ and their sum. Integrated neutral-current
fiducial cross sections are presented for the $Z$-peak region,
corresponding to $66 < \mll < 116\GeV$, where they are most precise.

The differential \Wpmlnu\ cross sections are measured as a function of
the absolute values of the charged-lepton pseudorapidity, \etal , in
bins with boundaries given by
\begin{equation}
  |\etal| = [0.00\,,\,0.21\,,\,0.42\,,\,0.63\,,\,0.84\,,\,1.05\,,\,1.37\,,\,1.52\,,\,1.74\,,\,1.95\,,\,2.18\,,\,2.50]\,.
\end{equation}
The differential $Z/\gamma^*$ cross sections are presented as a function of
dilepton rapidity, \yll , in three intervals of dilepton mass, \mll ,
with bin edges 
\begin{equation}
  \mll = [46\,,\,66\,,\,116\,,\,150]\gev\,.
\end{equation}
In the $Z$-peak region, the boundaries of the bins in dilepton
rapidity \yll\ are chosen to be
\begin{equation}
  |\yll| = [0.0\,,\,0.2\,,\,0.4\,,\,0.6\,,\,0.8\,,\,1.0\,,\,1.2\,,\,1.4\,,\,1.6\,,\,1.8\,,\,2.0\,,\,2.2\,,\,2.4]\,,
\end{equation}
while in the adjacent mass intervals, below and above the $Z$ peak,
the binning is twice as coarse and ranges also from $|\yll|=0$
to $2.4$.

A dedicated \Zgll\ analysis in the electron channel extends into the
forward region of \yll, covering the range from $|\yll| =1.2$ to
$3.6$. This analysis is only performed in the two higher
mass intervals, with the boundaries $\mll= [66\,,\,116\,,\,150]\GeV$,
as the region below $\mll<66\GeV$ cannot be measured with good
precision with the current lepton \pt\ acceptance in this channel. In
the $Z$-peak region of the forward \Zg\ analysis the boundaries of
the bins in dilepton rapidity \yll\ are chosen as
\begin{equation}
  |\yll| = [1.2\,,\,1.4\,,\,1.6\,,\,1.8\,,\,2.0\,,\,2.2\,,\,2.4\,,\,2.8\,,\,3.2\,,\,3.6]\,,
\end{equation}
while for the higher mass interval the same range is divided into six
bins of equal size.



\section{Electron channel measurements}
\label{sec:elana}

\newcommand{\totstat}{The sum of all expected background and signal
  contributions is shown as a solid line with a hashed band detailing
  the statistical uncertainty and labelled ``total (stat)''.}

\newcommand{\Etcone}{\ensuremath{E_\mathrm{T}^\mathrm{cone30}}}

\subsection{Event selection}
\label{sec:ElEvSel}
Events are required to have at least one primary vertex formed by at
least three tracks of $\pt >500\MeV$. If multiple vertices are
reconstructed, the one with the highest sum of squared transverse momenta
of associated tracks, $\sum \pTsq$, is selected as the primary vertex.

Central electron candidates are reconstructed from an ID track
matched to an energy deposit in the EM
calorimeter~\cite{egammapaper2011}. They are required
to be within the coverage of the ID and the precision region of the EM
calorimeter, $|\eta|< 2.47$. The transition region between the barrel
and end-cap calorimeters, $1.37<|\eta|<1.52$, is excluded, as the
reconstruction quality is significantly reduced compared to the rest
of the pseudorapidity range. The electron momentum vector is calculated by
combining the calorimeter measurement of the energy and the tracker
information on the direction. The electron is required to satisfy
``tight'' identification criteria~\cite{egammapaper2011} based on the
shower shapes of the cluster of energy in the calorimeter, the track
properties, and the track-to-cluster matching. The combined efficiency
for electrons from $W$ and $Z$ decays to be reconstructed and to meet
these ``tight'' identification criteria depends strongly on both
$\eta$ and $\pt$. In the most central region of the detector, at
$|\eta|<0.8$, this efficiency is about $65\%$ at $\pt=20\gev$ and
increases to about $80\%$ at $\pt=50\gev$. In the more forward region,
$2.0<|\eta|<2.47$, the corresponding efficiencies are in the range
$50$--$75\%$ for transverse momenta $\pt= 20$--$50\gev$.

The same ``tight'' requirements are imposed for all central electron
candidates to enable a coherent treatment across all $W^\pm$ and $\Zg$
analyses, even though the background rejection is less crucial for the
\Zg\ analysis with two central electrons. To improve the rejection of
background from non-isolated electrons, converted photons, or hadrons
misidentified as electrons, isolation criteria are imposed on
the electron candidates in the \Wenu\ and forward \Zgee\ analyses. The
isolation of central electron candidates in these channels is implemented
by setting an upper limit on both the energy measured in the calorimeter in a
cone of size $\Delta R = 0.2$ around the electron cluster and the
sum of transverse momenta of all tracks in a cone of size $\Delta R
= 0.4$ around the trajectory of the electron candidate.
The contribution from the electron candidate itself is 
excluded in both cases.
The specific criteria are optimized as a function of electron
$\eta$ and \pt\ to have a combined efficiency of about $95\%$ in the
simulation for isolated electrons from the decay of a $W$ or $Z$
boson.

Forward electron candidates are reconstructed in
the region $2.5 < |\eta| < 4.9$, excluding the transition region
between the end-cap and the FCAL calorimeter, 3.16$<|\eta|<$3.35, and  are
required to satisfy ``forward tight'' identification
requirements with a typical efficiency in the range 
of $65$--$85\%$~\cite{egammapaper2011}. As the forward region is not
covered by the ID, the electron identification has to
rely on calorimeter cluster shapes only. The forward electron momentum
is determined from the calorimeter cluster energy and position.

In an inclusive \Wln\ analysis, signal events can be
considered to consist of three contributions: the isolated charged
lepton, the undetected neutrino, and any further particles produced in
the hadronization of quarks and gluons produced in association with
the $W$ boson. This last contribution is referred to as the hadronic
recoil~\cite{Aad:2011fp}. The missing transverse momentum, \met, is
given by the negative vectorial sum of the transverse momentum components of
the charged lepton and the
hadronic recoil and  identified with the undetected
neutrino. The \met\ is reconstructed from energy deposits in the
calorimeters and muons reconstructed 
in the MS~\cite{Aad:2012re, ATLAS-CONF-2012-101}.
Calorimeter energy deposits associated to an electron candidate meeting the
``medium'' identification criteria~\cite{egammapaper2011} and 
exceeding $\pt>10\gev$ are calibrated to
the electron scale. Alternatively, if calorimeter energy deposits can
be associated to a jet reconstructed with the anti-$k_t$ algorithm
with radius parameter $R=0.6$ and $\pt >20 \GeV$, the calibrated jet is
used~\cite{Aad:2014bia}. Finally, identified combined and isolated
muons, as described in \Sec~\ref{sec:muana}, with $\pt >10\gev$, are
used in the \met\ reconstruction, removing the energy deposits of such
muons in the calorimeter. Any remaining energy deposits in the
calorimeters are added to the \met\ after calibration with the local
hadronic calibration~\cite{Aad:2014bia}.

During data collection, events with one central electron were selected
with a single-electron trigger with ``medium'' identification criteria
and a \pt\ threshold of $20\GeV$ or
$22\GeV$~\cite{ATLAS-CONF-2012-048}. The rise in threshold was
enforced by the increasing instantaneous luminosity delivered by the
LHC during 2011.  Events with two central electrons are furthermore
selected online by a dielectron trigger in which two electrons are
required to satisfy the ``medium'' identification criteria and a lower
\pt\ threshold of $12\GeV$.

To select $W$-boson events in the electron channel, exactly one
central identified and isolated electron is required with a transverse
momentum $\pt > 25\GeV$. This electron is also required to have passed
the single-electron trigger. Events with at least one additional
central electron meeting the ``medium'' identification
criteria~\cite{egammapaper2011} and $\pt > 20\gev$ are rejected to
reduce background from \Zgee\ events. The missing transverse momentum
is required to exceed $\met = 25\GeV$ and the transverse mass of the
electron--\met system, \mt, has to be larger than $40\GeV$.

The selection for the central \Zgee\ analysis requires exactly two
identified electrons with $\pt>20\gev$.  These two electrons must have
passed the dielectron trigger selection. No requirement is made on the
charge of the two electron candidates. The analysis examines the
invariant mass \mee\ interval from $46$ to $150\GeV$.

For the selection of forward \Zgee\ events over an extended range of
rapidity, a central identified and isolated electron is required as in
the \Wenu\ channel, but lowering the transverse momentum threshold to
the minimum $\pt = 23\GeV$ accessible with the single-electron trigger.
A second electron candidate with $\pt > 20\GeV$ has to be
reconstructed in the forward region. The invariant mass of the
selected pair is required to be between $66$ and $150\GeV$.

\subsection{Calibration and efficiencies}
\label{sec:eleeffcalib}

Comprehensive evaluations of the reconstruction of electrons
are described in Refs.~\cite{egammapaper2011, PERF-2013-05}. 
The energy of the electron
is calibrated using a multivariate algorithm trained on simulated
samples of single electrons to achieve an optimal response and
resolution. Residual corrections to the energy scale and resolution are
determined from data as a function of $\eta$ in the central and
forward regions by comparing the measured \Zee\ line shape to the one
predicted by the simulation~\cite{PERF-2013-05}. The energy-scale
corrections applied to the data are typically within a range of $\pm
2\%$ and the systematic uncertainty of the energy scale
is typically $0.1\%$. Resolution corrections of around $(1.0 \pm 0.3)\%$
are applied to the simulation to match the data, where the quoted
uncertainty corresponds to the precision of the correction.

The electron efficiencies are controlled in several steps
corresponding to the reconstruction and identification of electron
candidates as well as the isolation and trigger requirements described
above. All central electron efficiencies are measured as a function of
the electron pseudorapidity and electron transverse momentum, while
in the forward region $2.5<|\eta|<4.9$ the corrections are binned in
electron pseudorapidity only. All uncertainties in the electron
efficiency measurements are classified as being of
statistical or systematic origin, where the latter has components
correlated and uncorrelated across $\eta$ and
\pt~\cite{egammapaper2011}. This classification allows 
the corresponding systematics to be propagated correctly 
to the final measurement as described in \Sec~\ref{sec:combi}.

The efficiencies for electrons from $W$ or $Z$ decays in the central
region to satisfy the ``tight'' identification requirements are measured
using two different tag-and-probe methods performed with $W$ and $Z$
data samples~\cite{egammapaper2011}. The data-to-simulation ratios of
the efficiencies measured in these two samples are combined. They are
typically within $\pm 0.05$ of unity with significant
variations as a function of pseudorapidity. The total uncertainty in
these factors is $0.5 - 1.0\%$.

The central electron trigger, reconstruction and isolation efficiencies as well as
 the forward electron identification efficiencies are determined
using the $Z$ tag-and-probe method only. Corresponding correction
factors are derived in all cases and applied to the simulation. The
efficiencies for the reconstruction of central electrons are measured
with a precision of mostly better than $0.5\%$ and are found to be
described by the simulation within typically $\pm 1\%$. The efficiency
of the electron isolation requirement employed in the \Wenu\ and
forward \Zgee\ analysis is well described by the simulation within
$\pm 1\%$ variations and the corresponding correction factors have
typically $<0.3\%$ uncertainty. The electron trigger efficiencies are
measured separately for the single-electron and dielectron triggers and for
various different configurations employed during the data-taking. Most
data-to-simulation correction factors for the trigger selection are
within $\pm 1\%$ of unity and determined with a precision of
better than $1\%$.

The forward electron reconstruction efficiency has been found to be
nearly $100\%$ in the simulation. The
identification efficiencies are found to be lower in data than in the
simulation by about $10$\,\% 
and are measured with a precision of $3 - 8\%$.

The distinction between $W^+$ and $W^-$ events relies on the
measurement of the charge of the decay electron. The charge
misidentification probability as a function of $\eta$ is determined
in both data and simulation
from the fraction of \Zee\ events where both electrons are
reconstructed with the same sign. It depends on the identification
criteria and in general increases at large
$|\eta|$~\cite{egammapaper2011}. A correction is applied
to the simulation to match the rate observed in the data.
In the \Zgee\ analysis, the majority of 
dielectron events reconstructed with same charge, with an invariant
mass close to the $Z$-boson mass and satisfying the identification
requirements, are indeed signal events. The efficiency loss of an
opposite-charge selection through charge misidentification of either
electron incurs a non-negligible systematic uncertainty, which is
avoided by not applying the opposite-charge selection in the \Zgee\
analysis.

Uncertainties in the \met\ scale and resolution are determined by the
corresponding uncertainties for the electrons~\cite{PERF-2013-05},
muons~\cite{muPerfPaper}, and jets~\cite{Aad:2014bia} used in the
reconstruction. The uncertainties in the remaining ``soft'' part are
evaluated by reconstructing the hadronic recoil in \Zll\ events and
comparing the recoil response to the dilepton system in both data and
simulation~\cite{ATLAS-CONF-2012-101}.
 
\subsection{Backgrounds}
\label{sec:ElBkg}

The backgrounds contributing in the \Wen\ channel can be divided into
two categories: i) electroweak background processes and
top-quark production, which are estimated using MC
prediction, and ii) background from multijet production determined
with data-driven methods.
 
The largest electroweak background in the \Wen\ channel is due to the
\Wtau\ production where isolated electrons are produced in the decay
$\tau \to e\bar{\nu}\nu$. Relative to the number of all $W^\pm$
candidate events, this contribution is estimated to be between $1.6\%$
and $1.9\%$ for the different bins of the pseudorapidity with a
similar fraction in $W^+$ and $W^-$ events. The contamination of the
\Wen\ sample by \Zgee\ is determined to be between $0.7\%$ and
$1.3\%$. Further contributions, at the $0.1$--$0.5\%$ level, arise from
\ttbar, \Ztau, single top-quark and diboson production. The sum of
electroweak and top-quark backgrounds is between $3.3\%$ and $3.9\%$ in
the $W^-$ channel and between $2.8\%$ and $3.5\%$ in the $W^+$
channel. In contrast to the \Wtau\ background, the other electroweak
and top-quark background yields are of similar absolute size in $W^+$ and
$W^-$ channels.

Multijet production from QCD processes is a significant source of
background in the \Wen\ channel when non-isolated electrons, converted
photons or hadrons are misidentified as isolated electrons and
neutrinos from hadron decays or resolution effects cause a significant
measurement of missing transverse momentum in the event. This
background is estimated from the data using a template fit of the
\met\ distribution in a normalization region that differs from the signal
region by relaxed the \met\ and \mt\ requirements. A template to
represent the multijet background contribution is selected from data
using the same kinematic requirements as for signal electrons, but inverting
a subset of the electron identification criteria and requiring the
electron candidate not to be isolated. The isolation is estimated from
the energy deposited in the calorimeter in a cone of size $\Delta R =
0.3$ around the electron candidate, denoted by \Etcone, and the
condition $\Etcone/\pt >0.20$ is imposed. A second template that
combines the \Wen\ signal and electroweak and top-quark contributions
is taken from the simulation.

\begin{figure}[bth]
  \begin{center}
    \includegraphics[width=0.48\textwidth]{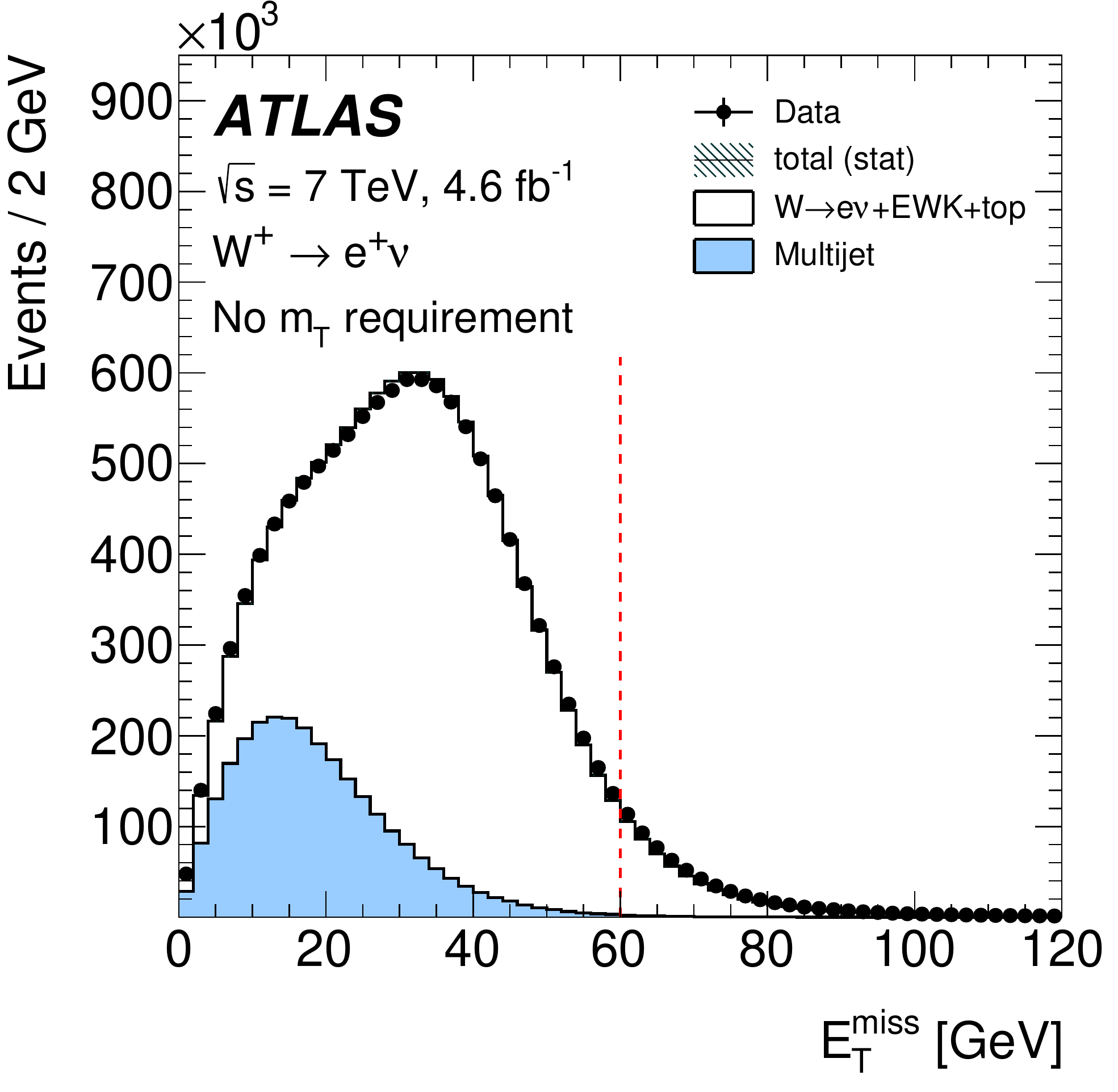}
    \includegraphics[width=0.48\textwidth]{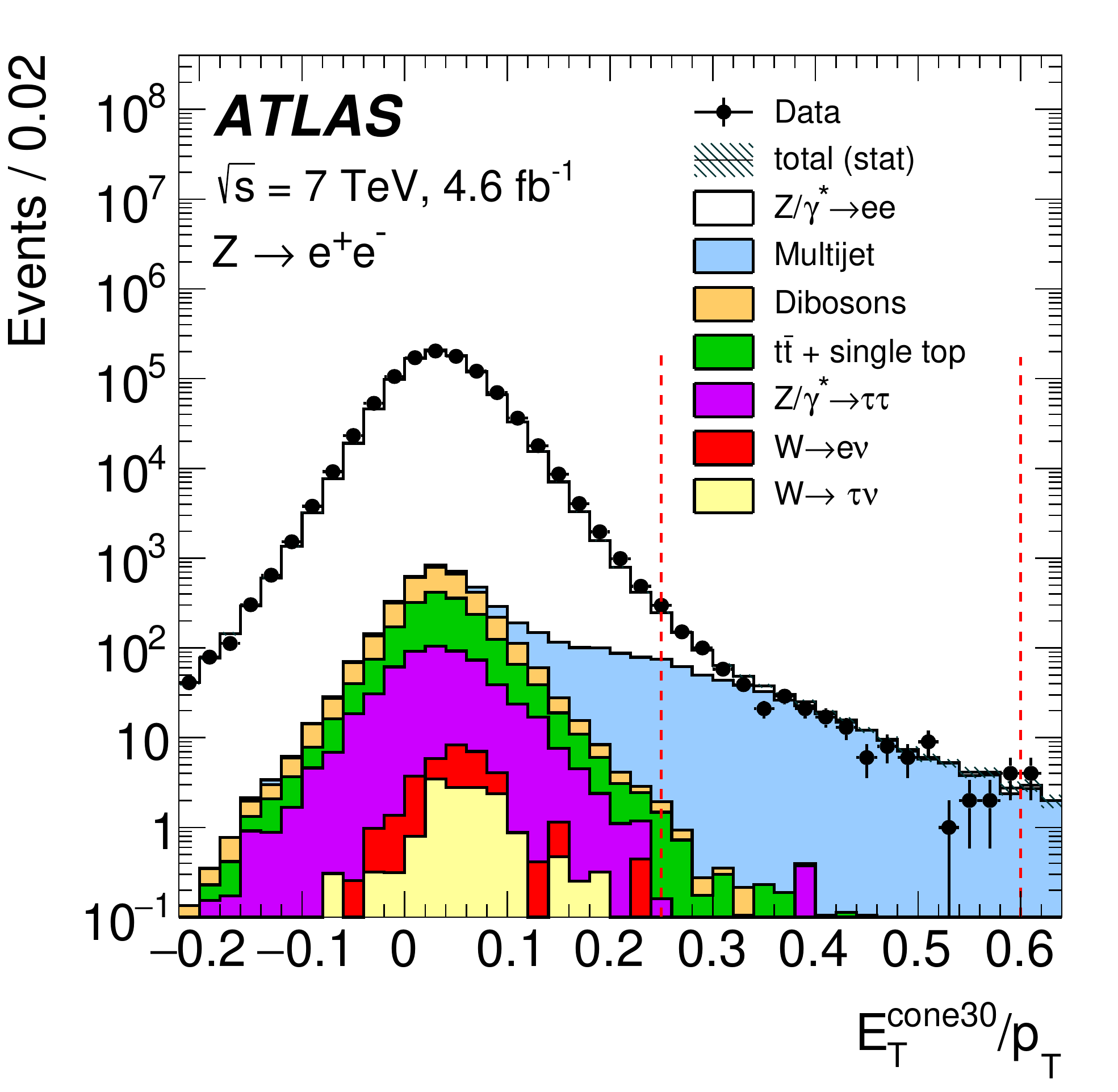}
    \caption{Distributions used for the estimation of the multijet
      background in the \Wenup\ channel (left) and \Zee\ channel
      (right). For the \Wenup\ channel, the result of the template fit
      in a multijet-enhanced region using the \MET\ distribution is
      shown. The vertical line indicates the upper boundary ($\met =
      60\gev$) of the region used in the fit. The label ``EWK+top''
      refers to the electroweak and top-quark background contributions
      estimated from MC simulation, which are here treated in a common
      template together with the \Wen\ signal. In the \Zee\ channel, the region
      of large isolation $\Etcone/\pt$,  between the two
      vertical lines, is used to normalize the multijet template from
      data. The shown distribution is taken from the central \Zee\ analysis in
      the region $66<\mee<116\gev$. \totstat}
    \label{fig:ele:qcdFit}
  \end{center}
\end{figure}

The relative fraction of the two components is determined by a fit to
the data in the normalization region. The normalization region
contains the signal region to constrain the signal contribution,
relaxes the lower \met\ and \mt\ requirements to increase the multijet
fraction and furthermore imposes $\met<60\gev$ to avoid a mismodelling
of the high \met\ region, which was established in a study of the \Zee\ sample.
No prior knowledge of either template's
normalization is assumed, and the fit is performed separately for the $W^+$ and
$W^-$ channels and also in each bin of electron pseudorapidity
to obtain the background for the differential analysis. The resulting
\met\ distribution for the case of the inclusive $W^+$ selection is
shown in the left panel of \Fig~\ref{fig:ele:qcdFit}. The background in the
signal region $\met > 25\gev$ and $\mt >40\gev$ is then obtained by
multiplying the multijet yield determined in the fit by the fraction of
events in the template sample that satisfy the signal region and
normalization region \met\ and \mt\ requirements, respectively. This
multijet estimate is found to change in a systematic way when the
\met\ and \mt\ requirements imposed for the normalization region are
progressively tightened to resemble more the \met\ and \mt\ 
requirements of the signal region. This
dependence is measured and linearly extrapolated to the point where the normalization
region has the same \met\ and
\mt\ thresholds as the signal region.
A corresponding correction of typically $10\%$ is applied to obtain an improved
multijet estimate, while the full size of this correction is
assigned as a systematic uncertainty. Further systematic uncertainties
are derived from variations of the background and signal template
shapes. Background shape uncertainties are obtained from varied
template selection criteria by changing the 
$\Etcone/\pt$ selection, requiring the electron-candidate track to
have a hit in the innermost layer of the ID, or changing the subset of
identification criteria that the electron is allowed to not satisfy
from the ``tight'' to the ``medium'' identification level.
The shape uncertainties on the signal template from
the detector systematic uncertainties discussed in
\Sec~\ref{sec:eleeffcalib} and using the alternative signal MC
simulation samples discussed in \Sec~\ref{sec:simulation} are considered as well.

The multijet background in the signal region ranges from $2.1\%$ in
the most central pseudorapidity bin to $6.9\%$ in the most forward
bin of the measurement for the $W^+$ and from $2.8\%$ to $11\%$
for the $W^-$ channel respectively. The total systematic uncertainty
is at the level of $15$--$25\%$ and the statistical uncertainty is
typically a factor of ten smaller. While this background is determined
separately for $W^+$ and $W^-$ samples, the resulting background
yields for the two charges are found to be compatible within their
statistical uncertainties. An alternative method for the determination
of the multijet fractions, following Ref.~\cite{Aad:2016naf}, gives an
estimate well within the systematic uncertainty assigned to the
baseline determination described above.

In the central \Zgee\ analysis, the relative background contributions 
due to electroweak processes with two isolated electrons, from \Ztau,
\ttbar, single top-quark, and diboson production are estimated using
the corresponding MC samples. That background is dominated by the
\Ztau\ process below the $Z$ peak and the \ttbar\ process above the
$Z$ peak, while it is very small in the $Z$-peak region
$\mee=66$--$116\gev$. The background from electroweak and top-quark
processes ranges from $6.2\%$ to $8.8\%$ for $\mee=46$--$66\gev$,
$0.23\%$ to $0.46\%$ for $\mee=66$--$116\gev$ and $2.0\%$ to $8.5\%$ for
$\mee=116$--$150\gev$, where a larger background contamination is
typically found at central rapidity.

To separate the central \Zgee\ signal from the multijet background,
the analysis relies on the same \Etcone\ quantity as described for the
\Wen\ case. The minimum of the value $\Etcone/\pt$ of the two electron
candidates is chosen to represent each event, as it was found to
provide optimal discrimination. The multijet fraction is then
estimated from data by fitting this distribution using a template
method similar to the \Wen\ analysis. The background template is
selected with inverted electron identification requirements and the
signal \Zgee , electroweak and \ttbar\ templates are taken from
simulation.  The non-isolated sample where the minimum of
$\Etcone/\pt$ of both electrons exceeds a certain value is found to be
dominated by multijet background and is used to adjust the
normalization of the background template, taking into account the small
signal contamination. The right panel of \Fig~\ref{fig:ele:qcdFit} shows the
isolation distribution used to obtain the multijet background in the
$Z$-peak region. This procedure yields a fraction of multijet
background decreasing towards larger rapidity with a typical size
between $1.9\%$ and $5.0\%$ in the low dielectron mass bin, between
$0.14\%$ and $1.6\%$ at high dielectron mass and between $0.02\%$ and
$0.15\%$ near the $Z$ peak. Uncertainties are dominated by the
statistical uncertainty of the sample containing
non-isolated electron candidates
and by the sensitivity of the procedure to the threshold applied to
the minimum of $\Etcone/\pt$ to select the non-isolated region and
amount to typically $20\%$ at and above the $Z$ peak ($66<\mll <
150\gev$) and $10\%$ below ($46<\mll < 66\gev$).

In the forward \Zgee\ analysis, the multijet background is estimated
with the same technique as described for the central \Zee\ analysis,
although only the isolation distribution of the central electron is
used. In total the multijet background is estimated to be $1.4$--$2.4\%$
in the $Z$-peak region and $18$--$26\%$ in the high-mass region. The
total relative uncertainties in these estimates are at the level of
$10\%$.

Furthermore, there is a significant contamination from $W(\to
e\nu)+$jets events in the forward \Zgee\ channel, where the electron
from the $W$ decay is detected in the central region and an associated
jet mimics the signature of an electron in the forward region. 
As the associated jet
production and fake-electron rates may be poorly modelled by the simulation,
the \Wen\ background component is determined by a data-driven
procedure. A control region is constructed starting from the nominal forward \Zgee\ 
event selection, but removing the $Z$-peak region $\mee = 80$--$100\GeV$ and
requiring \met\ and \mt\ selections similar to the \Wen\ signal
analysis. 
It is found that the \Powheg+\Pythia\ \Wen\ samples describe well all
relevant kinematic variables such as the invariant mass \mee\ or dielectron
rapidity \yee\ in the control region after applying an additional
normalization factor of $1.6 \pm 0.2$. This factor is
 then also applied 
 to the \Powheg+\Pythia\ \Wen\ samples in the forward \Zgee\ signal
region. The assigned uncertainty of this scale factor covers
systematic uncertainties induced by the extrapolation and is estimated
using variations of the control region with different 
\met\ or \mt\ selections. Other, smaller electroweak contributions from
\ttbar\ and diboson production are estimated using the corresponding MC
samples. The total \Wen\ and other electroweak backgrounds to the
forward \Zgee\ channel is about $1.9\%$ at the $Z$ peak and up to
$22$\,\% in the high-mass region. While the multijet background
fraction is found to be essentially independent of the dielectron
rapidity \yee, the \Wen\ and other electroweak backgrounds decrease
towards larger \yee.



\section{Muon channel measurements}
\label{sec:muana}
\newcommand{\ptcone}{\ensuremath{p_\mathrm{T}^\mathrm{cone40}}}

\subsection{Event selection}
\label{sec:MuEvSel}

The same requirement for a primary vertex is imposed as for the
electron channels. The analysis uses muon candidates that are defined
as ``combined muons'' in Ref.~\cite{muPerfPaper}. For combined muons
an independent track reconstruction is performed in the ID and the MS,
and a combined track is formed using a $\chi^2$ minimization
procedure.  In order to reject cosmic-ray background, the $z$ position
of the muon track extrapolated to the beam line has to match the $z$
coordinate of the primary vertex within $\pm 1$\,cm. The ID track is
required to satisfy the track-hit requirements described in
Ref.~\cite{muPerfPaper}; in addition, the ID track must include a
position measurement from the innermost layer of the pixel
detector. To reduce background from non-isolated muons produced in the
decay of hadrons within jets, muons are required to be isolated. This
is achieved with a track-based isolation variable defined as the sum
of transverse momenta of ID tracks with $\pt > 1\GeV$ within a cone
$\Delta R = 0.4$ around the muon direction and excluding the muon
track, denoted as \ptcone. The value of \ptcone\ is required to be
less than $10\%$ of the muon \pt. The efficiency of this isolation
requirement is about $92\%$ for signal muons with $\pt=20\gev$ and
increases to about $99\%$ for $\pt>40\gev$.

Events in the muon channels were selected during data-taking with a
trigger demanding the presence of a single muon with $\pt > 18\gev$.
The selection of $W$ events demands one muon with $\pt > 25\GeV$ and
$|\eta| < 2.4$, while a veto on any further muon with 
$\pt>20\gev$ is imposed to reduce contamination from the \Zgmumu\
process. The same missing transverse momentum $\met>25\gev$ and
transverse mass $\mt >40\gev$ requirements are imposed as in the \Wenu\ analysis.
Events for the \Zgmumu\ analysis are selected by requiring exactly two
muons with $\pt > 20\GeV$ and $|\eta| < 2.4$.  The two muons are
required to be of opposite charge, and the invariant mass of the
$\mu^+\mu^-$ pair, \mmumu, is required to be between $46$ and $150$\,GeV.

\subsection{Calibration and efficiencies}
\label{sec:mueffcalib}

Muon transverse momentum corrections and trigger and reconstruction
efficiencies are measured using the same methods as
applied in Ref.~\cite{Aad:2011dm} and documented in Refs.~\cite{muPerfPaper,
  ATLAS-CONF-2012-099}.
Muon transverse momentum resolution corrections are determined
comparing data and MC events as a function of $\eta$ in the barrel and 
end-cap regions \cite{muPerfPaper}.  They are derived by fitting the
 \Zmm\  invariant mass spectrum and the  distributions of
$1/\pt^\mathrm{ID} - 1/\pt^\mathrm{MS}$ for both $\mu^+$ and $\mu^-$,
where $\pt^\mathrm{ID}$ and $\pt^\mathrm{MS}$ are the muon
transverse momenta  in \Zmm\ and \Wmn\
events, measured in only the ID  and the muon spectrometer, respectively.
 Muon transverse momentum scale corrections are measured by
comparing the peak positions in the data and MC  \Zmm\ invariant mass distributions.
Further charge-dependent corrections are derived
by comparing the muon transverse momentum distributions in \Zmm\
events for positive and negative muons~\cite{muPerfPaper, Aad:2011yn}.
The momentum scale in the simulation is found to be higher than in the data by
about $0.1$--$0.2\%$ in the central region and $0.3$--$0.4\%$ in the forward
region.  An additional, momentum-dependent correction is applied to
account for charge-dependent biases. For a transverse momentum of
$40\GeV$ this correction is less than $0.1\%$ in the central region
and extends to $0.5$\,\% in the forward region. The muon momentum
resolution is found to be $2$--$5\%$ worse in the data than in the
simulation.  All scale and resolution corrections are applied to the
simulated event samples to match the characteristics of the data.

Muon trigger and reconstruction efficiencies are measured with a
tag-and-probe method in a sample of \Zmm\ events. Imposing tighter
selections on the invariant mass and on the angular correlation
between the two muons reduces the background contamination and
allows one of the muons to be selected with looser requirements to
measure the efficiencies~\cite{muPerfPaper}. The reconstruction
efficiencies are measured using a factorized approach: the efficiency
of the combined reconstruction is derived with respect to the ID
tracks, and the efficiency of reconstructing a muon in the inner
tracker is measured relative to the MS tracks.  The isolation
selection efficiency is estimated relative to combined tracks. Finally,
the trigger efficiency is measured relative to isolated combined
muons.

The measured data-to-simulation ratios of efficiencies are applied as
corrections to the simulation. In general, these factors are close to
unity, indicating that the simulation reproduces detector effects very
well. The corrections for the combined reconstruction efficiency are
$1$--$2\%$, except for a small region around $|\eta| \simeq 1.0$
where a larger correction of $6$--$7\%$ is applied to account for muon
chambers simulated but not installed. These correction factors are
parameterized in $\eta$ and $\phi$ and they are determined with a
$0.1$--$0.3\%$ relative uncertainty.  The efficiency of the isolation
requirement is also modelled well in the simulation. The correction is
derived as a function of the transverse momentum and is about $1\%$
for $\pt = 20\GeV$ and decreases as \pt\ increases to reach about $0.2\%$
for $\pt > 40\gev$. The relative uncertainty of the isolation efficiency
correction is about $0.1$--$0.3\%$.  A larger correction is needed to
account for the mismodelling of the trigger efficiency in simulation,
ranging from $5$--$10\%$. This is parameterized as a function of $\eta$
and \pt\ and known with a $0.1$--$0.8\%$ relative uncertainty.

\subsection{Backgrounds}
\label{sec:MuBkg}

The electroweak background in the \Wmn\ channel is dominated by \Wtau\
and \Zgmumu\ events and is estimated with the simulation.
Relative to the number
of all $W^\pm$ candidate events, the \Wtau\ contribution is determined
to be between $1.9\%$ and $2.1\%$ for the different bins of 
pseudorapidity and is a similar fraction of $W^+$ and $W^-$ events.
The \Zgmumu\ contribution is estimated to be between $1.1\%$ and
$5.7\%$.  Further contributions at the $0.1$--$0.8\%$ level arise from
\ttbar, \Ztau, single top-quark and diboson production. The sum of
electroweak and top-quark backgrounds ranges from $4.5\%$ to $9.6\%$ in the
$W^-$ channel and from $4.0\%$ to $7.0\%$ in the $W^+$ channel. In
contrast to \Wtau\ background, the other electroweak and
top-quark background yields are of similar absolute size in $W^+$ and $W^-$
events.

\begin{figure}[htbp]
  \begin{center}
    \includegraphics[width=0.48\textwidth]{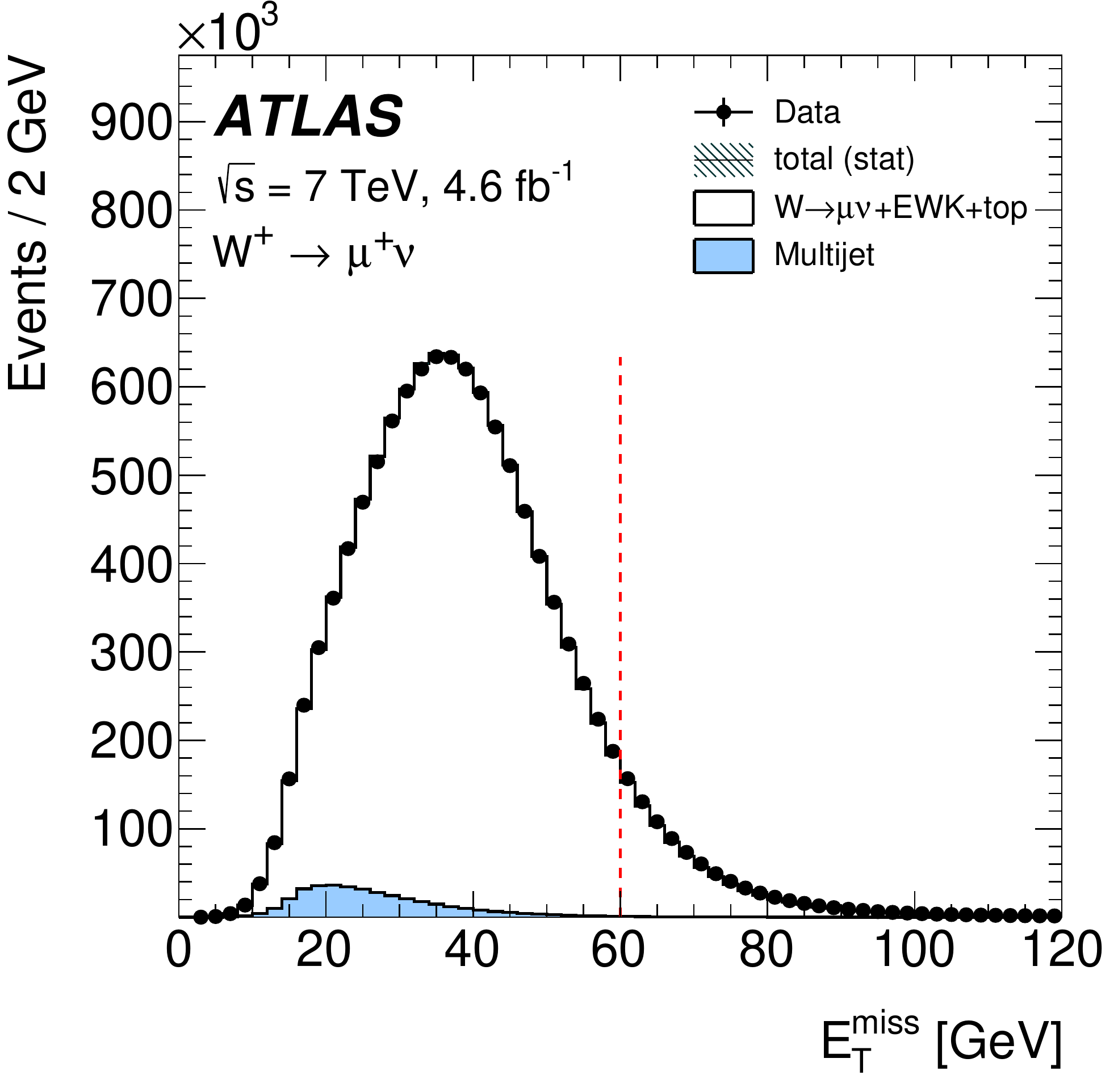}
    \includegraphics[width=0.48\textwidth]{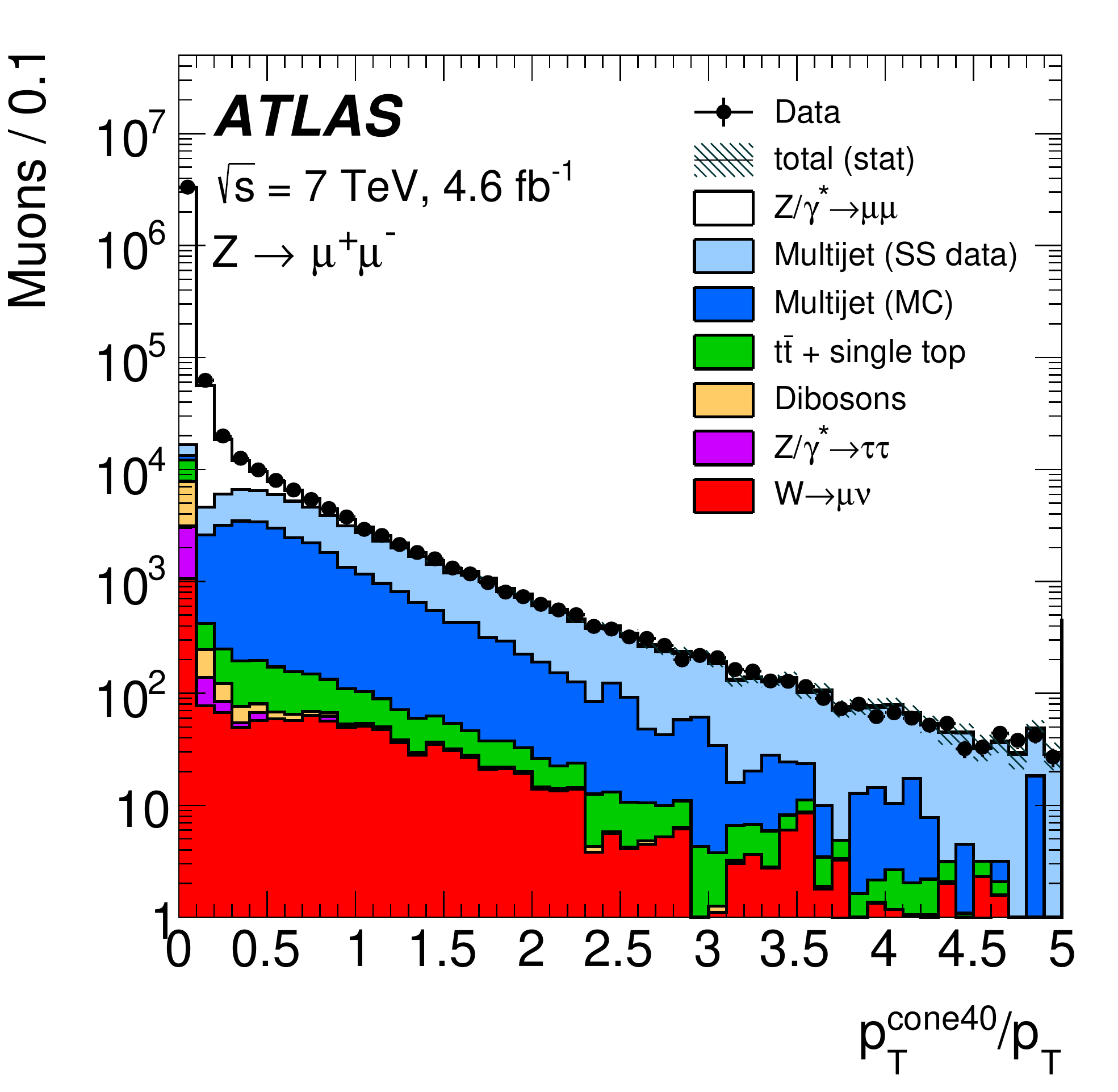}
    \caption{Distributions used for the estimation of the multijet
      background in the \Wmn\ channel (left) and \Zgmumu\ channel
      (right). For the \Wmn\ channel, the result of the template fit
      using the \MET\ distribution is shown.  The vertical line
      indicates the upper boundary ($\met = 60\gev$) of the region
      used in the fit.  The label ``EWK+top'' refers to the
      electroweak and top-quark background contributions estimated
      from MC simulation, which are here treated in a common template
      together with the \Wmn\ signal.  In the \Zgmumu\ channel, the
      full $\ptcone/\pt$ distribution is used to normalize the
      multijet template from data. \totstat}
    \label{WZmuon::fig:WZbkgFits}
  \end{center}
\end{figure}

The multijet background in the \Wmn\ channel originates primarily from
heavy-quark decays, with smaller contributions from pion and kaon
decays in flight and fake muons from hadrons that punch through the
calorimeter. Given the uncertainty in the dijet cross-section
prediction and the difficulty of properly simulating non-prompt muons,
the multijet background is derived from data. The number of background
events is determined from a binned maximum-likelihood template fit to
the \met\ distribution, as shown in the left panel of
\Fig~\ref{WZmuon::fig:WZbkgFits}. The fit is used to determine the
normalization of two components, one for the signal and electroweak
plus top-quark backgrounds, taken from simulation, and a second for
the multijet background, derived from data. No prior knowledge of the
normalization of the two components is assumed. The multijet template
is derived from a control sample defined by reversing the isolation
requirement imposed to select the signal and without applying any
requirement on \met.  The fits are done separately for $W^+$ and $W^-$
events and in each $\eta$ region of the differential cross-section
measurement.

This analysis yields a fraction of multijet background events between
$2.7\%$ in the most central pseudorapidity bin and $1.3\%$ in the most
forward bin of the measurement for the $W^+$ channel and between $3.5\%$ and
$2.6\%$ for the $W^-$ channel, respectively.  The systematic
uncertainty, dominated by the uncertainty in the \met\ modelling for
signal events in simulation, is estimated to be about $0.4$--$0.8\%$
relative to the number of background events.  While this background is
determined separately for $W^+$ and $W^-$ samples, the resulting
background yields are found to be compatible between both
charges within the statistical uncertainty. As in the electron
channel, the multijet background was also determined with an
alternative method following Ref.~\cite{Aad:2016naf}, which gives an
estimate well within the systematic uncertainty assigned to the
baseline determination described above.


The background contributions in the \Zgmumu\ channel due to isolated
muons from \ttbar, \Ztau, and diboson production behave 
similarly to those in the electron channel. In the $Z$-peak region,
$\mmumu=66$--$116\gev$, these are estimated to be \nttbarZmu,
\nZtautauZmu, and \nDiBosonZmu, respectively. The total background from
electroweak and top-quark processes outside the $Z$-peak region is around $6\%$ for
$\mmumu=46$--$66\gev$ and around $4\%$ for $\mmumu=116$--$150\gev$.

The multijet background in the \Zgmumu\ channel is estimated from data
using  various  methods. The first class of methods is based on
binned maximum-likelihood template fits using different discriminating
distributions: the isolation, transverse impact parameter and \pt\ of
the muon, and the dimuon invariant mass. The templates for the
multijet background are derived in most cases from data control
samples obtained by inverting the requirements on muon isolation or
the opposite-charge requirement on the muon pair, depending on the
quantity fitted. Alternative templates are also derived from
simulation of inclusive heavy-flavour production with semileptonic
decays of charm or bottom hadrons to muons. The right panel of
\Fig~\ref{WZmuon::fig:WZbkgFits} shows the result of the template fit
in the muon isolation distribution to determine the absolute scale of
the multijet background, which is then extrapolated to the isolated
region. For this particular method, the multijet template is modelled
by a combination of same-charge data events, used to represent the
background from light-quark production, and a contribution from simulated
heavy-flavour production, where the small same-charge fraction is
subtracted from the dominant opposite-charge dimuon contribution.

In addition to the template fits, a method extrapolating from control
regions defined by inverting the isolation, opposite charge, or both
requirements is employed. All methods, apart from the template fit
in \mmumu, are performed separately in the three mass regions of the
differential \Zgmumu\ cross-section measurements. The fraction of
background events is calculated as the weighted average of these
measurements and found to be \nQCDZmu\ in the $\mmumu=66$--$116\GeV$ mass
region. The relative statistical uncertainty is \nQCDZmuStatRel. A
relative systematic uncertainty of \nQCDZmuSystRel\ is assigned based
on the spread of the weighted measurements.  In the $\mmumu = 46$--$66$
($116$--$150$)\,\GeV\ mass region the fraction of multijet background
events is estimated to be $0.5$\,($0.2$)\,\% with relative
statistical and systematic uncertainties of $15\%$\,($14\%$) and
$80\%$\,($60\%$), respectively.

The shape of the multijet background as a function of \ymumu\ is
derived from a simulated sample of multijet events selected with a
looser muon isolation requirement to increase the statistical
precision. Systematic uncertainties in the shape of the multijet
background as a function of \ymumu\ are assessed by comparing the
shape in simulation obtained with the looser and nominal muon
selections as well as comparing the shape predicted by the
simulation to the shape in a data control region, where at least one muon fails
either the isolation or transverse impact parameter requirements. An
additional relative uncertainty of $22\%$ is obtained, treated as
uncorrelated in rapidity and mass bins.

Cosmic-ray muons overlapping in time with a collision event are
another potential source of background. From a study of non-colliding
bunches, this background contribution is found to be negligible.

%
\section{Cross-section results}
\label{sec:crossres}

\subsection{Analysis procedure}
\label{sec:anaproc}
The integrated and differential \Wpluslnu, \Wminuslnu, and \Zgll\ production cross sections times the branching ratio for decays into a single lepton
flavour ($\ell=e$ or $\mu$) are measured in fiducial volumes as
defined in \Sec~\ref{sec:fidureg}.
Integrated fiducial cross sections in the electron (muon) channel are
computed following the equation
\begin{equation} 
  \sigma^\mathrm{fid,e(\mu)}_{W \to e(\mu)\nu[Z\to ee(\mu\mu)]} = \frac{N_{W[Z]} - B_{W[Z]}}{\CWZ \cdot L_\mathrm{int}} \,,
\label{eq:WZxsecfidemu} 
\end{equation}
where $N_{W[Z]}$ is the number of observed signal candidates in data and $B_{W[Z]}$
is the number of background events expected in the selected sample.
The integrated luminosity of the sample is
$L_\mathrm{int}=(\lumiprecise\pm\dlumiabs)\ifb$ for all channels except
the \Wen\ analysis, where it is $L_\mathrm{int}=(4.51\pm\dlumiabs)\ifb$.
A correction for the event detection efficiency is applied with the
factor \CWZ\ , which is obtained from the simulation as
\begin{equation} \label{Eq:CWZ}
  \CWZ = \frac{N^\mathrm{MC, rec}_{W[Z]}}{N^\mathrm{MC, gen, fid}_{W[Z]}}\,.
\end{equation}
Here, $N^\mathrm{MC, rec}_{W[Z]}$ is the sum of event weights after
simulation, reconstruction and selection, adjusted for the observed
data-to-simulation differences such as in reconstruction, identification, and
trigger efficiencies. The denominator $N^\mathrm{MC, gen,
  fid}_{W[Z]}$ is computed with generator-level information after fiducial
requirements. To
correct the measurements for QED FSR effects,
the fiducial requirements at generator level are implemented
using the lepton momenta before photon radiation. The lepton pairs
($\ell^+\ell^-$, $\ell^+\nu$ or $\ell^-\bar{\nu}$) are required to
originate directly from the decay of the \Zg\ or $W^\pm$ bosons. The \CWZ\
correction is affected mostly by experimental uncertainties, which
are described in \Secs~\ref{sec:elana}~and~\ref{sec:muana}.

The following uncertainties in \CWZ\ of theoretical origin are
considered. PDF-induced uncertainties are determined by reweighting
the signal samples~\cite{Buckley:2014ana} to the 26 eigenvectors of
the CT10 set and scaling the resulting uncertainty to $68\%$
confidence level (CL). The effect of an imperfect description of the
boson transverse momentum spectra is estimated by an additional
reweighting of the $W^\pm$ and \Zg\ samples, beyond that discussed in
\Sec~\ref{sec:simulation}, by the data-to-simulation ratio observed in
the $Z$-peak region. 
Uncertainties related to the implementation of the NLO QCD matrix element and 
its matching to the parton shower are estimated from 
the difference between the \CWZ\ correction factors
 obtained from the \Powheg+\Herwig\ and \Mcatnlo+\Herwig\
  signal samples.
A similar systematic uncertainty related to the signal modelling is
estimated by changing the parton showering, hadronization, and
underlying event by comparing analysis results using \Powheg+\Pythia\
and \Powheg+\Herwig\ samples. When changing the signal generator, the
\CWZ\ correction factors vary by small amounts due to differences in
the simulated charged-lepton and neutrino kinematics, the detector
response to the hadronic recoil, and the electron and muon
identification and isolation efficiencies. The full data-driven estimate 
of multijet background in the \Wln\ channels is repeated 
when changing the signal samples, as the reconstructed \met\
and \mt\ shapes have a significant impact in the fit.

For the measurement of charge-separated $W^+$ and $W^-$ cross sections, the
\CW\ factor is modified to incorporate a correction for event
migration between the $W^+$ and $W^-$ samples as
\begin{equation}
  C_{W^+} = \frac{N^\mathrm{MC, rec+}_{W}}{N^\mathrm{MC, gen+, fid}_{W}}
  \:\:\:\:\:\:\mbox{and}\:\:\:\:\:\: C_{W^-} = \frac{N^\mathrm{MC, rec-}_{W}}{N^\mathrm{MC, gen-, fid}_{W}}\,,
\end{equation}
where $N^\mathrm{MC, rec+}_{W}$ and $N^\mathrm{MC, rec-}_{W}$
are sums of event weights reconstructed as $W^+$ or $W^-$, respectively,
regardless of the generated charge; similarly
$N^\mathrm{MC, gen+, fid}_{W}$ and $N^\mathrm{MC, gen-, fid}_{W}$
are sums of events generated as $W^+$ and $W^-$, respectively,
regardless of the reconstructed lepton charge. This charge
misidentification effect is only relevant for the electron channels
and  negligible in the muon channels.

The correction of the differential distributions follows a similar
methodology, but it is performed using the Bayesian Iterative
method~\cite{DAgostini:1994zf, DAgostini:2010}, as implemented in the
RooUnfold package~\cite{Adye:2011gm} using three iterations. The
differential distributions considered in this paper are constructed to
have bin purities of typically more than $90\%$, where the bin purity
is defined as the ratio of events generated and reconstructed in a
certain bin to all events reconstructed in that bin. Slightly lower
purities of $80$--$90\%$ are observed in the $Z/\gamma^*$ analyses below
the $Z$-peak region ($\mll=46$--$66\gev$) due to QED FSR effects and in
the forward \Zee\ analysis due to worse experimental
resolution. Because of the very high bin purities, the unfolding is to
a large extent reduced to an efficiency correction. Residual prior
uncertainties are covered by the variations of
theoretical origin as discussed for the \CWZ\ factors above.

Fiducial cross sections in the electron and muon channels, as reported
in \Secs~\ref{sec:cross_e}~and~\ref{sec:cross_mu}, are then
extrapolated 
to the common fiducial volume 
by applying a small correction $E^{e(\mu)}_{W[Z]}$
as mentioned in
\Sec~\ref{sec:fidureg}:
\begin{equation} 
  \sigma^\mathrm{fid}_{W \to \ell\nu[Z\to\ell\ell]} = \frac{\sigma^\mathrm{fid, e(\mu)}_{W \to e(\mu)\nu[Z\to ee(\mu\mu)]}}{E^{e(\mu)}_{W[Z]}}\,.
\label{eq:WZxsecfid} 
\end{equation} 
These $E^{e(\mu)}_{W[Z]}$ corrections account for the different
 $\eta$ acceptances for electrons and muons in both the 
CC and NC analyses and are calculated
from the nominal signal samples generated with \Powheg+\Pythia. These
correction factors are typically in the range of
$0.90$--$0.95$, but are as low as $0.65$ in a few bins at high
lepton pseudorapidity or dilepton rapidity. Uncertainties in these
extrapolation factors  account for PDF uncertainties
as well as further signal modelling uncertainties obtained by comparing 
samples generated with \Powheg+\Herwig\ and \Mcatnlo. These
uncertainties are found to be small, $\sim 0.1\%$, and are
always well below the experimental precision of the measurements.

The total \Wpmlnu\ and \Zgll\ cross sections, times leptonic
branching ratio, are calculated using the relation
\begin{equation} 
  \sigma^\mathrm{tot}_{W \to \ell\nu [Z \to \ell\ell]} = \frac{\sigma^\mathrm{fid}_{W \to \ell\nu [Z \to \ell\ell]}}{\AWZ}\,, 
\label{eq:WZxsectot} 
\end{equation} 
where the acceptance \AWZ\ extrapolates the cross section for the
$W^+$, $W^-$ and the \Zg\ channels, measured in the fiducial volume,
$\sigma^\mathrm{fid}_{W \to \ell\nu [Z \to \ell\ell]}$, to the full
kinematic region. It is given by
\begin{equation}
  \AWZ = \frac{N^\mathrm{MC, gen, fid}_{W[Z]}}{N^\mathrm{MC, gen, tot}_{W[Z]}}\,,
\end{equation}
where $N^\mathrm{MC, gen, tot}_{W[Z]}$ is the total sum of weights of all generated
MC events. Uncertainties in the acceptance from the theoretical uncertainties in
the process modelling and in the PDFs  are
estimated as indicated above and amount 
to typically $\pm (1.5$--$2.0)\%$. This therefore
significantly increases the uncertainty in the total cross sections
with respect to the fiducial cross sections.

\subsection{Cross-section measurements}
\label{sec:cross}

\subsubsection{Electron channels}
\label{sec:cross_e}
\newcommand{\visible}{The legend lists only background sources with a
  visible contribution.}

\newcommand{\luminorm}{The simulated samples are normalized to the
  data luminosity.}

\newcommand{\mjplot}{The multijet background shape is taken from a
  data control sample and normalized to the estimated yield of
  multijet events.}

To ensure an adequate description of important kinematic variables in
the electron channels,
\Figs~\ref{wenu:fig:candWpt}~to~\ref{zee:fig:candZYCF} compare several
distributions of the data to the signal simulation and estimated
backgrounds. The signal and electroweak background distributions are
taken from the simulation and normalized to the corresponding data
luminosity. The distributions of the background from multijet
production are obtained from data and normalized as described in
\Sec~\ref{sec:ElBkg}. \FFigs~\ref{wenu:fig:candWpt},
\ref{wenu:fig:candWeta}, \ref{wenu:fig:candWmet} and
\ref{wenu:fig:candWmt} show the distributions of the electron
transverse momentum, the electron pseudorapidity, the missing
transverse momentum, and the transverse mass of candidate $W$ events,
respectively. The invariant mass distribution of electron pairs,
selected by the \Zgee\ analyses, and the dilepton rapidity
distributions are shown in \Figs~\ref{zee:fig:candZm},
\ref{zee:fig:candZYCC} and \ref{zee:fig:candZYCF}, respectively.  
Good agreement between data and the predictions is observed in general
for all kinematic distributions. Small disagreements in the shapes of
the \met\ and \mt\ distributions of $W$-boson candidates are visible at
the level of $2$--$10\%$. These deviations are covered by
uncertainties on the multijet background and on the signal modelling,
for the latter specifically the variations related to
the hadronic recoil response and $W$-boson \pt\ spectrum.
In the forward \Zgee\ distributions, small disagreements at low
\mee\ and localised in particular \yee\ bins of the high mass region
$\mee=116$--$150\GeV$ are covered by the systematic uncertainties on
the electron energy scale and resolution, and background yields,
respectively.

\begin{figure}[tbp]
  \begin{center}
    \includegraphics[width=0.48\textwidth]{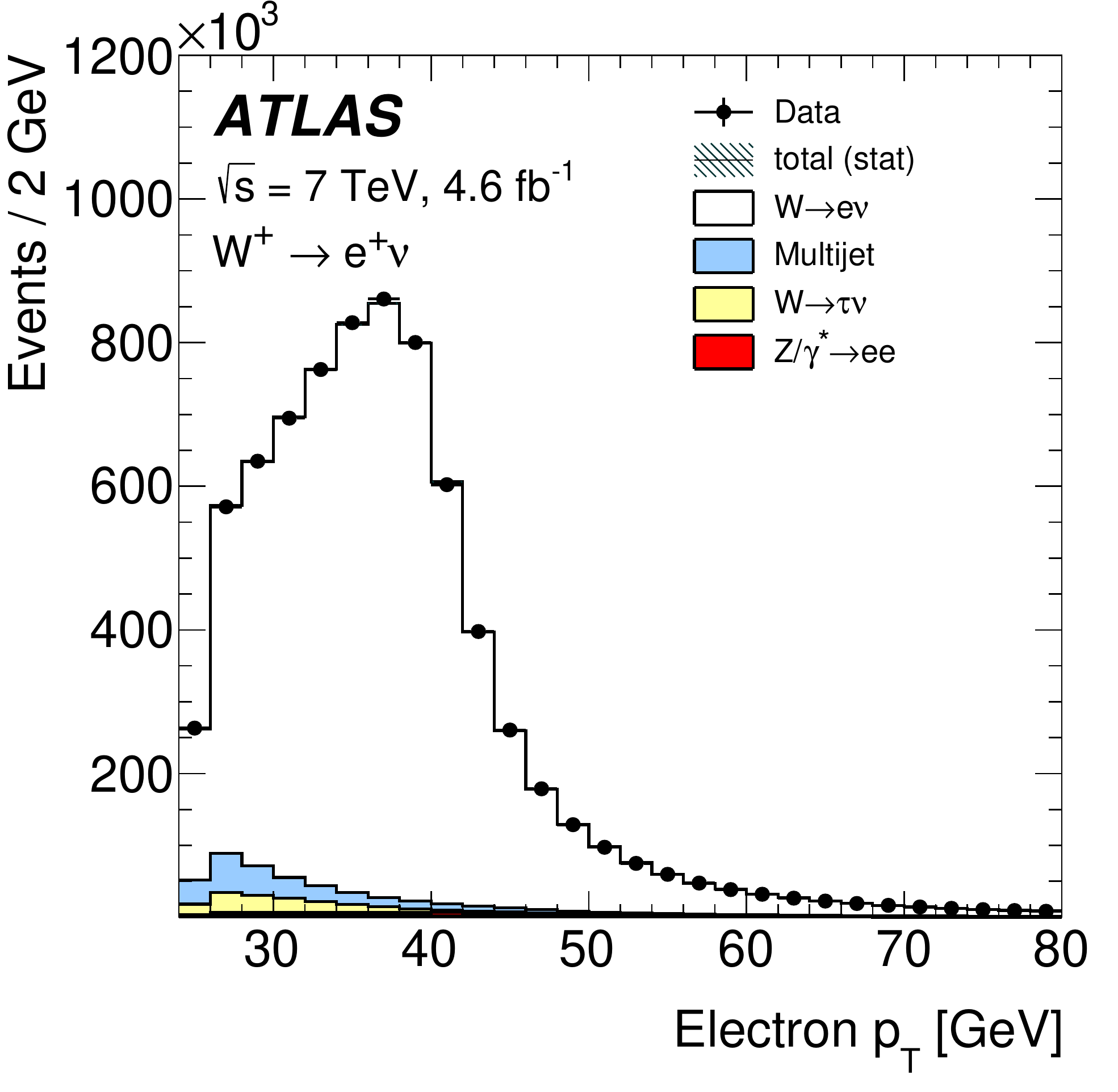}
    \includegraphics[width=0.48\textwidth]{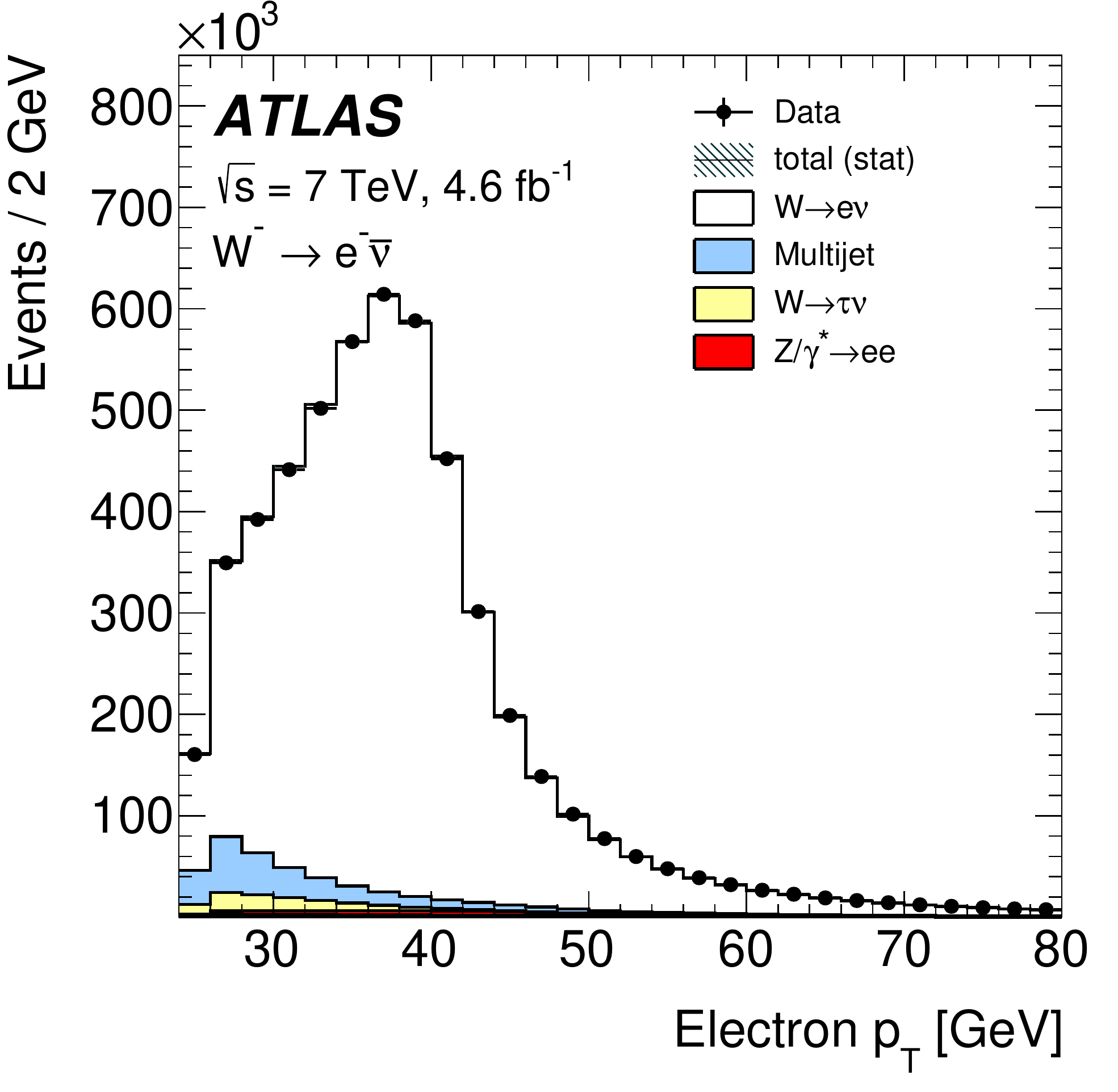}
    \caption{The transverse momentum distribution of electrons for
      \Wenup\ candidates (left) and \Wenum\ candidates (right). 
      \luminorm\  \mjplot\ \totstat\ \visible }
    \label{wenu:fig:candWpt}
  \end{center}
\end{figure}

\begin{figure}[tbp]
  \begin{center}
    \includegraphics[width=0.48\textwidth]{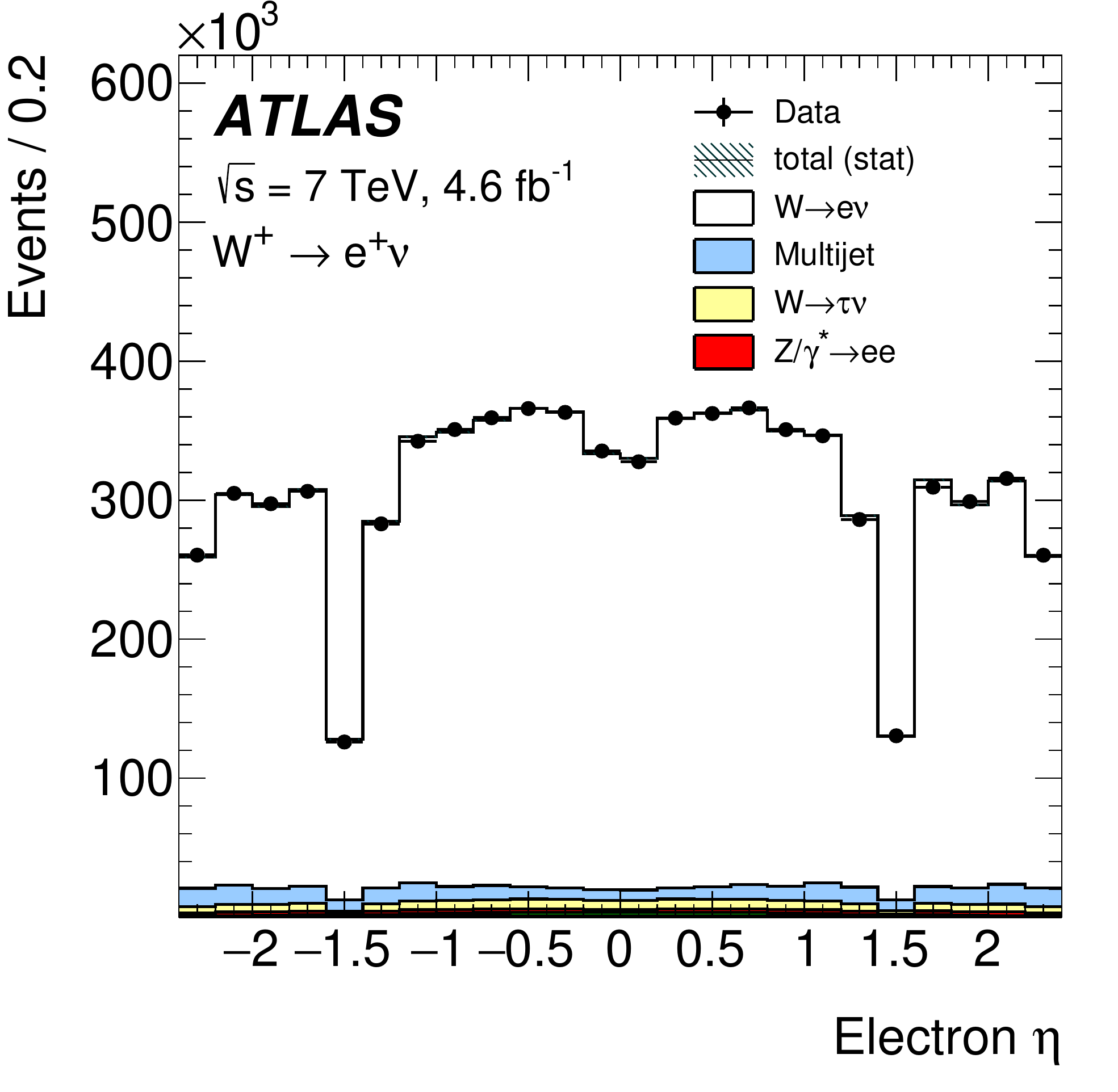}
    \includegraphics[width=0.48\textwidth]{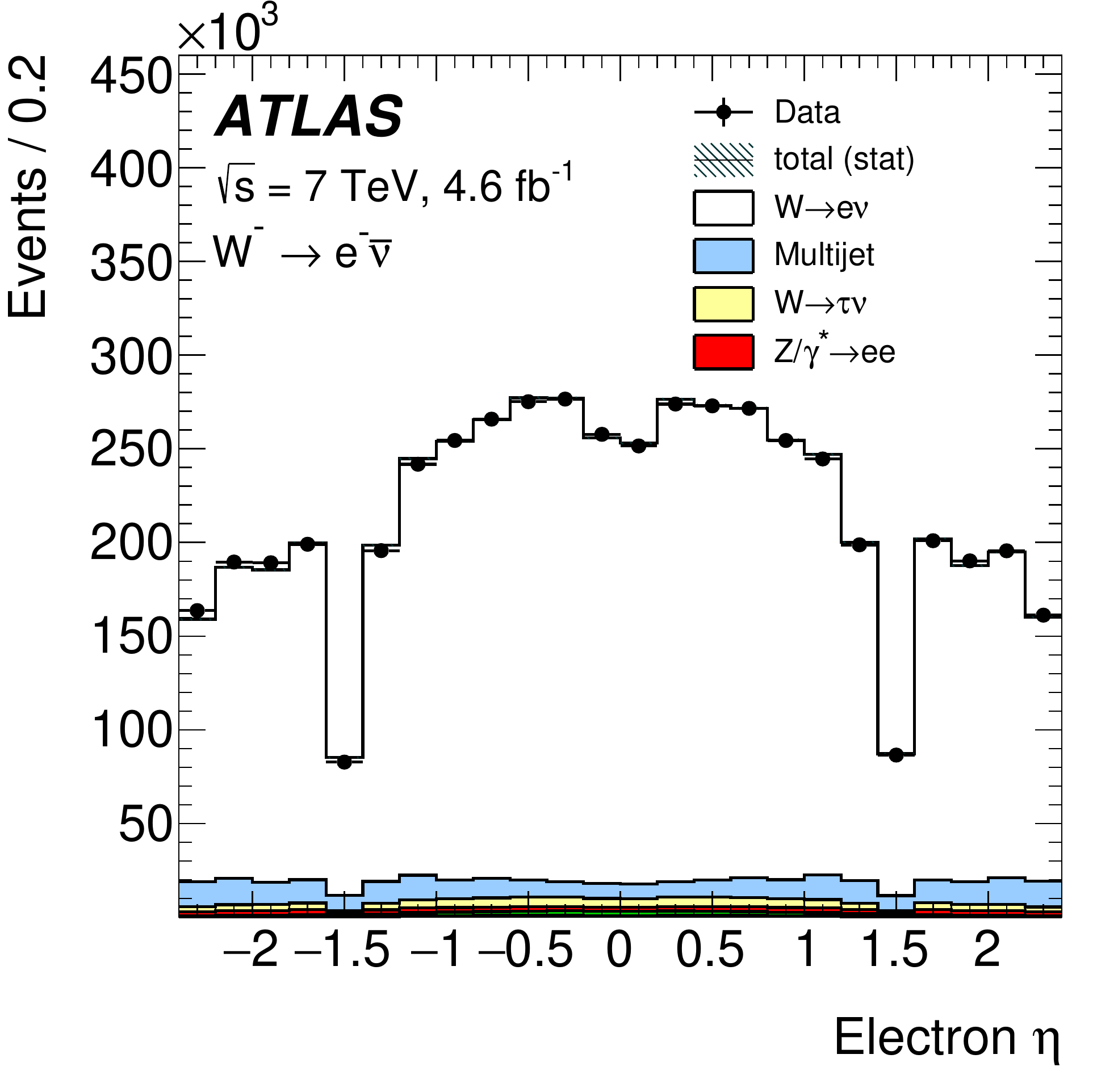}
    \caption{The pseudorapidity distribution of electrons for
      \Wenup\ candidates (left) and \Wenum\ candidates (right). 
      \luminorm\  \mjplot\ \totstat\ \visible }
    \label{wenu:fig:candWeta}
  \end{center}
\end{figure}

\begin{figure}[tbp]
  \begin{center}
    \includegraphics[width=0.48\textwidth]{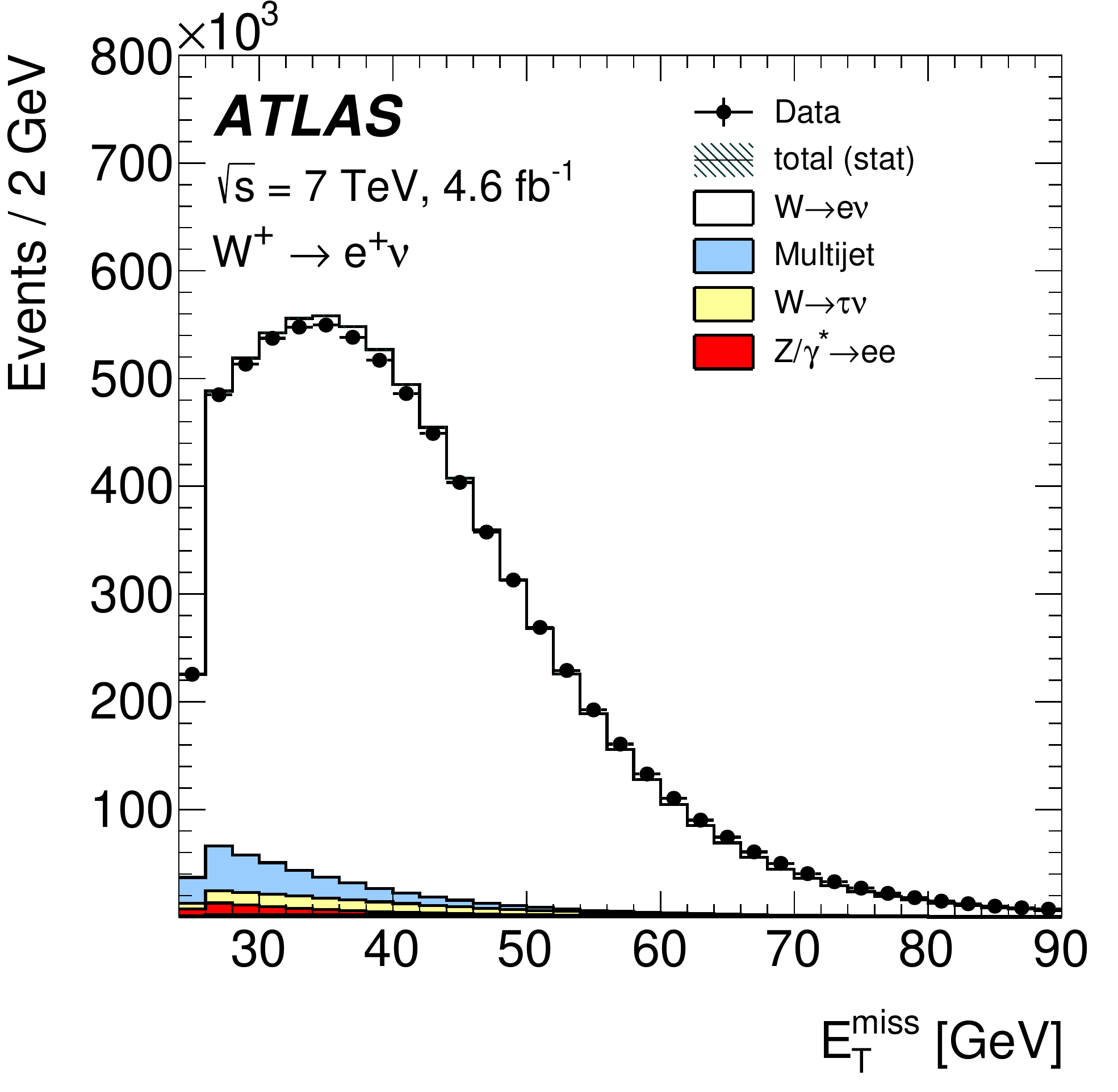}
    \includegraphics[width=0.48\textwidth]{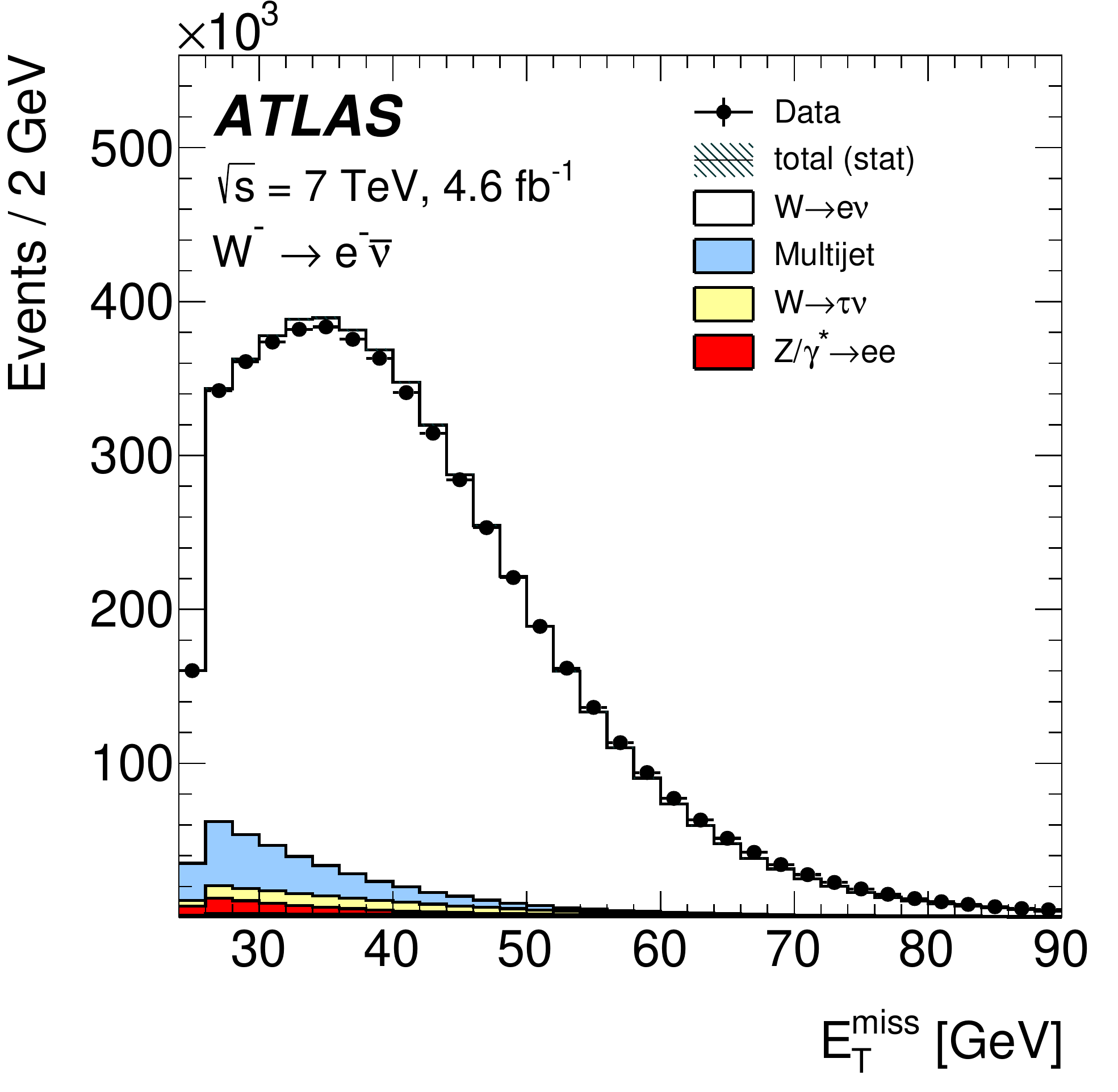}
    \caption{The missing transverse momentum distribution for \Wenup\
      candidates (left) and \Wenum\ candidates (right).  \luminorm\
      \mjplot\ \totstat\ \visible }
    \label{wenu:fig:candWmet}
  \end{center}
\end{figure}

\begin{figure}[tbp]
  \begin{center}
    \includegraphics[width=0.48\textwidth]{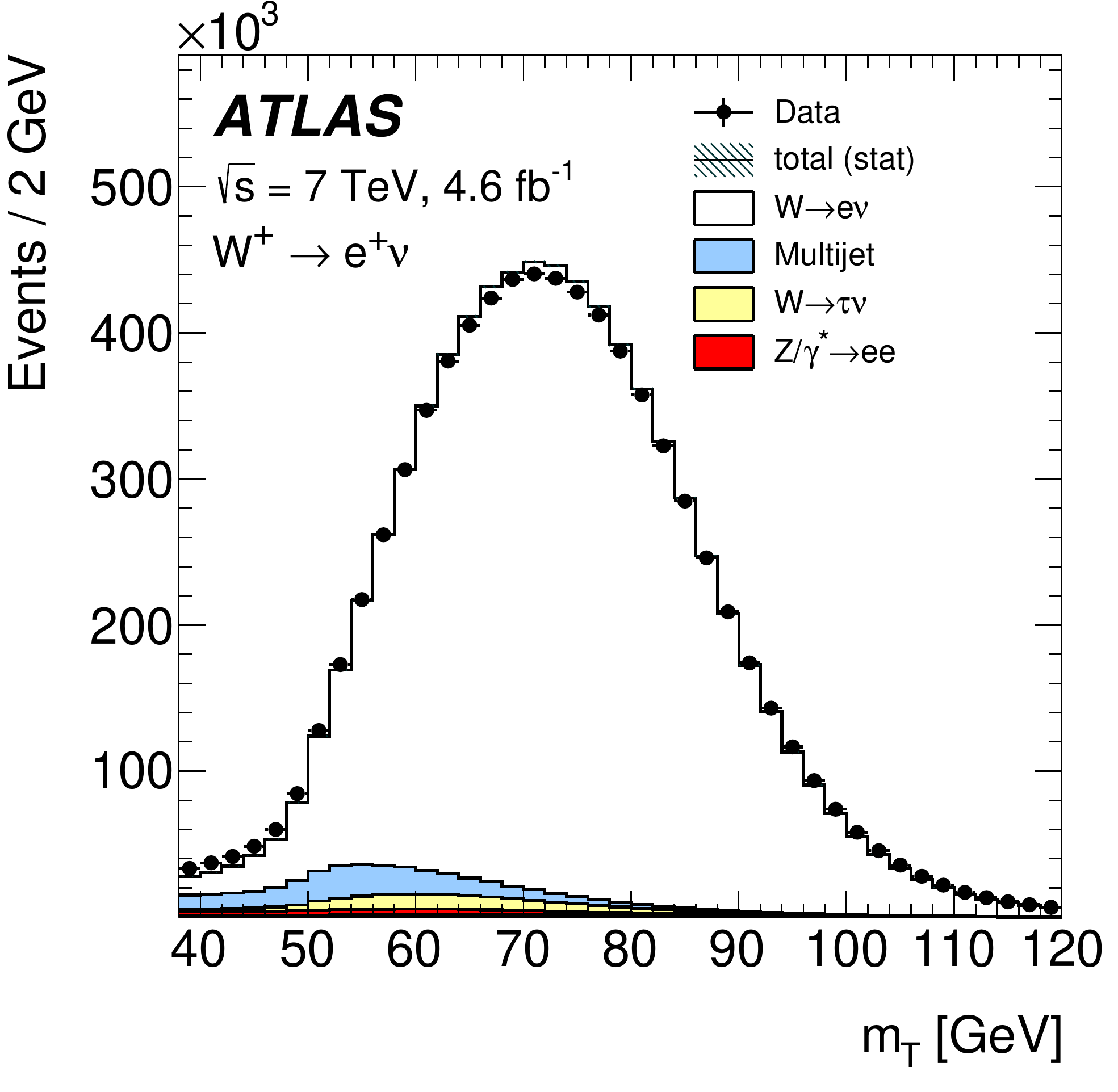}
    \includegraphics[width=0.48\textwidth]{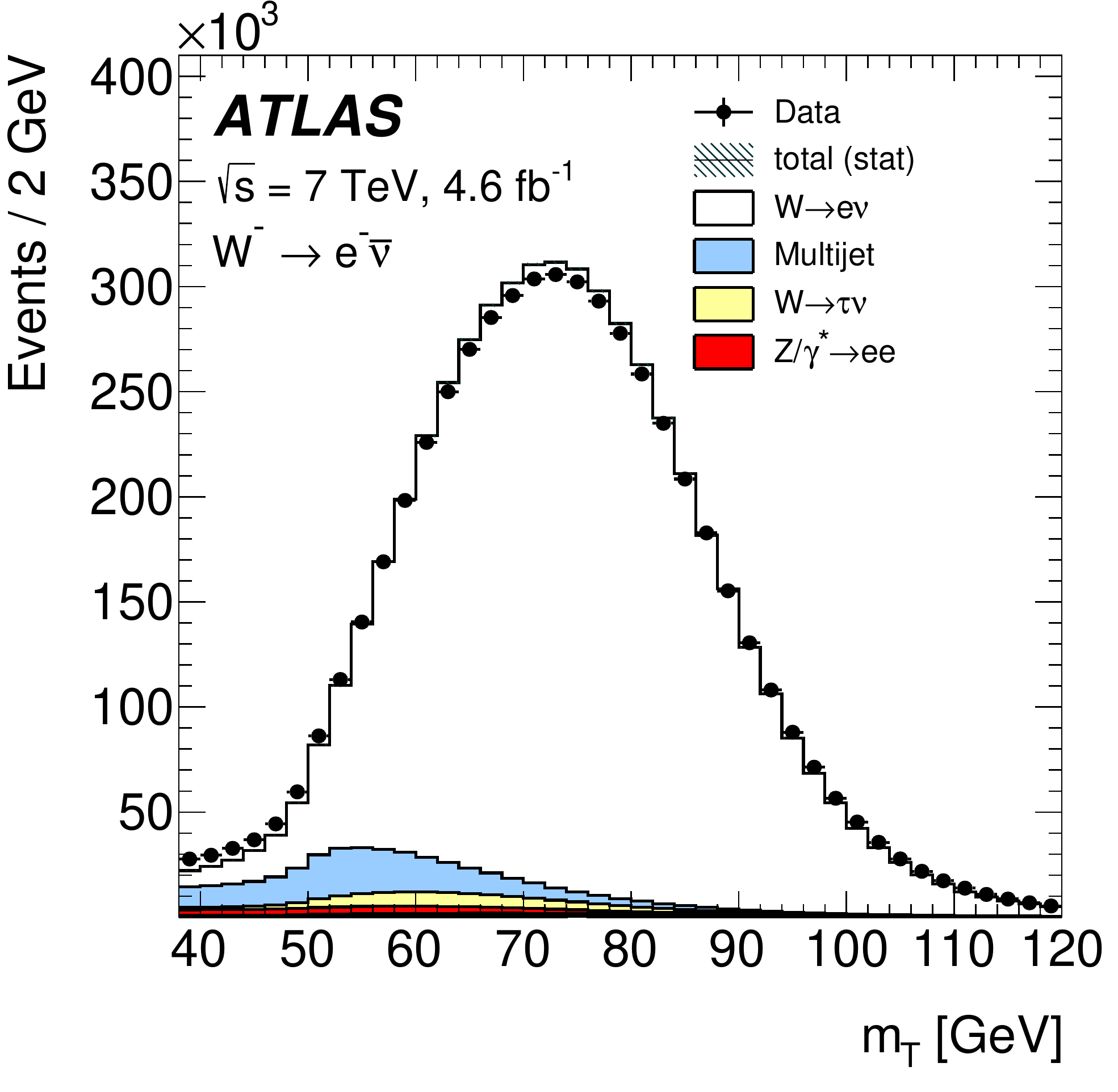}
    \caption{The transverse mass distribution for \Wenup\ candidates
      (left) and \Wenum\ candidates (right).  \luminorm\ \mjplot\
      \totstat\ \visible}
    \label{wenu:fig:candWmt}
  \end{center}
\end{figure}

\begin{figure}[tbp]
  \begin{center}
    \includegraphics[width=0.48\textwidth]{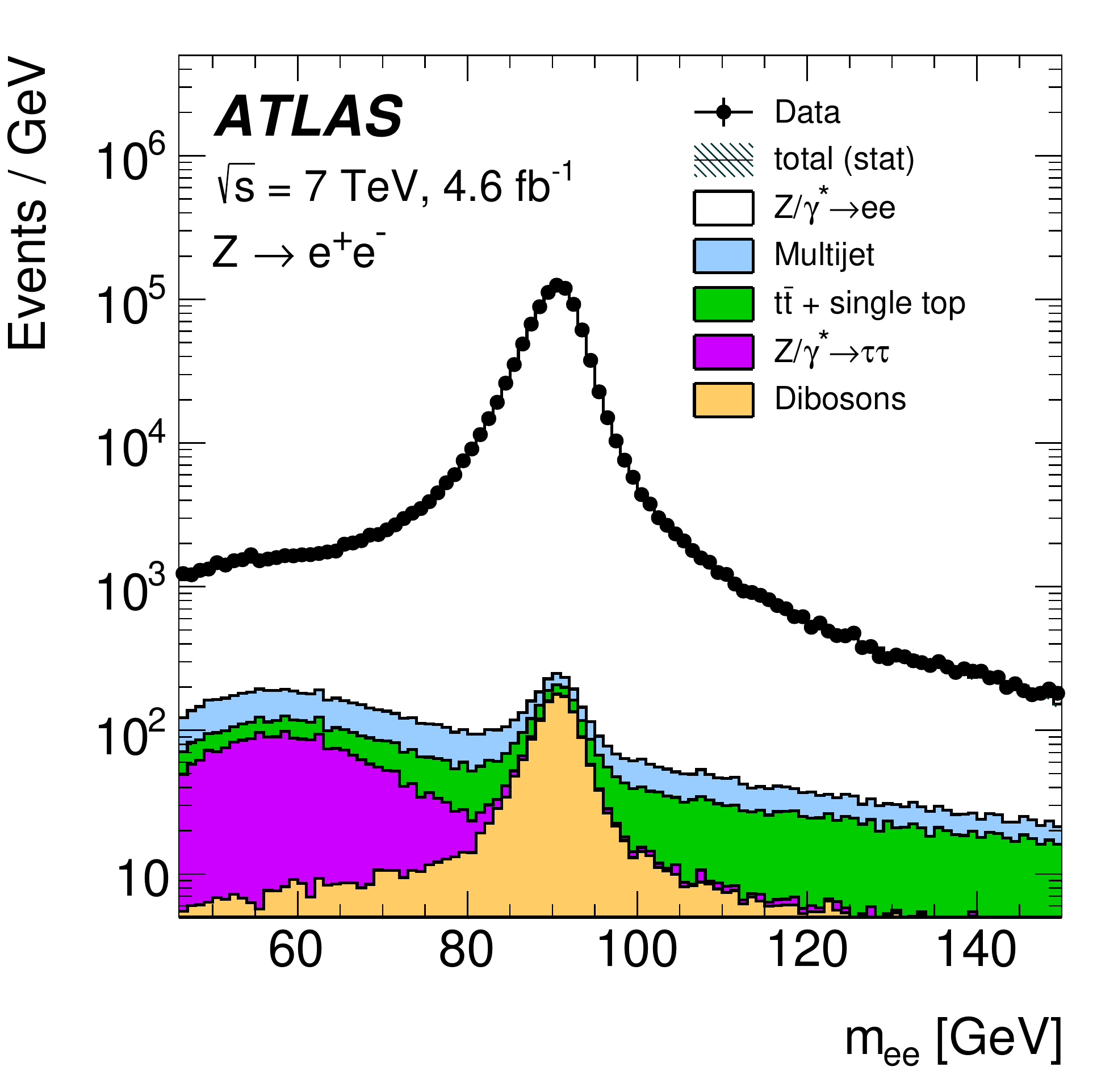}
    \includegraphics[width=0.48\textwidth]{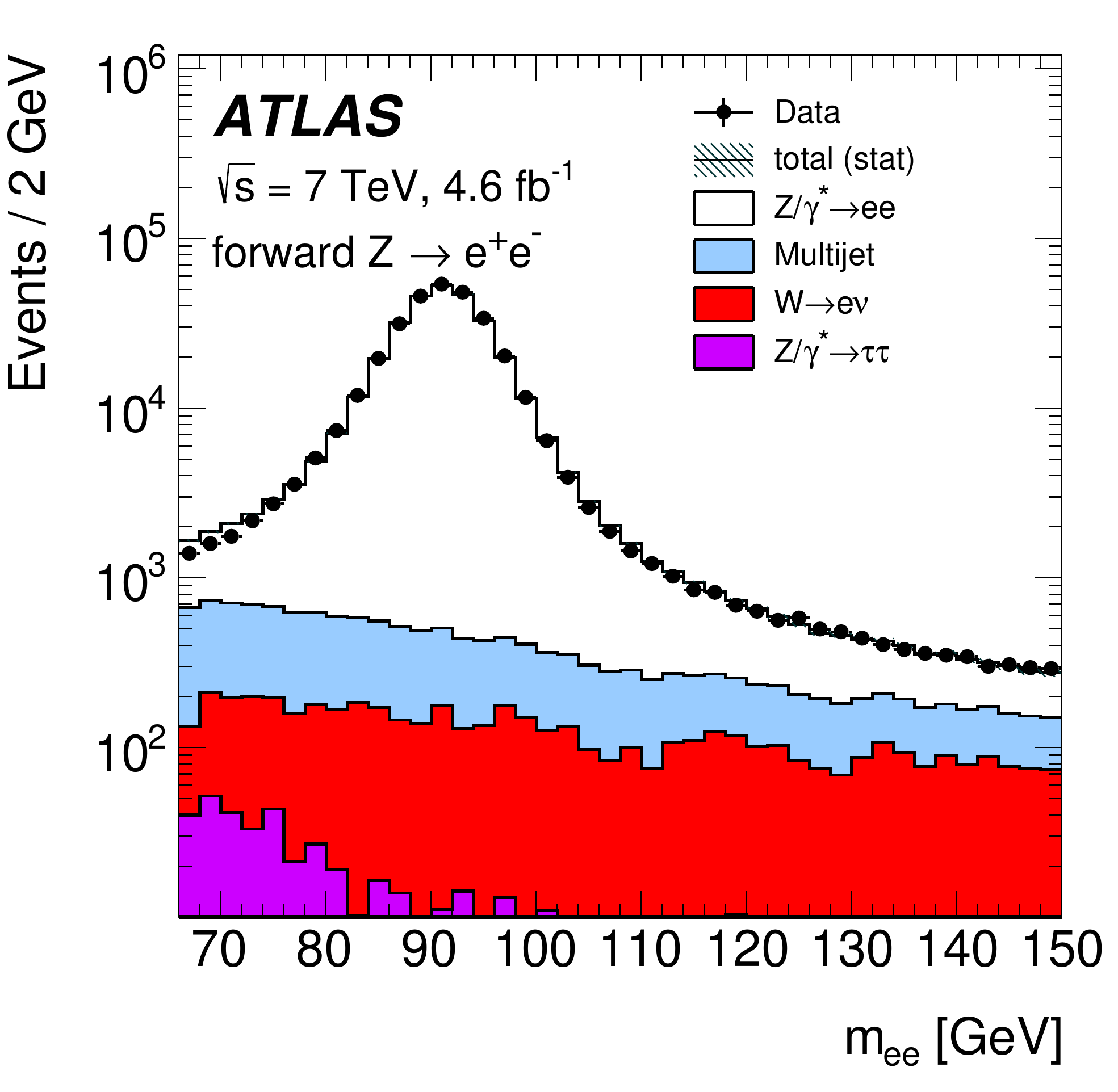}
    \caption{The dilepton invariant mass distributions for \Zgee\
      candidates with two central electrons (left) and one central and
      one forward electron (right). \luminorm\ \mjplot\ \totstat\
      \visible }
    \label{zee:fig:candZm}
  \end{center}
\end{figure}

\begin{figure}[tbp]
  \begin{center}
    \includegraphics[width=0.32\textwidth]{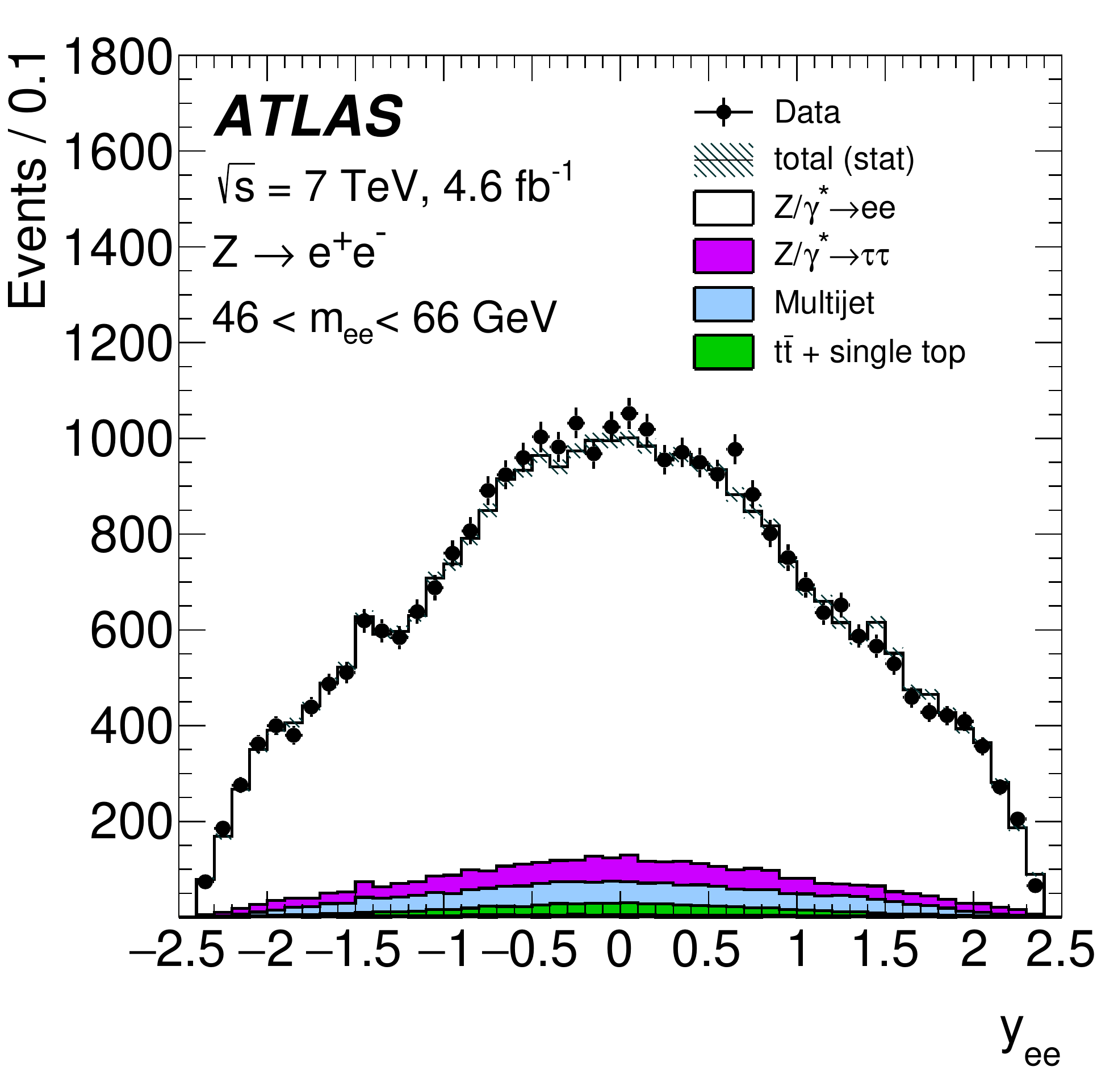}
    \includegraphics[width=0.32\textwidth]{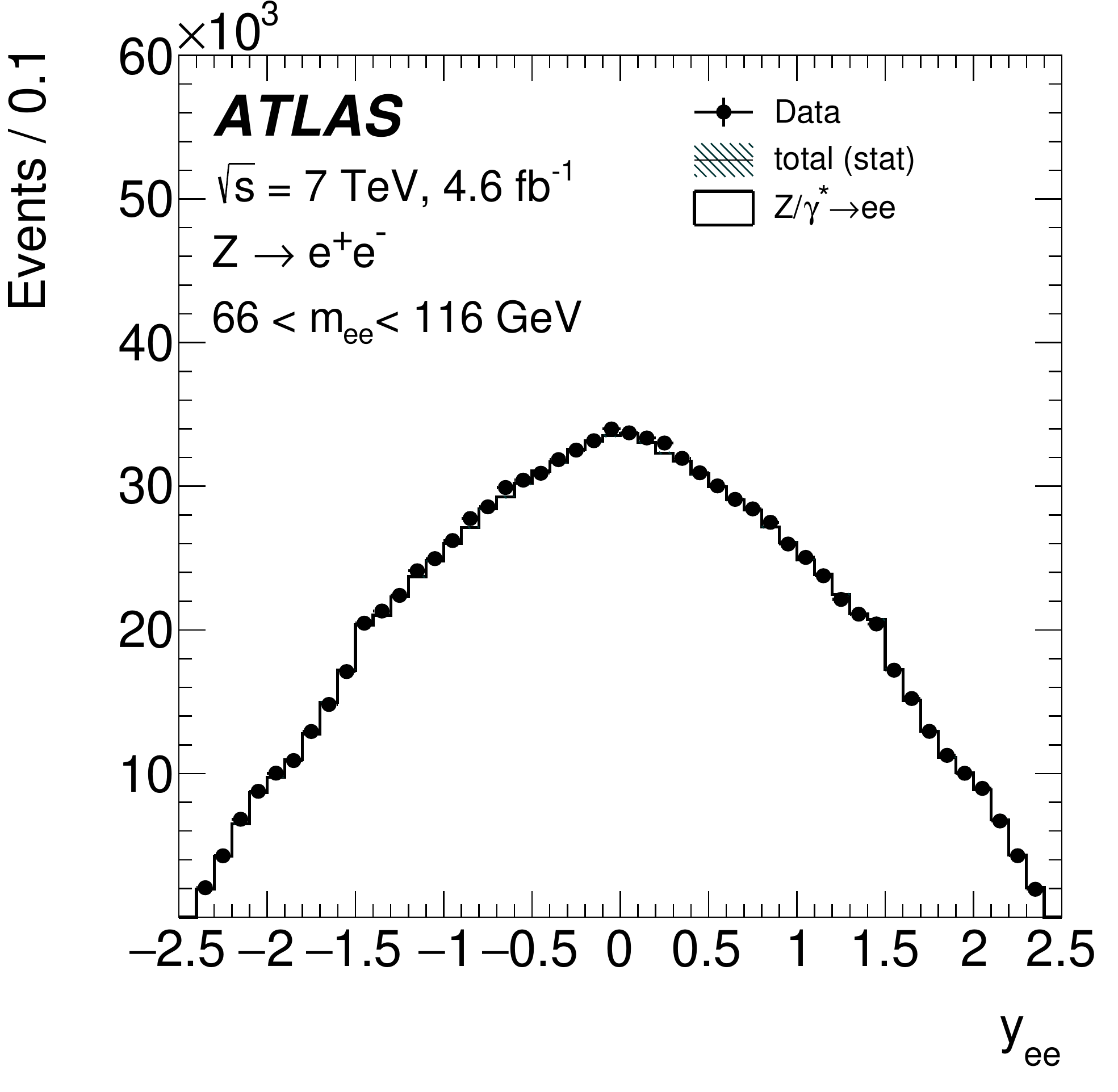}
    \includegraphics[width=0.32\textwidth]{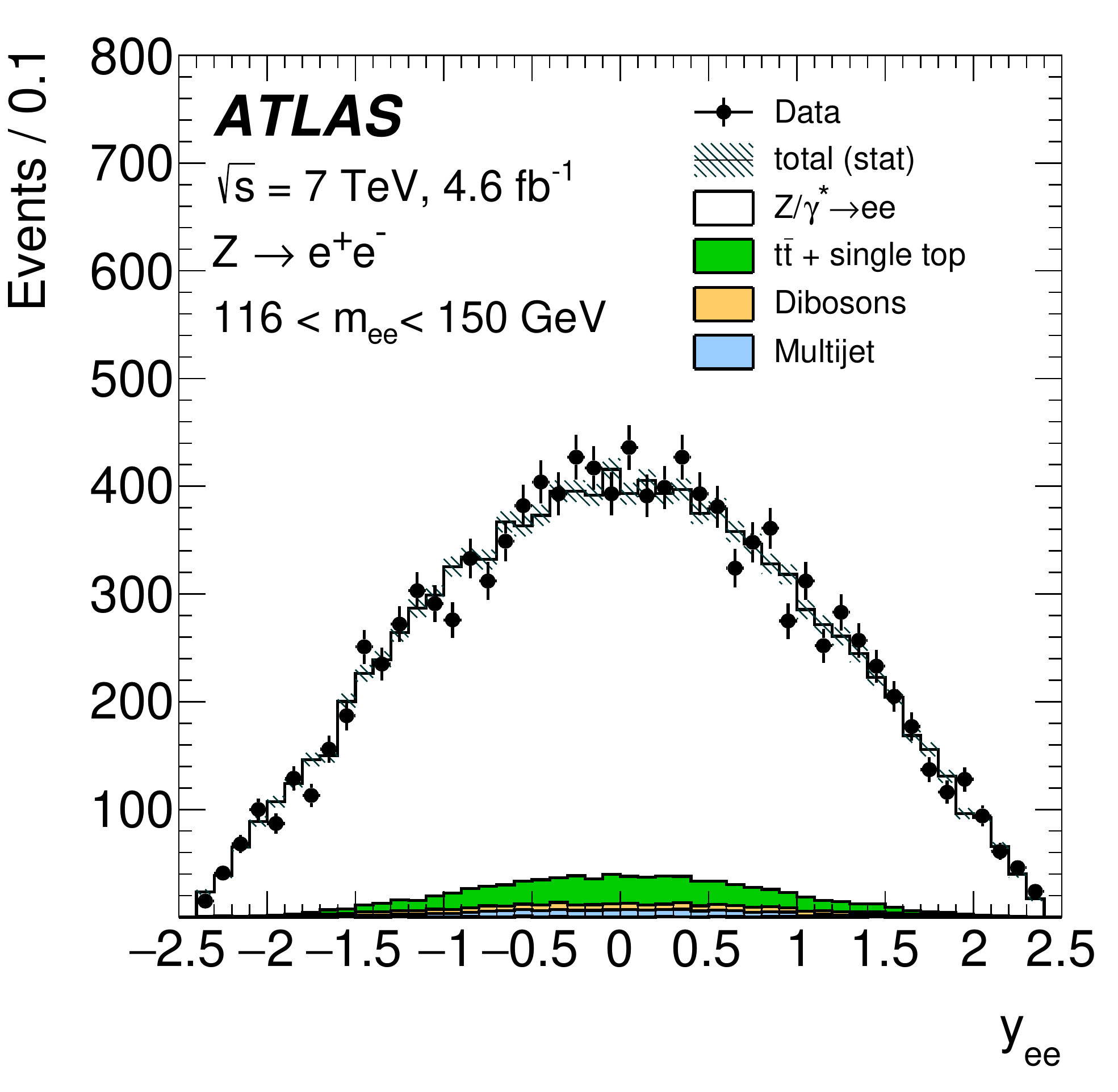}
    \caption{The dilepton rapidity distributions for \Zgee\
      candidates with two central electrons in the mass regions $46 <
      \mee < 66\GeV$ (left), $66 < \mee < 116\GeV$ (middle) and $116 <
      \mee < 150\GeV$ (right). \luminorm\ \mjplot\ \totstat\ \visible
    }
    \label{zee:fig:candZYCC}
  \end{center}
\end{figure}

\begin{figure}[tbp]
  \begin{center}
    \includegraphics[width=0.48\textwidth]{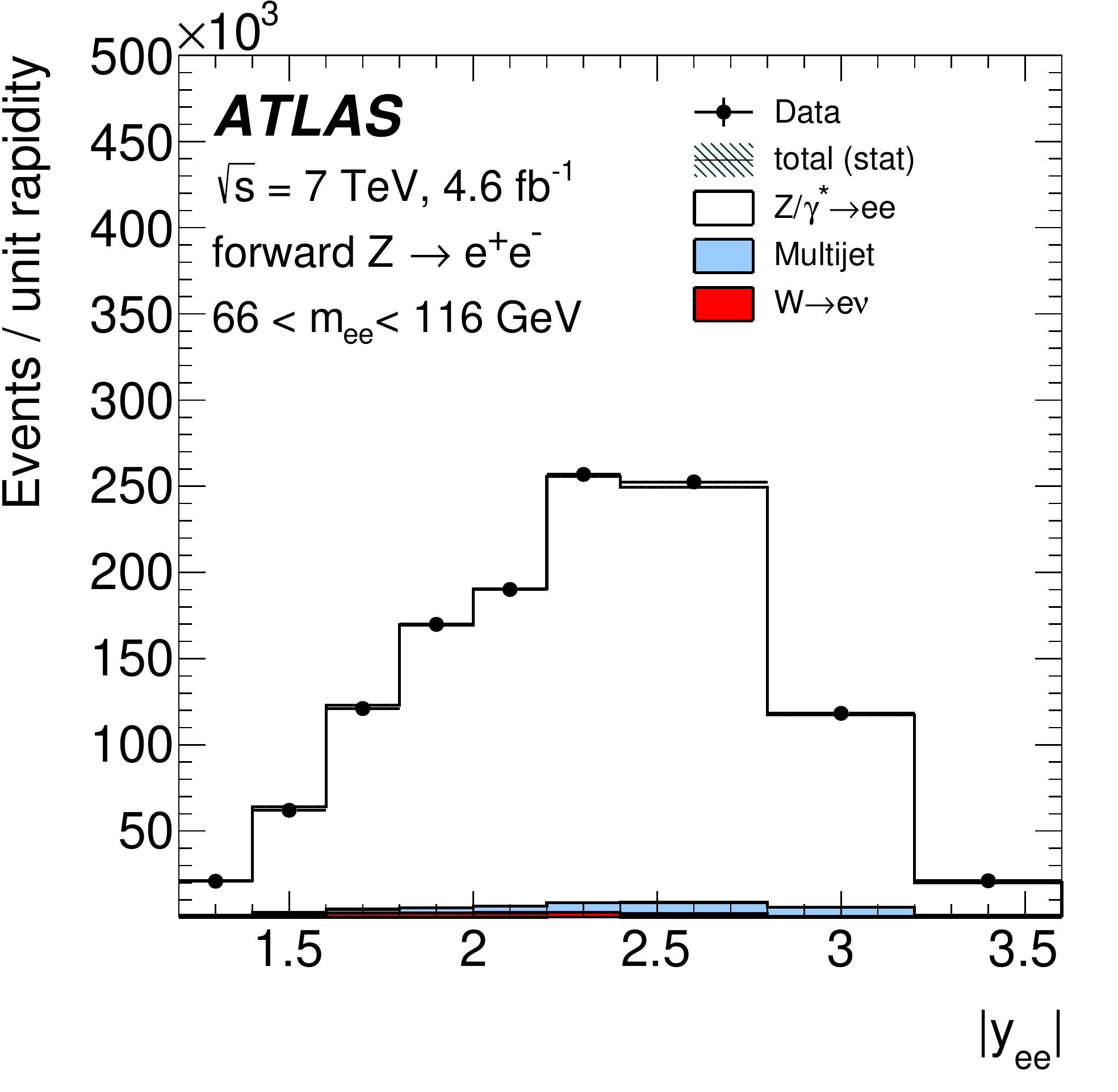}
    \includegraphics[width=0.48\textwidth]{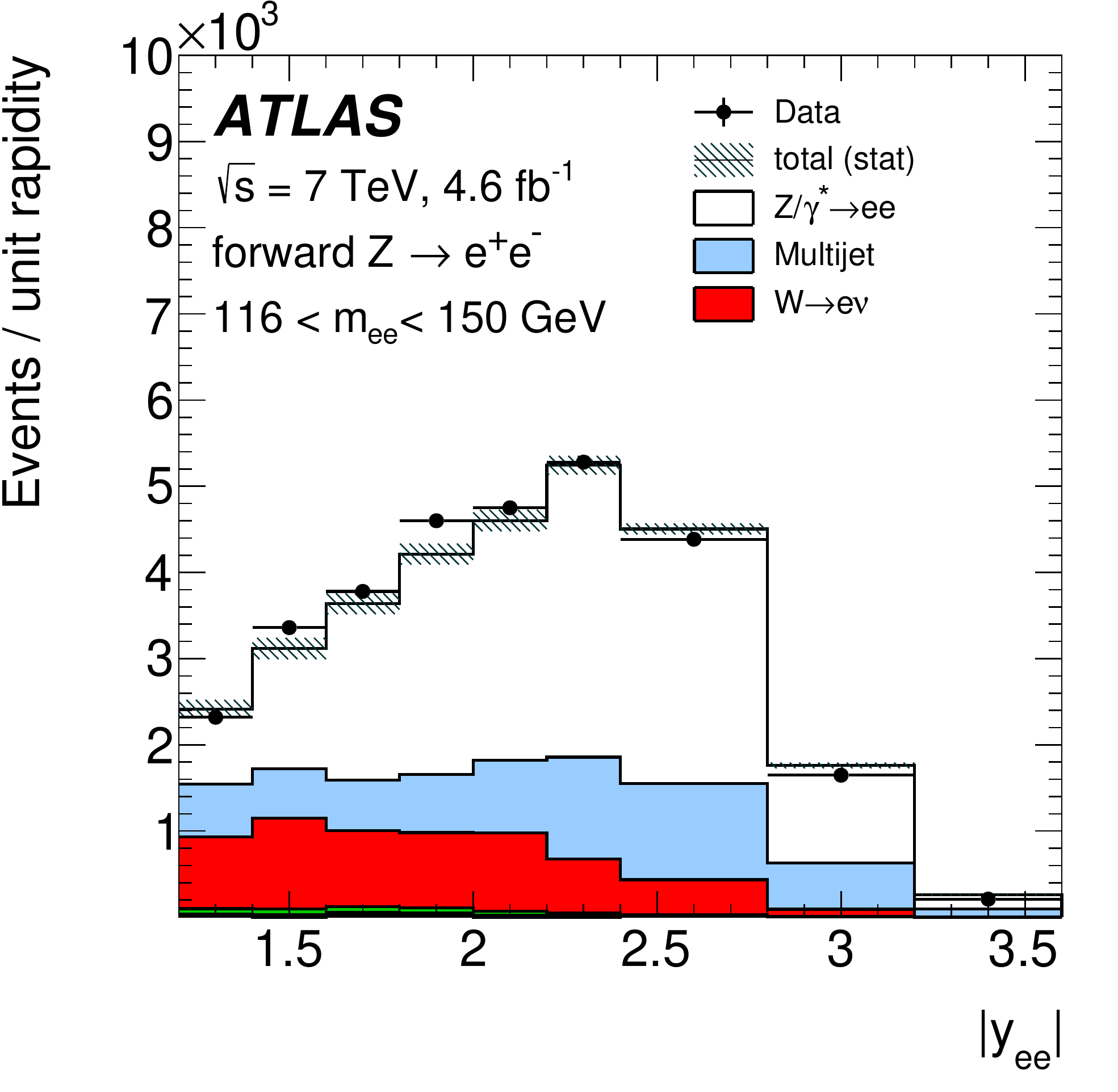}
    \caption{The dilepton rapidity distributions for \Zgee\ candidates
      with one central and one forward electron in the mass region $66
      < \mee < 116\GeV$ (left) and $116\GeV < \mee < 150\GeV$
      (right). \luminorm\ \mjplot\ \totstat\ \visible }
    \label{zee:fig:candZYCF}
  \end{center}
\end{figure}

\TTab~\ref{tab:WZaccEle} summarizes the number of selected candidates,
estimated background events and the \CWZ\ correction factors used for
the four different integrated electron channel measurements: $W^+$,
$W^-$, central \Zg, and forward \Zg\ analyses, both \Zg\ analyses
in the $Z$-peak region of $66<\mee<116\gev$. The
corresponding four integrated cross sections in the fiducial phase
space specific to the electron channels are reported in
\Tab~\ref{tab:elexsec} with their uncertainties due to data statistics,
luminosity, and further experimental systematic uncertainties. 

The systematic uncertainties split into their different components are
shown in \Tab~\ref{tab:elesyst}. Apart from the luminosity
contribution of \dlumi\%, the \Wenu\ cross section is measured with an
experimental uncertainty of $0.9\%$ for the $W^+$ channel and $1.1\%$
for the $W^-$ channel. The central \Zgee\ cross section in the
$Z$-peak region is measured with an uncertainty of $0.35\%$. The
extended forward rapidity \Zgee\ cross section is measured with an
uncertainty of $2.3\%$. 

The uncertainties of the data-driven determinations of 
the electron and hadronic recoil responses,
discussed in \Sec~\ref{sec:eleeffcalib},
are propagated to the measurements. These comprise uncertainties in the electron
detection efficiencies, separated into contributions from the trigger,
reconstruction, identification, and isolation, which are relatively
small for the \Wenu\ channel,  about $0.2\%$ in total, but constitute the
dominant systematic uncertainties in the central $Z$ data and
amount to $0.25\%$. In the forward
$Z$ analysis the dominant systematic uncertainty, of about $1.5\%$,
comes from 
the forward electron identification. The effects from
charge misidentification only affect the \Wpmenu\ cross sections and
are very small, $<0.1\%$. Both the central and forward electron \pt\
resolution and scale uncertainties are in general subdominant, amounting to
about $0.2\%$. The \Wenu\ analyses are also affected by
uncertainties in the hadronic recoil response, decomposed into soft
\met\ and jet energy scale and resolution uncertainties, which add up to
a total contribution of about $0.2\%$.

Signal modelling variations using different event generators, as
discussed in \Sec~\ref{sec:anaproc}, contribute significant
uncertainties of $0.6$--$0.7\%$ to the \Wenu\ analysis and $1.1\%$ to the
forward $Z$ analysis, while the effect on the central $Z$ analysis is
smaller with $0.2\%$. 
This source of uncertainty 
comprises effects from the lepton
efficiencies and, for the \Wen\ analysis, effects from the multijet
background determination, which relies on \met\ and \mt\ shapes, and
the hadronic recoil response. Other theoretical modelling
uncertainties, due to PDFs and boson \pt\ effects, are at the level of
$0.1$--$0.2\%$.

Uncertainties in the background subtraction are discussed in
\Sec~\ref{sec:ElBkg}. The contribution from the electroweak and
top-quark backgrounds is small and $<0.2\%$ for all channels. The
multijet background to the \Wenu\ channel, however, represents one of
the dominant uncertainties with $0.6$--$0.7\%$.

\begin{table}[htbp]
  \centering    
  \begin{tabular}{lrcc}
    \hline
    \hline
              &      \multicolumn{1}{c}{$N$}     &    $B$           & $C$ \\
    \hline
    \Wenup     & \NWeplusCands  & \nWeplusBkg    & \CWeplus   \\
    \Wenum     & \NWeminusCands & \nWeminusBkg   & \CWeminus  \\
    Central \Zgee      & \NZeeCCCandspm   & \nZeeCCBkgpm        & \CZeeCCpm \\
    Forward \Zgee\       & \NZeeCFCandspm   & \nZeeCFBkgpm        & \CZeeCFpm \\
    \hline 
    \hline
  \end{tabular}
  \caption{Number of observed event candidates $N$, of estimated background
    events $B$, and the  correction factors $C$ for the $W^+$, $W^-$,
    central  and forward \Zg\ ($66 < \mee <116\GeV$) electron channels.
    The correction factors $C$ were defined in Eq.~\eqref{Eq:CWZ}.
    The charge asymmetry in the background to the $W^{\pm}$ channels stems from the
    \Wtau\ contribution, which is proportional to the signal yield.
    The given uncertainties are the sum in quadrature of
    statistical and systematic components.
    The statistical uncertainties in $C$ are negligible.}
  \label{tab:WZaccEle}
\end{table}

\begin{table}[tbph]
\small
\begin{center}
\begin{tabular}{lc}
\hline
\hline
  & $\sigma^\mathrm{fid,e}_{W \to e\nu}$ [pb] \\[.25ex]
\hline
\Wenup\  &   $~\WeplusSigmaFid$  \\
\Wenum   &   $~\WeminusSigmaFid$  \\
\hline
\hline
 & $\sigma^\mathrm{fid,e}_{Z/\gamma^* \to ee}$ [pb] \\[.25ex]
\hline
Central \Zgee      &    \ZeeCCSigmaFidpm   \\
Forward \Zgee      &  \ZeeCFSigmaFidpm   \\
\hline
\hline
\end{tabular}
\caption{Fiducial cross sections times branching ratios for $W^+$,
  $W^-$, central  and forward \Zg\ ($66 <\mee<116\GeV$) production
  in the electron decay channels. The fiducial regions used for the
  measurement are those defined for the combined fiducial regions in
  \Sec~\ref{sec:fidureg}, except that the central electron
  pseudorapidity is restricted to be $|\eta|<2.47$ and
  excludes $1.37<|\eta|<1.52$, and the forward electron pseudorapidity
  excludes the region $3.16<|\eta|<3.35$. The uncertainties denote the
  statistical (stat), the systematic (syst) and the
  luminosity (lumi) uncertainties.}
\label{tab:elexsec}
\end{center}
\end{table}

\begin{table}[tbph]
\small
\sisetup{round-mode=places}
\begin{center}  
\begin{tabular}{lS[round-precision=2]S[round-precision=2]S[round-precision=2]S[round-precision=2]}
\hline 
\hline 
                      & $\delta \sigma_{W+}$ & $\delta \sigma_{W-}$ & $\delta \sigma_{Z}$ & $\delta \sigma_{\mathrm{forward}\,Z}$    \\
                      & $[\%]$ & $[\%]$ & $[\%]$ & $[\%]$    \\
\hline 
Trigger efficiency                    & 0.03         & 0.03         & 0.045        & 0.05  \\
Reconstruction efficiency             & 0.12         & 0.12         & 0.204        & 0.128  \\
Identification efficiency             & 0.09         & 0.09         & 0.163        & 0.121 \\
Forward identification efficiency     & $\mathrm{-}$ & $\mathrm{-}$ & $\mathrm{-}$ & 1.514 \\
Isolation efficiency                  & 0.03         & 0.03         & $\mathrm{-}$ & 0.039 \\
Charge misidentification              & 0.04         & 0.06         & $\mathrm{-}$ & $\mathrm{-}$    \\
Electron \pt resolution               & 0.02         & 0.03         & 0.009        & 0.011 \\
Electron \pt scale                    & 0.22         & 0.18         & 0.075        & 0.115 \\
Forward electron \pt scale + resolution & $\mathrm{-}$ & $\mathrm{-}$ & $\mathrm{-}$ & 0.179 \\
\met\ soft term scale              & 0.14         & 0.13         & $\mathrm{-}$ & $\mathrm{-}$    \\
\met\ soft term resolution         & 0.06         & 0.04         & $\mathrm{-}$ & $\mathrm{-}$     \\
Jet energy scale                      & 0.04         & 0.02         & $\mathrm{-}$ & $\mathrm{-}$    \\
Jet energy resolution                 & 0.11         & 0.15         & $\mathrm{-}$ & $\mathrm{-}$    \\
Signal modelling (matrix-element generator)     & 0.57         & 0.64         & 0.027        & 1.12  \\
Signal modelling (parton shower and hadronization)       & 0.24         & 0.25         & 0.18         & 1.25  \\
PDF                                  & 0.10         & 0.12         & 0.09         & 0.059 \\
Boson \pt                & 0.22         & 0.19         & 0.006        & 0.043 \\
Multijet background                   & 0.55         & 0.72         & 0.028        & 0.052 \\
Electroweak+top background        & 0.17         & 0.19         & 0.022        & 0.139 \\
Background statistical uncertainty    & 0.02         & 0.03         & <0.01        & 0.044 \\
Unfolding statistical uncertainty     & 0.03         & 0.04         & 0.038        & 0.134 \\
\hline
Data statistical uncertainty          & 0.04         & 0.05         & 0.099        & 0.18  \\
\hline
Total experimental uncertainty        & 0.94         & 1.08         & 0.346        & 2.288 \\
\hline 
Luminosity                       &                \multicolumn{4}{c}{$\dlumi$}   \\
\hline 
\hline 
\end{tabular}
\caption{Relative uncertainties $\delta\sigma$ in the measured
  integrated fiducial cross sections times branching ratios of $W^+$, $W^-$,
  central  and forward \Zg\ ($66 < \mee <116\GeV$) in the electron
  channels.}
\label{tab:elesyst}       
\end{center}
\end{table}

The differential cross-section measurements as a function of the
$W^{\pm}$ electron pseudorapidity and the dielectron rapidity and
mass for the \Zg\ channel are summarized in the
Appendix~\ref{sec:simple_xsectable} in the
\Tabs~\ref{tab:w_eta_e}--\ref{tab:zhigh_yll_e}.
The statistical uncertainties in the \Wen\ differential cross sections
are about $0.1$--$0.2\%$, and the total uncertainties are in the range of
$0.9$--$2.2\%$, excluding the luminosity uncertainty.

The differential \Zgee\ cross sections in the central region are
measured in the $\mee=66$--$116\GeV$ invariant mass region with a
statistical uncertainty of about $0.3$--$0.5\%$ up to $|\yll|=2.0$ and of
$0.9\%$ for $|\yll| = 2.0$--$2.4$. The total uncertainty, excluding the
luminosity uncertainty, is $0.5$--$0.7\%$ up to $|\yll|=2.0$ and $1.4\%$
for $|\yll|= 2.0$--$2.4$. The statistical uncertainties of the differential
\Zgee\ cross sections measured in the regions $\mee=46$--$66\gev$ and
$116$--$150\GeV$ are in the range  $1.5$--$5\%$, dominating the total
uncertainties of $2$--$6\%$.

The uncertainties in the forward \Zgee\ differential cross sections
are  dominated by systematic uncertainties. At the $Z$ peak, the
total uncertainty is  $3$--$8\%$, while in the high-mass
region it is about $10$--$20\%$.


\clearpage

\subsubsection{Muon channels}
\label{sec:cross_mu}
The description of important kinematic variables in the muon-channel
data by the signal simulation and the estimated backgrounds is
illustrated in
\Figs~\ref{wmunu:fig:candWpt}~to~\ref{zmumu:fig:candZmPt}. The signal
and electroweak background distributions are taken from MC simulation
and normalized to the corresponding data luminosity. The distributions
for the background from multijet production are obtained from data and
normalized as described in \Sec~\ref{sec:MuBkg}.
\FFigs~\ref{wmunu:fig:candWpt}, \ref{wmunu:fig:candWeta} and
\ref{wmunu:fig:candWmet} show the distributions of muon transverse
momentum, muon pseudorapidity and the missing transverse momentum of
candidate $W$ events for positive and negative charges. The transverse
mass distributions are shown in \Fig~\ref{wmunu:fig:candWmt}. The
dimuon mass distribution of muon pairs selected by the \Zgmumu\
analysis are shown in \Fig~\ref{zmumu:fig:candZmmass}, while
\Fig~\ref{zmumu:fig:candZmPt} shows the dimuon rapidity distributions
for the three invariant mass regions. The level of agreement between data and
simulation is good in all cases. Small disagreements in the shapes of
the \met\ and \mt\ distributions of W-boson candidates are visible in a similar way as in the electron channel and are covered by the systematic uncertainties.

\begin{figure}[htbp]
  \begin{center}
    \includegraphics[width=0.48\textwidth]{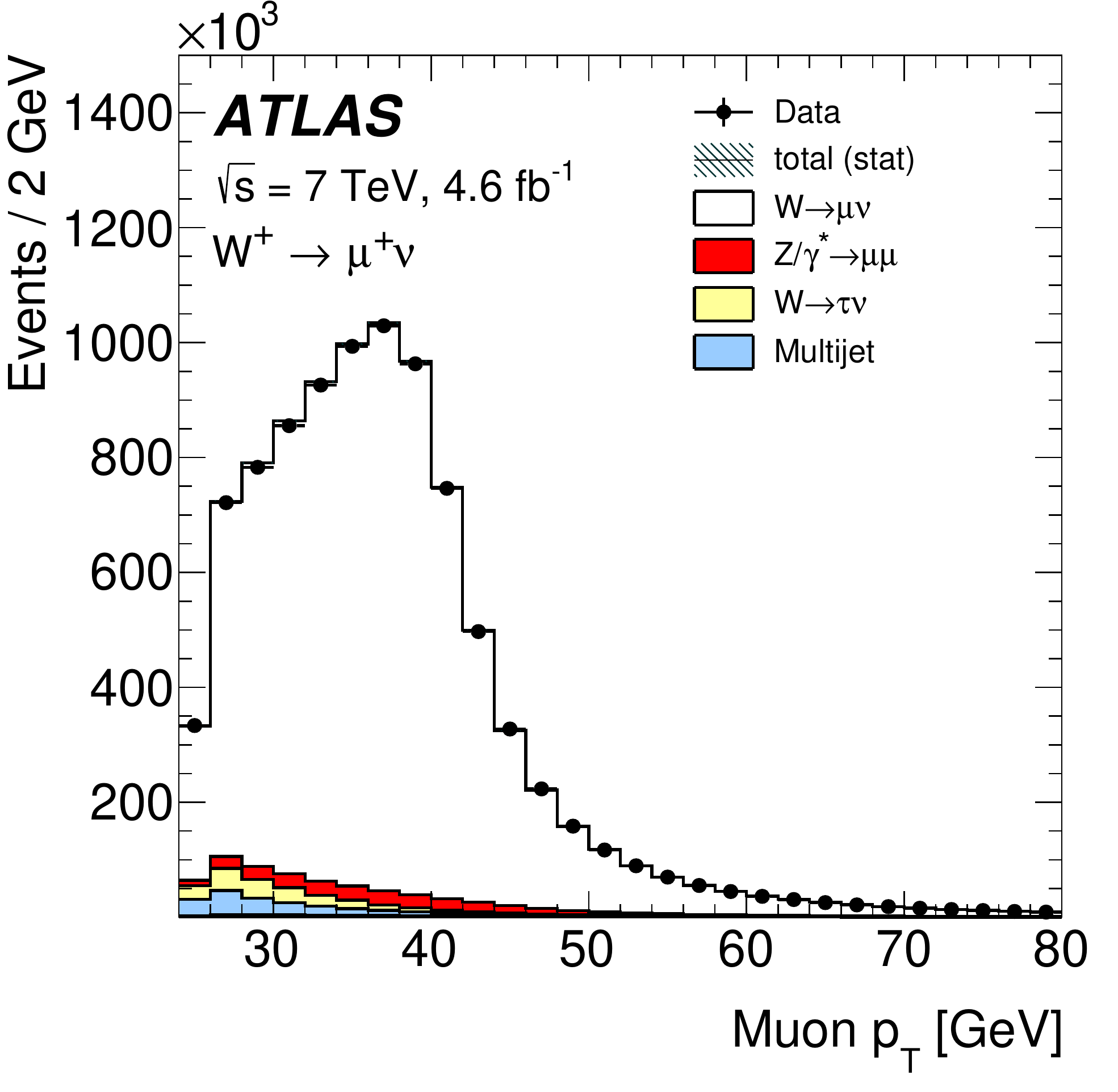}
    \includegraphics[width=0.48\textwidth]{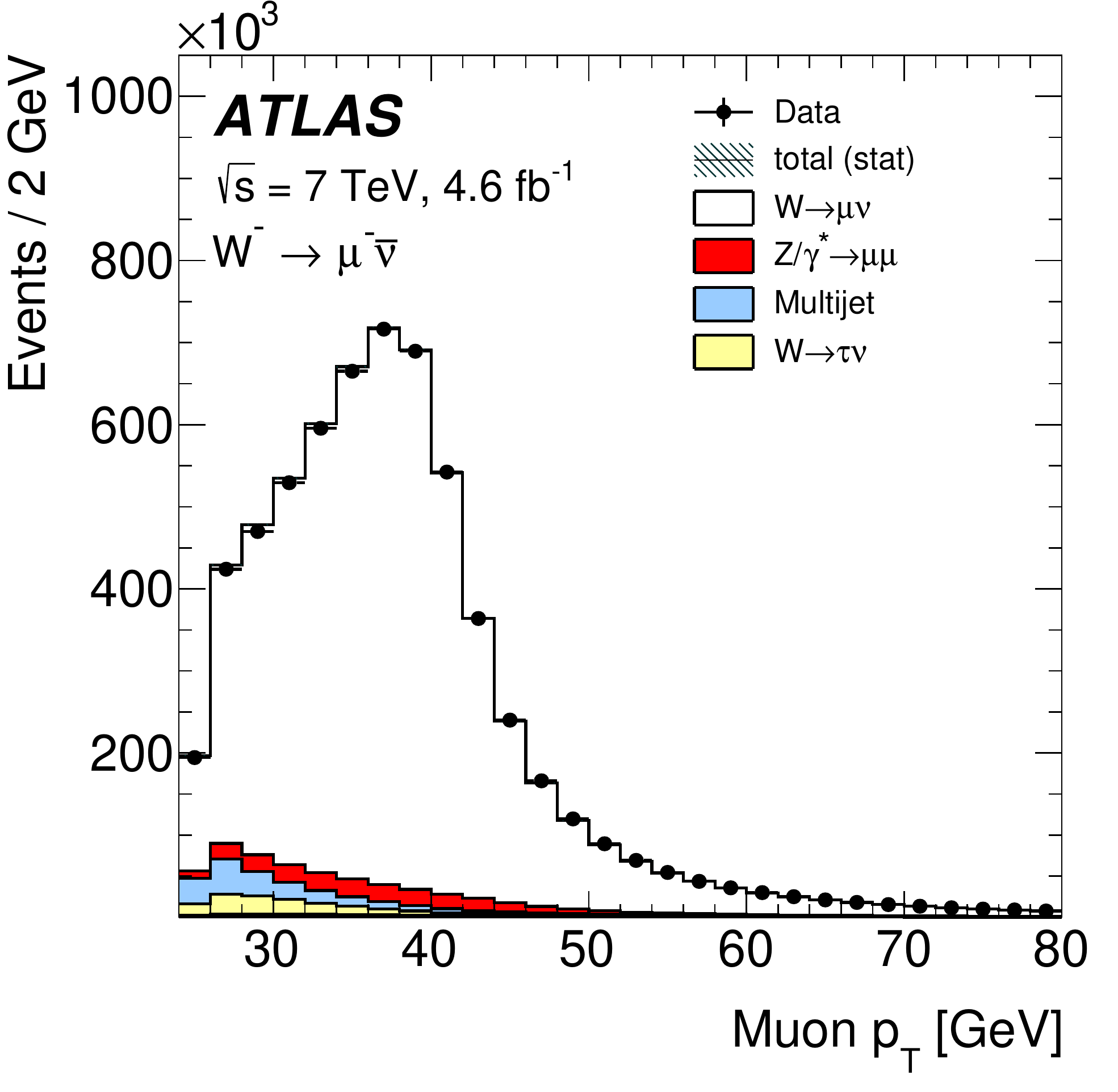}
    \caption{The transverse momentum distribution of muons for
      \Wmunup\ candidates (left) and \Wmunum\ candidates (right). 
      \luminorm\  \mjplot\ \totstat\ \visible }
    \label{wmunu:fig:candWpt}
  \end{center}
\end{figure}

\begin{figure}[htbp]
  \begin{center}
    \includegraphics[width=0.48\textwidth]{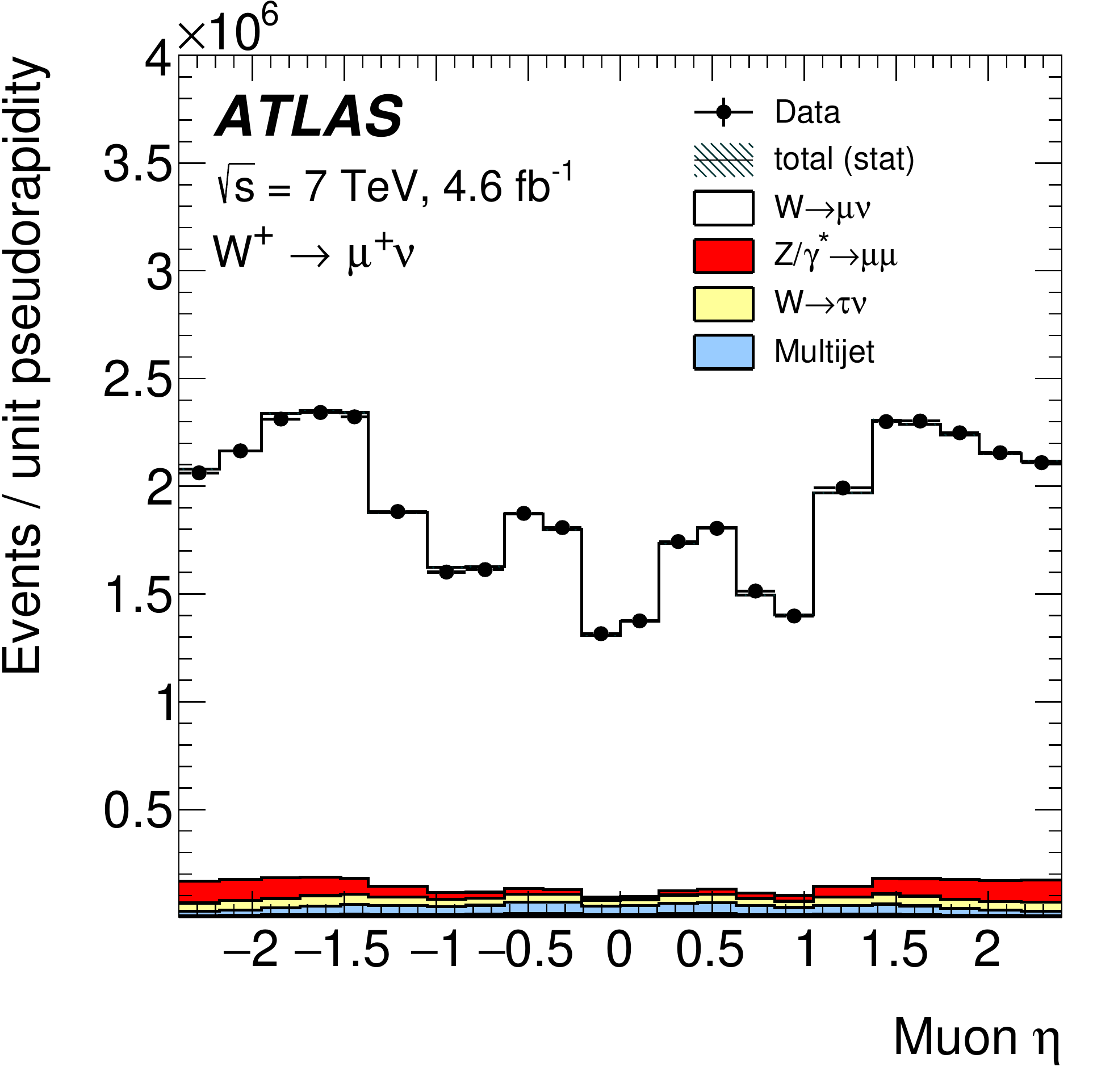}
    \includegraphics[width=0.48\textwidth]{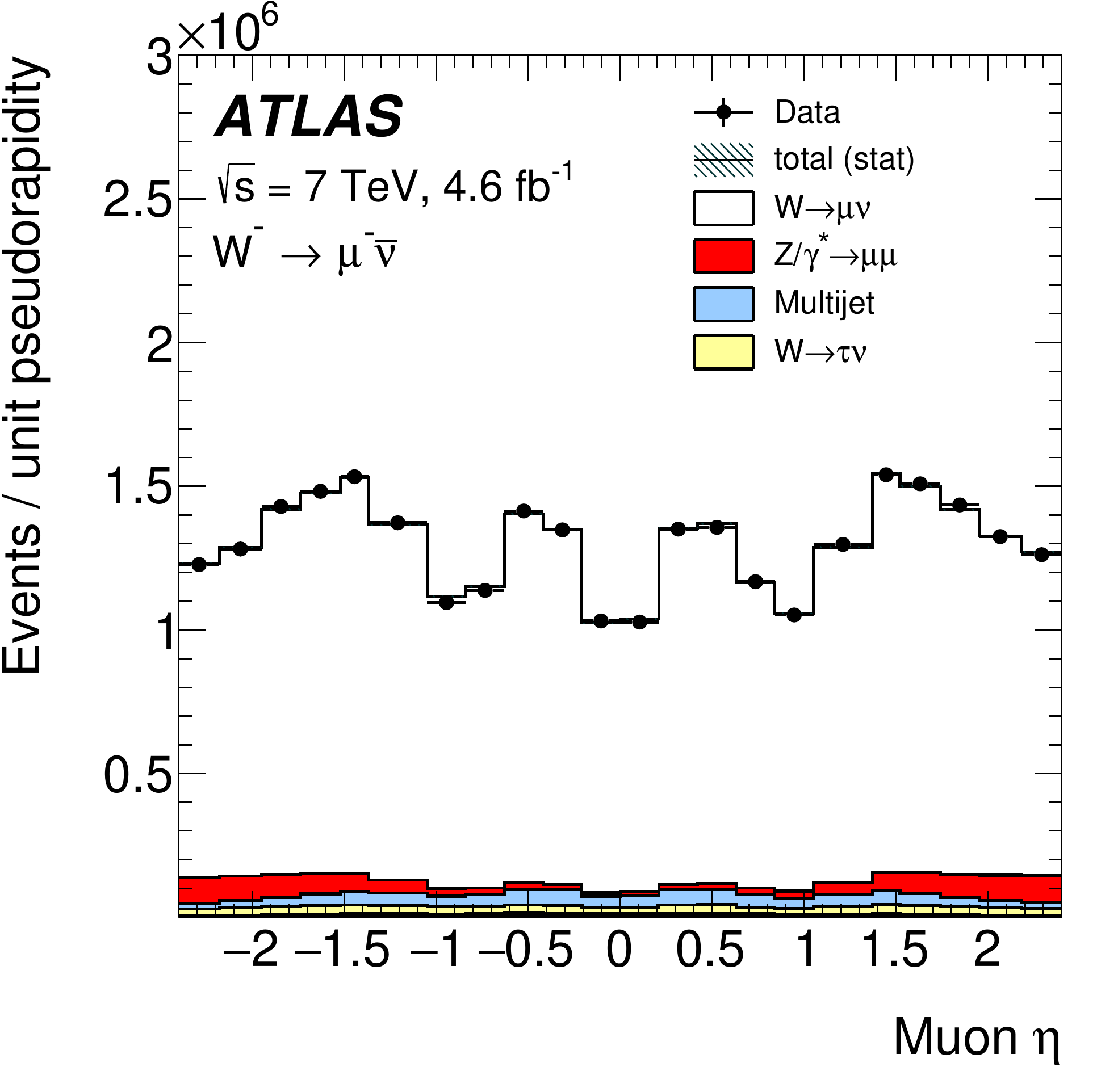}
    \caption{The pseudorapidity distribution of muons for
      \Wmunup\ candidates (left) and \Wmunum\ candidates (right). 
      \luminorm\  \mjplot\ \totstat\ \visible }
    \label{wmunu:fig:candWeta}
  \end{center}
\end{figure}

\begin{figure}[htbp]
  \begin{center}
    \includegraphics[width=0.48\textwidth]{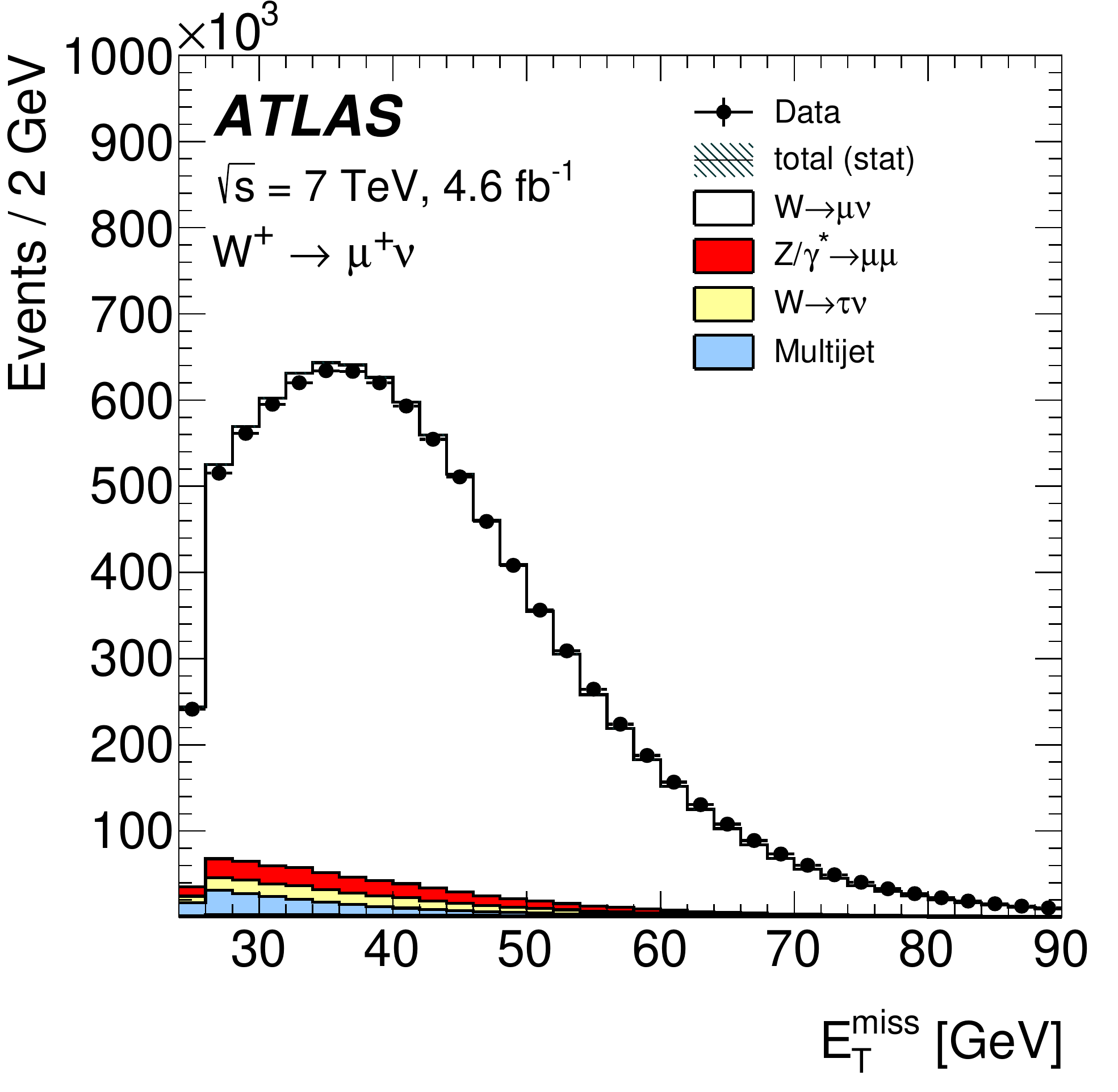}
    \includegraphics[width=0.48\textwidth]{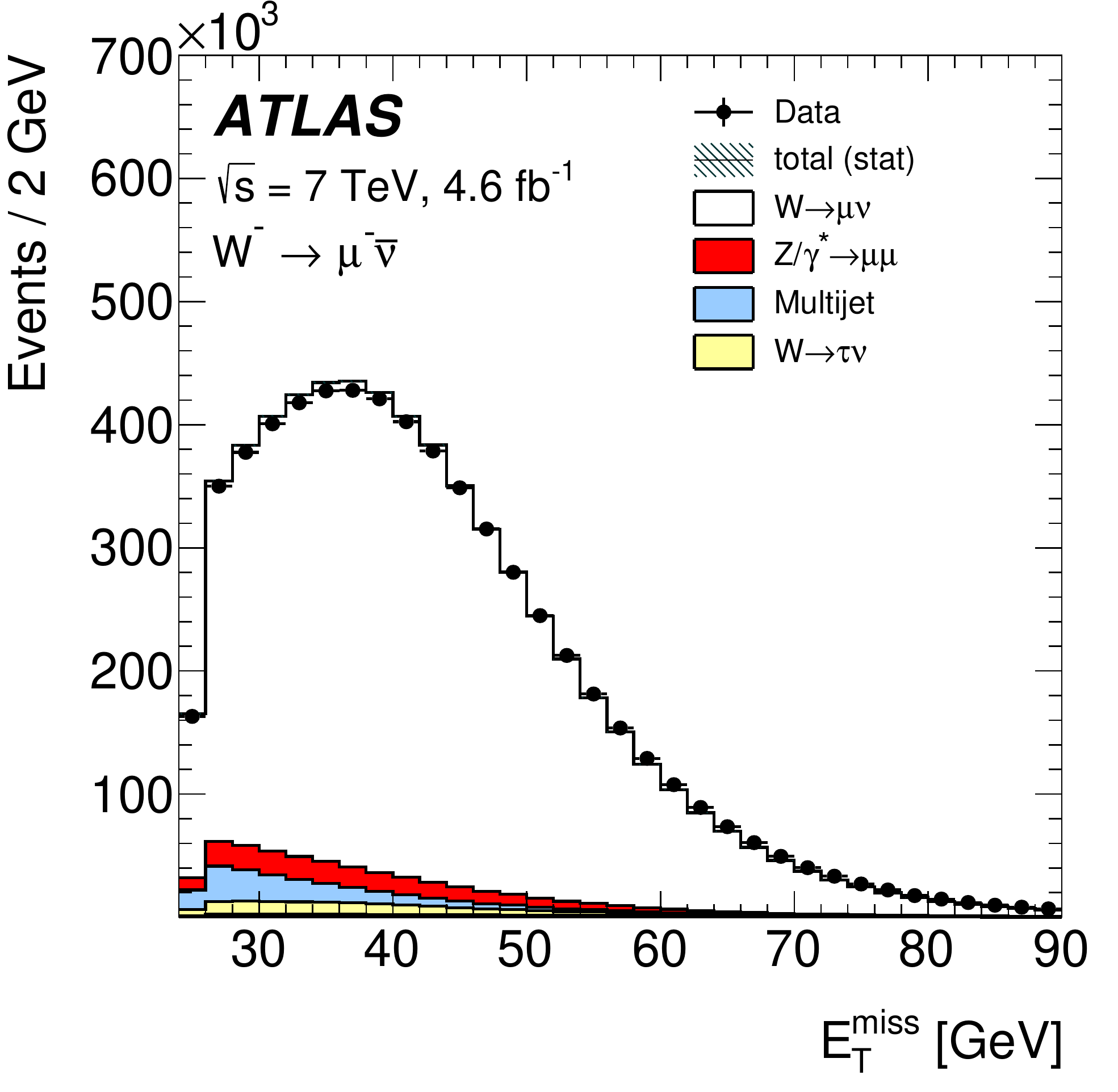}
    \caption{The missing transverse momentum distribution for \Wmunup\
      candidates (left) and \Wmunum\ candidates (right).  \luminorm\
      \mjplot\ \totstat\ \visible }
      \label{wmunu:fig:candWmet}
  \end{center}
\end{figure}

\begin{figure}[htbp]
  \begin{center}
    \includegraphics[width=0.48\textwidth]{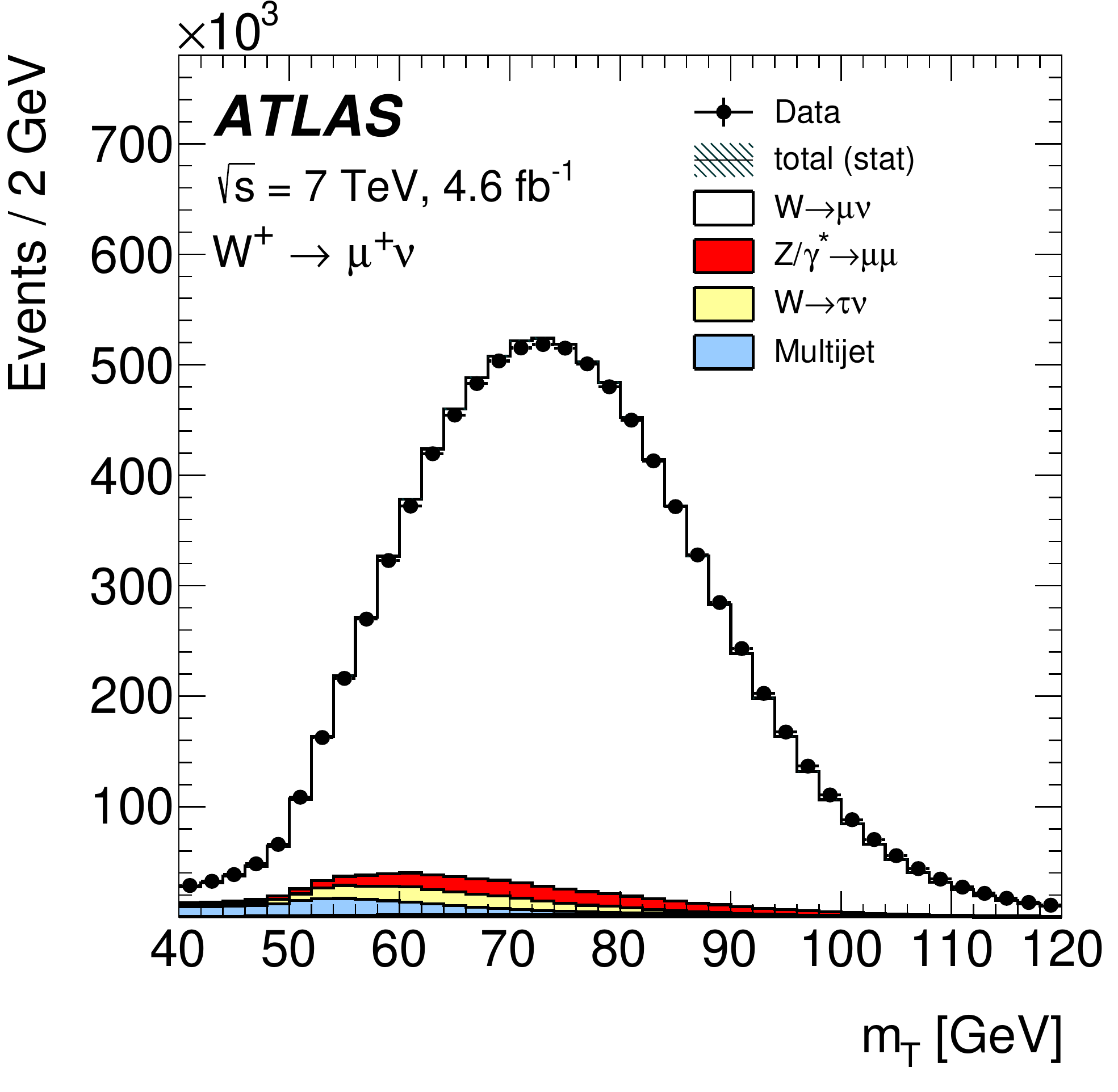}
    \includegraphics[width=0.48\textwidth]{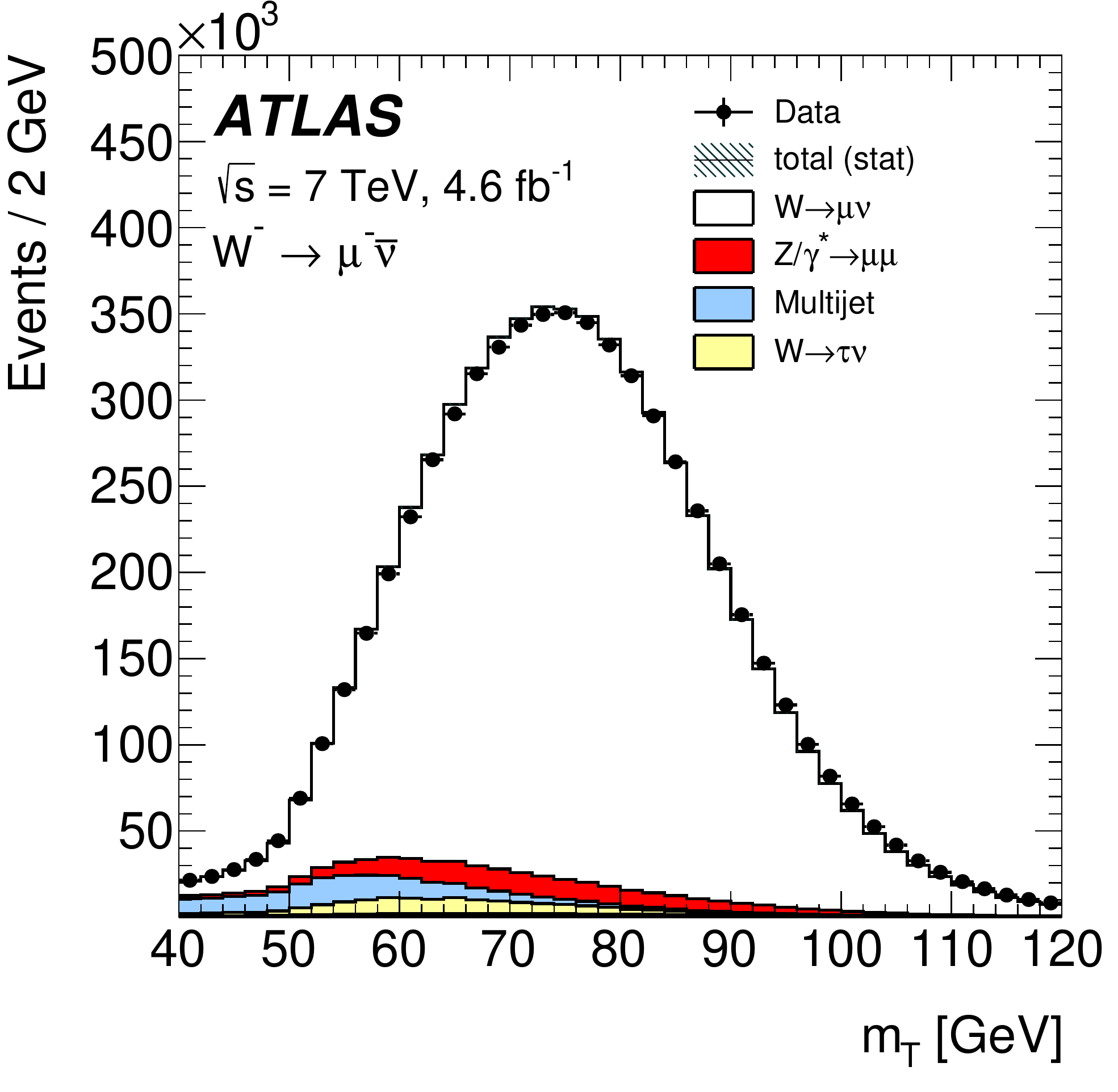}
    \caption{The transverse mass distribution for \Wmunup\ candidates
      (left) and \Wmunum\ candidates (right).  \luminorm\ \mjplot\
      \totstat\ \visible}
    \label{wmunu:fig:candWmt}
  \end{center}
\end{figure}

\begin{figure}[htbp]
  \begin{center}
    \includegraphics[width=0.48\textwidth]{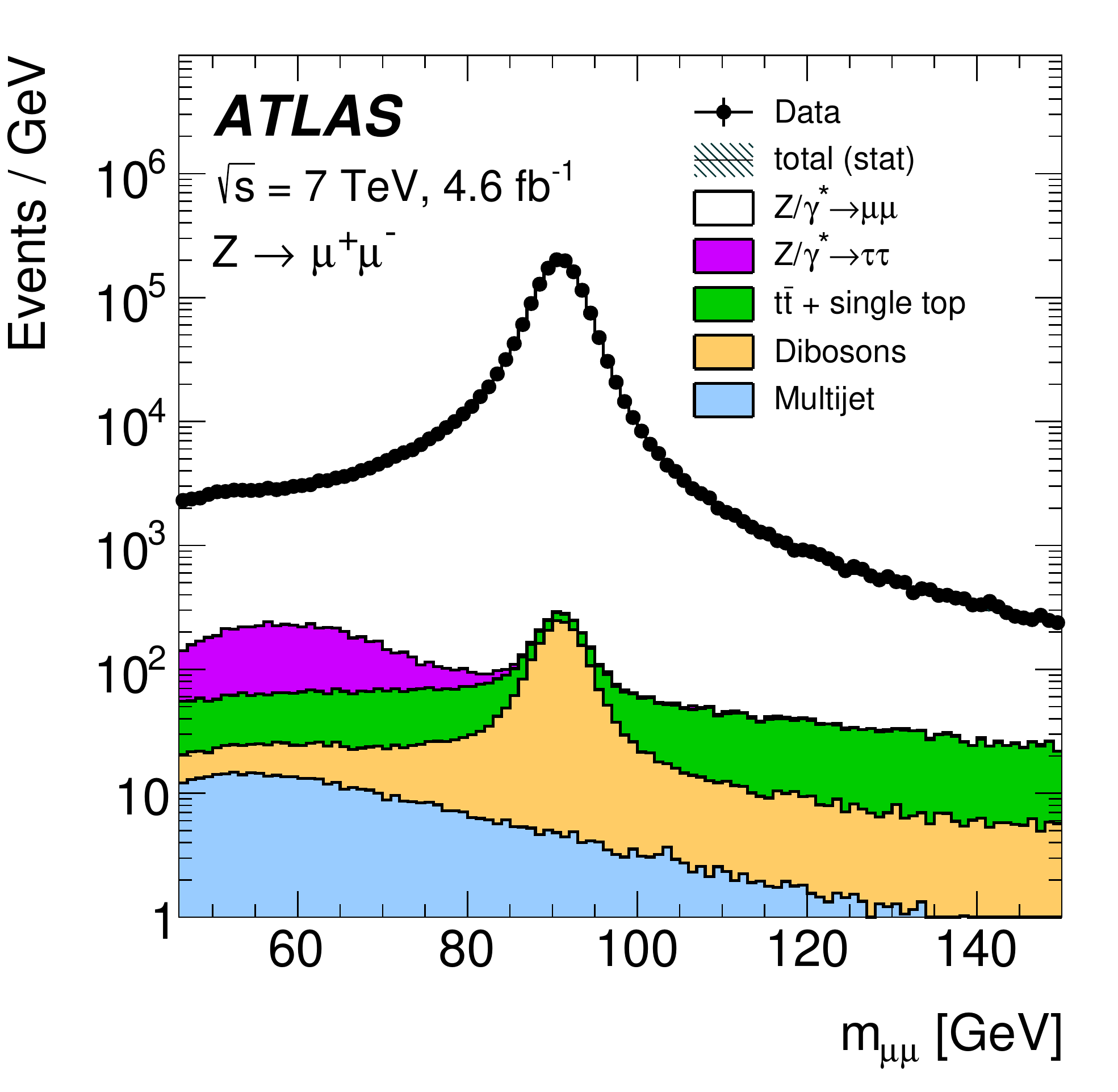}
    \caption{The dilepton invariant mass distributions for \Zgmumu\
      candidates. \luminorm\ \mjplot\ \totstat\ \visible}
    \label{zmumu:fig:candZmmass}
  \end{center}
\end{figure}

\begin{figure}[htbp]
  \begin{center}
    \includegraphics[width=0.32\textwidth]{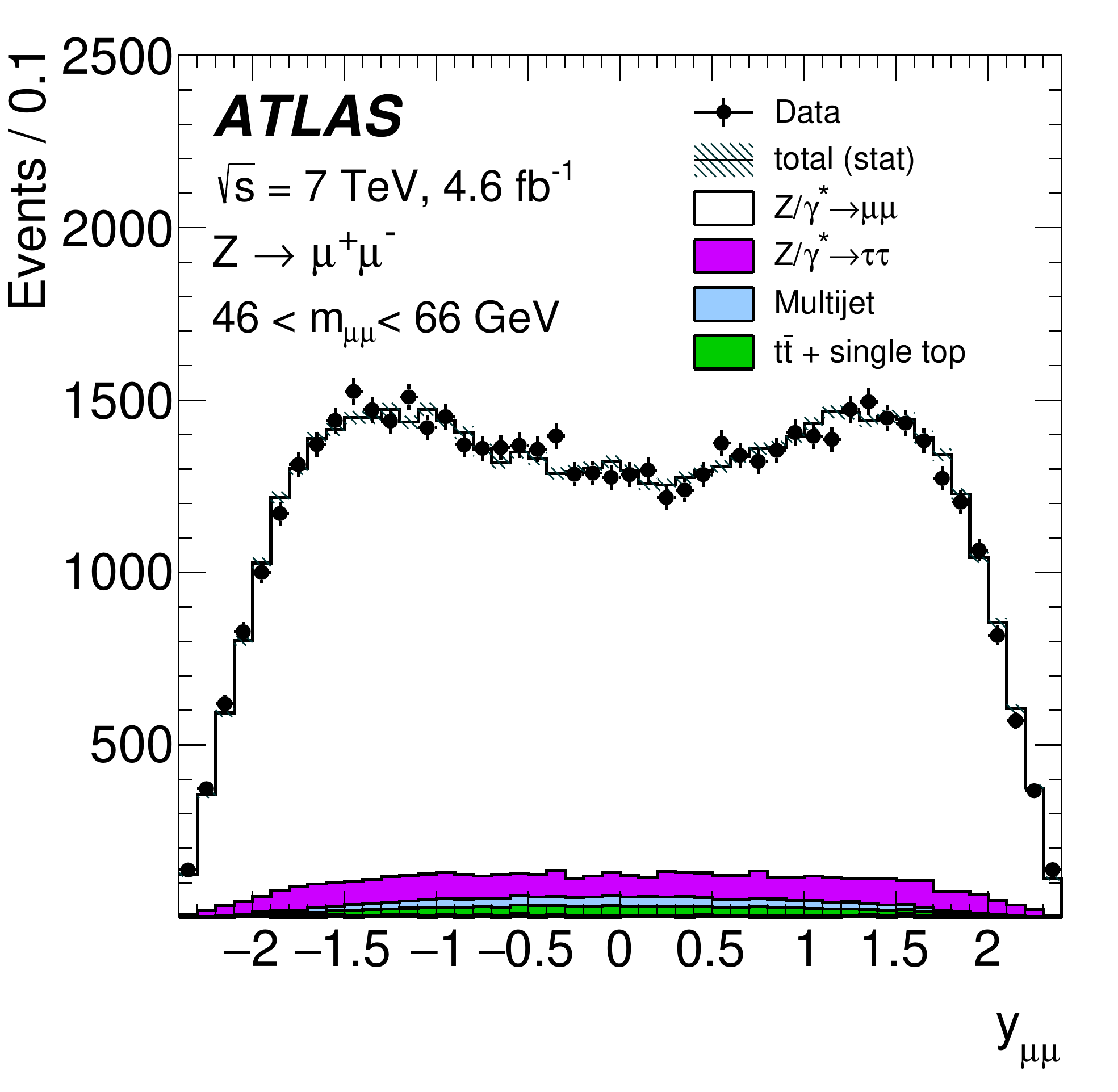}
    \includegraphics[width=0.32\textwidth]{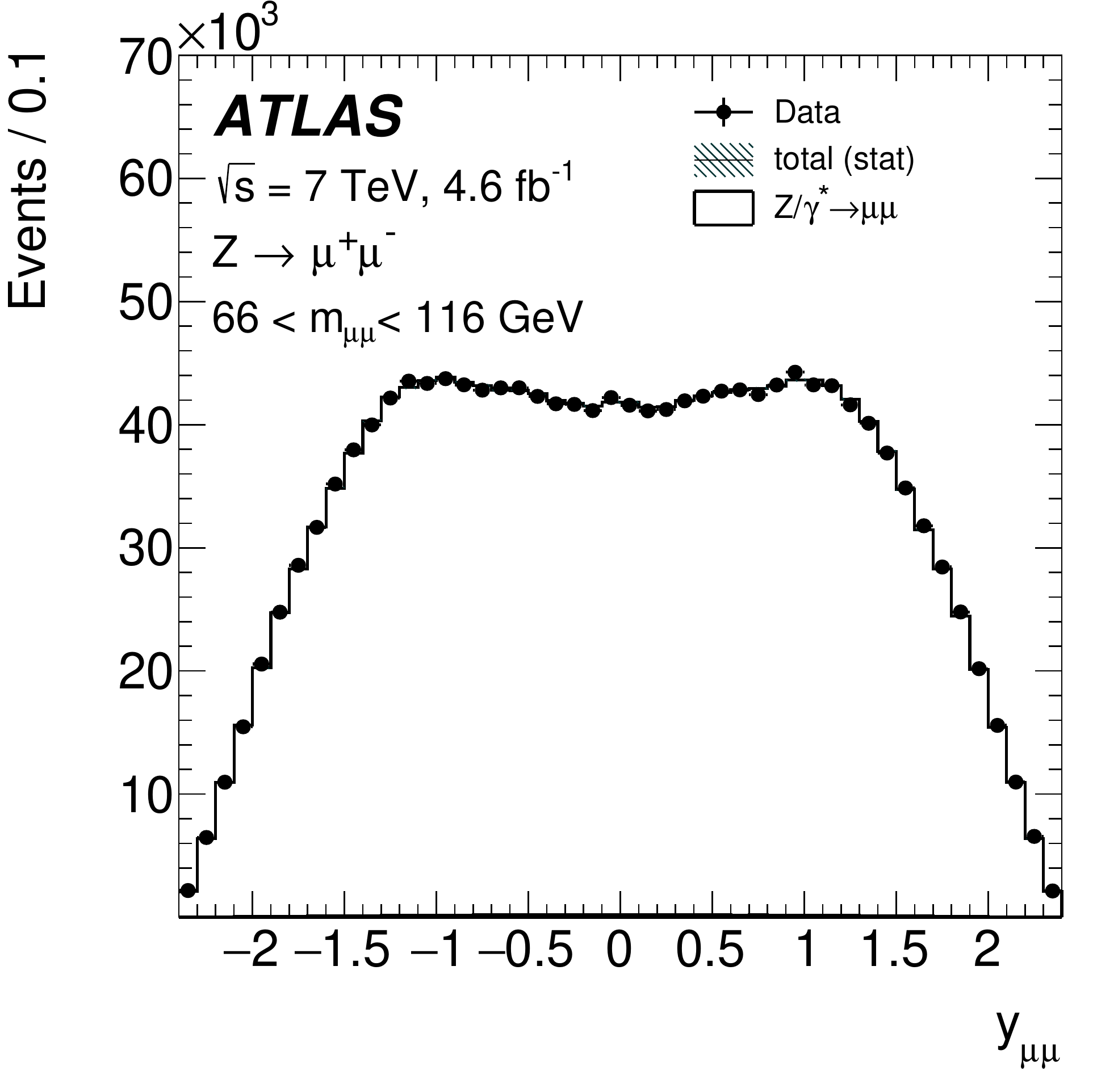}
    \includegraphics[width=0.32\textwidth]{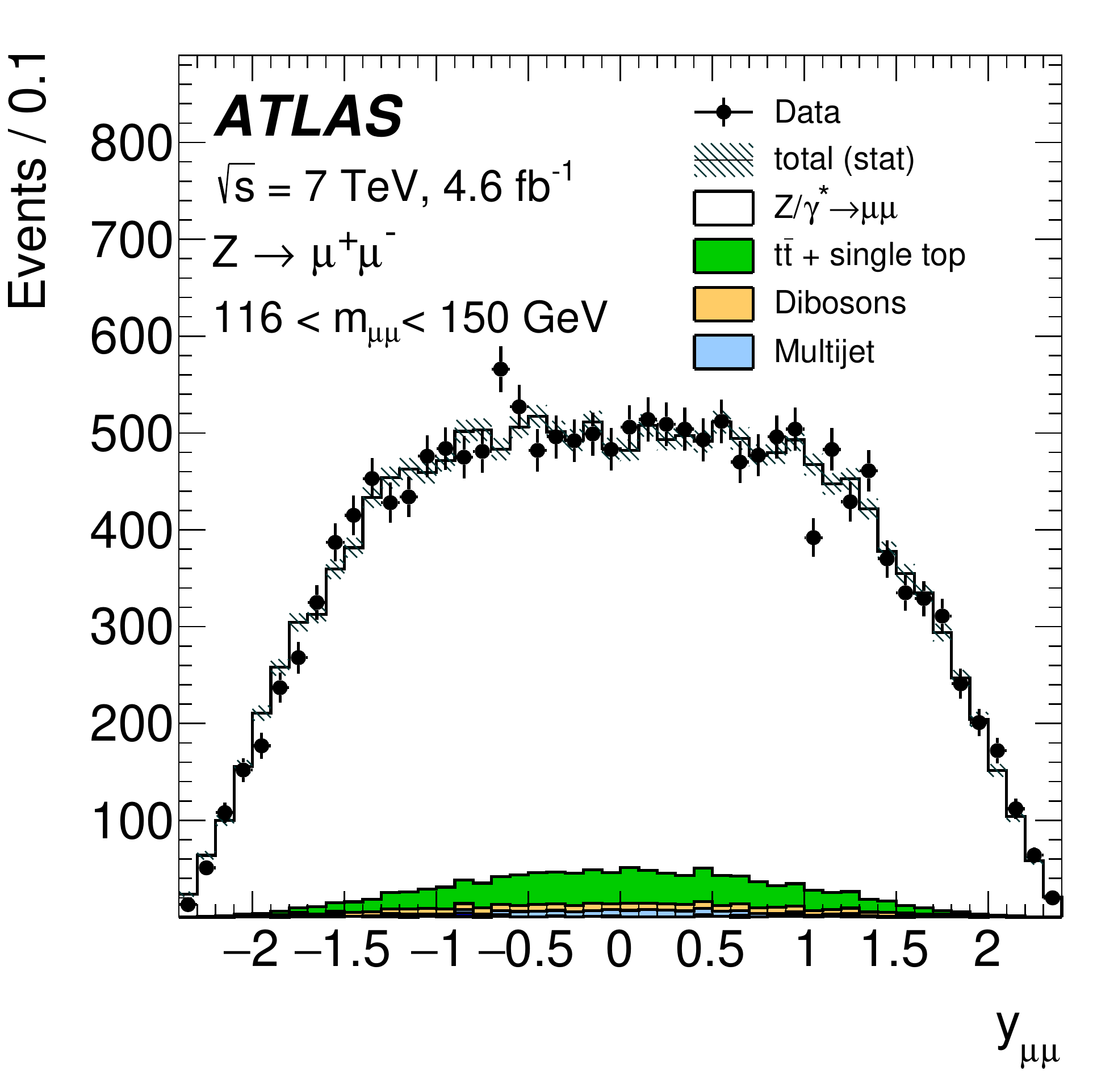}
    \caption{The dilepton rapidity distributions for \Zgmumu\
      candidates in the mass regions $46 < \mmumu < 66\GeV$ (left),
      $66 < \mmumu < 116\GeV$ (middle) and $116 < \mmumu < 150\GeV$
      (right). \luminorm\ \mjplot\ \totstat\ \visible}
      \label{zmumu:fig:candZmPt}
  \end{center}
\end{figure}

\TTab~\ref{tab:WZaccMuon} reports the number of candidates, the
estimated background events and the \CWZ\ correction factors used for
the three different integrated muon channel measurements of the $W^+$,
$W^-$, and $Z/\gamma^*$ cross sections, the latter in the $Z$-peak
region of $66<\mmumu<116\gev$.  The corresponding three integrated
cross sections in the fiducial phase space specific to the muon
channels are reported in \Tab~\ref{tab:muonxsec} with their
uncertainties due to data statistics, luminosity, and further
experimental systematic uncertainties.

\begin{table}[htbp]
  \centering    
  \begin{tabular}{lccc}
    \hline
    \hline
              &      $N$         &    $B$           & $C$           \\
    \hline
    \Wmunup     & \NWmuplusCands  & \nWmuplusBkg    & $0.656 \pm 0.003$   \\
    \Wmunum     & \NWmuminusCands & \nWmuminusBkg   & $0.649 \pm 0.003$   \\
    \Zgmumu       & \NZmuCands      & ~~\nZmuBkg        & $0.734 \pm 0.003$   \\
    \hline 
    \hline
  \end{tabular}
  \caption{Number of observed candidates $N$, of expected background
    events $B$, and the  correction factors $C$ for the $W^+$, $W^-$, 
    and \Zg\ ($66<\mmumu<116\gev$) muon channels.
    The correction factors $C$ were defined in Eq.~\eqref{Eq:CWZ}.
    The charge asymmetry in the background to the $W^{\pm}$ channels stems from the
    \Wtau\ contributions, which is proportional to the signal yield.
    The uncertainties are the quadratic sum of
    statistical and systematic components.
    The statistical uncertainties in $C$ are negligible.}
  \label{tab:WZaccMuon}
\end{table}

\begin{table}[tbhp]
\small
\begin{center}
\begin{tabular}{l c}
\hline
\hline
  & $\sigma^\mathrm{fid,\mu}_{W \to \mu\nu}$ [pb] \\[.25ex]
\hline
\Wmunup      &    $~\sigfidWmuplus$  \\
\Wmunum      &    $~\sigfidWmuminus$  \\
\hline
\hline
 & $\sigma^\mathrm{fid,\mu}_{Z/\gamma^* \to \mu\mu}$ [pb] \\[0.25ex]
\hline
\Zgmumu      &   $~\sigfidZmu$   \\
\hline
\hline
\end{tabular}
\caption{Fiducial cross sections times branching ratios for $W^+$,
  $W^-$, and $\Zg$ ($66<\mmumu<116\gev$) production in the muon decay
  channel. The fiducial regions used for the measurement are those
  defined for the combined fiducial regions in \Sec~\ref{sec:fidureg},
  except that the muon pseudorapidity is restricted to be within
  $|\eta|<2.4$.  The uncertainties denote the statistical (stat), the
   systematic (syst), and the luminosity (lumi)
  uncertainties.}
\label{tab:muonxsec}
\end{center}
\end{table}

The breakdown of the systematic uncertainty in all channels is shown
in \Tab~\ref{muonsyst}. Apart from the luminosity contribution of
\dlumi\,\%, the \Wmn\ cross sections are measured with an experimental
uncertainty of $0.6\%$ and the \Zmm\ cross section is measured with an
experimental uncertainty of $0.4\%$. 

The uncertainties of the data-driven determinations of muon and
hadronic recoil responses, discussed in \Sec~\ref{sec:mueffcalib},
are propagated to the measurements. This comprises the uncertainties in the muon detection
efficiencies, separated into contributions from the trigger,
reconstruction, and isolation, which are relatively small for the
\Wmunu\ channels and about $0.2\%$ in total, but constitute the dominant
systematic uncertainties
in the \Zmumu\ case with $0.34\%$. The muon \pt\ resolution and
scale uncertainties are very small for $Z$ and subdominant for the
\Wmunu\ channels at about $0.2\%$. The \Wmunu\ analyses are furthermore
affected by uncertainties in the hadronic recoil response, decomposed
into soft \met\ and jet energy scale and resolution uncertainties,
which add up to a total uncertainty contribution of about $0.2\%$.

Signal modelling variations with different event generators as
discussed in \Sec~\ref{sec:anaproc} contribute uncertainties of about
$0.1\%$ to both the \Wmunu\ and \Zmumu\ analyses. The high precision
is achieved after a dedicated re-evaluation of
the data-to-simulation correction factor for the muon isolation using
alternative signal samples, which  is
especially relevant for the \Zmumu\ peak data analysis, where the
overlap of the samples used for efficiency calibration and
cross-section analysis is very large. For the \Wmn\ analysis, smaller
effects from the multijet background determination and the hadronic
recoil response remain. Other theoretical modelling uncertainties from
PDFs and boson \pt\ sources are also at the level of $0.1$--$0.2\%$.

The determination of 
uncertainties in the background subtraction follows the discussion in
\Secs~\ref{sec:MuBkg}. The contribution of  electroweak and
top-quark backgrounds is about $0.2\%$ for the \Wmn\ analyses
and much smaller for the $Z$ analysis. With 
a contribution of about $0.3\%$ 
the multijet background dominates the systematic uncertainty for the 
 \Wmunup\ and  \Wmunum\ channels.

\begin{table}[tbh]
  \small
  \begin{center}  
    \begin{tabular}{lccc}
      \hline 
      \hline 
      & $\delta \sigma_{W+}$    &  $\delta \sigma_{W-}$   & $\delta \sigma_Z$  \\
      & $[\%]$ & $[\%]$ & $[\%]$    \\
      \hline 
      Trigger efficiency                       & 0.08 & 0.07 & 0.05 \\ 
      Reconstruction efficiency                & 0.19 & 0.17 & 0.30 \\ 
      Isolation efficiency                     & 0.10 & 0.09 & 0.15 \\ 
      Muon \pt resolution                     & 0.01 & 0.01 & $<$0.01 \\ 
      Muon \pt scale                          & 0.18 & 0.17 & 0.03 \\
      \met\ soft term scale                  & 0.19 & 0.19 & $-$ \\
      \met\ soft term resolution             & 0.10 & 0.09 & $-$ \\
      Jet energy scale                         & 0.09 & 0.12 & $-$ \\
      Jet energy resolution                    & 0.11 & 0.16 & $-$ \\
      Signal modelling (matrix-element generator)      & 0.12 & 0.06 & 0.04 \\
      Signal modelling (parton shower and hadronization)          & 0.14 & 0.17 & 0.22 \\
      PDF                                      & 0.09 & 0.12 & 0.07 \\
      Boson \pt                                & 0.18 & 0.14 & 0.04  \\
      Multijet background                      & 0.33 & 0.27 & 0.07 \\
      Electroweak+top background        & 0.19 & 0.24 & 0.02 \\
      Background statistical uncertainty      & 0.03 & 0.04 & 0.01 \\
      Unfolding statistical uncertainty       & 0.03 & 0.03 & 0.02 \\
      \hline                                                                                    
      Data statistical uncertainty            & 0.04 & 0.04 & 0.08 \\
      \hline                                                                                        
      Total experimental uncertainty           & 0.61 & 0.59 & 0.43 \\
      \hline 
      Luminosity                       &     \multicolumn{3}{c}{$\dlumi$}   \\
      \hline 
      \hline 
    \end{tabular}
    \caption{Relative uncertainties $\delta\sigma$ in the measured
      integrated fiducial cross sections times branching ratios in the muon
      channels.  The efficiency  uncertainties
      are partially correlated between the trigger, reconstruction and
      isolation terms. This is taken into account in the computation
      of the total uncertainty quoted in the table.}
    \label{muonsyst}       
  \end{center}
\end{table}

The differential cross-section measurements, as a function of the
$W^+$ and $W^-$ muon pseudorapidity and of
the dimuon rapidity and mass for the \Zg\ channel, are summarized in 
Appendix~\ref{sec:simple_xsectable} in the
\Tabs~\ref{tab:w_eta_mu}--\ref{tab:zlowhigh_yll_mu}. The statistical
uncertainties in the \Wmn\ differential cross sections are about
$0.1$--$0.2\%$, and the total uncertainties are 
$0.6$--$0.9\%$, excluding the luminosity uncertainty.

The differential \Zgmumu\ cross sections are measured in the
$\mmumu=66$--$116\GeV$ invariant mass region with a statistical
uncertainty of about $0.3\%$ up to $|\yll|<2.0$ and of $0.8\%$ for larger
$|\yll| < 2.4$. The total uncertainty, excluding the luminosity
uncertainty, is $0.5\%$ up to $|\yll|<2.0$ and $1.0\%$ for
$|\yll|=2.4$. The statistical uncertainties of the differential
\Zgmumu\ cross sections measured in the $\mmumu=46$--$66\gev$ and
$116$--$150\GeV$ invariant mass regions are  $1.3$--$4\%$,
and the total uncertainties amount to $2$--$5\%$.

\clearpage

\subsection{Test of electron--muon universality}
\label{emuuni}
Ratios of the measured $W$ and $Z$ production cross sections in the
electron and muon decay channels are evaluated from the corresponding
measurements minimally extrapolated to the common fiducial phase space
according to Eq.~\eqref{eq:WZxsecfid}. These $e/\mu$ cross-section
ratios represent direct measurements of the corresponding relative
branching fractions, which are predicted to be unity in the SM given
that lepton mass effects are negligible. Considering the case of the
$W$ boson, the ratio $R_W$ is obtained from the sum of $W^+$ and $W^-$
cross sections as:
\begin{eqnarray*}
  R_{W} &=& \frac{\sigma^\mathrm{fid,e}_{W \to e\nu} /E_W^\mathrm{e}}{\sigma^\mathrm{fid,\mu}_{W \to \mu\nu}/E_W^\mathrm{\mu}} = \frac{\sigma^\mathrm{fid}_{W \to e\nu}}{\sigma^\mathrm{fid}_{W \to \mu\nu}} =\frac{BR(W \rightarrow e\nu)}{BR(W \rightarrow \mu \nu)} \\
  &=& 0.9967 \pm 0.0004\,\mathrm{(stat)} \pm 0.0101\,\mathrm{(syst)} \\
  &=& 0.997 \pm 0.010 \,.
\end{eqnarray*}
This measurement is more precise than the combination of LEP
results from $e^+e^-\to W^+W^-$ data of $1.007 \pm
0.019$~\cite{Schael:2013ita}. It also significantly improves on the
previous ATLAS measurements of $1.006 \pm 0.024$ with the
2010 data~\cite{Aad:2011dm} and of $1.036 \pm 0.029$ with the 2015
data~\cite{Aad:2016naf}. Related measurements were published by the
CDF Collaboration with $R_W = 1.018 \pm 0.025$~\cite{Abulencia:2005ix}
and recently by the LHCb Collaboration with $R_W = 1.020 \pm
0.019$~\cite{Aaij:2016qqz}.

Similarly, the $e/\mu$ ratio of the $Z$-boson cross sections is extracted:
\begin{eqnarray*}
R_{Z} &=& \frac{\sigma^\mathrm{fid, e}_{Z \to ee}/E_Z^\mathrm{e}}{\sigma^\mathrm{fid, \mu}_{Z \to \mu\mu}/E_Z^\mathrm{\mu}} = \frac{\sigma^\mathrm{fid}_{Z \to ee}}{\sigma^\mathrm{fid}_{Z \to \mu\mu}} = \frac{BR(Z \rightarrow ee)}{BR(Z \rightarrow \mu \mu)} \\
 &=& 1.0026 \pm 0.0013\,\mathrm{(stat)} \pm 0.0048\,\mathrm{(syst)} \\
 &=& 1.0026 \pm 0.0050\,.
\end{eqnarray*}
The result agrees well with the value obtained from the
combination of $e^+e^-\to Z$ LEP and SLC data of $0.9991 \pm
0.0028$~\cite{ALEPH:2005ab}. It is significantly more precise than
the previous
ATLAS measurements:  $1.018 \pm 0.031$ with the 2010
data~\cite{Aad:2011dm} and  $1.005 \pm 0.017$ with the 2015
data~\cite{Aad:2016naf}.

The $R_W$ and $R_Z$ measurements therefore confirm lepton ($e$--$\mu$)
universality in the weak vector-boson decays. The result, taking into
account the correlations between the $W$ and $Z$ measurements, is illustrated in
\Fig~\ref{Fig:lUniFid} as an ellipse. For comparison, bands are shown
representing the above cited combined measurements from $e^+e^-$
colliders.

For the leptonic $W$ branching fraction, $BR(\Wln)$, precise
constraints are also derived from off-shell $W$ bosons in
$\tau$-lepton, $K$-meson, and $\pi$-meson decays. For $\tau$ decays
the HFAG group~\cite{Amhis:2014hma} obtains $R_W= (g_e/g_\mu)^2 =
0.9964 \pm 0.0028$, where $g_e$ and $g_\mu$ are the couplings of the
$W$ boson to $e$ and $\mu$, respectively. The K\TeV\ measurement of
$K\to\pi^\pm\ell^\mp\nu$ decays results in $R_W = 1.0031 \pm
0.0048$~\cite{Alexopoulos:2004sy}. The measurement of
$K^\pm\to\ell^\pm\nu$ decays
by NA62 corresponds to an equivalent of $R_W = 1.0044 \pm
0.0040$~\cite{Lazzeroni:2012cx}. Finally, measurements of
$\pi^\pm\to\ell^\pm\nu$ decays may be translated to a value of $R_W =
0.9992 \pm 0.0024$~\cite{Aguilar-Arevalo:2015cdf}.

\begin{figure}
  \begin{center}
    \includegraphics[width=0.55\textwidth]{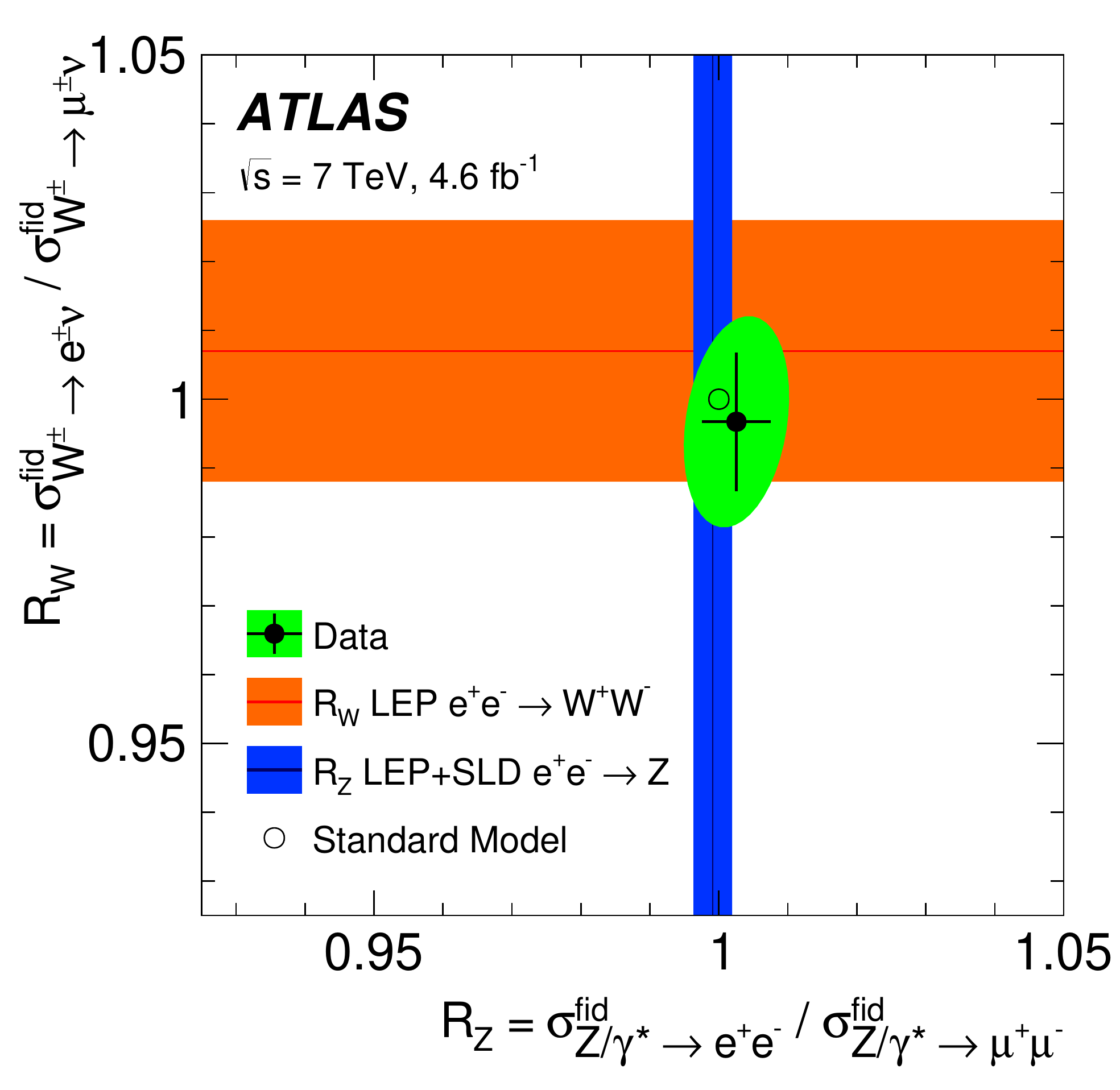}
  \end{center}
  \caption{Measurement of the electron-to-muon cross-section ratios
    for the $W$ and $Z$ production, $R_W$ and $R_Z$. The orange and blue, 
    shaded bands
    represent the combination of the ratios of electron and muon
    branching fractions for on-shell $W$ and $Z$ production as
    obtained at the $e^+e^-$ colliders LEP and SLC~\cite{Schael:2013ita, ALEPH:2005ab}.
    The green shaded ellipse represents the $68\%$
    CL for the correlated measurement of $R_{W}$ and $R_{Z}$, while
    the black error bars give the one-dimensional standard
    deviation. The SM expectation of $R_W=R_Z=1$ is indicated with an
    open circle.}
\label{Fig:lUniFid}
\end{figure}

\subsection{Combination of cross sections}
\label{sec:combi}
\subsubsection{Combination procedure}
\label{sec:CombiProc}

The \Wpmlnu\ and \Zgll\ cross-section measurements are performed in
both the electron and muon decay channels. Assuming lepton universality,
this provides a cross-check of
experimental consistency and, as described later in this section, a
means to improve the measurements when accounting for correlated and
uncorrelated experimental uncertainties in the combination of the $e$
and $\mu$ channel measurements. Correlations arise from the use of 
electrons, muons, or \MET\ reconstructed in the same way
for different channels, but also
due to similar or identical analysis techniques, e.g. in the
background estimation. 
The method used to combine the cross-section
data was also applied  in the previous
inclusive $W,~Z$ cross-section measurement~\cite{Aad:2011dm}. It was
introduced for the combination of HERA cross-section measurements in
Refs.~\cite{Glazov:2005rn,Aaron:2009bp}.

The combination procedure minimizes the deviation of the combined
measurement $\sigma_{\mathrm{comb}}^i$ in a kinematic interval $i$
from the input measurements $\sigma^i_k$, where $k=1,2$ denotes the
electron and muon measurements. This is achieved by allowing the
contributions $b_j$ of the correlated uncertainty sources $j$ to
shift, where $b_j$ is expressed in units of standard deviations. The
procedure requires as input a list of $\gamma_{j,k}^i$ values that
specify the influence of the correlated uncertainty source $j$ on the
measurement $i$ in the data set $k$.  The relative data statistical and
uncorrelated systematic uncertainties are given by
$\delta^i_{\mathrm{sta},k}$ and $\delta^i_{\mathrm{unc},k}$,
respectively.  The resulting $\chi^2$ function

\begin{equation}
  \chi^2(\vec{\sigma}_{\mathrm{comb}},\vec{b}) = \sum_{k,i} \frac{\left[\sigma^i_k - \sigma_{\mathrm{comb}}^i (1 - \sum_j \gamma^i_{j,k}b_j) \right]^2}{(\Delta^i_k)^2} + \sum_{j}b_j^2
  \label{Eq:CorrChiSq}
\end{equation}
with
\begin{equation}
  (\Delta^i_k)^2=\left(\delta^i_{\mathrm{sta},k}\right)^2 \sigma^i_k \sigma_{\mathrm{comb}}^i  +\left(\delta^i_{\mathrm{unc},k} \sigma_{\mathrm{comb}}^{i} \right)^{2}
  \label{Eq:DeltaiComb}
\end{equation}

includes a penalty term for the systematic shifts $b_j$. The
definition of $\Delta^i_k$ ensures the minimization of biases due to
statistical fluctuations, affecting the estimate of the statistical
uncertainty, and treats systematic uncertainties in a multiplicative
way~\cite{Aaron:2009bp}. Given the size of the statistical and
systematic uncertainties for the data considered here, the differences
between $\Delta^i_k$ as used here and the simpler form without scaling
are very small.

The uncertainties due to electron and muon momentum scales and
resolutions are treated as fully correlated between the \Wpmlnu\ and
\Zgll\ channels of a specific decay channel. Uncertainties in the
hadronic recoil response, separated into jet and soft \MET\ scales and
resolutions, only affect the $W^\pm$ channels and are treated in a
correlated way between the $W^+$ and $W^-$ measurements and the $e$ and $\mu$ channels.

The accurate determination of lepton selection efficiencies for online
selection, reconstruction, identification, and isolation is an
important input to the analysis. The efficiencies are measured in data
and applied as correction factors to the simulation. These correction
factors have statistical and procedural uncertainties, which are
propagated to the measurements using pseudo-experiments for all
channels in a consistent way. A covariance matrix is constructed from
typically $1000$ pseudo-experiments and then decomposed into
a list of fully correlated uncertainty sources $\gamma$ and bin-to-bin
uncorrelated uncertainties in the measurements.

The following theoretical uncertainties are largely correlated between
all channels: i) uncertainties in the measurements due to signal
modelling, such as the boson transverse momentum spectrum; ii)
theoretical uncertainties in signal modelling and hadronic recoil
simulation, estimated with alternative signal samples, and iii)
extrapolations applied to the measurements to account for the small
differences in experimental fiducial phase spaces.

The uncertainties due to background estimation from simulated MC
samples are treated as fully correlated between all channels, but 
separately for each background source. Data-driven background
estimates are uncorrelated between channels and often
contain significant statistical components, especially in the
low-background \Zgll\ analyses. There is, however, a significant
correlated part between $W^+$ and $W^-$ of a given lepton decay
channel as the employed procedures are the same. 

\subsubsection{Integrated cross sections}
\label{sec:IntCrossSec}

The combination of fiducial integrated \Zgll, \Wpluslnu, and
\Wminuslnu\ cross sections, including the full information contained
in 66 correlated sources of uncertainty, gives a $\chi^2$ per number
of degrees of freedom (\chindf) of $0.5/3$, indicating  that
 the measurements are compatible. \TTab~\ref{tab:fidint} summarizes
the separate electron and muon channel measurements in the common
fiducial volume and gives the final integrated fiducial cross-section
results. Apart from the luminosity uncertainty of \dlumi\%, a fiducial cross-section
measurement precision of $0.32\%$ is reached for the NC channel
and of $0.5\%~(0.6)\%$ for the $W^+$ ($W^-$) channels. The new $Z~(W)$ fiducial cross-section measurements are $10~(3.5)$ times more precise
than the previous ATLAS measurements~\cite{Aad:2011dm} when
considering the statistical and systematic uncertainties added in
quadrature.

\begin{table}[ptbh]
  \small
  \begin{center}
    \begin{tabular}{l c}
      \hline
      \hline
      & $\sigma^\mathrm{fid}_{W \to \ell\nu}$ [pb] \\[.25ex]
      \hline
      \Wenup    & $2939 \pm 1 \,\mathrm{(stat)} \pm 28 \,\mathrm{(syst)} \pm 53 \,\mathrm{(lumi)}$  \\
      \Wmunup   & $2948 \pm 1 \,\mathrm{(stat)} \pm 21 \,\mathrm{(syst)} \pm 53 \,\mathrm{(lumi)}$  \\
      \Wpluslnu & $2947 \pm 1 \,\mathrm{(stat)} \pm 15 \,\mathrm{(syst)} \pm 53 \,\mathrm{(lumi)}$     \\
      \hline
      \Wenum    & $1957 \pm 1 \,\mathrm{(stat)} \pm 21 \,\mathrm{(syst)} \pm 35 \,\mathrm{(lumi)}$  \\
      \Wmunum   & $1964 \pm 1 \,\mathrm{(stat)} \pm 13 \,\mathrm{(syst)} \pm 35 \,\mathrm{(lumi)}$   \\
      \Wminuslnu& $1964 \pm 1 \,\mathrm{(stat)} \pm 11 \,\mathrm{(syst)} \pm 35 \,\mathrm{(lumi)}$     \\
      \hline
      \Wenu     & $4896 \pm 2\,\mathrm{(stat)} \pm 49 \,\mathrm{(syst)} \pm 88 \,\mathrm{(lumi)}$     \\
      \Wmunu    & $4912 \pm 1\,\mathrm{(stat)} \pm 32 \,\mathrm{(syst)} \pm 88 \,\mathrm{(lumi)}$   \\
      \Wlnu     & $4911 \pm 1\,\mathrm{(stat)} \pm 26 \,\mathrm{(syst)} \pm 88 \,\mathrm{(lumi)}$ \\
      \hline
      \hline
      & $\sigma^\mathrm{fid}_{Z/\gamma^* \to \ell\ell}$ [pb] \\[.25ex]
      \hline
      \Zgee     & $502.7 \pm 0.5 \,\mathrm{(stat)} \pm 2.0 \,\mathrm{(syst)} \pm 9.0 \,\mathrm{(lumi)}$ \\
      \Zgmumu   & $501.4 \pm 0.4 \,\mathrm{(stat)} \pm 2.3 \,\mathrm{(syst)} \pm 9.0\,\mathrm{(lumi)}$ \\
      \Zgll     & $502.2 \pm 0.3 \,\mathrm{(stat)} \pm 1.7 \,\mathrm{(syst)} \pm 9.0 \,\mathrm{(lumi)}$ \\
      \hline
      \hline
    \end{tabular}
    \caption{Integrated fiducial cross sections times
      leptonic branching ratios in the electron and muon channels and
      their combination with statistical and systematic
      uncertainties, for $W^+$, $W^-$, their sum and the \Zg\
      process measured at $\sqrt{s}=7\TeV$. The \Zg\ cross section is defined for the
      dilepton mass window $66<\mll<116\gev$. 
      The common fiducial regions are defined in
      \Sec~\ref{sec:fidureg}. The uncertainties denote the statistical
      (stat), the experimental systematic (syst), and the luminosity
      (lumi) contributions.}
    \label{tab:fidint}
    \end{center}
\end{table}

Excluding the common luminosity uncertainty, the correlation
coefficients of the $W^+$ and $Z$, $W^-$ and $Z$, and $W^+$
and $W^-$ fiducial
cross-section measurements are $0.349,~0.314,$ and $0.890$,
respectively. Including the luminosity, all three measurements are
highly correlated, with coefficients of $0.964,~0.958$ and $0.991$,
respectively. \TTab~\ref{tab:fidratio} presents four ratios that may
be obtained from these fiducial integrated \Zg\ and $W^\pm$ cross
sections, where the luminosity uncertainty as well as other correlated
uncertainties are eliminated or strongly reduced. The precision of
these ratio measurements is very high with a total experimental
uncertainty of $0.4\%$ for the $W^{+}/W^{-}$ ratio and $0.5\%$ for the
$W^{\pm}/Z$ ratio.

\begin{table}[ptbh]
  \small
  \begin{center}
    \begin{tabular}{l l}
      \hline
      \hline
      $R^\mathrm{fid}_{W^{+}/W^{-}}$ &
      $1.5006 \pm 0.0008\,\mathrm{(stat)} \pm 0.0037\,\mathrm{(syst)}$\\
      $R^\mathrm{fid}_{W/Z}$ & 
      $9.780 \pm 0.006\,\mathrm{(stat)} \pm 0.049\,\mathrm{(syst)}$\\
      $R^\mathrm{fid}_{W^{+}/Z}$ & 
      $5.869 \pm 0.004\,\mathrm{(stat)} \pm 0.029\,\mathrm{(syst)}$\\
      $R^\mathrm{fid}_{W^{-}/Z}$ &
      $3.911 \pm 0.003\,\mathrm{(stat)} \pm 0.021\,\mathrm{(syst)}$\\
      \hline
      \hline
    \end{tabular}
    \caption{Ratios of integrated fiducial CC and NC
      cross sections obtained from the combination of electron and
      muon channels with statistical (stat) and systematic (syst)
      uncertainties. The common fiducial regions are defined in
      \Sec~\ref{sec:fidureg}.}
    \label{tab:fidratio}
    \end{center}
\end{table}

In order to obtain the total cross sections,
the combined integrated fiducial cross sections are also
extrapolated to the full phase space with the procedure discussed in
\Sec~\ref{sec:anaproc}. Results are provided in
\Tab~\ref{tab:totint}. The uncertainties in these total cross sections
receive significant contributions from PDF and signal modelling
uncertainties, which are similar in size to the luminosity
uncertainty. Ratios of these total cross sections are provided in
\Tab~\ref{tab:totratio}. While for these ratios the luminosity
uncertainty and a large part of the signal modelling uncertainties in
the extrapolation are found to cancel, a significant uncertainty
remains from PDF uncertainties.

\begin{table}[ptbh]
  \small
  \begin{center}
    \begin{tabular}{l c}
      \hline
      \hline
      & $\sigma^\mathrm{tot}_{W \to \ell\nu}$ [pb] \\[.25ex]
      \hline
      \Wpluslnu &
      $ 6350 \pm 2\,\mathrm{(stat)} \pm 30\,\mathrm{(syst)} \pm 110\,\mathrm{(lumi)} \pm 100\,\mathrm{(acc)}$\\
      \Wminuslnu&
      $ 4376 \pm 2\,\mathrm{(stat)} \pm 25\,\mathrm{(syst)} \pm 79\,\mathrm{(lumi)} \pm 90\,\mathrm{(acc)}$\\
      \Wlnu     &
      $ 10720 \pm 3\,\mathrm{(stat)} \pm 60\,\mathrm{(syst)} \pm 190\,\mathrm{(lumi)} \pm 130\,\mathrm{(acc)}$\\
      \hline
      \hline
      & $\sigma^\mathrm{tot}_{Z/\gamma^* \to \ell\ell}$ [pb] \\[.25ex]
      \hline
      \Zgll &
      $990 \pm 1\,\mathrm{(stat)} \pm 3\,\mathrm{(syst)} \pm 18\,\mathrm{(lumi)} \pm 15\,\mathrm{(acc)}$\\
      \hline
      \hline
    \end{tabular}
    \caption{Total  cross sections times
      leptonic branching ratios obtained from the combination of
      electron and muon channels with statistical and systematic
      uncertainties,  for $W^+$, $W^-$, their sum and the \Zg\
      process measured at $\sqrt{s}=7\TeV$. The \Zg\ cross section is defined for the
      dilepton mass window $66<\mll<116\gev$. The uncertainties
      denote the statistical (stat), the experimental systematic
      (syst), the luminosity (lumi), and acceptance extrapolation
      (acc) contributions.}
    \label{tab:totint}
    \end{center}
\end{table}

\begin{table}[ptbh]
  \small
  \begin{center}
    \begin{tabular}{l l}
      \hline
      \hline
      $R^\mathrm{tot}_{W^{+}/W^{-}}$ &
      $1.450 \pm 0.001\,\mathrm{(stat)} \pm 0.004\,\mathrm{(syst)} \pm 0.029\,\mathrm{(acc)}$\\
      $R^\mathrm{tot}_{W/Z}$ & 
      $10.83 \pm 0.01\,\mathrm{(stat)} \pm  0.05\,\mathrm{(syst)}  \pm 0.09\,\mathrm{(acc)}$\\
      $R^\mathrm{tot}_{W^{+}/Z}$ & 
      $6.407 \pm 0.004\,\mathrm{(stat)} \pm 0.032\,\mathrm{(syst)} \pm 0.062\,\mathrm{(acc)}$\\
      $R^\mathrm{tot}_{W^{-}/Z}$ &
      $4.419  \pm 0.003\,\mathrm{(stat)} \pm 0.024\,\mathrm{(syst)} \pm 0.082\,\mathrm{(acc)}$\\
      \hline
      \hline
    \end{tabular}
    \caption{Ratios of total  CC and NC cross
      sections obtained from the combination of electron and muon
      channels with statistical and systematic uncertainties.    The \Zg\
      cross section is defined for the dilepton mass window
      $66<\mll<116\gev$. The uncertainties denote the statistical
      (stat), the experimental systematic (syst), the luminosity
      (lumi), and acceptance extrapolation (acc) contributions.}
    \label{tab:totratio}
    \end{center}
\end{table}

\subsubsection{Differential cross sections}

For the combination of the rapidity-dependent differential cross
sections, a simultaneous averaging of $105$ data points, characterized
by more than one hundred correlated sources from all channels, is
performed leading to $61$ combined measurement points. As the phase
space regions of the central and forward \Zgll\ analyses are disjoint, and
there is no \Zmm\ analysis in the forward region, the combination in
this region is based solely on the \Zee\ analysis. The forward \Zee\
analysis is nevertheless included in the $e$--$\mu$ averaging to account
for possible shifts and reductions of correlated uncertainties in a
consistent way. Similarly, $W^\pm$ measurements in the bin $|\etal|
\in [1.37, 1.52]$ are covered only by the muon channel.

The combination of the differential cross sections measured in the
electron and muon channels is illustrated in \Figs~\ref{Fig:WCombi}
and \ref{Fig:ZCombi} for the \Wpmlnu\ and \Zgll\ channels.  The top
panels show the measured muon and electron cross sections together
with their combination. The central panel illustrates the $e/\mu$
ratio. The lowest panel shows the \textit{pulls}, which are the
deviations of the input measurements from the combination in terms of their uncorrelated uncertainties when fixing the systematic shifts
$b_j$ at the values leading to the total $\chi^2$ minimum.

The  measurements in  the electron and muon decay channels
are compatible. This can be
quantified with the total combination \chindf\ of $47.2/44$
and  be inferred from the pulls displayed with 
\Figs~\ref{Fig:WCombi} and \ref{Fig:ZCombi}. The
partial $\chi^2$ values are listed in
\Tab~\ref{tab:chitab} 
as well as the contribution of the penalty term
constraining the shifts of correlated uncertainties .

\begin{table}
  \begin{center}
    \begin{tabular}{lc}
      \hline
      \hline
      Channel & \chindf\\\hline

      \Wpluslnu & $6.7/10$  \\
      \Wminuslnu & $4.5/10$  \\
      $\Zgll\ (46<\mll<66\gev)$ & $3.3/6$ \\
      $\Zgll\ (66<\mll<116\gev)$ & $15.2/12$ \\
      $\Zgll\ (116<\mll<150\gev)$ & $1.8/6$ \\
      Correlated & 15.7 \\\hline
      Total &  $47.2/44$\\
      \hline
      \hline
    \end{tabular}
  \end{center}
  \caption{Partial and total \chindf\ for the combination
    of the differential $\rd\sigma/\rd|\etal|$ and $\rd\sigma/\rd|\yll|$ cross sections.
    The contribution of the penalty term
    constraining the shifts of correlated uncertainties is listed separately
    in the row labelled ``Correlated'', see Eq.~\eqref{Eq:CorrChiSq}.}
  \label{tab:chitab}
\end{table}

\begin{figure}
  \begin{center}
    \includegraphics[width=0.48\textwidth]{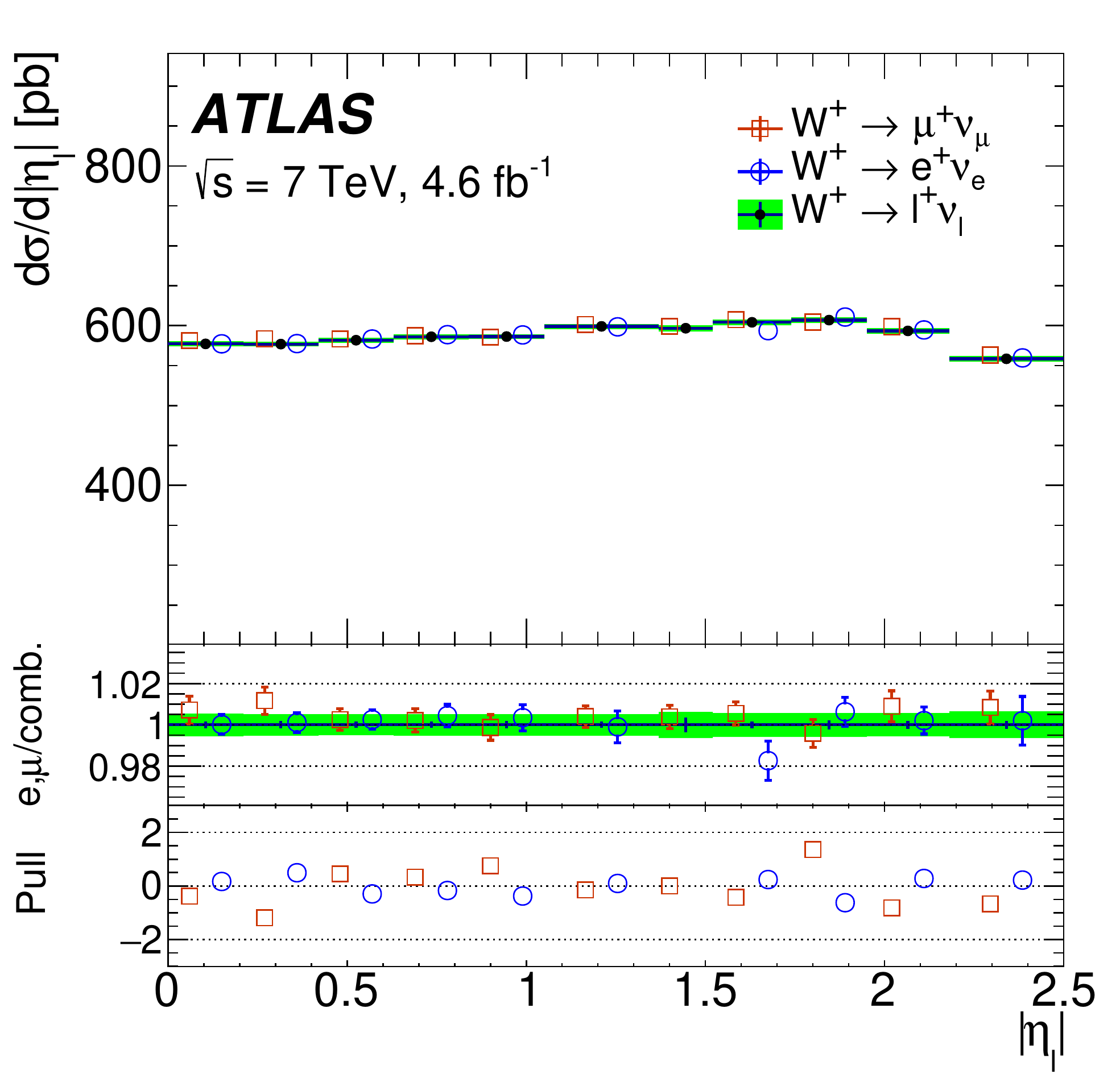}
   \includegraphics[width=0.48\textwidth]{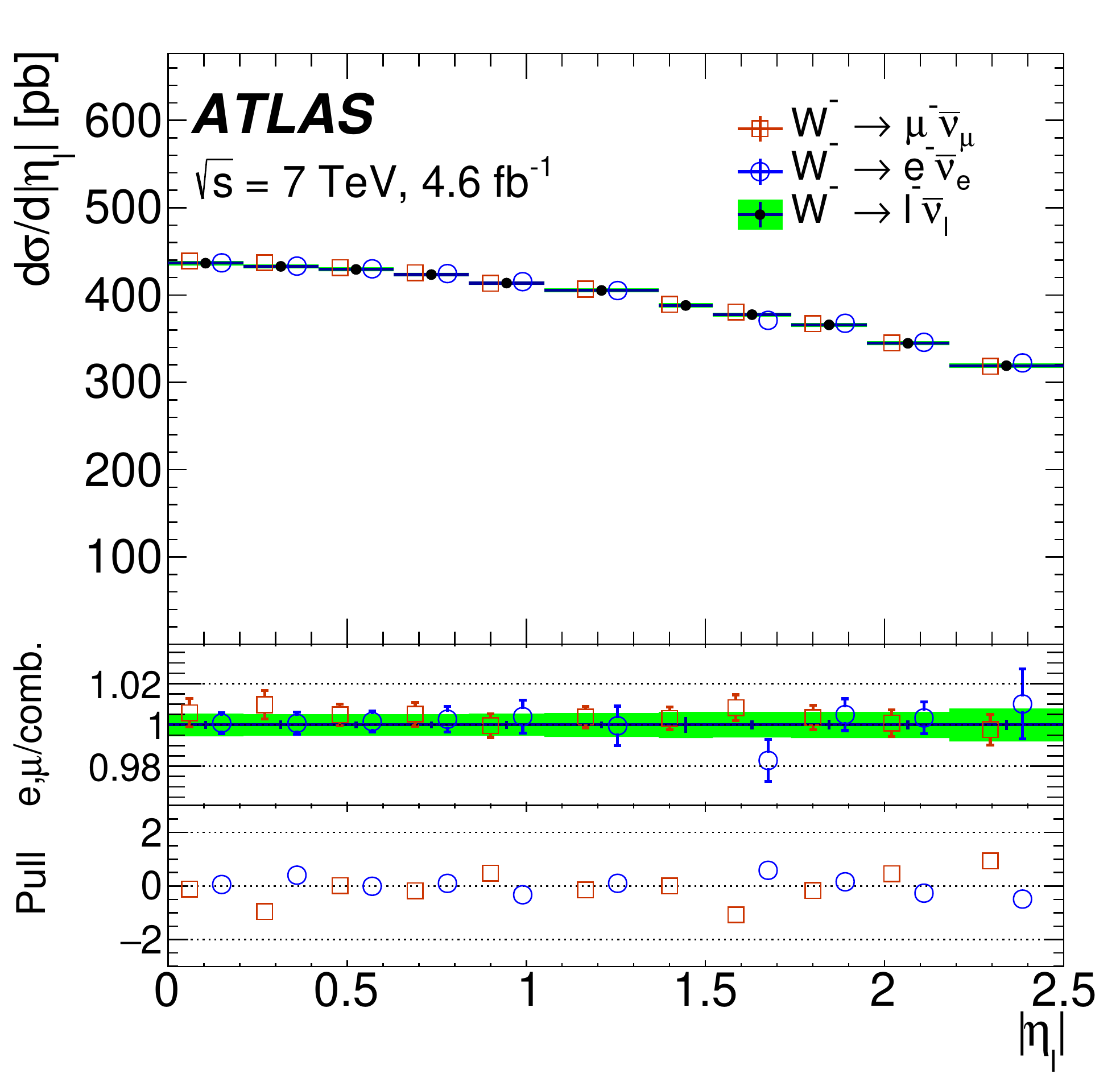}
  \end{center}
  \caption{Differential $\rd\sigma/\rd|\etal|$ cross-section measurements
    for $W^+$ (left) and $W^-$ (right), for the electron channel (open
    circles), the muon channel (open squares) and their combination
    with uncorrelated uncertainties (crosses) and the total
    uncertainty, apart from the luminosity error (green band).  Also
    shown are the ratios of the $e$ and $\mu$ measurements to the
    combination and the pulls of the individual measurements in terms
    of their uncorrelated uncertainties, see text.}
  \label{Fig:WCombi}
\end{figure}

\begin{figure}
  \begin{center}
    \includegraphics[width=0.325\textwidth]{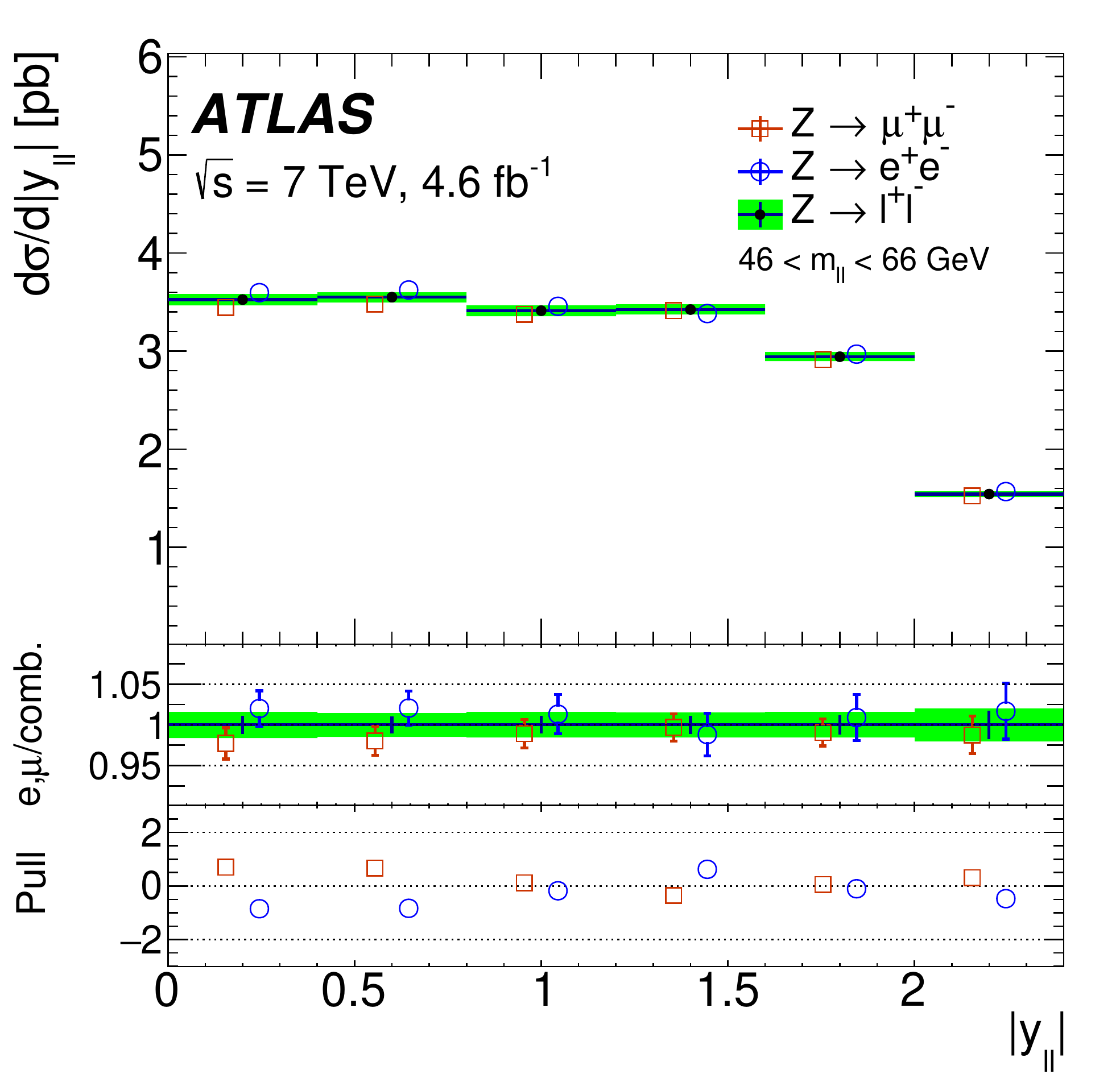}%
    \includegraphics[width=0.325\textwidth]{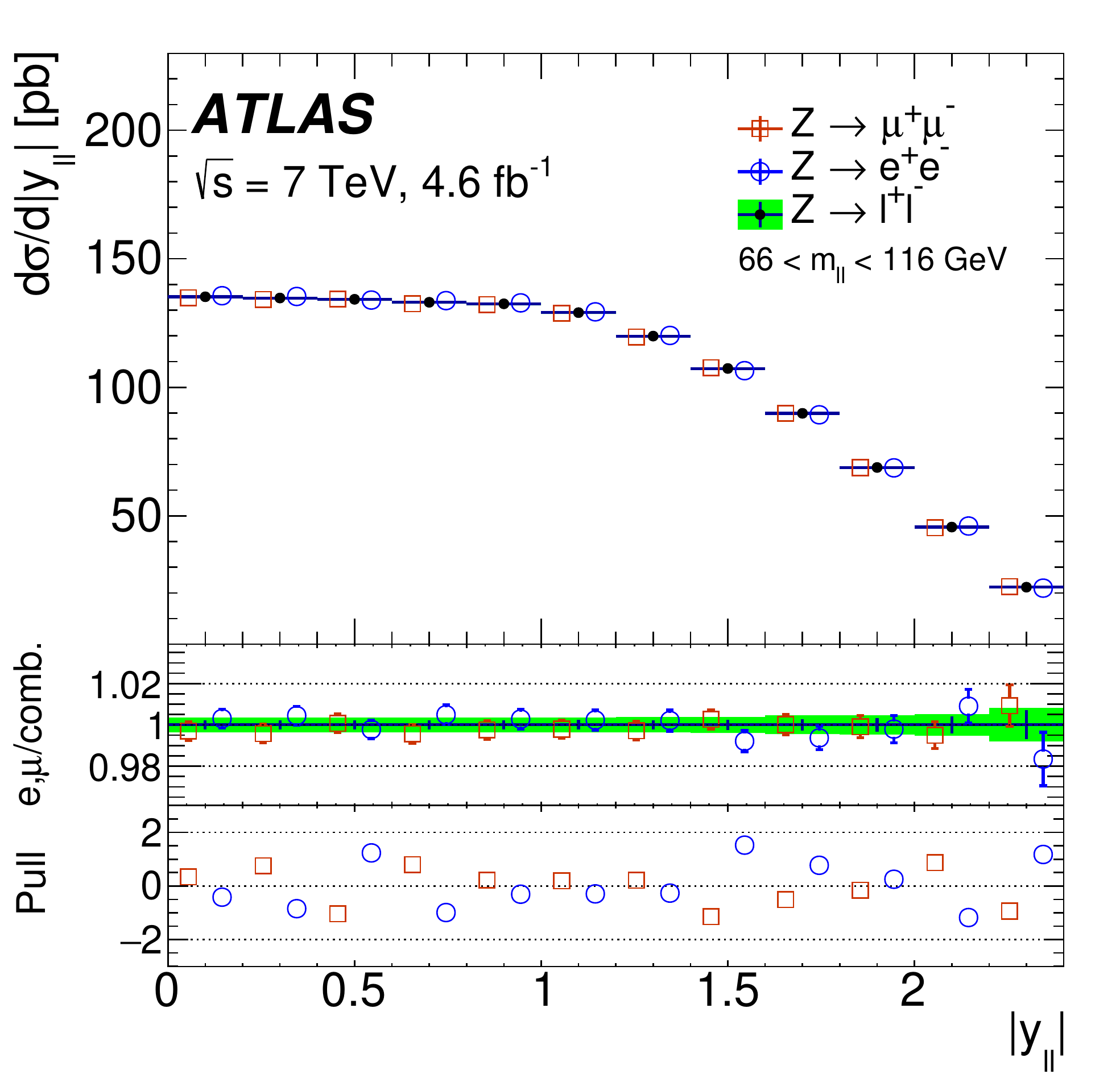}%
    \includegraphics[width=0.325\textwidth]{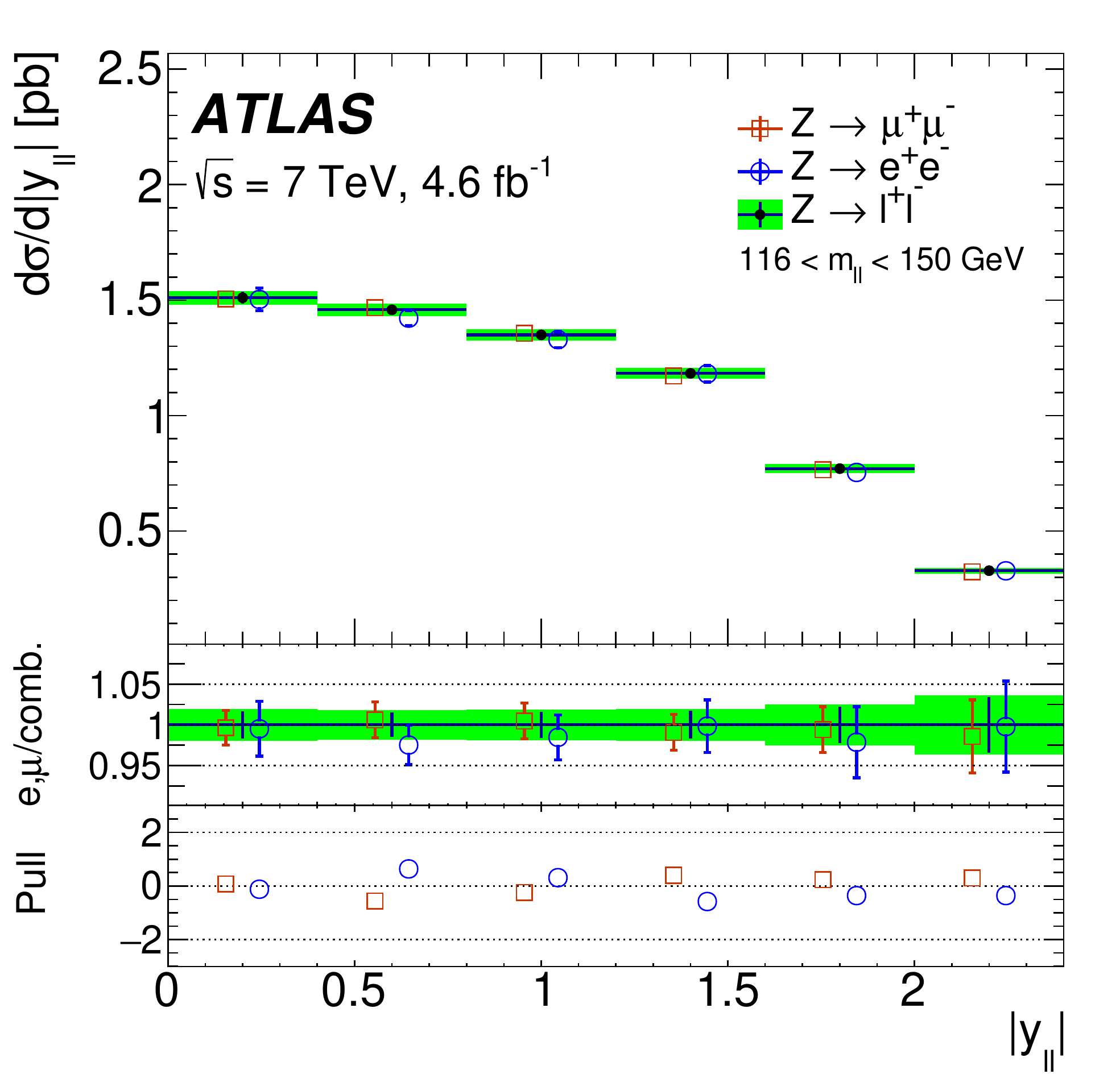}%
  \end{center}
  \caption{Differential $\rd\sigma/\rd|\yll|$ cross-section measurements
    for $Z/\gamma^{*} \rightarrow \ell\ell$ in the three \mll\
    regions, for the electron channel (open circles), the muon channel
    (open squares) and their combination with uncorrelated
    uncertainties (crosses) and the total uncertainty, apart from the
    luminosity error (green band). Also shown are the ratios of the
    $e$ and $\mu$ measurements to the combination and the pulls of the
    individual measurements in terms of their uncorrelated
    uncertainties, see text.}
  \label{Fig:ZCombi}
\end{figure}

Apart from the common luminosity uncertainty of \dlumi\%, the precision
of the combined differential cross sections reaches $0.4$--$0.6\%$ for
the $W^+$ and $W^-$ as well as the central $Z$ peak measurements. Off-peak and forward
measurements have significantly larger uncertainties of typically a
few percent but reaching as high as $20\%$. The differential combined
measurement results are summarized in \Tabs~\ref{tab:w_eta_comb} to
\ref{tab:zfwd_yll_comb}. The full measurement information is provided
in HEPDATA.  The measurements presented here supersede the results
published in Ref.~\cite{Aad:2011dm} because of their significantly
higher precision and extended kinematic coverage.

\begin{table}[p]
  \begin{center}

    \begin{tabular}{cc|c|cccc}
      \hline
      \hline
      \multicolumn{7}{c}{\Wpluslnu} \\\hline
      $|\etal|^\mathrm{min}$ &
      $|\etal|^\mathrm{max}$ &
      $\mathrm{d}\sigma/\mathrm{d}|\etal|$ &
      $\delta_\mathrm{sta}$ &
      $\delta_\mathrm{unc}$ &
      $\delta_\mathrm{cor}$ &
      $\delta_\mathrm{tot}$ \\
      & & $[\mathrm{pb}]$ & $[\%]$ & $[\%]$ & $[\%]$ & $[\%]$ \\\hline

      0.0 & 0.21 & 577.15 &   0.11 &   0.13 &   0.52 &   0.55\\
0.21 & 0.42 & 576.87 &   0.11 &   0.15 &   0.49 &   0.52\\
0.42 & 0.63 & 581.75 &   0.09 &   0.12 &   0.49 &   0.51\\
0.63 & 0.84 & 586.07 &   0.10 &   0.11 &   0.50 &   0.52\\
0.84 & 1.05 & 586.33 &   0.10 &   0.14 &   0.50 &   0.53\\
1.05 & 1.37 & 599.07 &   0.08 &   0.13 &   0.51 &   0.53\\
1.37 & 1.52 & 596.75 &   0.13 &   0.33 &   0.52 &   0.63\\
1.52 & 1.74 & 604.17 &   0.11 &   0.13 &   0.55 &   0.57\\
1.74 & 1.95 & 606.93 &   0.12 &   0.18 &   0.54 &   0.58\\
1.95 & 2.18 & 593.40 &   0.11 &   0.14 &   0.53 &   0.56\\
2.18 & 2.5 & 558.46 &   0.12 &   0.14 &   0.62 &   0.64\\

      \hline
      \multicolumn{7}{c}{\Wminuslnu} \\\hline
      $|\etal|^\mathrm{min}$ &
      $|\etal|^\mathrm{max}$ &
      $\mathrm{d}\sigma/\mathrm{d}|\etal|$ &
      $\delta_\mathrm{sta}$ &
      $\delta_\mathrm{unc}$ &
      $\delta_\mathrm{cor}$ &
      $\delta_\mathrm{tot}$ \\
      & & $[\mathrm{pb}]$ & $[\%]$ & $[\%]$ & $[\%]$ & $[\%]$ \\\hline

      0.0 & 0.21 & 436.45 &   0.12 &   0.14 &   0.52 &   0.55\\
0.21 & 0.42 & 432.78 &   0.12 &   0.16 &   0.48 &   0.52\\
0.42 & 0.63 & 429.29 &   0.11 &   0.13 &   0.49 &   0.52\\
0.63 & 0.84 & 423.38 &   0.12 &   0.13 &   0.50 &   0.53\\
0.84 & 1.05 & 413.64 &   0.11 &   0.15 &   0.50 &   0.54\\
1.05 & 1.37 & 405.26 &   0.10 &   0.14 &   0.56 &   0.59\\
1.37 & 1.52 & 388.02 &   0.17 &   0.34 &   0.52 &   0.64\\
1.52 & 1.74 & 377.51 &   0.14 &   0.16 &   0.58 &   0.62\\
1.74 & 1.95 & 365.82 &   0.12 &   0.20 &   0.58 &   0.63\\
1.95 & 2.18 & 344.70 &   0.13 &   0.17 &   0.59 &   0.63\\
2.18 & 2.5 & 319.04 &   0.14 &   0.19 &   0.75 &   0.79\\

      \hline\hline
    \end{tabular}
  \end{center}
  \caption{Differential cross section for the \Wpluslnu\ (top) and \Wminuslnu\ (bottom) processes,
    extrapolated to the common fiducial region. The relative statistical
    ($\delta_\mathrm{sta}$), uncorrelated systematic
    ($\delta_\mathrm{unc}$), correlated systematic
    ($\delta_\mathrm{cor}$), and total ($\delta_\mathrm{tot}$)
    uncertainties are given in percent of the cross-section
    values. The overall \dlumi\% luminosity uncertainty is not
    included.}
  \label{tab:w_eta_comb}
\end{table}

\begin{table}[p]
  \begin{center}  
    \begin{tabular}{cc|c|cccc}
      \hline
      \hline
      \multicolumn{7}{c}{Central \Zgll, $46<\mll<66\gev$} \\\hline
      $|\yll|^\mathrm{min}$ &
      $|\yll|^\mathrm{max}$ &
      $\mathrm{d}\sigma/\mathrm{d}|\yll|$ &
      $\delta_\mathrm{sta}$ &
      $\delta_\mathrm{unc}$ &
      $\delta_\mathrm{cor}$ &
      $\delta_\mathrm{tot}$ \\
      & & $[\mathrm{pb}]$ & $[\%]$ & $[\%]$ & $[\%]$ & $[\%]$ \\\hline

      0.0 & 0.4 &   3.524 &   0.97 &   0.52 &   1.14 &   1.58\\
0.4 & 0.8 &   3.549 &   0.95 &   0.47 &   1.05 &   1.49\\
0.8 & 1.2 &   3.411 &   0.97 &   0.48 &   1.13 &   1.57\\
1.2 & 1.6 &   3.423 &   1.00 &   0.48 &   1.03 &   1.52\\
1.6 & 2.0 &   2.942 &   1.09 &   0.47 &   1.02 &   1.57\\
2.0 & 2.4 &   1.541 &   1.64 &   0.60 &   1.02 &   2.03\\

      \hline
      \multicolumn{7}{c}{Central \Zgll, $66<\mll<116\gev$} \\\hline
      $|\yll|^\mathrm{min}$ &
      $|\yll|^\mathrm{max}$ &
      $\mathrm{d}\sigma/\mathrm{d}|\yll|$ &
      $\delta_\mathrm{sta}$ &
      $\delta_\mathrm{unc}$ &
      $\delta_\mathrm{cor}$ &
      $\delta_\mathrm{tot}$ \\
      & & $[\mathrm{pb}]$ & $[\%]$ & $[\%]$ & $[\%]$ & $[\%]$ \\\hline

      0.0 & 0.2 & 135.22 &   0.19 &   0.10 &   0.29 &   0.36\\
0.2 & 0.4 & 134.74 &   0.19 &   0.10 &   0.28 &   0.35\\
0.4 & 0.6 & 134.24 &   0.19 &   0.09 &   0.28 &   0.35\\
0.6 & 0.8 & 133.08 &   0.20 &   0.09 &   0.28 &   0.36\\
0.8 & 1.0 & 132.48 &   0.20 &   0.10 &   0.28 &   0.36\\
1.0 & 1.2 & 129.06 &   0.20 &   0.11 &   0.28 &   0.36\\
1.2 & 1.4 & 119.92 &   0.21 &   0.09 &   0.29 &   0.37\\
1.4 & 1.6 & 107.32 &   0.23 &   0.12 &   0.29 &   0.39\\
1.6 & 1.8 &  89.87 &   0.25 &   0.11 &   0.36 &   0.45\\
1.8 & 2.0 &  68.80 &   0.29 &   0.15 &   0.32 &   0.46\\
2.0 & 2.2 &  45.62 &   0.36 &   0.22 &   0.31 &   0.52\\
2.2 & 2.4 &  22.23 &   0.59 &   0.37 &   0.41 &   0.81\\

      \hline
      \multicolumn{7}{c}{Central \Zgll, $116<\mll<150\gev$} \\\hline
      $|\yll|^\mathrm{min}$ &
      $|\yll|^\mathrm{max}$ &
      $\mathrm{d}\sigma/\mathrm{d}|\yll|$ &
      $\delta_\mathrm{sta}$ &
      $\delta_\mathrm{unc}$ &
      $\delta_\mathrm{cor}$ &
      $\delta_\mathrm{tot}$ \\
      & & $[\mathrm{pb}]$ & $[\%]$ & $[\%]$ & $[\%]$ & $[\%]$ \\\hline

      0.0 & 0.4 &   1.510 &   1.41 &   0.90 &   1.03 &   1.97\\
0.4 & 0.8 &   1.458 &   1.37 &   0.61 &   1.03 &   1.82\\
0.8 & 1.2 &   1.350 &   1.45 &   0.73 &   0.95 &   1.88\\
1.2 & 1.6 &   1.183 &   1.54 &   0.75 &   0.92 &   1.95\\
1.6 & 2.0 &   0.7705 &   2.03 &   0.99 &   1.06 &   2.49\\
2.0 & 2.4 &   0.3287 &   3.17 &   1.31 &   1.25 &   3.65\\

      \hline\hline
    \end{tabular}
    \caption{Differential cross section for the \Zgll\ process in the
      central region in three dilepton invariant mass regions,
      extrapolated to the common fiducial region. The relative
      statistical ($\delta_\mathrm{sta}$), uncorrelated systematic
      ($\delta_\mathrm{unc}$), correlated systematic
      ($\delta_\mathrm{cor}$), and total ($\delta_\mathrm{tot}$)
      uncertainties are given in percent of the cross-section
      values. The overall \dlumi\% luminosity uncertainty is not
      included.}
    \label{tab:zcen_yll_comb}
  \end{center}
\end{table}

\begin{table}
  \begin{center}  
    \begin{tabular}{cc|c|cccc}
      \hline
      \hline
      \multicolumn{7}{c}{Forward \Zgll, $66<\mll<116\gev$} \\\hline
      $|\yll|^\mathrm{min}$ &
      $|\yll|^\mathrm{max}$ &
      $\mathrm{d}\sigma/\mathrm{d}|\yll|$ &
      $\delta_\mathrm{sta}$ &
      $\delta_\mathrm{unc}$ &
      $\delta_\mathrm{cor}$ &
      $\delta_\mathrm{tot}$ \\
      & & $[\mathrm{pb}]$ & $[\%]$ & $[\%]$ & $[\%]$ & $[\%]$ \\\hline

      1.2 & 1.4 &   7.71 &   1.76 &   1.84 &   3.10 &   4.01\\
1.4 & 1.6 &  17.93 &   1.02 &   1.11 &   2.93 &   3.30\\
1.6 & 1.8 &  32.52 &   0.73 &   0.70 &   2.68 &   2.87\\
1.8 & 2.0 &  50.55 &   0.59 &   1.77 &   2.52 &   3.14\\
2.0 & 2.2 &  68.88 &   0.58 &   2.66 &   2.14 &   3.46\\
2.2 & 2.4 &  86.59 &   0.50 &   1.90 &   1.90 &   2.73\\
2.4 & 2.8 &  86.21 &   0.34 &   3.03 &   1.68 &   3.48\\
2.8 & 3.2 &  40.69 &   0.49 &   0.64 &   5.49 &   5.55\\
3.2 & 3.6 &  10.95 &   1.23 &   3.69 &   6.40 &   7.48\\

      \hline
      \multicolumn{7}{c}{Forward \Zgll, $116<\mll<150\gev$} \\\hline
      $|\yll|^\mathrm{min}$ &
      $|\yll|^\mathrm{max}$ &
      $\mathrm{d}\sigma/\mathrm{d}|\yll|$ &
      $\delta_\mathrm{sta}$ &
      $\delta_\mathrm{unc}$ &
      $\delta_\mathrm{cor}$ &
      $\delta_\mathrm{tot}$ \\
      & & $[\mathrm{pb}]$ & $[\%]$ & $[\%]$ & $[\%]$ & $[\%]$ \\\hline
      1.2 & 1.6 &   0.300 &   6.84 &   6.58 &   8.96 &  13.06\\
1.6 & 2.0 &   0.548 &   5.21 &   7.78 &   7.20 &  11.81\\
2.0 & 2.4 &   0.925 &   3.99 &  13.52 &   4.26 &  14.72\\
2.4 & 2.8 &   0.937 &   3.87 &  20.86 &   3.87 &  21.57\\
2.8 & 3.2 &   0.437 &   5.30 &  14.40 &   6.59 &  16.70\\
3.2 & 3.6 &   0.0704 &  14.49 &  11.60 &   7.04 &  19.85\\

      \hline\hline
    \end{tabular}
    \caption{Differential cross section for the \Zgll\ process in the
      forward region in two dilepton invariant mass ranges,
      extrapolated to the common fiducial region. The relative
      statistical ($\delta_\mathrm{sta}$), uncorrelated systematic
      ($\delta_\mathrm{unc}$), correlated systematic
      ($\delta_\mathrm{cor}$), and total ($\delta_\mathrm{tot}$)
      uncertainties are given in percent of the cross-section
      values. The overall \dlumi\% luminosity uncertainty is not
      included.}
    \label{tab:zfwd_yll_comb}
  \end{center}
\end{table}

\clearpage

\section{Comparison with theory}
\label{sec:comnnlo}

\subsection{Theoretical framework and methodology}
\label{sec:thyframe}
\subsubsection{Drell--Yan cross-section predictions}
\label{subho}

Predictions for Drell--Yan production in proton--proton collisions in
this paper are calculated at fixed order in perturbative QCD using the
programs \DYNNLO~1.5~\cite{Catani:2007vq,Catani:2009sm} and \FEWZ~
3.1.b2~\cite{Gavin:2010az,Gavin:2012sy,Li:2012wn}. Both programs
calculate $W$ and \Zg\ boson production up to
next-to-next-to-leading order  in the strong coupling constant,
$\mathcal{O}(\alphas^2)$, and include the boson decays to leptons
($\ell^+\nu$, $\ell^-\bar{\nu}$, or $\ell^+\ell^-$) with full spin
correlations, finite width, and interference effects.  They allow
kinematic phase-space requirements to be implemented for a direct
comparison with experimental data.  In addition, the programs
\textsc{ZWPROD}~\cite{Hamberg:1991} and
\textsc{VRAP}~\cite{Anastasiou:2003ds} are available for total
cross-section calculations enabling  cross-checks or fast estimates
of factorization and renormalization scale uncertainties.

At leading order (LO) in the electroweak (EW) couplings, there is a significant
dependence of the cross-section predictions
on the electroweak parameter scheme. For all calculations
the $G_\mu$ scheme~\cite{Dittmaier:2009cr} is chosen, in which 
the primary parameters are the Fermi constant and the particle masses. Corrections for
NLO EW effects reduce the dependence on the EW scheme and are
important at the precision level required
 for the present measurements.
 These NLO EW corrections, however, require a separate
treatment, discussed in \Sec~\ref{QCDandEW}, as they are currently not
provided by the NNLO QCD programs, with the exception of the NC
Drell--Yan calculation in \FEWZ~\cite{Li:2012wn}.

\begin{table}
  \begin{center}
\begin{tabular}{lrlr}
\hline
\hline
    $m_Z$                       & 91.1876 \GeV                        & $|V_{ud}|$      & 0.97427  \\
    $\Gamma_Z$                  & 2.4949 \GeV                            & $|V_{us}|$      & 0.22534  \\
    $\Gamma(\Zll)$  & 0.08400 \GeV                             & $|V_{ub}|$      & 0.00351  \\
    $m_W$                       & 80.385 \GeV                        & $|V_{cd}|$      & 0.22520  \\
    $\Gamma_W$                  & 2.0906 \GeV                            & \Vcs      & 0.97344  \\
    $\Gamma(\Wlnu)$ & 0.22727 \GeV                          & $|V_{cb}|$      & 0.0412   \\
    $m_H$                       & 125     \GeV                        & $|V_{td}|$      & 0.00867  \\
    $m_t$                       & 173.5   \GeV                        & $|V_{ts}|$      & 0.0404   \\
    $G_\mathrm{F}$                       & $1.1663787 \times 10^{-5}$ \GeV$^{-2}$ & $|V_{tb}|$      & 0.999146 \\
    \hline 
    $\sin^2\theta_\mathrm{W}$             & 0.222897                   &               &       \\
    $\alpha_{G_{\mu}}$                  & $7.562396 \times 10^{-3}$  &               &       \\
    $v_{u}$                  & 0.405607                   &               &       \\
    $v_{d}$                  & $-$0.702804                 &               &       \\
    $v_{\ell}$                  & $-$0.108411                  &               &       \\
    \hline
    \hline
    \end{tabular}
    \caption{Electroweak input parameters, in the $G_{\mu}$ scheme, for
      the NC and CC Drell--Yan $pp$ and deep inelastic $ep$ scattering
      cross-section calculations, see text. Standard Model parameters
      are taken from~Refs.~{\mbox{\cite{Beringer:1900zz, Agashe:2014kda}}},
      except
      $\Gamma(\Wlnu)$. The $V_{ij}$ symbols denote the elements of the
      CKM matrix. The parameters below the line, the weak mixing angle
      $\sin^2\theta_\mathrm{W}$, the fine-structure constant
      $\alpha_{G_{\mu}}$, and the vector couplings of up-type quarks
      $v_{u}$, down-type quarks $v_{d}$, and charged
      leptons $v_{\ell}$ to the $Z$ boson, are calculated at
      tree level from the ones above.}
    \label{ewqcd:tabEWpar}
      \end{center}
\end{table}

The QCD analysis of the $ep$ and $pp$ data presented below assumes
that the SM electroweak parameters are known. Their values are taken
from the PDG~\cite{Agashe:2014kda}, and are listed for reference in
\Tab~\ref{ewqcd:tabEWpar}. The leptonic decay width of the $W$ boson,
$\Gamma(\Wln)$, is an exception. The predicted
value of $\Gamma(\Wln)=226.36\MeV$ quoted in the PDG effectively
includes higher-order EW effects. For consistency with the
higher-order EW corrections, provided by
\textsc{MCSANC}~\cite{Bondarenko:2013nu}, however, the leading-order
partial width value, $\Gamma(\Wln)=227.27\MeV$, is used in both the
QCD and EW calculations. It was verified that consistent results were
obtained by using the PDG value and omitting the extra NLO EW
corrections. For the leptonic decay width of the Z boson, the
predicted value of $\Gamma(\Zll)=84.00\mev$ differs only by $0.1\%$
from the leading-order value of $\Gamma(\Zll)=83.92\MeV$ and this
difference is of no practical relevance for the NC Drell--Yan cross-section calculation. The values of the magnitudes of the CKM matrix elements, listed
in \Tab~\ref{ewqcd:tabEWpar}, are taken from
Ref.~\cite{Beringer:1900zz}.  The \Vcs\ matrix parameter is accessible
through $cs \to W$ production and thus related to the fraction of
strange quarks in the proton, which is of special interest in this
analysis.  In \Sec~\ref{sec:vcs} a dedicated QCD fit analysis is
presented, where no prior knowledge is assumed on the magnitude of the CKM matrix element \Vcs\ , which instead is determined from the data together with the PDF parameters.

The nominal theoretical predictions of the differential, fiducial and
total cross sections at NNLO in QCD are computed with \DYNNLO\,1.5
using the default program parameters.\footnote{Using the default
  parameters of this program, with an intrinsic \textsc{xqtcut}
  parameter chosen to be $0.008$, the fiducial NNLO QCD predictions
  are found to behave in a continuous way with respect to small
  variations in the minimum lepton \pt\ requirements around the choice
  of equal threshold values chosen for all fiducial regions of this
  paper.}  For an estimate of the current uncertainties of
fixed-order perturbative QCD NNLO calculations, the \DYNNLO\,
predictions are compared with predictions using \FEWZ\,3.1.b2. For the
total cross sections,  agreement to better than $0.2\%$ is
observed. For the fiducial and differential cross-section measurements
with additional kinematic requirements on the lepton transverse
momenta and rapidities, however, poorer agreement is found:
for the integrated fiducial $W^+,~W^-,~Z/\gamma^*$ cross sections, the
differences between \FEWZ\ and \DYNNLO\ predictions calculated with
the \epWZ12 PDF set amount to $(+1.2,\,+0.7,\,+0.2)\%$, which may be
compared to the experimental uncertainties of
$\pm(0.6,\,0.5,\,0.32)\%$, respectively.\footnote{ The \FEWZ\ and
  \DYNNLO\ programs differ in the subtraction schemes used, which leads
  to small differences in the boson \pt\ distributions at low
  values. This effect on the fiducial cross-section predictions is
  significant compared to the present experimental precision. Further
  efforts will be needed to understand this effect and the role of
  boson \pt\ in fiducial cross-section predictions and to reduce the
  impact on the extracted PDFs. See Ref.~\cite{Alioli:2016fum} for a
  further discussion of this effect.}

In the calculation of the Drell--Yan cross sections,
the renormalization and factorization scales, $\mur$ and $\muf$, are
chosen to be the dilepton invariant mass, \mll\ , at the
centre of the respective cross-section bin in the NC case and the
$W$-boson mass, $m_W$, in the CC case. Variations of the scales by 
factors of $2$ and $1/2$ are conventionally used as an estimate of the
approximation represented by NNLO as compared to still unknown higher-order
corrections. 
The numerical implication of the scale choices, termed
scale uncertainties, is considered in the evaluation of the QCD fit
results on the strange-quark fraction and the CKM element \Vcs.
The DIS cross sections are calculated in all cases at
the scale of  $\mur =\muf = \sqrt{Q^2}$, where $Q^2$ denotes the
negative square of the four-momentum transfer in NC and CC $ep$ scattering. 

The relative uncertainty of the LHC proton beam energy of $\pm
0.1\%$~\cite{Wenninger:2254678} induces an uncertainty of the
cross-section predictions of typically $\pm 0.1\%$, which is negligible
compared to the other theoretical uncertainties discussed above.

\subsubsection{Electroweak corrections and combination with QCD predictions}\label{QCDandEW}

In Drell--Yan production, the dominant part of the higher-order
electroweak corrections is the QED radiation from the final-state
leptons. This contribution is included in the Drell--Yan MC
samples using \Photos~\cite{Golonka:2005pn} and then passed through
the detailed ATLAS detector simulation as described in
\Sec\,\ref{sec:simulation}. The data are unfolded for QED FSR effects
at the same time as for other detector effects.
The calculations of the QED FSR effects by
\Photos\ and \textsc{MCSANC}~1.20~\cite{Bardin:2012jk} agree very
well~\cite{Arbuzov:2012dx}. The remaining NLO EW corrections are then
calculated using \textsc{MCSANC}, excluding the QED FSR contributions,
for both the NC and CC Drell--Yan processes. These terms include NLO
contributions from initial-state photon radiation, EW loop
corrections, and initial-state--final-state photon interference.

The NLO EW corrections calculated with \textsc{MCSANC} need to be
combined with the NNLO QCD predictions, calculated with \DYNNLO, to
obtain complete predictions.\footnote{Combined higher-order $\alpha
  \cdot \alphas$ corrections to resonant $W,~Z$ production were
  recently considered in Ref.~\cite{Dittmaier:2015rxo}. Another
  approach to combine NLO QCD and NLO EW effects, using the Powheg method,
  has been presented in Refs.~\cite{Bernaciak:2012hj, Barze:2012tt,
    Barze:2013yca}.  } This combination may be achieved using either a
factorized or an additive approach~\cite{Butterworth:2014efa}. A
common PDF set at NNLO, \epWZ12, is used for the calculation of both
the absolute NNLO QCD and NLO EW cross sections. The combination of
QCD and EW calculations in the factorized approach may be expressed
using \kfactor\ corrections as
\begin{equation}
  \label{ewqcd:eqf}
  \sigma_\mathrm{NNLO\;QCD}^\mathrm{NLO\;EW}   = 
  \sigma_\mathrm{NNLO\;QCD}^\mathrm{LO\;EW} \cdot K^\mathrm{EW} 
  = \sigma_\mathrm{LO\;QCD}^\mathrm{LO\;EW} \cdot K_\mathrm{QCD} \cdot K^\mathrm{EW}
\end{equation}
with the electroweak $K^\mathrm{EW}$ and QCD
$K_\mathrm{QCD}$ correction factors defined as
\begin{equation}\label{ewqcd:eqf1}
 K_\mathrm{QCD } =  \frac {\sigma_\mathrm{NNLO\;QCD}^\mathrm{LO\;EW}}{\sigma_\mathrm{LO\;QCD}^\mathrm{LO\;EW}} \;\;\;\;\mbox{and} \;\;\;\;
  K^\mathrm{EW}  =  \frac {\sigma^\mathrm{NLO\;EW}_\mathrm{LO\;QCD} }{\sigma_\mathrm{LO\;QCD}^\mathrm{LO\;EW}} \,. 
\end{equation}
This assumes that the fractional higher-order EW corrections,
quantified by $K^\mathrm{EW}$, are the same for all orders of QCD. They
thus can be determined based on LO QCD Drell--Yan cross-section
calculations.

The alternative additive approach assumes the absolute contribution of
the EW correction to be independent of the order of the
underlying QCD calculation. Thus the relative fraction of the
higher-order EW corrections is different for each order of
QCD by $(K^\mathrm{EW}-1)/K_\mathrm{QCD}$. The combination
of QCD and EW calculations then proceeds as
\begin{equation}\label{ewqcd:eqa}
\sigma_\mathrm{NNLO\;QCD}^\mathrm{NLO\;EW}   =  \sigma_\mathrm{NNLO\;QCD}^\mathrm{LO\;EW} + \left( \sigma^\mathrm{NLO\;EW}_\mathrm{LO\;QCD} -\sigma_\mathrm{LO\;QCD}^\mathrm{LO\;EW} \right)=\sigma_\mathrm{NNLO\;QCD}^\mathrm{LO\;EW}  \cdot \left(1+\frac{K^\mathrm{EW}-1}{K_\mathrm{QCD}}  \right) \,.
\end{equation}

The central value of the combined NNLO QCD and NLO EW prediction is
taken from the additive approach, which is also implemented in
\FEWZ~\cite{Li:2012wn}. The corrections to be applied to the NNLO QCD
fiducial cross sections according to Eq.~\eqref{ewqcd:eqa} are about
$-0.4\%$ and $-0.3\%$ for the $W^+$ and $W^-$ channels,
respectively.  For the neutral-current channels, those corrections are
$+6\%$, $-0.3\%\,(-0.4\%)$ and $-0.5\%\,(-1.2\%)$ for the
central (forward) selection in the low-mass, $Z$-peak
and high-mass regions of \mll, respectively. The corrections are
calculated separately for each measurement bin, but they depend
only weakly on \etal\ and \yll\ for the CC and NC case,
respectively.

The differences between the additive and factorized approaches are in
general found to be small and significantly smaller than the
experimental uncertainty of the results presented in this paper. They
are at most $0.3-0.9\%$ for the low-mass $\mll = 46$--$66\gev$ region
for the NC case with larger effects observed at central rapidity. In
the forward $Z$-peak phase space, they extend to $0.4\%$. 
In all other
regions of phase space, the effect is  $<0.1\%$. These
differences are taken as a systematic uncertainty applied
symmetrically to the central value obtained using the additive
approach.

Additional two-loop EW corrections for the leading contributions are
calculated using \textsc{MCSANC} for the NC case~\cite{Arbuzov:2015yja}.
This correction is found to be $<0.1\%$
everywhere except for the region $\mll= 46$--$66\GeV$, where it reaches
$(-0.62 \pm 0.15)\%$.

The radiation of real (on-shell) $W$ and $Z$ bosons is very small for the
considered phase space~\cite{Baur:2006sn} and neglected.
An important background to the NC process
outside the $Z$-boson mass region arises from photon-induced dileptons,
$\gamma\gamma \to \ell\ell$. This contribution is calculated including
NLO effects for the fiducial phase space with the
MCSANC~\cite{Bardin:2012jk} program and subtracted from the unfolded
data. The calculation uses the average of the two available
MRST2004qed~\cite{Martin:2004dh} predictions for the photon PDF as the
central value and half the difference as an uncertainty estimate. The
size of the photon-induced contribution is about $1.5\%$ in
the low and high \mll\ bins, while it is negligible 
($<0.1\%$) at the $Z$ peak. Due to large uncertainties on the 
photon PDF, the fractional uncertainties are at the level
of $30$--$50\%$.

\subsubsection{Methodology of PDF profiling}
\label{subprof}

The impact of new data on a given PDF set can be estimated in a
quantitative way with a profiling
procedure~\cite{Paukkunen:2014zia,Camarda:2015zba}. The profiling is
performed using a $\chi^2$ function which includes both the
experimental uncertainties and the theoretical ones arising from PDF
variations:

\begin{eqnarray}
\nonumber \chi^2(\vec{b}_{\mathrm{exp}},\vec{b}_{\mathrm{th}}) &=& \sum_{i=1}^{N_\mathrm{data}} \frac{\textstyle \left[ \sigma^\mathrm{exp}_i -  \sigma^\mathrm{th}_i (1 - \sum_j \gamma^\mathrm{exp}_{ij} b_{j,\mathrm{exp}} - \sum_k \gamma^\mathrm{th}_{ik}b_{k,\mathrm{th}}) \right]^2}{\Delta_i^2} \\
&+&  \sum_{j=1}^{N_\mathrm{exp. sys}} b_{j,\mathrm{exp}}^2 + \sum_{k=1}^{N_\mathrm{th. sys}} b_{k,\mathrm{th}}^2\,.   \label{eq:chi2prof}
\end{eqnarray}

This $\chi^2$ function resembles the one used for the combination,
described in Section\,\ref{sec:combi}.  The index $i$ runs over all
$N_\mathrm{data}$ data points. The measurements and the theory
predictions are given by $\sigma^\mathrm{exp}_i$ and
$\sigma_i^\mathrm{th}$, respectively. The correlated experimental and
theoretical uncertainties are included using the nuisance parameter
vectors $\vec{b}_{\mathrm{exp}}$ and $\vec{b}_{\mathrm{th}}$,
respectively. Their influence on the data and theory predictions is
described by the matrices $\gamma^\mathrm{exp}_{ij}$ and
$\gamma^\mathrm{th}_{ik}$, where the index $j$ ($k$) corresponds to
the $N_\mathrm{exp. sys}$ experimental ($N_\mathrm{th. sys}$
theoretical) nuisance parameters.  Both the correlated and
uncorrelated systematic uncertainties are treated as
multiplicative. The estimation of the statistical uncertainties is
protected against statistical fluctuations in data using the expected
rather than the observed number of events and the denominator is hence
calculated as
\begin{equation}
  \Delta_i^2=\delta_{i,\mathrm{sta}}^2\sigma^\mathrm{exp}_i \sigma^\mathrm{th}_i + \left(\delta_{i,\mathrm{unc}}\sigma^\mathrm{th}_i\right)^2\,.
  \label{eq:deltai}
\end{equation}
The contribution to the $\chi^2$ from the two sums over $b_{j,k}^2$,
which implement the $\pm 1\sigma$ constraints of the nuisance
parameters, is later also referred to as the ``correlated''
contribution. The $\chi^2$ function of Eq.~\eqref{eq:chi2prof} can be
generalized to account for asymmetric uncertainties, as described in
Ref.~\cite{Camarda:2015zba}.

The value of the $\chi^2$ function at its minimum provides a
compatibility test of the data and theory.
In addition, the values of the nuisance parameters at this minimum,
$b^\mathrm{min}_{k,\mathrm{th}}$, can be interpreted as an optimization
(``profiling'') of PDFs to describe the data~\cite{Paukkunen:2014zia}. The profiled
central PDF set $f'_0$ is given by
\begin{equation}
   f'_0 = f_0 + \sum_k  \left[ b^\mathrm{min}_{k, \mathrm{th}} \left( \frac{f^{+}_k - f^{-}_k}{2} \right)  + \left(b^\mathrm{min}_{k, \mathrm{th}}\right)^2  
  \left(\frac{f^{+}_k + f^{-}_k - 2f_0}{2} \right)^2\right]\,, \label{eq:prof}
\end{equation}
where $f_0$ is a short notation for the original central PDFs of each
parton flavour, $f_0 = xf(x,Q^2)$, and $f^{\pm}_k$ represent the
eigenvector sets corresponding to up and down variations.  For the
LHAPDF6~\cite{Buckley:2014ana} parameterizations, $f_0$ and
$f^{\pm}_k$ are given as data tables at fixed $x,Q^2$ grid points.
Equation~\eqref{eq:prof} is a parabolic approximation of the PDF
dependence close to the central value, e.g. for a single nuisance
parameter, taking the values $b_\mathrm{th} = +1,~-1,~0$, the values
of $f'_0$ are $f'_0 = f^+,~f^-,~f_0$, respectively.

The profiled PDFs $f'_0$ have reduced uncertainties. In general, the
shifted eigenvectors are no longer orthogonal and are transformed to
an orthogonal representation using a standard
procedure~\cite{Aaron:2009bp}, which can be extended to asymmetric
uncertainties.  The profiling procedure used in this analysis is
implemented in the xFitter package~\cite{Alekhin:2014irh}.
The $\chi^2$ function used in the analysis takes into account
asymmetric PDF uncertainties.

The profiling procedure quantifies the compatibility of a data set
with the predictions based on a PDF set and estimates the PDF
sensitivity of the data set.  However, the results of profiling are
only reliable when the prediction is broadly consistent with the data
within the PDF uncertainties because of the approximation involved in
Eq.~\eqref{eq:prof}, and the profiling cannot act as a substitute 
for a full QCD fit
analysis. A second caveat is that the $\chi^2$ tolerance criteria, which 
many global PDF analyses use~\cite{Pumplin:2009bb}, are different 
from the
$\Delta \chi^2=1$ employed in the profiling. Thus the impact of the
data in a full PDF fit pursued by those groups may  differ
from the result of a profiling analysis as outlined
here.  Profiling results are presented below for the PDF sets ABM12,
CT14, MMHT2014, NNPDF3.0 (Hessian representation~\cite{Carrazza:2015aoa}), and \epWZ12.

\subsection{Integrated cross sections and their ratios}
\label{sec:crorat}
\begin{figure}[tb]
  \begin{center}
    \includegraphics[width=0.45\textwidth]{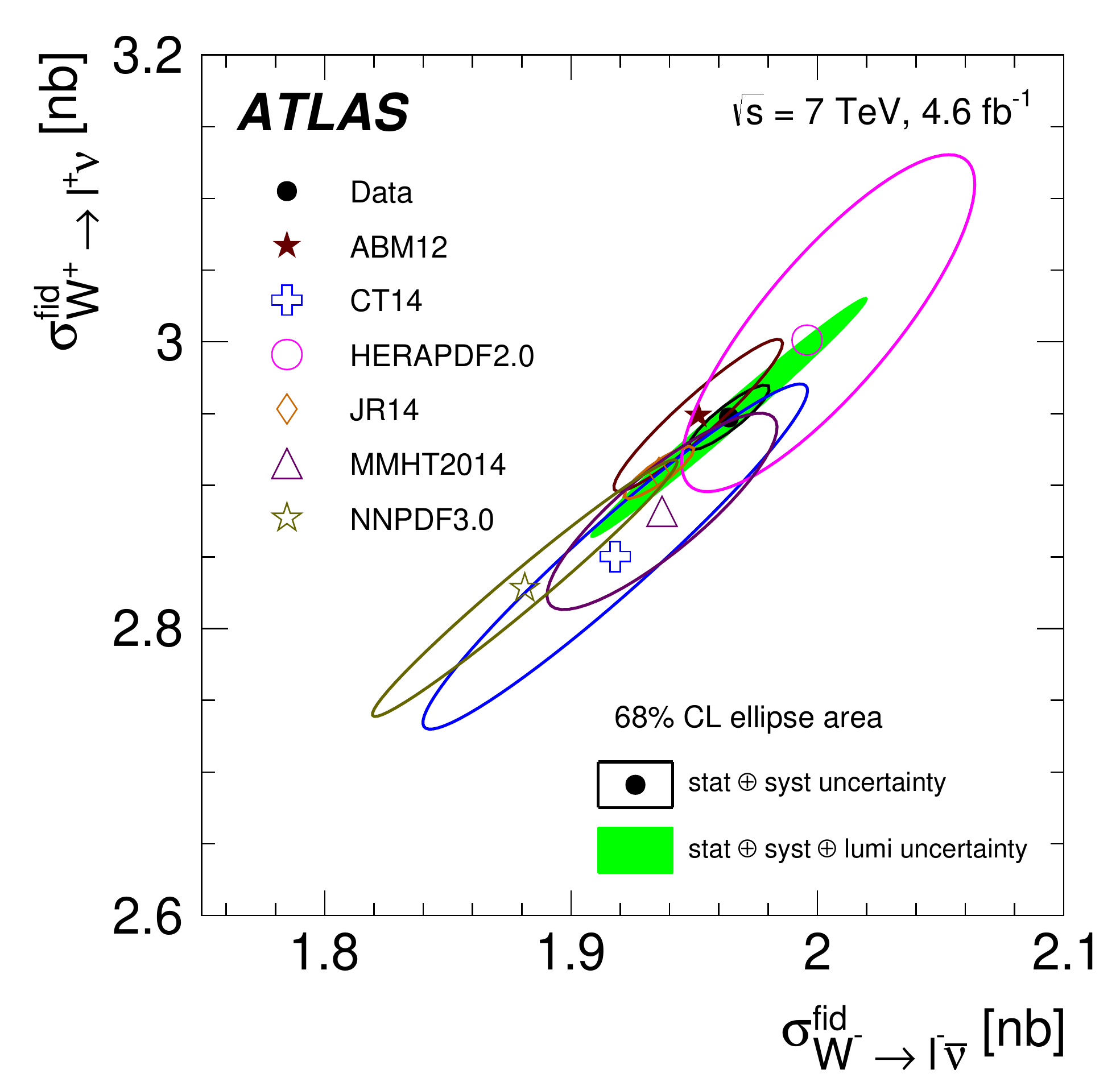}
    \includegraphics[width=0.45\textwidth]{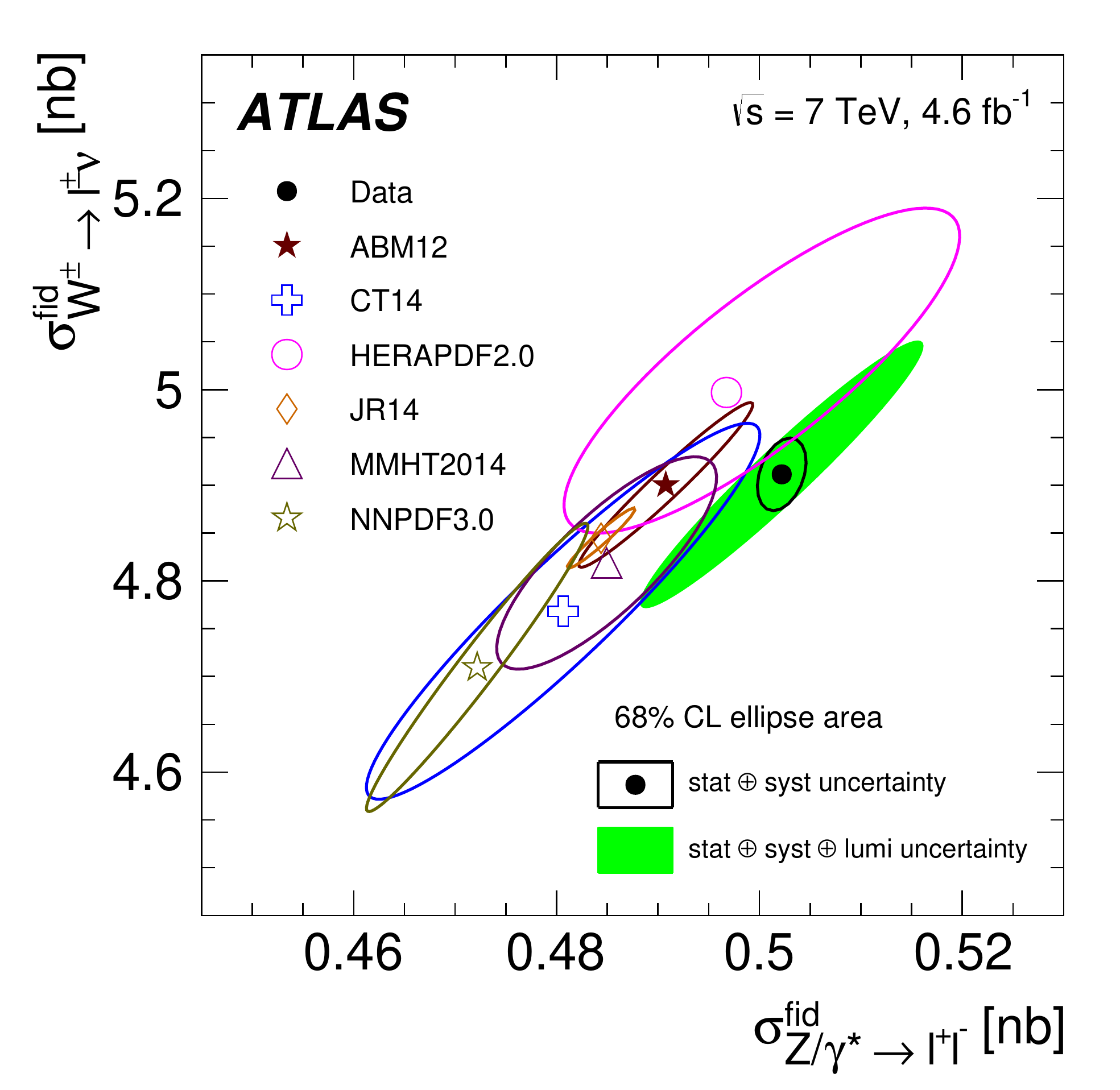}
  \end{center}
  \caption{Integrated fiducial cross sections times leptonic branching ratios of
    $\sigma^\mathrm{fid}_{W^+\to\ell^+\nu}$
    vs. $\sigma^\mathrm{fid}_{W^-\to\ell^-\bar{\nu}}$ (left) and
    $\sigma^\mathrm{fid}_{W^\pm\to\ell^\pm\nu}$
    vs. $\sigma^\mathrm{fid}_{Z/\gamma^{*}\to\ell^+\ell^-}$ (right). The data
    ellipses illustrate the $68\%$ CL coverage for the total
    uncertainties (full green) and total excluding the luminosity
    uncertainty (open black). Theoretical predictions based on various
    PDF sets are shown with open symbols of different colours. The
    uncertainties of the theoretical calculations correspond to the
    PDF uncertainties only.}
  \label{Fig:IntFidCrossSections}
\end{figure}

The combined integrated cross sections in the fiducial phase space are
shown in \Fig~\ref{Fig:IntFidCrossSections}. NNLO QCD predictions with
NLO EW corrections based on the ABM12, CT14, HERAPDF2.0, JR14,
MMHT2014, NNPDF3.0 PDF sets are compared to the data. 
The central values and their uncertainties for these PDF sets are
provided in \Tab~\ref{tab:fidxsec_theory} together with the combined measurements
reported before in \Tab~\ref{tab:fidint}.

\begin{table}[htbp]
  \begin{center}  
    \begin{tabular}{lcccc}
      \hline
      \hline
      PDF set 
      & $\sigma^\mathrm{fid}_{W^+\to\ell^+\nu}$ [pb]
      & $\sigma^\mathrm{fid}_{W^-\to\ell^-\bar{\nu}}$ [pb]
      & $\sigma^\mathrm{fid}_{W^\pm\to\ell^\pm\nu}$ [pb]
      & $\sigma^\mathrm{fid}_{Z/\gamma^*\to\ell\ell}$ [pb]\\
      \hline
      ABM12 &
      $2949 \pm 35$ &
      $1952 \pm 23$ &
      $4900 \pm 57$ &
      $490.8 \pm 5.7$ 
      \\
      CT14&
      $2850^{+77}_{-82}$&
      $1918^{+46}_{-57}$&
      $4770^{+120}_{-140}$&
      $481^{+11}_{-14}$
      \\
      HERAPDF2.0 &
      $3001^{+89}_{-66}$ &
      $1996^{+48}_{-31}$ &
      $5000^{+140}_{-90}$ &
      $497^{+16}_{-9}$
      \\
      JR14 &
      $2909^{+13}_{-11}$&
      $1936^{+10}_{-9}$ &
      $4845^{+23}_{-19}$&
      $484.4 \pm 2.2$
      \\
      MMHT2014 &
      $2882^{+49}_{-42}$&
      $1937^{+30}_{-32}$&
      $4819^{+75}_{-72}$&
      $485^{+7.4}_{-6.9}$
      \\
      NNPDF3.0 &
      $2828 \pm 59$&
      $1881 \pm 41$&
      $4709 \pm 99$&
      $472.2 \pm 7.2$
      \\
      \hline 
      Data &
      $2947 \pm 55$&
      $1964 \pm 37$&
      $4911 \pm 92$&
      $502.2 \pm 9.2$
      \\
      \hline 
      \hline
    \end{tabular}
    \caption{Predictions at NNLO QCD and NLO EW as obtained with
      \DYNNLO~1.5 for the integrated fiducial cross sections. The given
      uncertainties correspond to PDF uncertainties only and are
      evaluated following the different prescriptions of the  PDF
      groups. The measured cross sections as reported before in \Tab~\ref{tab:fidint}
      are shown in the last row with their total uncertainties.}
    \label{tab:fidxsec_theory}
  \end{center}
\end{table}

The two-dimensional presentation is particularly instructive, as it
conveys both the values and correlations of both the measurements
and predictions.
The cross-section calculations are performed with the DYNNLO program
as described in \Sec~\ref{sec:thyframe}. All experimental and
theoretical ellipses are defined such that their area corresponds to
$68\%$ CL.\footnote{This implies that the projections onto the axes
  correspond to $1.52$ times the one-dimensional uncertainty. This is
  the same convention as chosen in Refs.~\cite{Aad:2011dm,
    Aad:2016naf}. However, in the literature one may find an
  alternative definition, where the size of ellipses reflect
  the one-dimensional uncertainties when projected on the
  axes~\cite{Watt:2011kp}.}

Correlations between the predicted cross sections are evaluated from
individual error eigenvectors in each PDF set. The spread of the
predictions as well as the size of the individual PDF uncertainties
are significantly larger than the uncertainty of the data. The
measurements are seen to discriminate between different PDF choices
and to provide information to reduce PDF uncertainties.  As seen in
\Fig~\ref{Fig:IntFidCrossSections}, the PDF sets  CT14, MMHT2014
and NNPDF3.0 give predictions that are lower for both the $W^+$ and the
$W^-$ cross sections, a trend that is also observed for the
\Zg\ cross section.

The ratios of the combined fiducial cross sections, presented before in
\Tab~\ref{tab:fidratio}, are compared in \Fig~\ref{Fig:IntFidCrossSectionRatios}
to NNLO QCD predictions based on various PDF sets. It is
observed that the measured $W^+/W^-$ ratio is well reproduced, but,
as already seen in the correlation plots above, all
PDF sets predict a higher $W/Z$ ratio than measured in the data. 

\begin{figure}[tb]
  \begin{center}
    \includegraphics[width=0.47\textwidth]{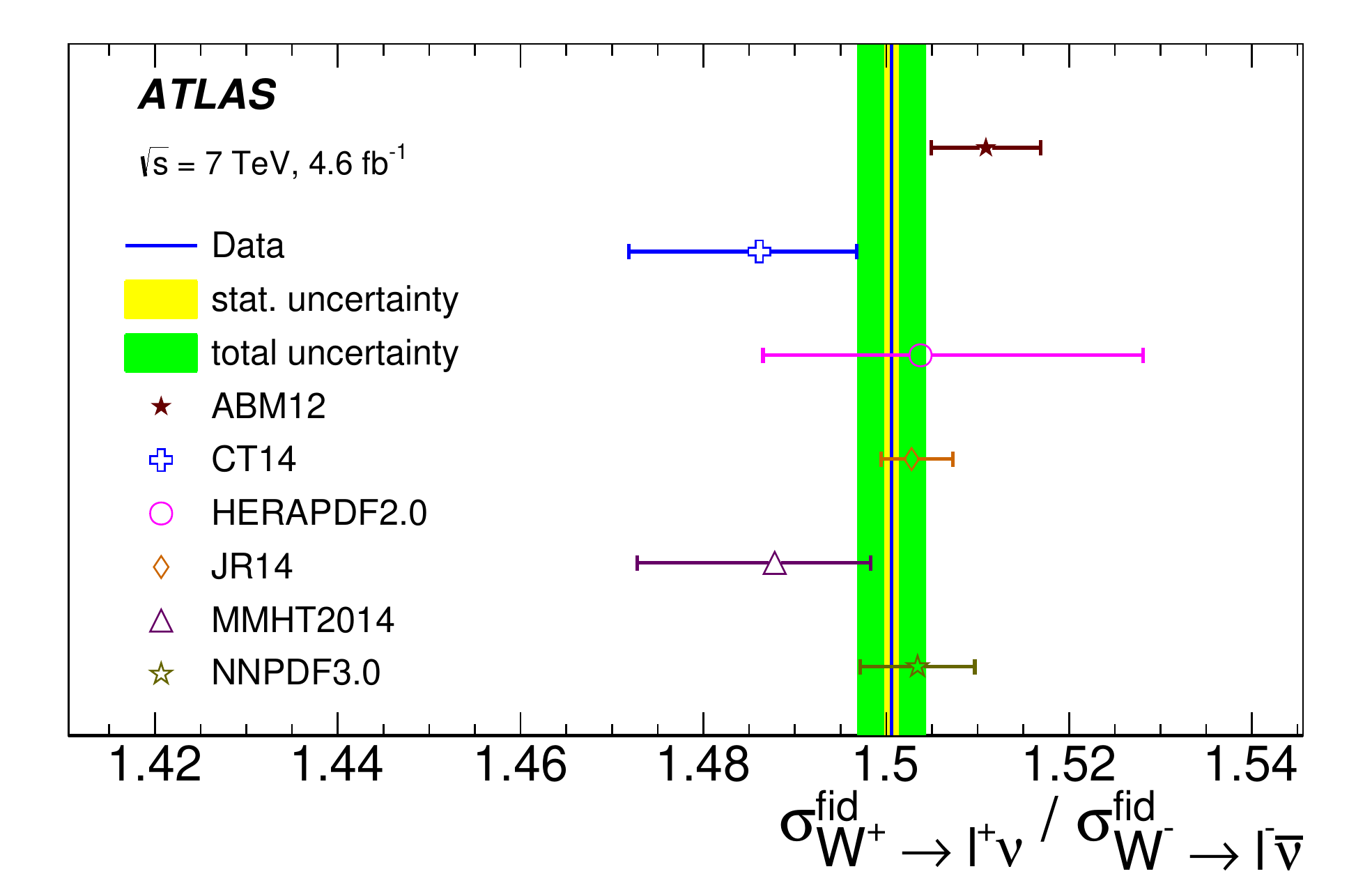}
    \includegraphics[width=0.47\textwidth]{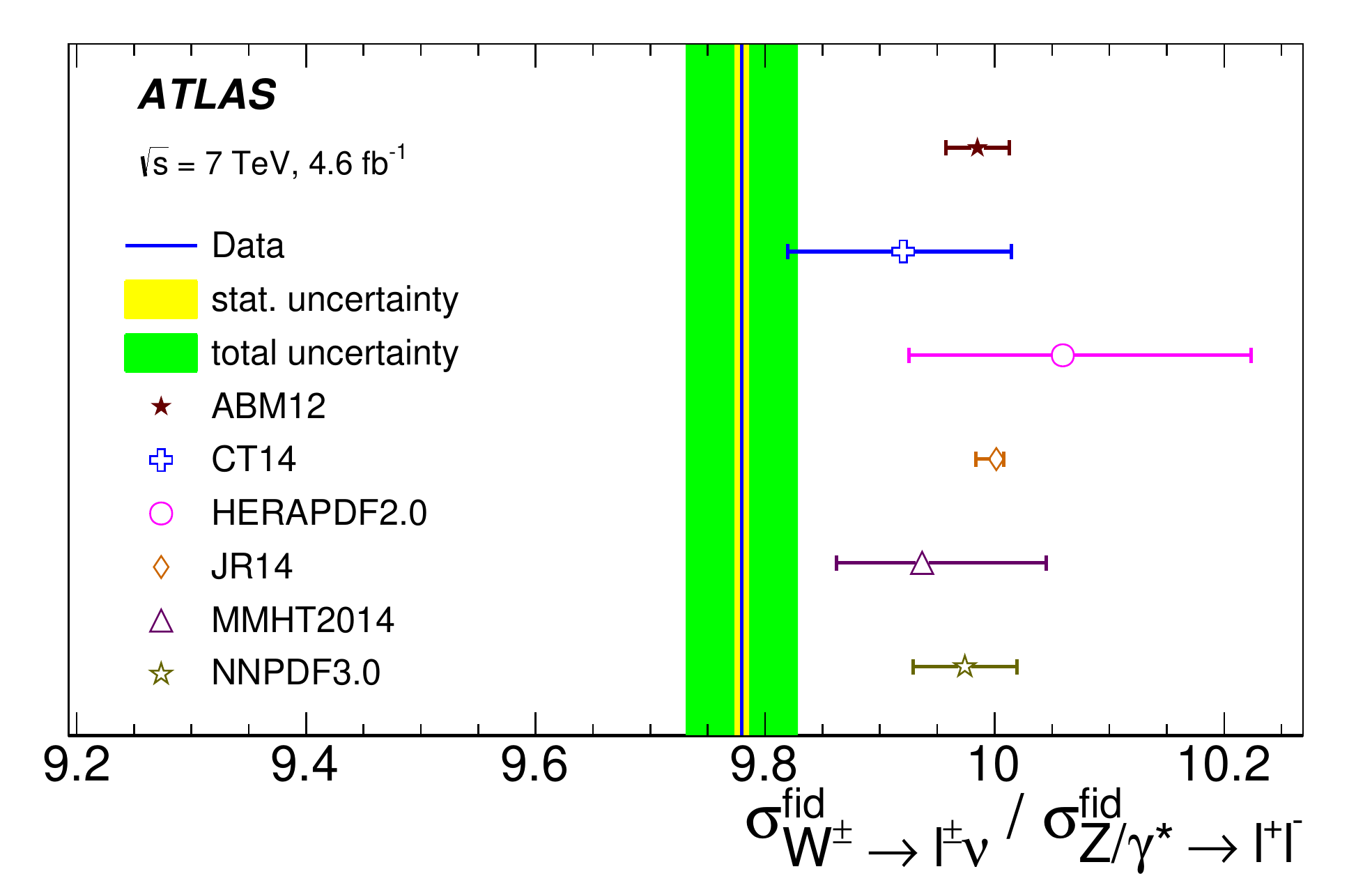}
  \end{center}
  \caption{Ratios of the fiducial cross sections times leptonic
    branching ratios of
    $\sigma^\mathrm{fid}_{W^+\to\ell^+\nu}/\sigma^\mathrm{fid}_{W^-\to\ell^-\bar{\nu}}$
    (left) and
    $\sigma^\mathrm{fid}_{W^\pm\to\ell^\pm\nu}/\sigma^\mathrm{fid}_{Z/\gamma^{*}\to\ell^+\ell^-}$
    (right). The data (solid blue line) are shown with the statistical
    (yellow band) and the total uncertainties (green
    band). Theoretical predictions based on various PDF sets are shown
    with open symbols of different colours. The uncertainties of the
    theoretical calculations correspond to the PDF uncertainties
    only.}
  \label{Fig:IntFidCrossSectionRatios}
\end{figure}

\subsection{Rapidity distributions}
\label{sec:sigdiff}

\newcommand{\diffnnlocap}{Predictions computed at NNLO QCD with NLO EW
  corrections using various PDF sets (open symbols) are compared to
  the data (full points). The ratio of theoretical predictions to the
  data is also shown. The predictions are displaced within each bin
  for better visibility. The theory uncertainty corresponds to the
  quadratic sum of the PDF uncertainty and the statistical uncertainty
  of the calculation.}

\subsubsection{$W^+$ and $W^-$ cross sections}

Differential cross sections as a function of lepton pseudorapidity in
\Wln\ decays, for both $W^+$ and $W^-$, are shown in \Fig~\ref{fig:Comb:NNLOW} and compared to
NNLO perturbative QCD predictions, including NLO EW corrections. The
predictions with the ABM12 PDF set match the data particularly well,
while the predictions of NNPDF3.0, CT14, MMHT14 and JR14, tend to be
below and the HERAPDF2.0 set slightly above
the $W$ cross-section data. For many PDF sets, the differences, however, do not exceed the luminosity uncertainty of $1.8\%$ by a significant amount. Different groups producing PDF sets make different choices in their evaluation of uncertainties. For
example, the JR14 set is less consistent with these data even though it
is somewhat closer to the data than the NNPDF3.0 set, which quotes
much larger uncertainties than JR14.

\begin{figure}
  \begin{center}
    \includegraphics[width=0.49\textwidth]{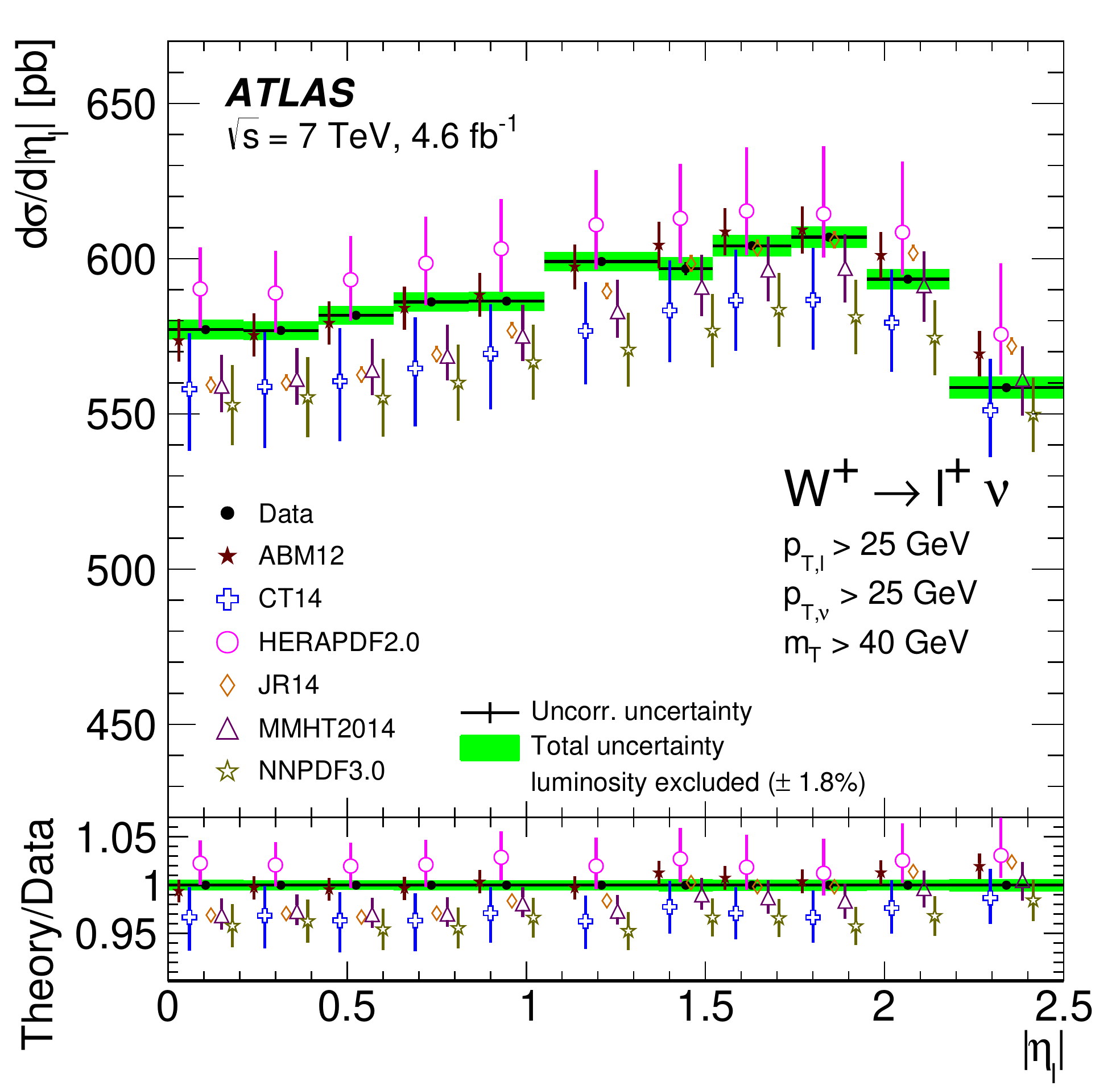}
    \includegraphics[width=0.49\textwidth]{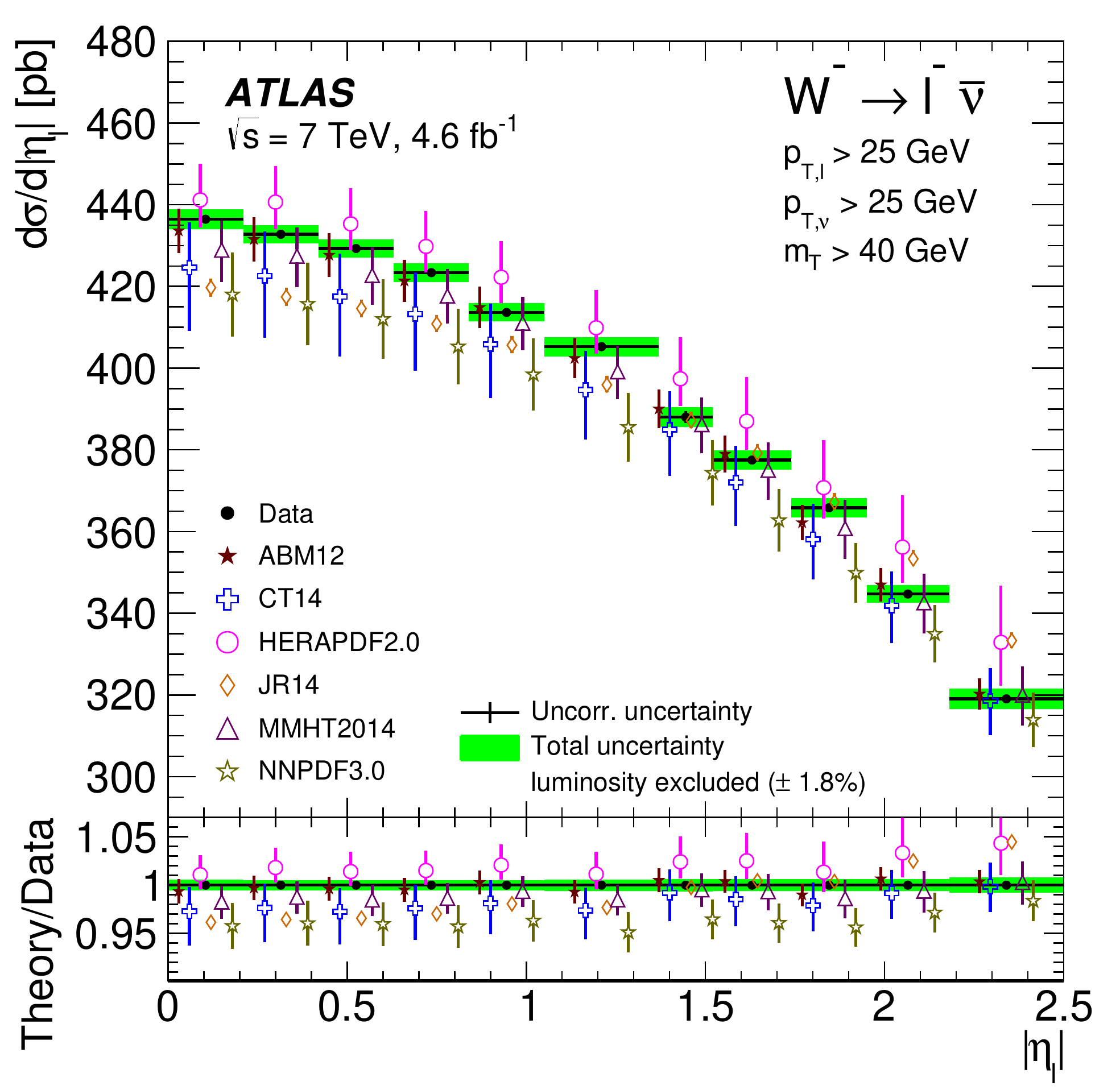}
  \end{center}
  \caption{ Differential $\rd\sigma_{W+}/\rd|\etal|$ (left) and
    $\rd\sigma_{W-}/\rd|\etal|$ (right) cross-section measurement for
    \Wln. \diffnnlocap}
  \label{fig:Comb:NNLOW}
\end{figure}

The measurements of $W^{+}$ and $W^{-}$ cross sections as a function
of $\etal$ are used to extract the lepton charge asymmetry
\begin{equation}
  A_{\ell} = \frac{\rd\sigma_{W+}/\rd|\etal| - \rd\sigma_{W-}/\rd|\etal|}
  {\rd\sigma_{W+}/\rd|\etal| + \rd\sigma_{W-}/\rd|\etal|} \,,
\end{equation}
taking into account all sources of correlated and uncorrelated
uncertainties.

\FFig~\ref{fig:Comb:NNLO:Wasym25} shows the measured charge asymmetry
and the predictions based on various PDF sets. The experimental
uncertainty ranges from $0.5\%$ to $1\%$.  Most of the predictions
agree well with the asymmetry measurement, only CT14 somewhat
undershoots the data. The NNPDF3.0 set, which uses $W^{\pm}$
asymmetry data from the CMS Collaboration~\cite{Chatrchyan:2012xt,
  Chatrchyan:2013mza}, matches the ATLAS data very well, even within
its very small uncertainties. On the other hand, these
predictions are in general $3$--$5\%$ below both the measured $W^+$ and $W^-$
differential cross sections. This highlights the additional information provided by precise, absolute differential measurements with full
uncertainty information, including the correlations, as compared to an asymmetry measurement.

\begin{figure}
  \begin{center}
    \includegraphics[width=0.49\textwidth]{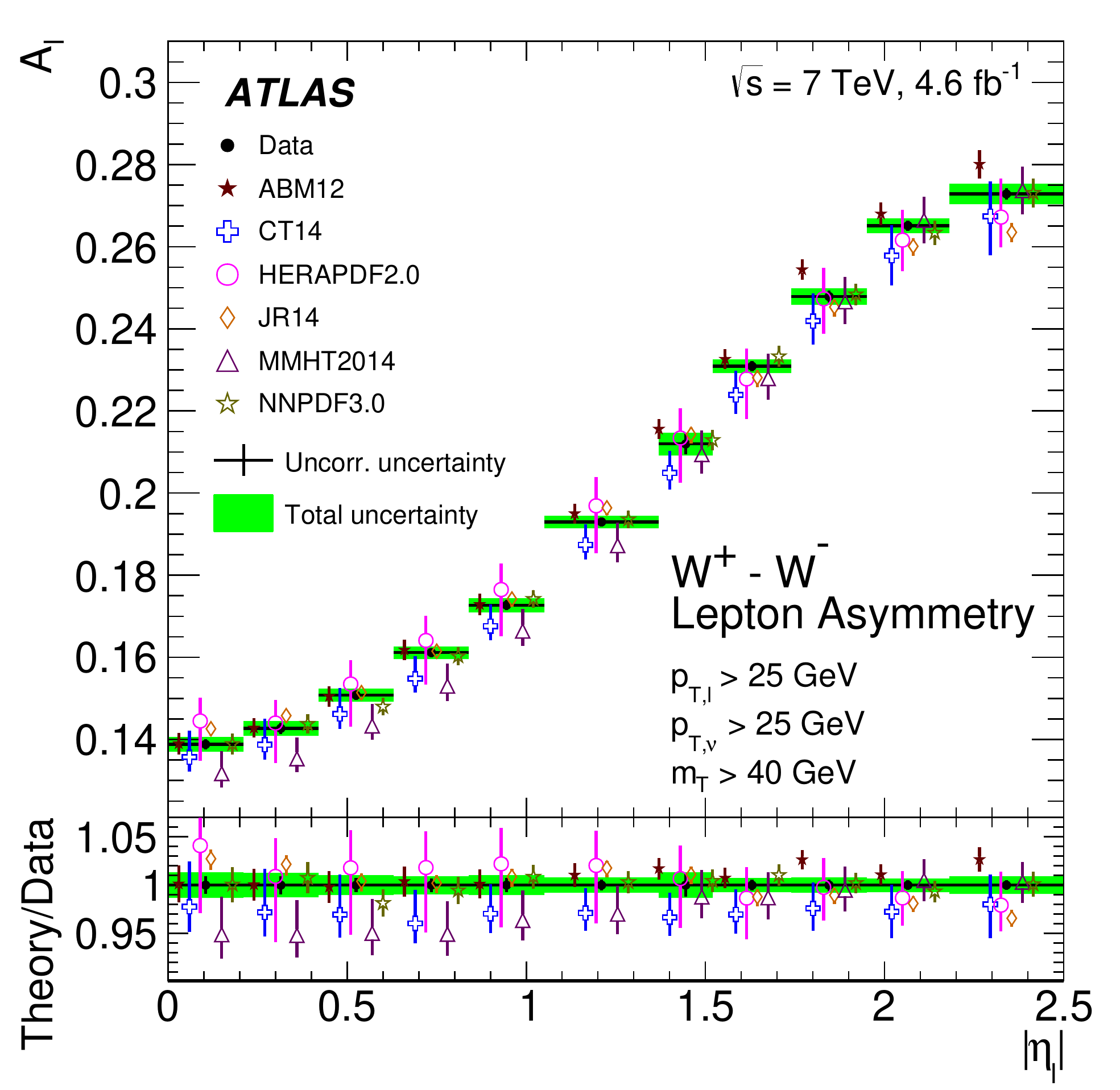}
  \end{center}
  \caption{Lepton charge asymmetry $A_{\ell}$ in \Wlnu\ production as a
    function of the lepton pseudorapidity $|\etal|$. \diffnnlocap}
  \label{fig:Comb:NNLO:Wasym25}
\end{figure}

\subsubsection{$Z/\gamma^*$ cross sections}

Differential \Zgll\ cross-sections, as a function of the dilepton
rapidity, are shown in \Figs~\ref{fig:Comb:NNLO:Zpeak} and
\ref{fig:Comb:NNLOZoffpeak}, and compared to NNLO perturbative QCD
predictions, including NLO EW corrections. The predictions are
evaluated with various PDF sets. At the $Z$ peak, where the highest
precision is reached for the data, all predictions are below
the data at central rapidity, $|\yll| < 1$, but least for the HERAPDF2.0
set, which quotes the largest uncertainties. In the forward region,
the PDFs agree well with the measurement, which, however, is
only precise to the level of a few percent and thus not very sensitive
to differences between PDFs. In the low mass \Zgll\ region,
\Fig~\ref{fig:Comb:NNLOZoffpeak}, several of the PDF sets exhibit a
different rapidity dependence than the data although being mostly
consistent with the measurement. This also holds for the central
rapidity region at high mass, $116 < \mll < 150\GeV$. The precision of
the data in the forward region at high mass is too low to allow 
discrimination between the various PDF sets, all of which reproduce the 
measured rapidity dependence within the quoted uncertainties.

\begin{figure}   
  \begin{center}   
    \includegraphics[width=0.49\textwidth]{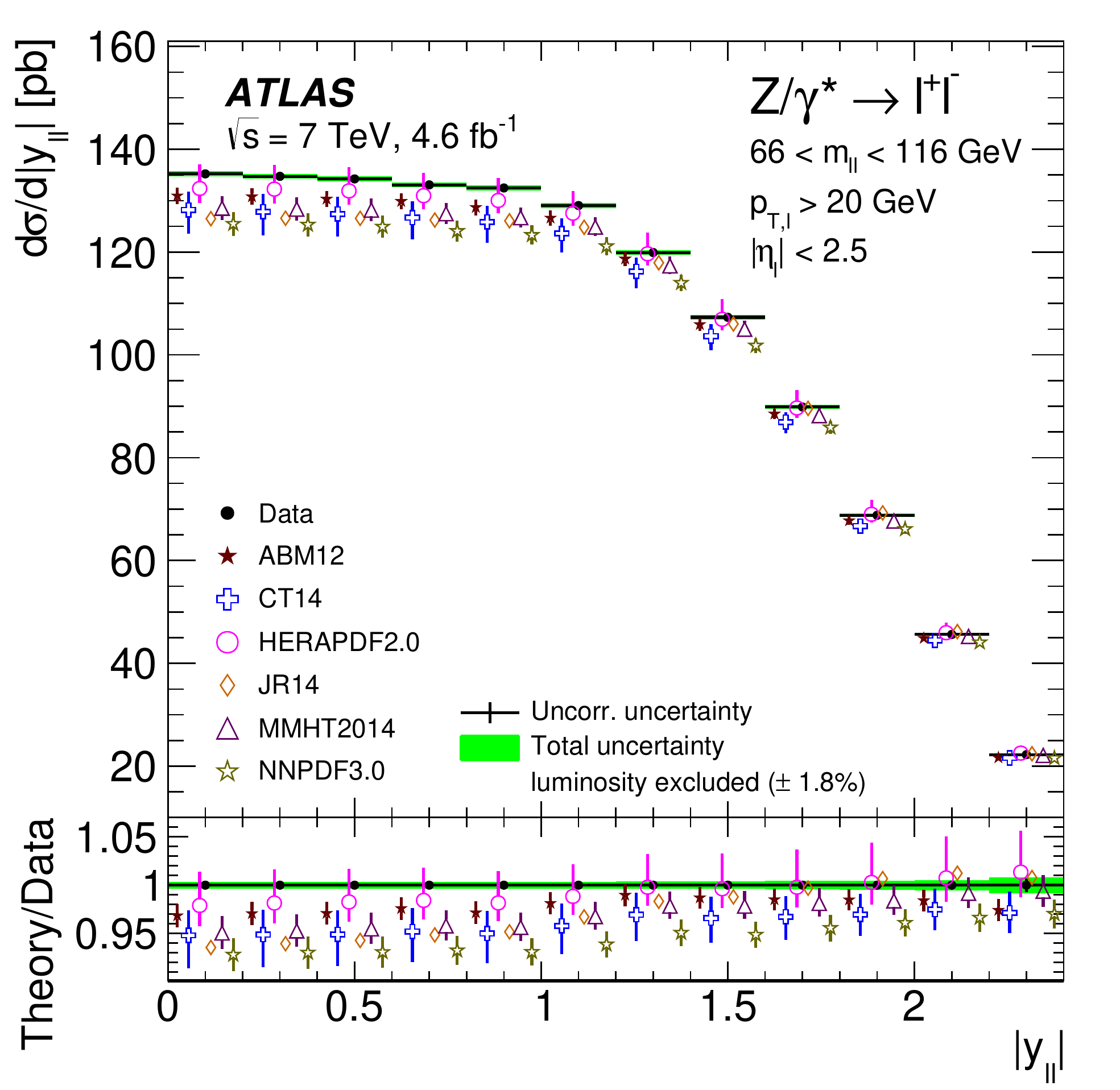}
    \includegraphics[width=0.49\textwidth]{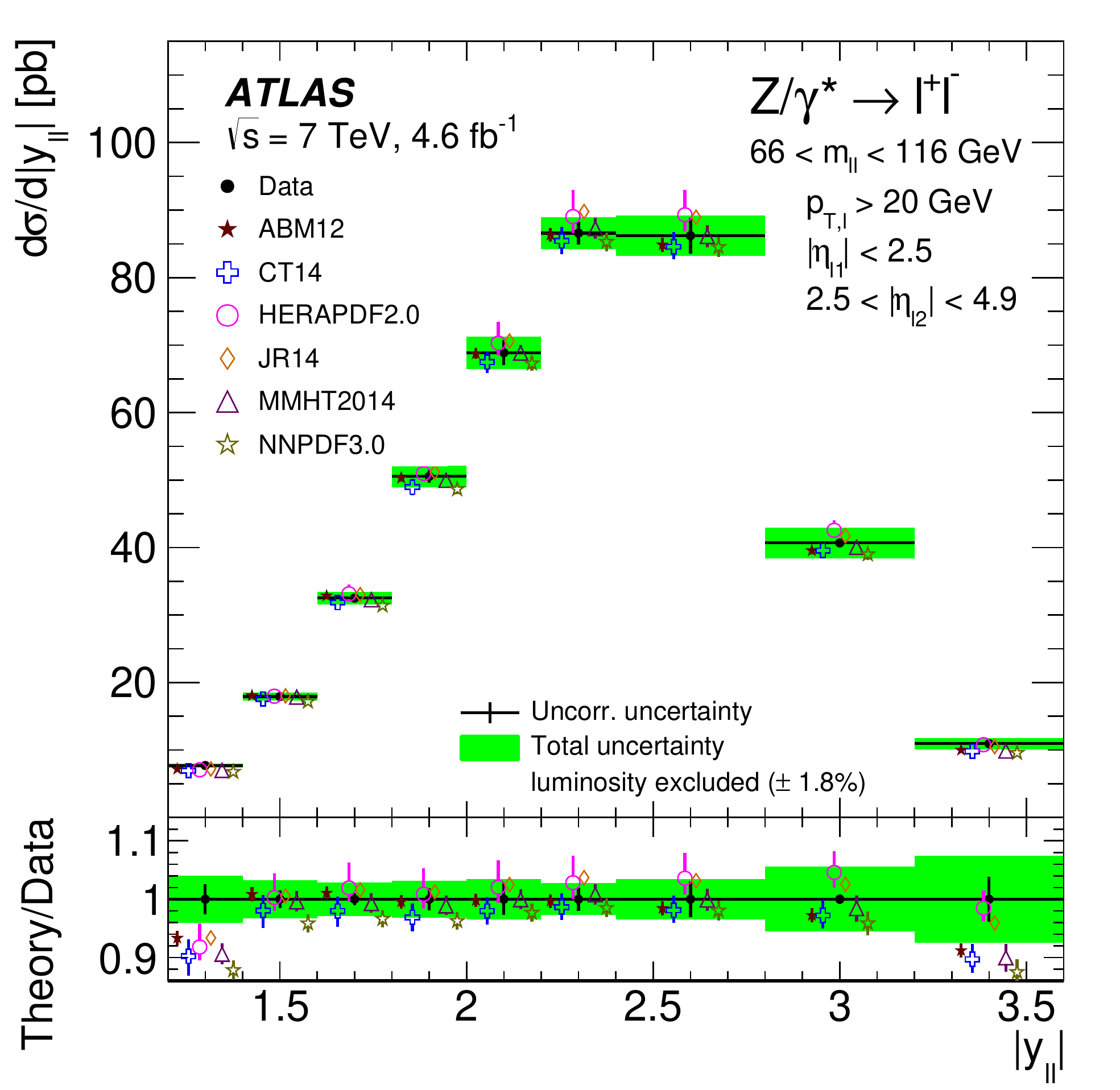}
  \end{center}
  \caption{Differential cross-section measurement $\rd\sigma/\rd|\yll|$
    for \Zgll\ in the $Z$-peak region, $66 < \mll < 116\gev$, for
    central (left) and forward rapidity values (right). \diffnnlocap}
  \label{fig:Comb:NNLO:Zpeak}
\end{figure}

\begin{figure}
  \begin{center}
    \includegraphics[width=0.325\textwidth]{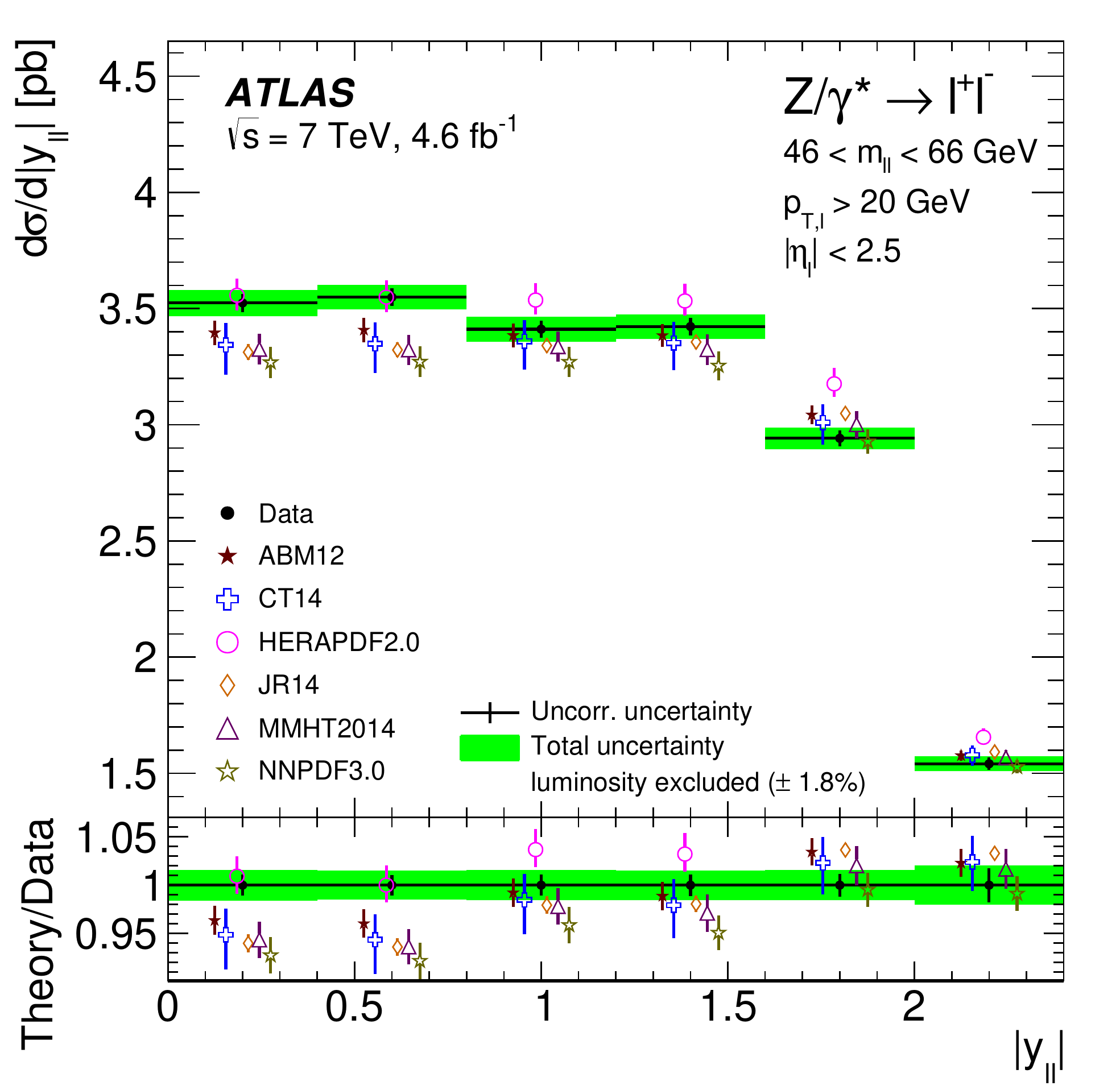}
    \includegraphics[width=0.325\textwidth]{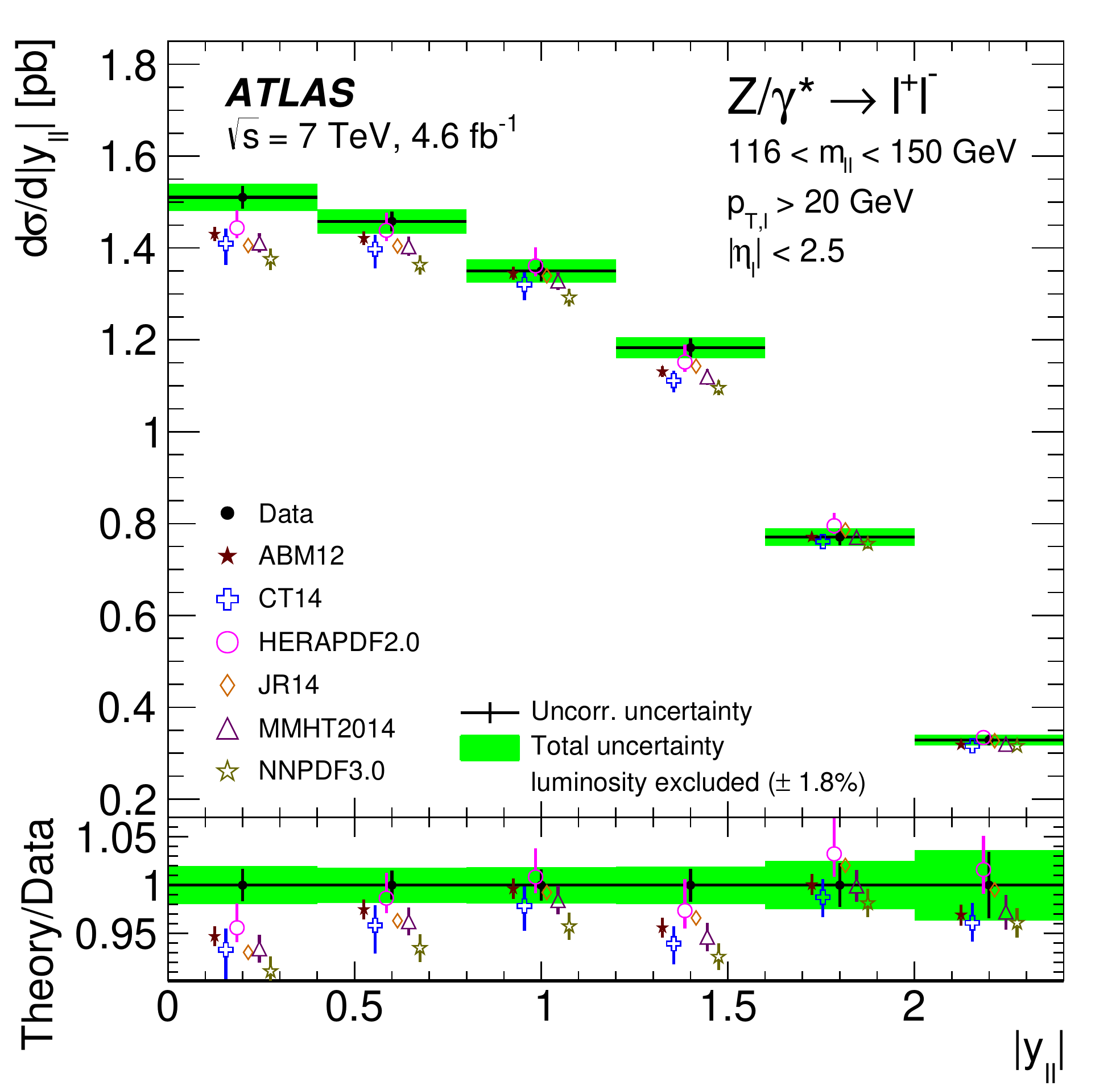}
    \includegraphics[width=0.325\textwidth]{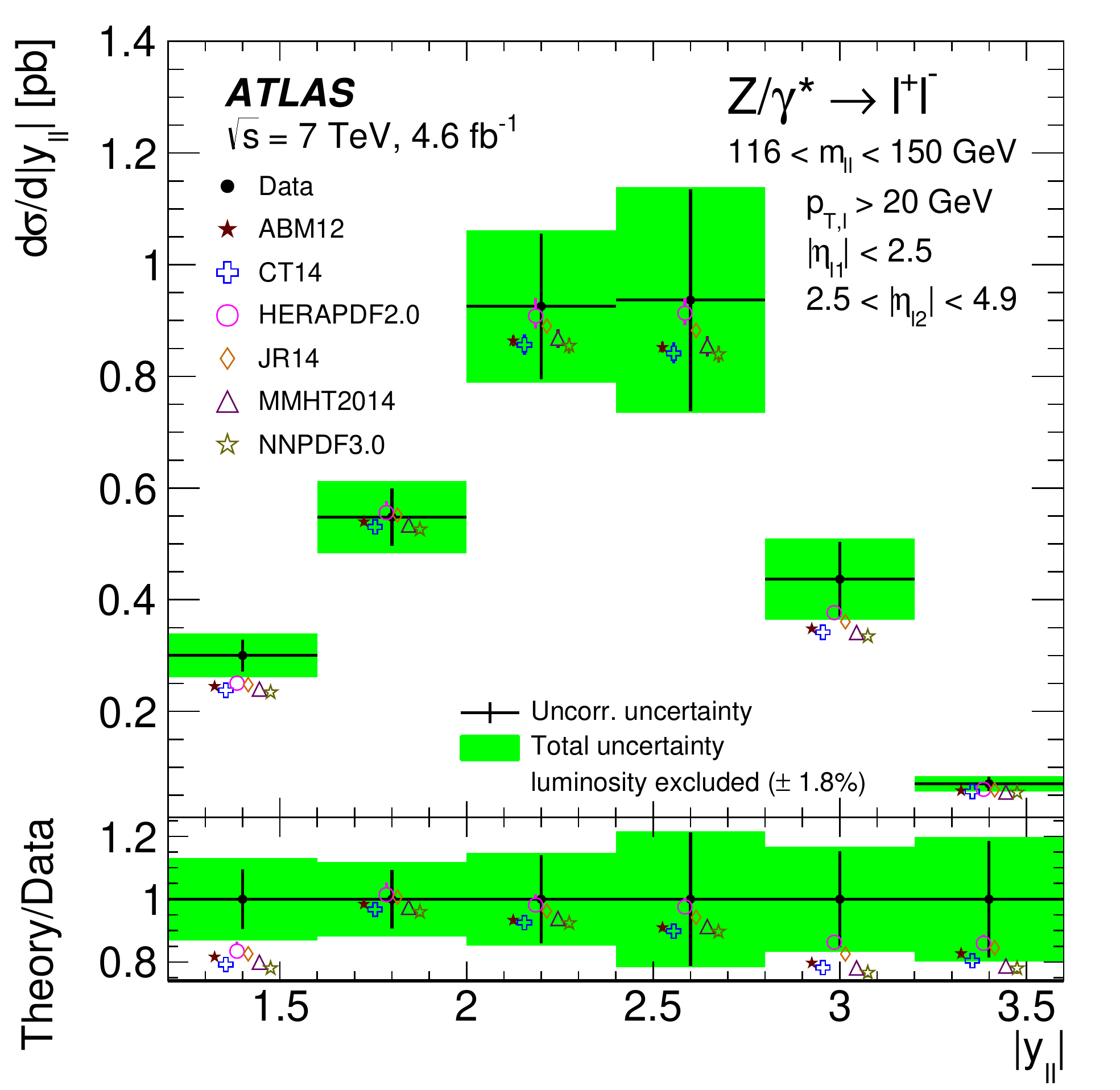}
  \end{center}
  \caption{Differential cross-section measurement $\rd\sigma/\rd|\yll|$
    for \Zgll\ in the central-rapidity low-mass region (left), the
    central-rapidity high-mass region (middle), and the
    forward-rapidity high-mass region (right). \diffnnlocap}
  \label{fig:Comb:NNLOZoffpeak}
\end{figure}

\subsection{PDF profiling results}
\label{sec:prof}
Using the profiling technique introduced in \Sec~\ref{sec:thyframe},
the agreement between data and predictions can be quantitatively
assessed. \TTab~\ref{tab:ch2} provides \chindf\ values for each
Drell--Yan data set and a number of PDFs, taking into account the
experimental uncertainties, and also including  the uncertainties
provided by the individual PDF sets. Including the full PDF
uncertainties, a satisfactory description of the data is achieved with
the CT14 PDFs, where the \chindf\ is similar to the dedicated PDF
analysis presented in \Sec~\ref{sec:qcdana}.~\footnote{The $\chi^2$ for the CT10 NNLO PDF set~\cite{Gao:2013xoa} is similar to that of CT14.}
The predictions with the
MMHT14 and \epWZ12 sets have a total $\chi^2$ increased by about ten
units compared to CT14, while the ABM12 and NNPDF3.0 predictions
exhibit a larger tension with the data.
The poorer description of the \Zgll\ data in the low mass region
$\mll=46$--$66\gev$ may reflect the enhanced theoretical uncertainties
below the $Z$ peak, which are not included in the $\chi^2$ calculation.

\begin{table}
  \begin{center}
    \scriptsize
    \begin{tabular}{lcccccc}
      \hline
      \hline
      Data set     & n.d.f. & ABM12  & CT14  & MMHT14  & NNPDF3.0   & \epWZ{}12  \\ 
      \hline
      \Wpluslnu & 11 & 11$|$21 & 10$|$26 & 11$|$37 & 11$|$18 & 12$|$15 \\ 
      \Wminuslnu & 11 & 12$|$20 & 8.9$|$27 & 8.1$|$31 & 12$|$19 & 7.8$|$17  \\ 
      \Zgll\ $(\mll=46-66\gev)$ & 6 & 17$|$21 & 11$|$30 & 18$|$24 & 21$|$22 & 28$|$36  \\ 
      \Zgll\ $(\mll=66-116\gev)$ & 12 & 24$|$51 & 16$|$66 & 20$|$116 & 14$|$109 & 18$|$26  \\ 
      Forward \Zgll\ $(\mll=66-116\gev)$ & 9 & 7.3$|$9.3 & 10$|$12 & 12$|$13 & 14$|$18 & 6.8$|$7.5  \\ 
      \Zgll\ $(\mll=116-150\gev)$ & 6 & 6.1$|$6.6 & 6.3$|$6.1 & 5.9$|$6.6 & 6.1$|$8.8 & 6.7$|$6.6  \\ 
      Forward \Zgll\ $(\mll=116-150\gev)$ & 6 & 4.2$|$3.9 & 5.1$|$4.3 & 5.6$|$4.6 & 5.1$|$5.0 & 3.6$|$3.5  \\ 
  Correlated $\chi^2$  & & 57$|$90& 39$|$123 & 43$|$167 & 69$|$157& 31$|$48  \\ 
  \hline
  Total $\chi^2$  & 61 & 136$|$222 & 103$|$290 & 118$|$396 & 147$|$351 & 113$|$159  \\ 
  \hline
  \hline
  \end{tabular}
  \caption{Values of $\chi^2$ for the predictions using various PDF
    sets split by data set with the respective number of degrees of
    freedom ($\mathrm{n.d.f.}$). The contribution of the penalty term
    constraining the shifts of experimental and theoretical correlated
    uncertainties is listed separately in the row labelled
    ``Correlated $\chi^2$'', see Eq.~\eqref{eq:chi2prof}.  The values
    to the left (right) of the vertical line refer to $\chi^2$ when
    the PDF uncertainties are included (excluded) in the evaluation.}
    \label{tab:ch2}
  \end{center}
\end{table}

Profiling PDFs, by introducing the data presented
here, provides a shifted set of parton distributions with generally
reduced uncertainties. Given the previous
observation~\cite{Aad:2012sb} of an enlarged strangeness fraction of
the light sea, the effect of the data on the strange-quark
distribution is examined. This is illustrated in
\Fig~\ref{fig:profCT10MMHTrs}, where the ratio
$R_s(x)=(s(x)+\bar{s}(x))/(\bar{u}(x)+\bar{d}(x))$ is shown for two selected PDF
sets, MMHT14 and CT14, before and after profiling, at a scale of $Q^2=1.9\gev^2$.
The uncertainties
of $R_s$ are seen to be significantly reduced and the central values,
at $x \simeq 0.023$, increased towards unity, supporting the
hypothesis of an unsuppressed strange-quark density at low $x$.

\begin{figure}[tbp]
  \begin{center}
    \includegraphics[width=0.45\textwidth]{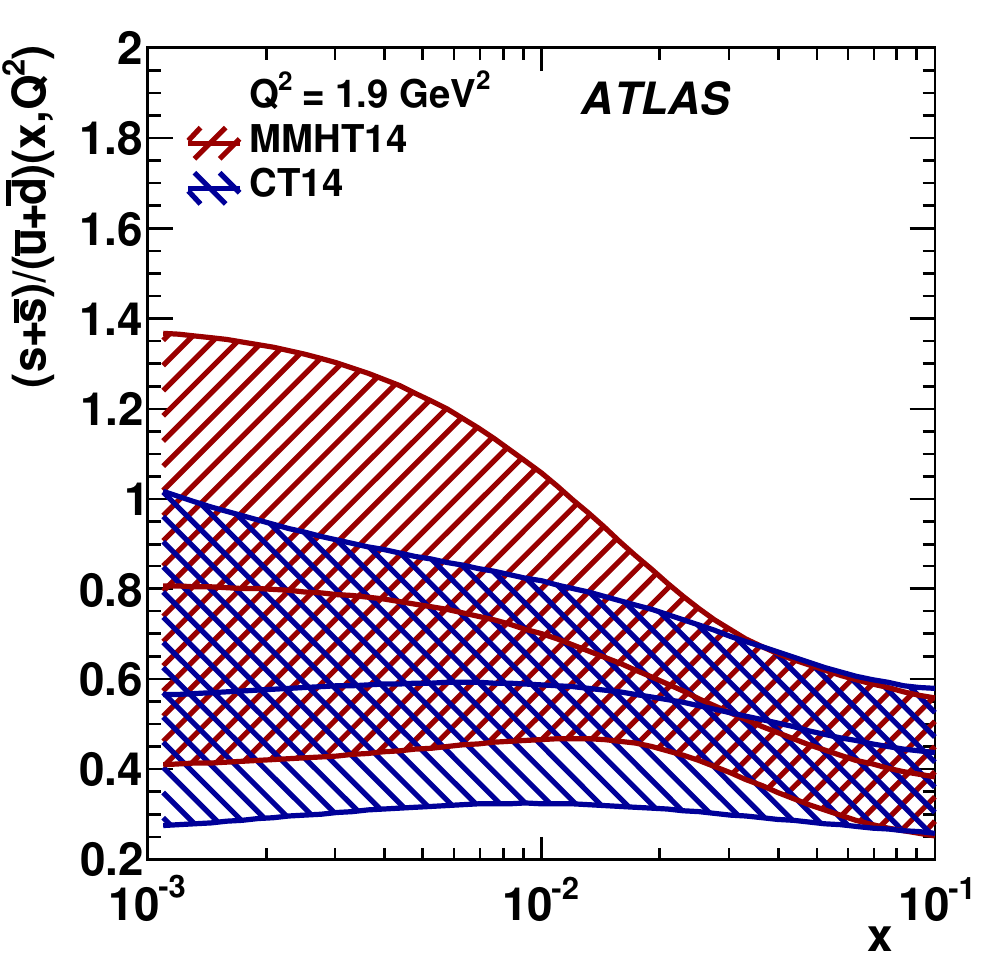}
    \includegraphics[width=0.45\textwidth]{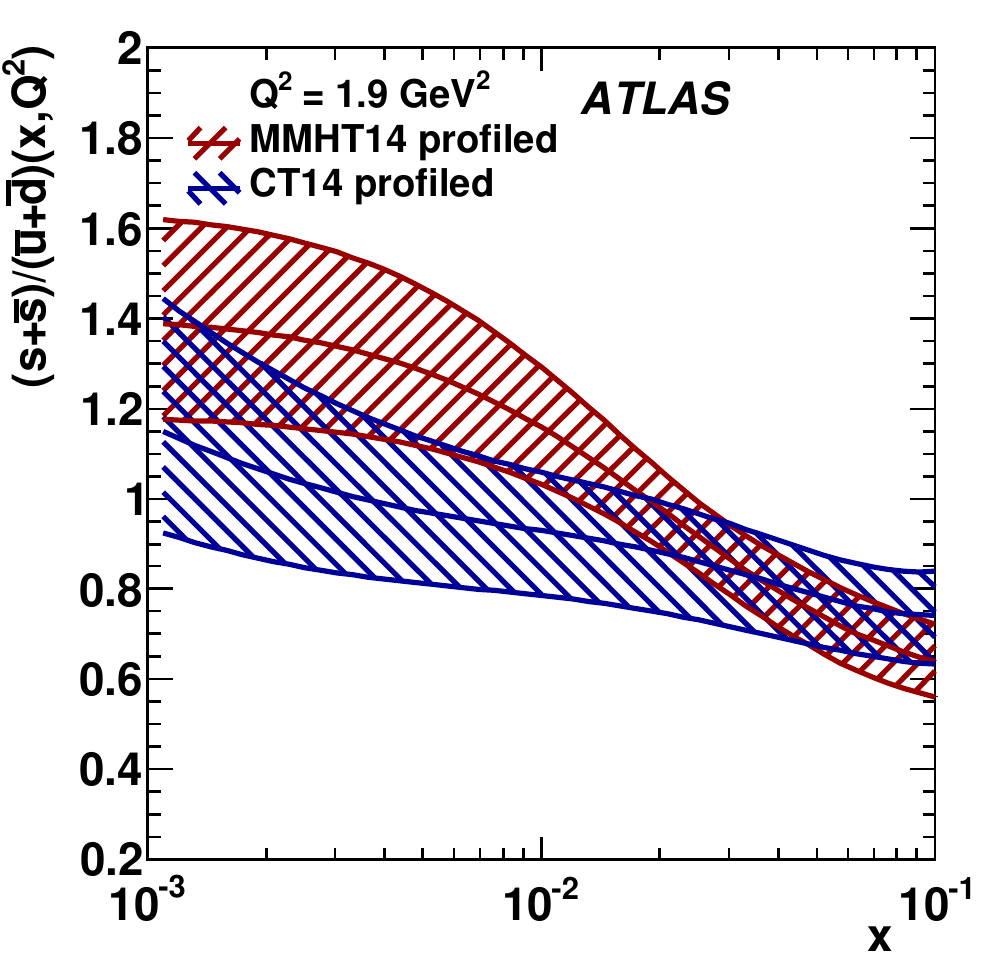}  
    \caption{Ratio $R_s(x) = (s(x)+\bar{s}(x))/(\bar{u}(x)+\bar{d}(x))$ as a function
      of Bjorken-$x$ at a scale of $Q^2=1.9\gev^2$ for the original
      MMHT14 and CT14 PDF sets (left) and for the MMHT14 and CT14 sets
      when profiled with the new $W,~Z$ differential cross-section
      data (right).}
    \label{fig:profCT10MMHTrs}
  \end{center}
\end{figure}

The sea-quark distributions, $x\bar{u}$, $x\bar{d}$ and $x\bar{s}$,
before and after profiling with the MMHT14 set, are shown in
\Fig~\ref{fig:udsmmht}. The strange-quark distribution is
significantly increased and the uncertainties are reduced. This in
turn leads to a significant reduction of the light sea,
$x\bar{u}+x\bar{d}$, at low $x$, resulting from the tight constraint
on the sum $4 \bar{u} + \bar{d} + \bar{s}$ from the precise
measurement of the proton structure function $F_2$ at HERA.  Some
reduction of the uncertainty is also observed for the valence-quark
distributions, $x\uv$ and $x\dv$, as is illustrated in
\Fig~\ref{fig:profCTMMHTval} for the CT14 and MMHT14 sets.

\begin{figure}[tbp]
  \begin{center}
    \includegraphics[width=0.325\textwidth]{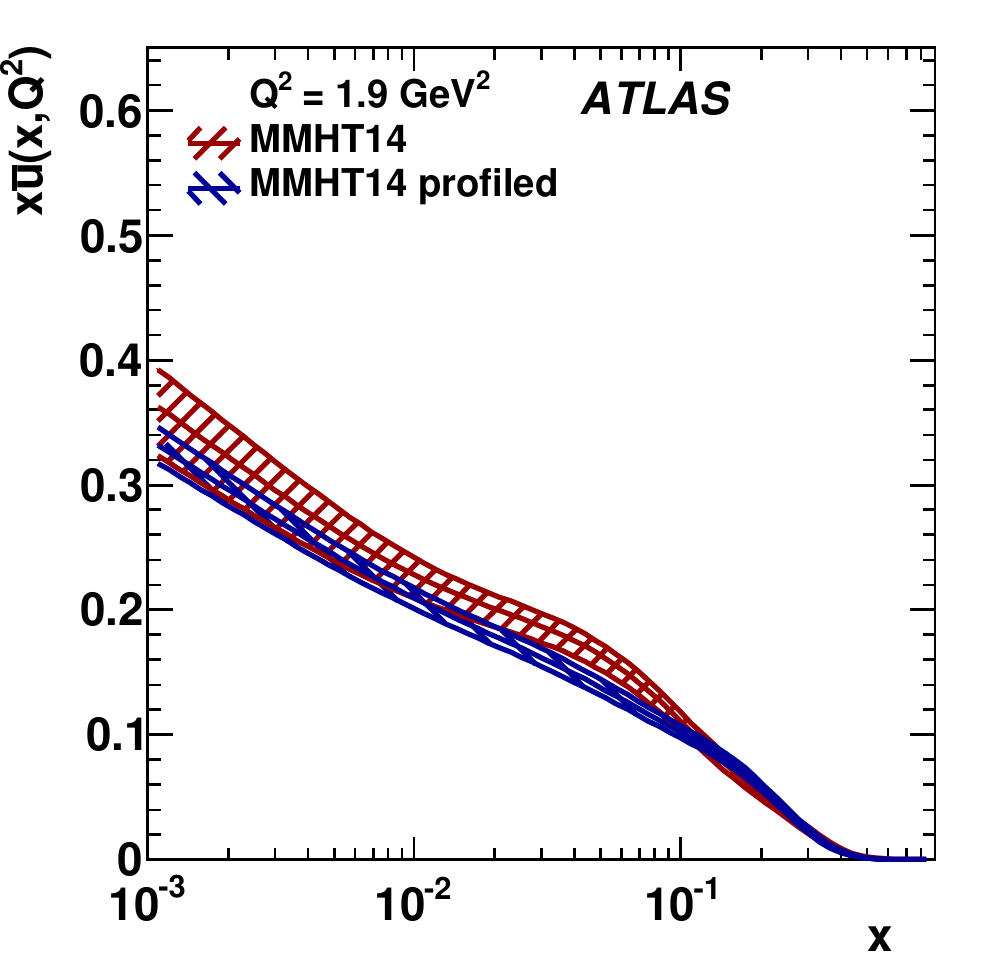}
    \includegraphics[width=0.325\textwidth]{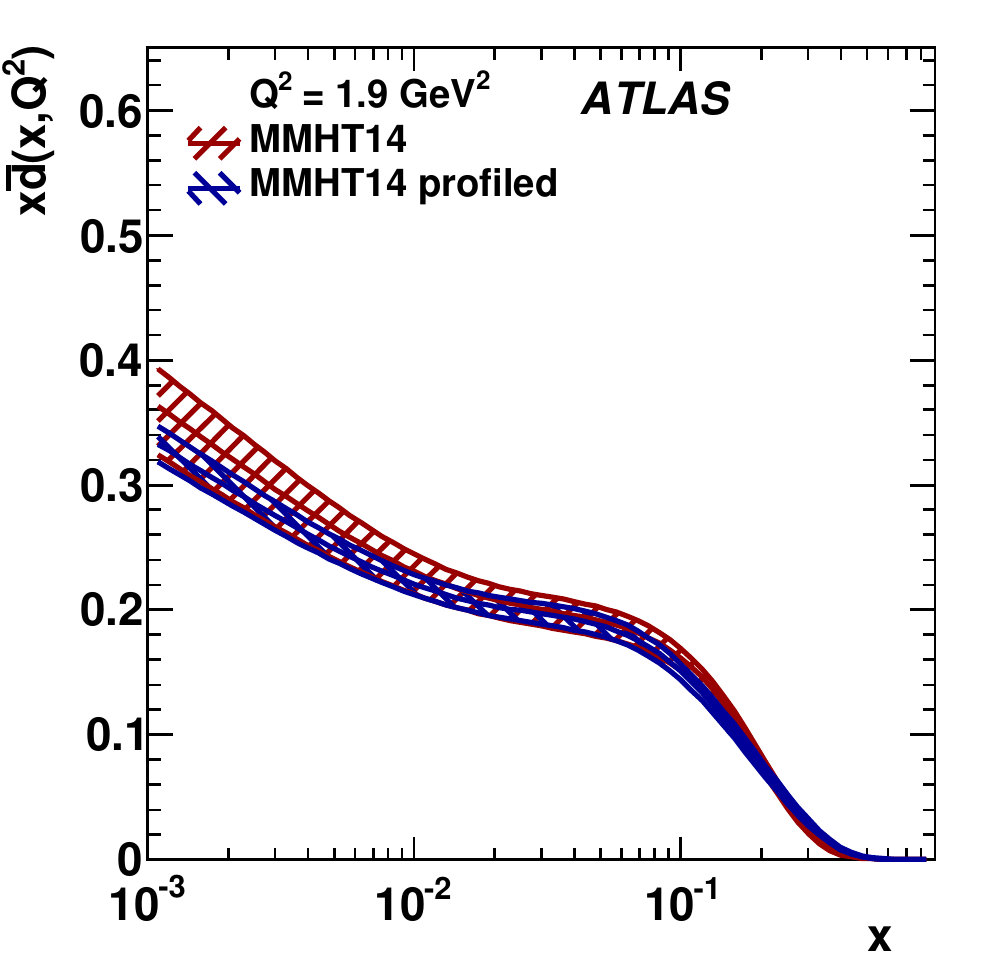}
    \includegraphics[width=0.325\textwidth]{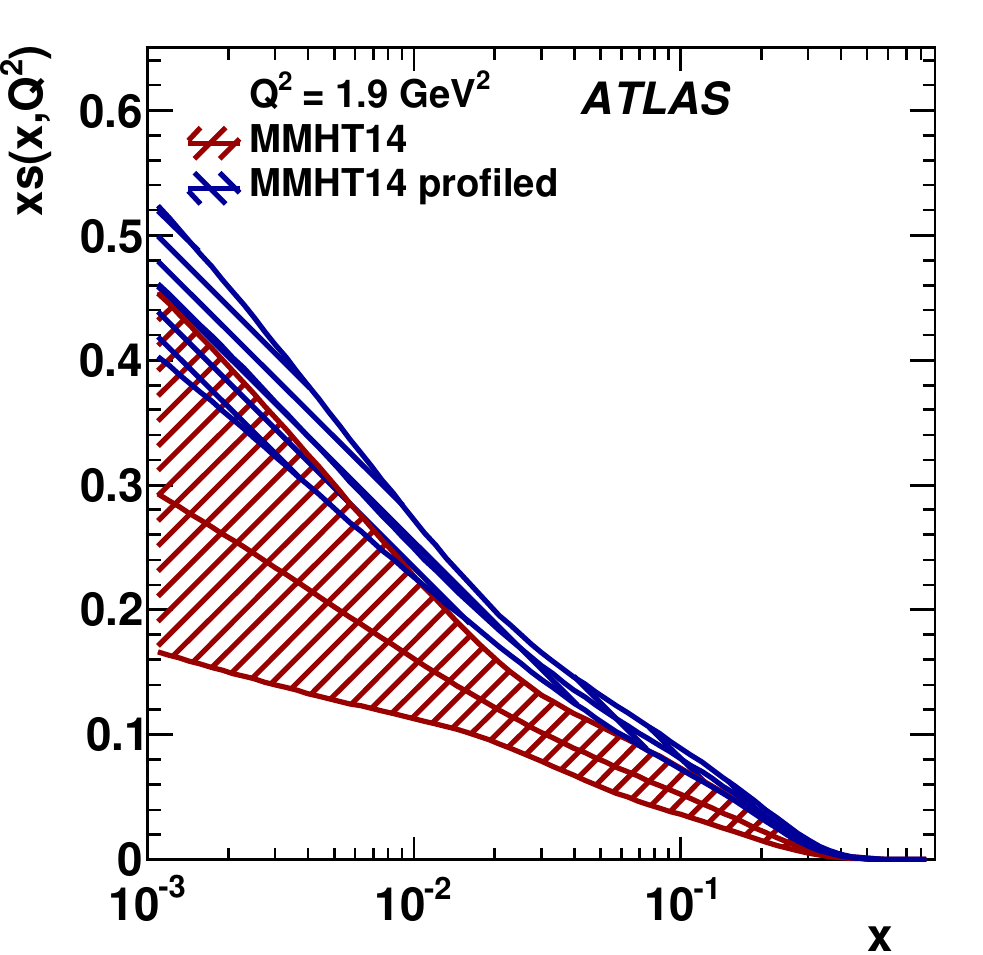}  
    \caption{Distribution of $x\bar{u}$ (left), $x\bar{d}$ (middle)
      and $xs$ (right) PDFs as a function of Bjorken-$x$ at a scale of
      $Q^2=1.9\gev^2$ for the MMHT14 PDF set before and after
      profiling.}
    \label{fig:udsmmht}
  \end{center}
\end{figure}

\begin{figure}[tbp]
  \begin{center}
    \includegraphics[width=0.4\textwidth]{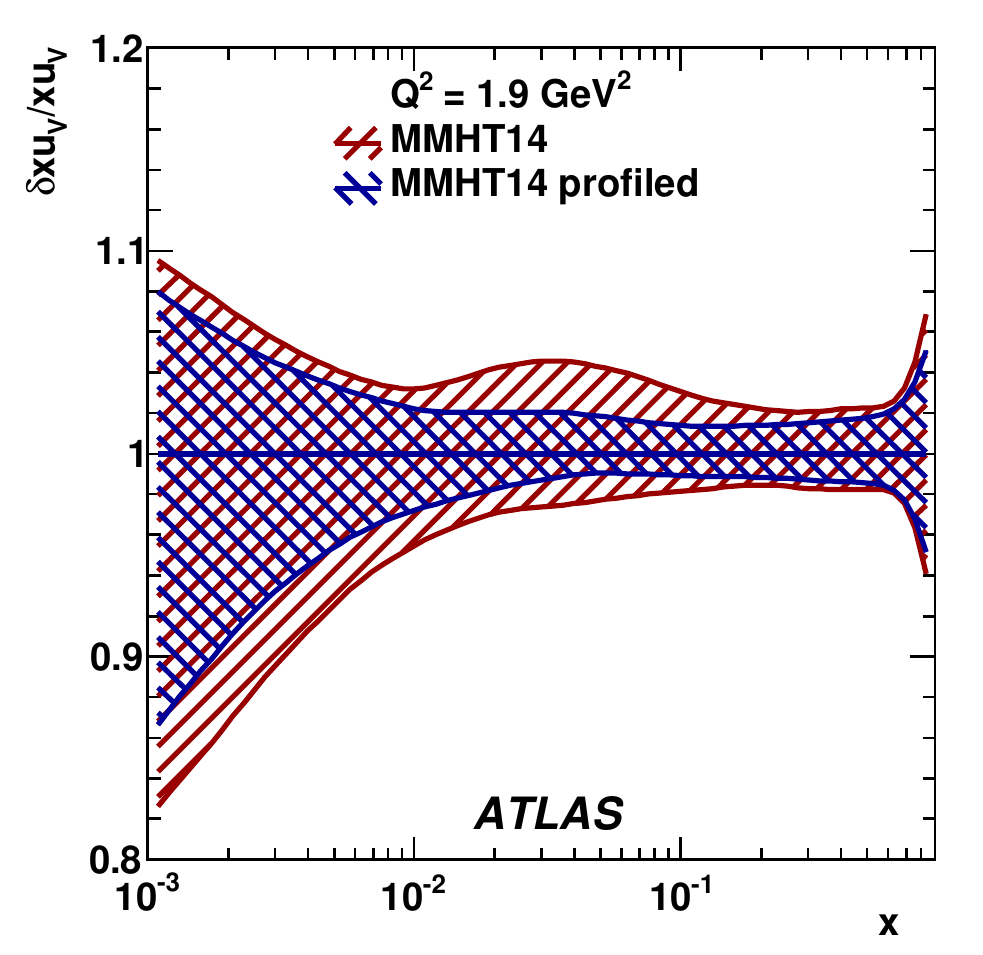}
    \includegraphics[width=0.4\textwidth]{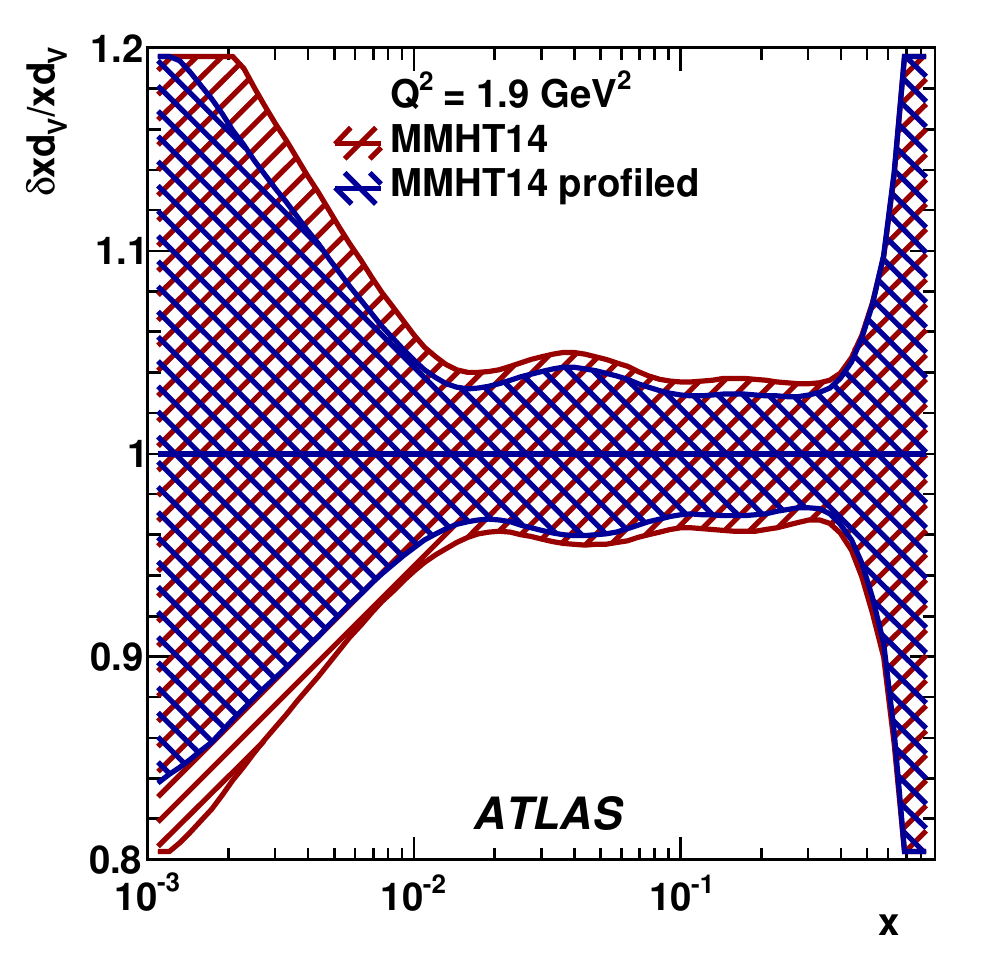}
    \includegraphics[width=0.4\textwidth]{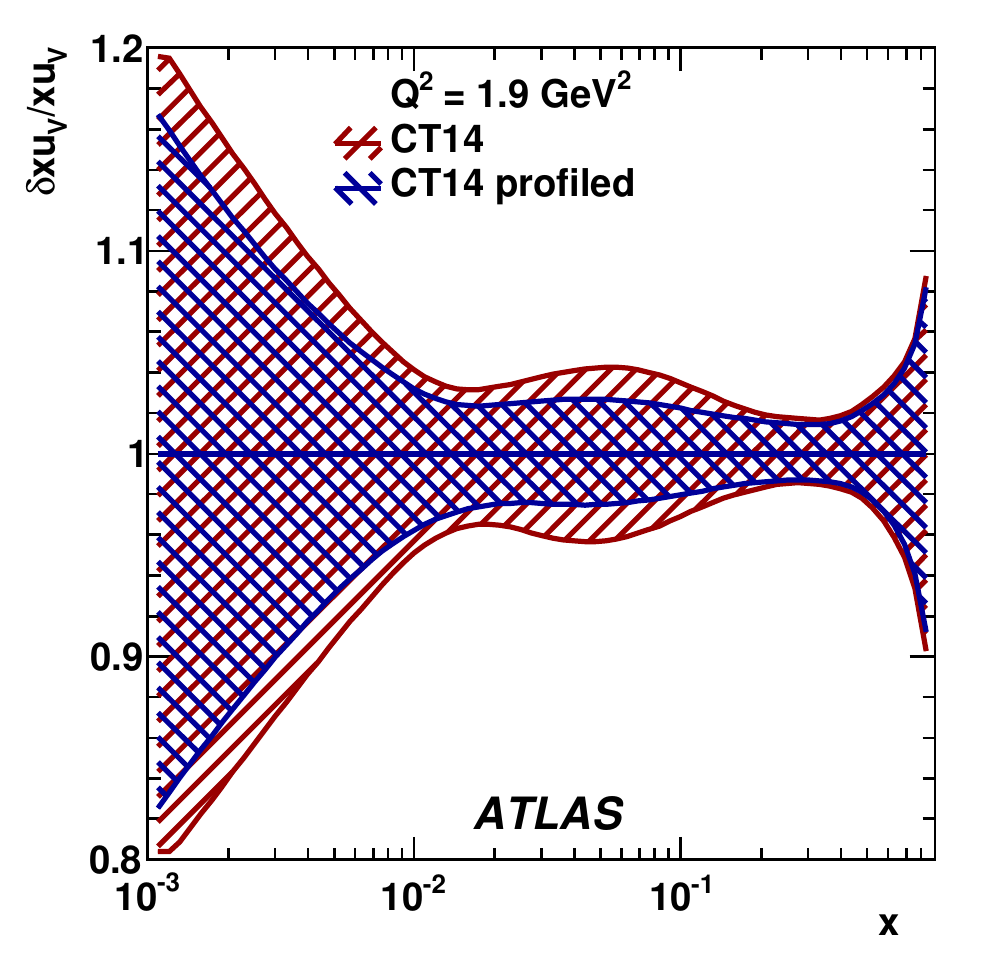}
    \includegraphics[width=0.4\textwidth]{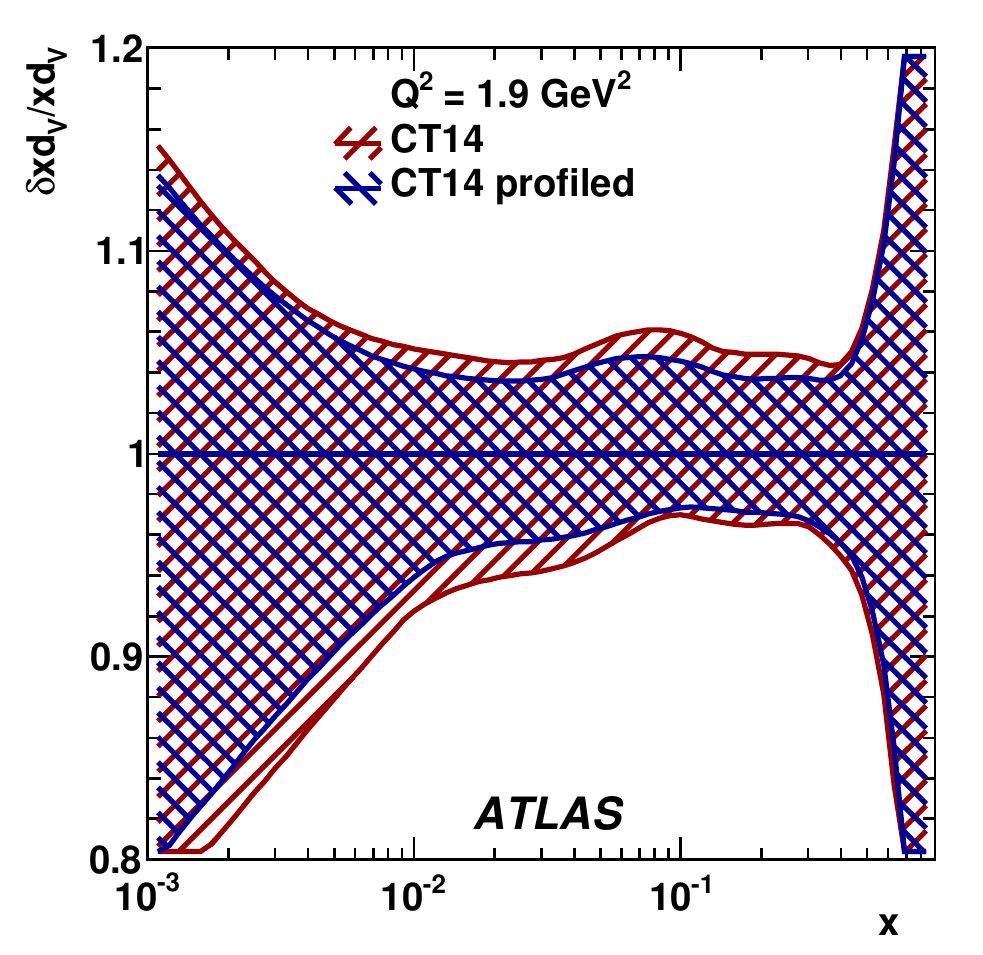} 
    \caption{Effect of profiling on the relative uncertainties of the
      valence up-quark distribution $\delta x\uv(x)/x\uv(x)$ (left)
      and the valence down-quark distribution $\delta x\dv(x)/x\dv(x)$
      (right) as a function of Bjorken-$x$ at a scale of
      $Q^2=1.9\gev^2$. The top row shows the MMHT14 PDF set and the
      bottom row shows the CT14 PDF set.}
    \label{fig:profCTMMHTval}
  \end{center}
\end{figure}


\section{QCD analysis}
\label{sec:qcdana}
In this section, the  differential
Drell--Yan production cross sections of \Wpmlnu\ and
\Zgll~$(\ell=e,\mu)$ are studied in combination with the final NC and
CC deep inelastic scattering (DIS) HERA I+II
data~\cite{Abramowicz:2015mha} within the framework of perturbative QCD. The
Drell--Yan and DIS reactions are theoretically very well understood
processes for such an analysis, and $ep$ and $pp$ collider data are
particularly suitable because of the absence of nuclear corrections
and negligible higher-twist effects. The HERA data alone can provide a
full set of PDFs with certain assumptions~\cite{Abramowicz:2015mha}. Adding the
ATLAS data provides more sensitivity to the flavour
composition of the quark sea as well as to the valence-quark
distributions at lower $x$. The HERA and ATLAS data are used to obtain a new set
of PDFs, termed \epWZ16. Following the previous, similar QCD fit
analysis in Ref.~\cite{Aad:2012sb}, special attention is given to the
evaluation of the strange-quark distribution, which was found to be
larger than previous expectations based on dimuon data in DIS
neutrino--nucleon scattering. The enhanced precision of the present
data also permits a competitive determination of the magnitude of the CKM  matrix element \Vcs.

\subsection{Fit framework}
\label{sec:fitpro}
The present QCD fit analysis is performed using the xFitter
platform~\cite{HERAFitter,Alekhin:2014irh} which uses
QCDNUM~\cite{Botje:2010ay} for PDF evolution and
MINUIT~\cite{James:1975dr} for minimization. Each step is
cross-checked with an independent fit program as also used in
Ref.~\cite{Abramowicz:2015mha}.

Predictions for the differential CC and NC Drell--Yan cross sections
calculated at fixed order in QCD at NNLO accuracy and with NLO
electroweak corrections are described in
\Sec~\ref{sec:thyframe}. These calculations, however, cannot be used
directly in an iterative fit because of the large computational effort required to
produce even a single prediction. Therefore, the xFitter package
uses the APPLGRID~\cite{Carli:2010rw} code interfaced to the
predictions of MCFM~\cite{Campbell:2010ff} for the fast
calculation at fixed-order NLO accuracy in QCD. The improved
NNLO QCD and NLO EW predictions 
discussed above are incorporated in the fit with
additional \kfactors\ defined as
\begin{equation}
  K_\mathrm{f} = \frac{\sigma_\mathrm{NNLO\;QCD}^\mathrm{NLO\;EW}(\mathrm{DYNNLO})}{\sigma_\mathrm{NLO\;QCD}^\mathrm{LO\;EW} (\mathrm{APPLGRID})}\,.
\end{equation}
All predictions are calculated in the respective
fiducial phase space of the experimental data. The
\kfactors\ are applied bin-by-bin and estimated using the same PDF,
\epWZ12, in both the numerator and denominator. They are typically
close to unity within $1$--$2$\,\%, but are up to $6\%$ in the low-mass
region, $\mll = 46$--$66\GeV$. These higher-order corrections are
calculated using DYNNLO 1.5 and cross-checked with FEWZ3.1.b2 as
detailed in \Sec~\ref{sec:thyframe}.  The \kfactors\ are available
as xFitter format files.

The QCD analysis uses the full set of ATLAS \Wpmln\ and \Zgll\ data,
as described in the preceding sections, together with the combined
H1 and ZEUS $ep$ data~\cite{Abramowicz:2015mha}.  There are $131$ sources of
experimental correlated systematic uncertainty for the ATLAS data and
$167$ sources of experimental correlated systematic uncertainty for
the HERA data. The statistical precision of the \kfactors\, is
typically $<0.1\%$ per measurement bin and is accounted for as an extra uncorrelated systematic uncertainty.

The nominal fit analysis is performed using the variable flavour
number scheme from
Refs~\cite{Thorne:1997ga,Thorne:2006qt}.\footnote{The choice of the
  heavy-flavour scheme is especially relevant for the HERA
  measurements at lower $Q^2$, see
  Ref.~\cite{Abramowicz:2015mha}.} The heavy-quark distributions are
generated dynamically above the respective thresholds chosen as $m_c =
1.43\GeV$ for the charm quark and as $m_b=4.5\GeV$ for the bottom
quark, corresponding to the recent heavy-quark differential
cross-section measurements at HERA~\cite{Abramowicz:2015mha}. The PDFs
are parameterized at the starting scale $Q_0^2=1.9\GeV^2$, chosen to
be below the charm-mass threshold as required by QCDNUM. The
strong coupling constant at the $Z$ mass is set to be
$\alphas(m_Z)=0.118$, a value conventionally used by recent PDF
analyses.

Besides the gluon distribution, $xg$, the valence and anti-quark
distributions $x\uv$, $x\dv$, $x\bar{u}$, $x\bar{d}$, $x\bar{s}$, are
parameterized at the starting scale $Q_0^2$, assuming that the sea
quark and anti-quark distributions are the same. These distributions
are evolved to the scale of the measurements and convolved with
hard-scattering coefficients to obtain the theoretical cross-section
predictions.  The prediction is then confronted with the data through the
$\chi^2$ function,
\begin{eqnarray}
\nonumber \chi^2(\vec{b}_{\mathrm{exp}})& = &  \nonumber \sum_{i=1}^{N_\mathrm{data}} \frac{\textstyle \left[ \sigma^\mathrm{exp}_i  -  \sigma^\mathrm{th}_i (1 - \sum_j \gamma^\mathrm{exp}_{ij} b_{j,\mathrm{\exp}}) \right]^2}{\Delta_i^2}\\
& + &  \sum_{j=1}^{N_\mathrm{exp. sys.}} b_{j,\mathrm{exp}}^2 + \sum_{i=1}^{N_\mathrm{data}} \ln \frac {\Delta_i^2}{\left(\delta_{i,\mathrm{sta}}\sigma^\mathrm{exp}_i\right)^2 + \left(\delta_{i,\mathrm{unc}}\sigma^\mathrm{exp}_i\right)^2} \,,   \label{eq:chi2fit}
\end{eqnarray}
which is defined similarly to Eq.~\eqref{eq:chi2prof} and accounts for the various sources of correlated and uncorrelated uncertainties. The
definition of $\Delta_i^2$ with scaled uncertainties is given
by Eq.~\eqref{eq:deltai} and
discussed there. This particular form is of higher
importance in this context, as the relative uncertainties of the HERA
data points can be large in parts of the phase space. The use of this
form of $\Delta_i^2$ leads to a logarithmic term, introduced in
Ref.~\cite{Aaron:2012qi}, arising from the likelihood transition to
$\chi^2$. The contribution to the $\chi^2$ from the last two sums
related to the nuisance parameter constraints and the logarithmic term
is referred to as ``correlated + Log penalty'' later.

The optimal functional form for the parameterization of each parton
distribution is found through a parameter scan requiring $\chi^2$
saturation~\cite{Adloff:2000qk,Aaron:2009aa}. The general form is of
the type $A_i x^{B_i}(1-x)^{C_i} P_i(x)$ for each parton flavour
$i$. The scan starts with the contribution of the factors
$P_i(x)=(1+D_i x+E_ix^2)e^{F_i x}$ set to unity by fixing the parameters $D_i=E_i=F_i=0$
for all parton flavours. The parameter $A_g$ is
constrained by the momentum sum rule relating the sum of the quark and
gluon momentum distribution integrals, while the parameters $A_{\uv}$
and $A_{\dv}$ are fixed by the up and down valence-quark number sum
rules. The assumption that $\bar u=\bar d$ as $x\to 0$ implies that
$A_{\bar{u}} = A_{\bar{d}}$ and $B_{\bar{u}} = B_{\bar{d}}$.  The
procedure thus starts with ten free parameters and, subsequently,
additional parameters are introduced one at a time.\footnote{An
  exception is the introduction of a negative term in the gluon
  parameterization, $-A'_g x^{B'_g} (1-x)^{C'_g}$, for which two
  parameters, $A'_g$ and $B'_g$, are introduced simultaneously. As in
  Ref.~\cite{Abramowicz:2015mha}, the parameter $C'_g$ is fixed to a
  large value, chosen to be $C'_g=25 \gg C_g$ to suppress the
  contribution at large $x$.} A parameterization with $15$ variables
is found to be sufficient to saturate the $\chi^2$ value after minimization, i.e. no
further significant $\chi^2$ reduction is observed when adding further
parameters. The final parameterization used to describe the parton
distributions at $Q^2=Q_0^2$ is:
\begin{eqnarray}
  x \uv(x) &=&  A_{\uv} x^{B_{\uv}} (1-x)^{C_{\uv}} ( 1 + E_{\uv} x^2)\, ,\nonumber \\
  x \dv(x) &=&  A_{\dv} x^{B_{\dv}} (1-x)^{C_{\dv}}              \, ,\nonumber \\
  x \bar{u} (x) &=& A_{\bar{u}} x^{B_{\bar{u}}} (1-x)^{C_{\bar{u}}}\, , \nonumber \\  
  x \bar{d} (x) &=& A_{\bar{d}} x^{B_{\bar{d}}} (1-x)^{C_{\bar{d}}}\, , \nonumber \\
  x g(x)   &=& A_g x^{B_g} (1-x)^{C_g} - A'_gx^{B'_g}(1-x)^{C'_g}\, , \nonumber \\
  x \bar{s}(x) &=& A_{\bar{s}} x^{B_{\bar{s}}} (1-x)^{C_{\bar{s}}} \, \label{eq:pdf},  
\end{eqnarray}  
where $A_{\bar{u}}=A_{\bar{d}}$ and
$B_{\bar{s}}=B_{\bar{d}}=B_{\bar{u}}$. Given the enhanced sensitivity
to the strange-quark distribution through the ATLAS data,
$A_{\bar{s}}$ and $C_{\bar{s}}$ appear as free parameters, assuming $s
= \bar{s}$.  The experimental data uncertainties are propagated to the
extracted QCD fit parameters using the asymmetric Hessian method based
on the iterative procedure of Ref.~\cite{Pumplin:2000vx}, which
provides an estimate of the corresponding PDF uncertainties.

\subsection{Fit results}
\label{sec:fitres}
The $\chi^2$ values characterizing the NNLO QCD fit to the ATLAS
Drell--Yan and HERA DIS data are listed in
\Tab~\ref{tab:nnlo_results}. The fit describes both the HERA and
the ATLAS data well. Most of the correlated systematic uncertainties
are shifted by less than one standard deviation and none are shifted
by more than twice their original size in the fit. The overall
normalization is shifted by less than half of the luminosity
uncertainty of $1.8\%$. The only significant departure from a partial
$\chindf \sim 1$ is seen for the low-mass \Zgll\ data.  Here
the \kfactors\ are large, and the theoretical uncertainties, such as
the FEWZ-DYNNLO difference, are sizable.  As described below, this
part of the data has little influence on the extracted PDFs.

\begin{table}
  \begin{center}
    \begin{tabular}{lc}
      \hline
      \hline
      Data set     & \epWZ{}16  \\ 
      & \chindf  \\ 
      \hline
      ATLAS \Wpluslnu & 8.4  / 11  \\ 
      ATLAS \Wminuslnu & 12.3  / 11  \\ 
      ATLAS \Zgll\ $(\mll=46$--$66\gev)$ & 25.9 / 6  \\ 
      ATLAS \Zgll\ $(\mll=66$--$116\gev)$ & 15.8 / 12  \\ 
      ATLAS forward \Zgll\ $(\mll=66$--$116\gev)$ & 7.4  / 9  \\ 
      ATLAS \Zgll\ $(\mll=116$--$150\gev)$ & 7.1 / 6  \\ 
      ATLAS forward \Zgll\ $(\mll=116$--$150\gev)$ & 4.0  / 6  \\ 
      ATLAS Correlated + Log penalty & 27.2  \\ 
      \hline
      ATLAS Total &  108 / 61 \\
      \hline 
      HERA I+II CC $e^+p$ & 44.3 / 39  \\ 
      HERA I+II CC $e^-p$ & 62.7 / 42  \\ 
      HERA I+II NC $e^-p$ & 222 / 159  \\ 
      HERA I+II NC $e^+p$ &      838 / 816 \\
      HERA Correlated + Log penalty & 45.5  \\ 
      \hline
      HERA Total &  1213 / 1056  \\
      \hline
      Total  & 1321 / 1102  \\ 
      \hline
      \hline
    \end{tabular}
  \end{center}
  \caption{Quality of the QCD fit, expressed as the \chindf,
    to the final DIS HERA data and the
    ATLAS differential \Wlnu\ and \Zgll\ cross-section measurements.
    This NNLO fit is the base for the new \epWZ{}16 set of PDFs.}
  \label{tab:nnlo_results}
\end{table}

\FFig~\ref{fig:nnlo_resultsW} shows the \Wpluslnu\ and \Wminuslnu\ lepton pseudorapidity distributions, which are well described by the
fit. The fit results are presented before (solid) and after (dashed)
application of the shifts accounting for the correlated systematic
uncertainties of the data.
\FFig~\ref{fig:nnlo_resultsZ} presents the new ATLAS \Zgll\ measurements
in the three different mass bins, further subdivided into the central
and forward measurements. Also these data are well described by the QCD fit.

\begin{figure}[ptb]
  \begin{center}
    \includegraphics[width=0.42\textwidth]{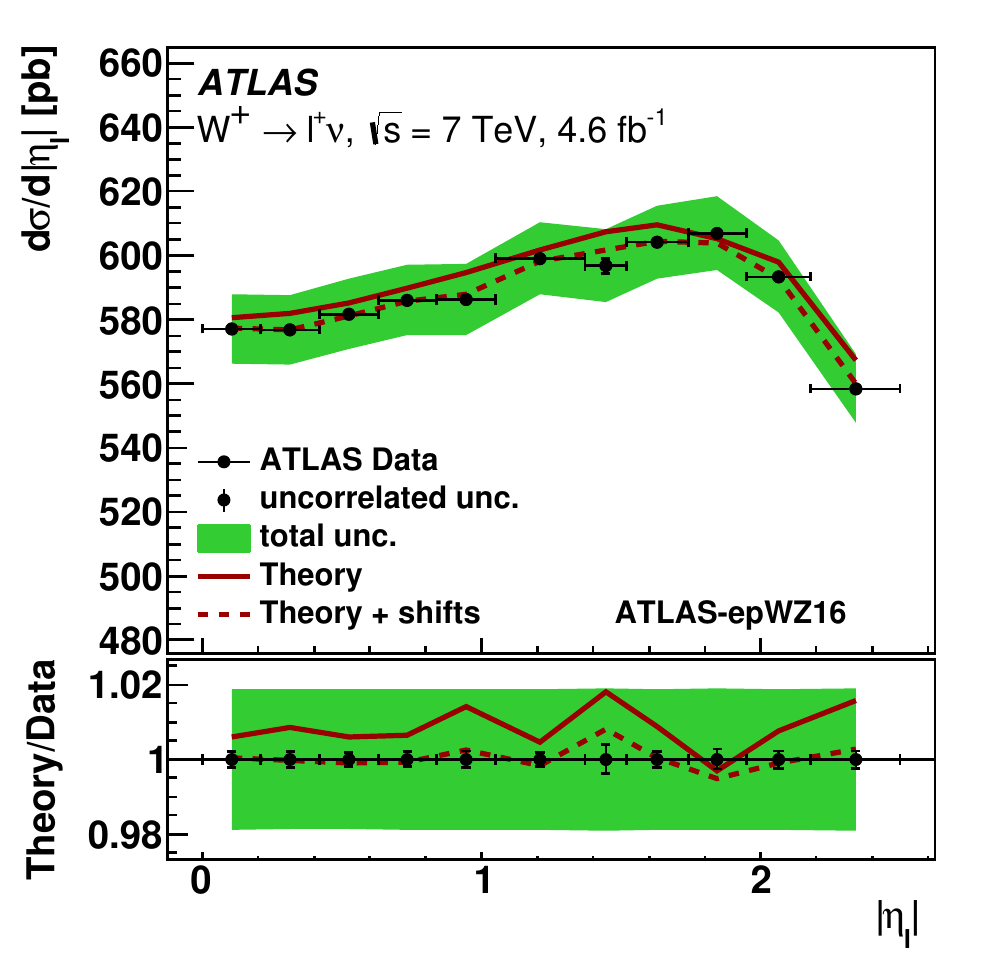}
    \includegraphics[width=0.42\textwidth]{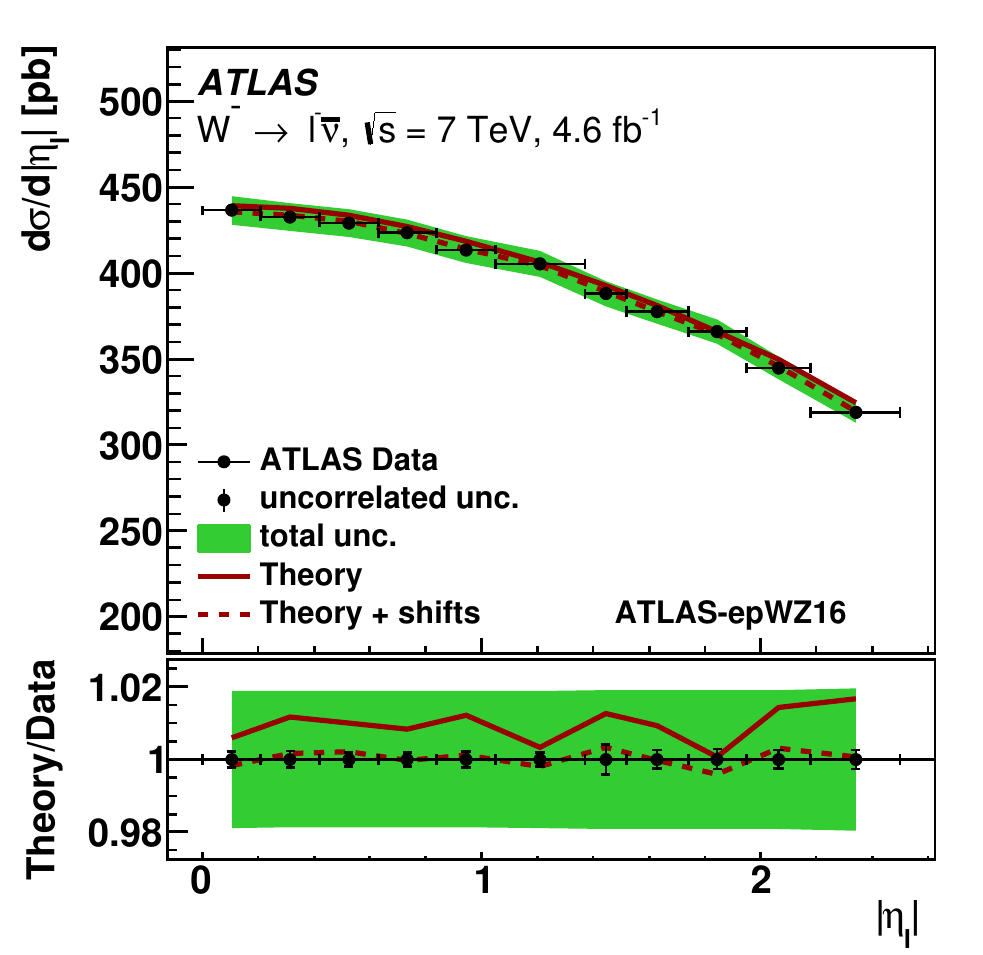}
  \end{center}
  \caption{Differential cross-section measurements for \Wpluslnu\
    (right) and \Wminuslnu\ (left) compared to the predictions of the
    QCD fit. The predictions are shown before (solid lines) and after
    (dashed lines) the shifts due to the correlated uncertainties are
    applied. The lower box of each plot shows the ratio of the
    theoretical calculations
    to the data.}
  \label{fig:nnlo_resultsW}
\end{figure}

\begin{figure}[ptb]
  \begin{center}
    \includegraphics[width=0.42\textwidth]{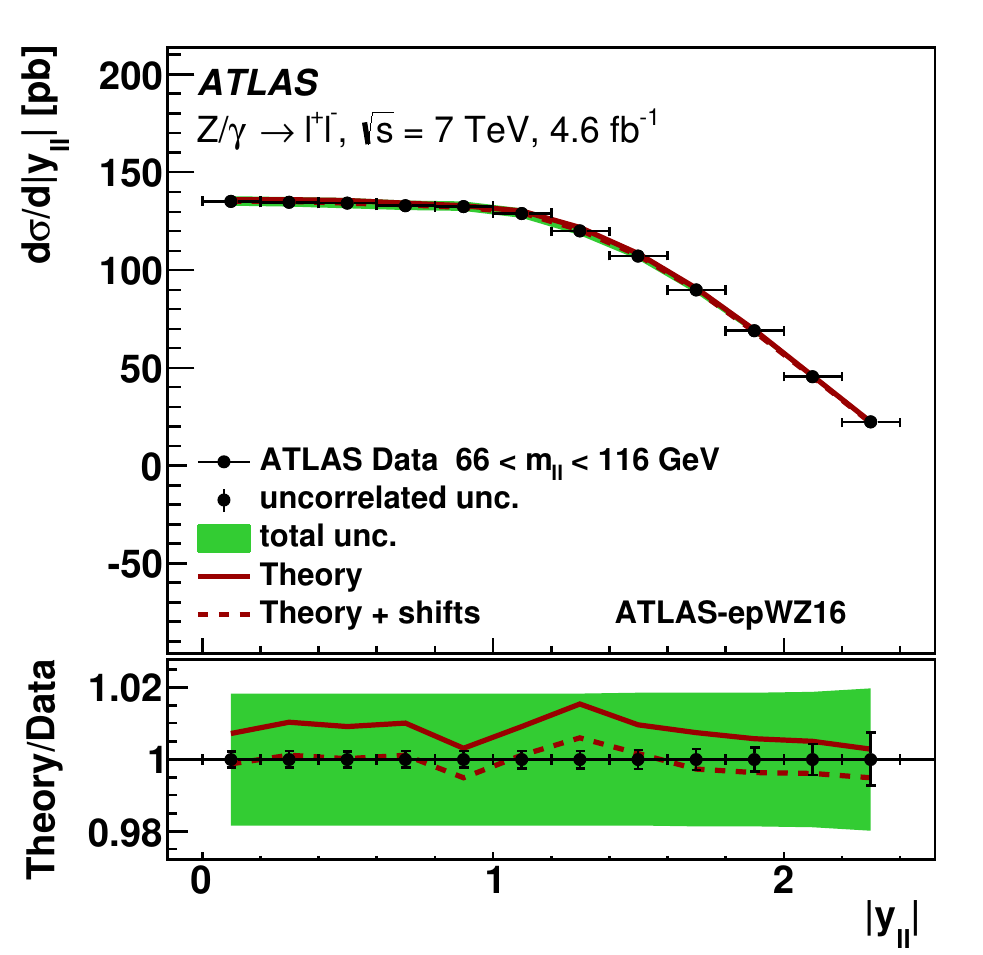}%
    \includegraphics[width=0.42\textwidth]{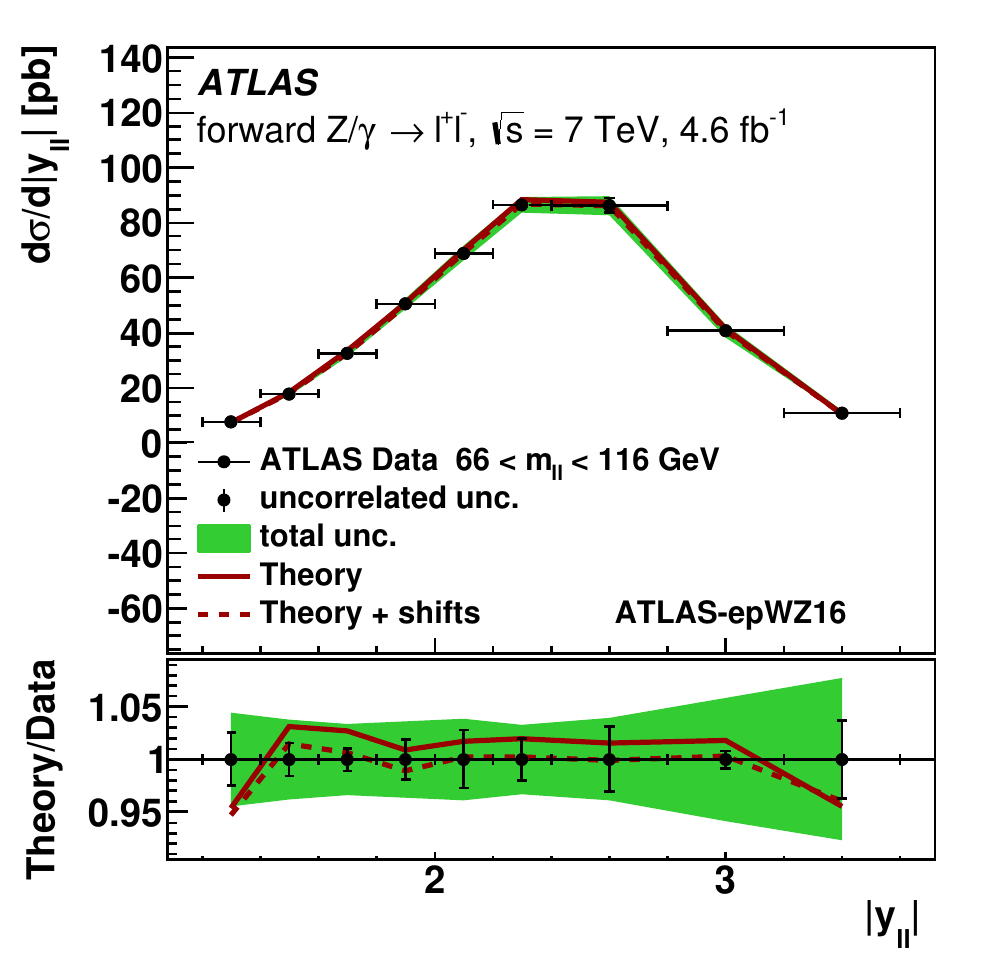}\\
    \includegraphics[width=0.42\textwidth]{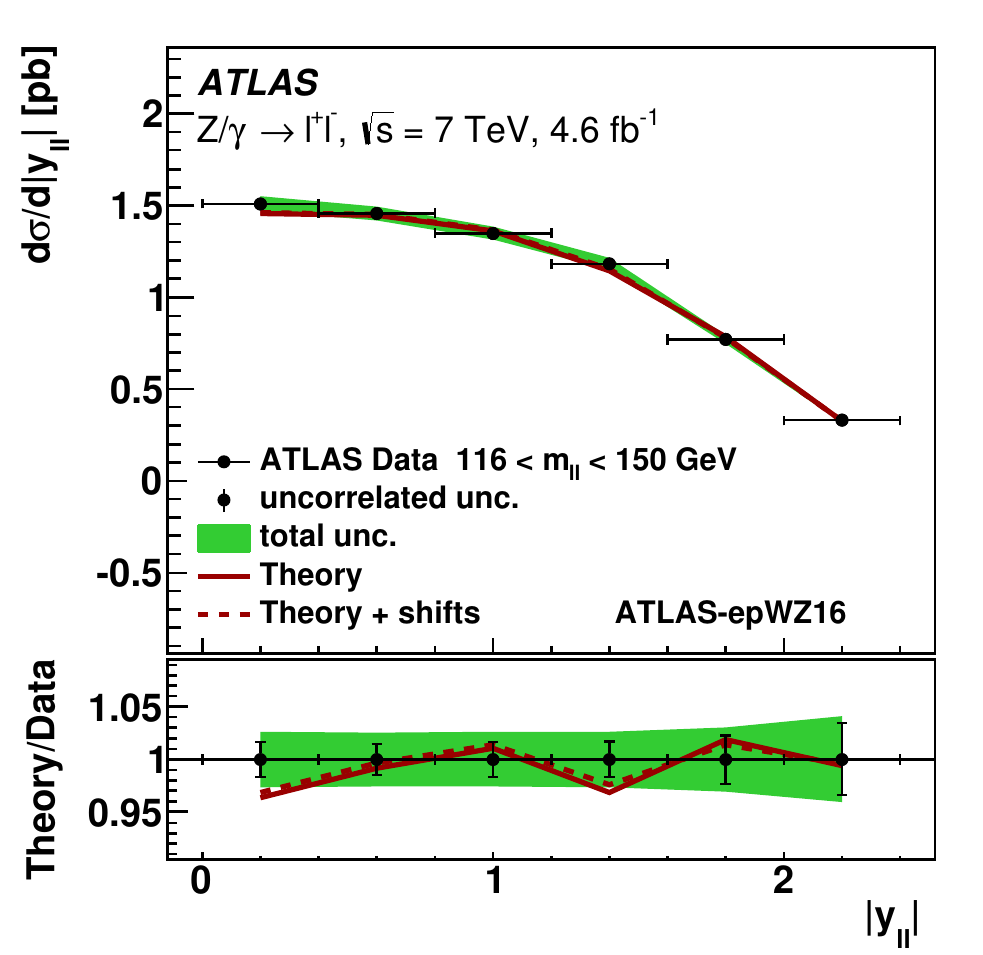}%
    \includegraphics[width=0.42\textwidth]{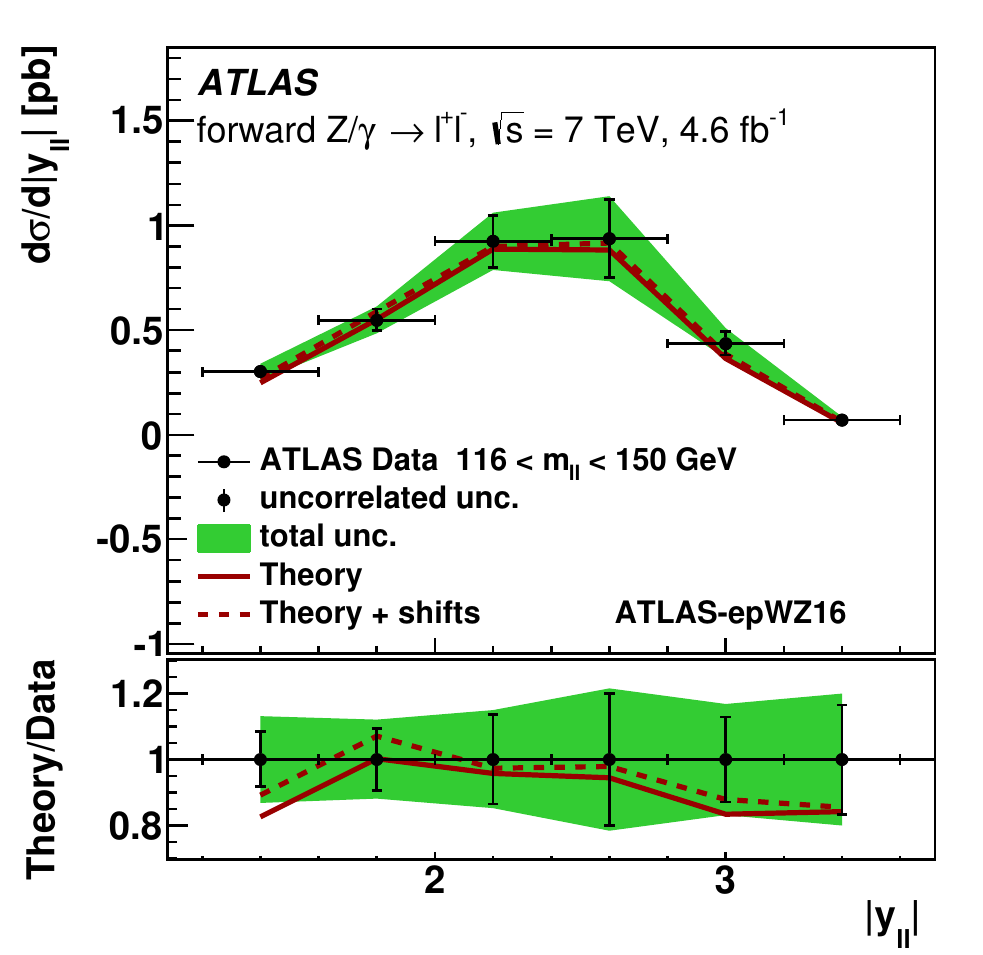}\\
    \includegraphics[width=0.42\textwidth]{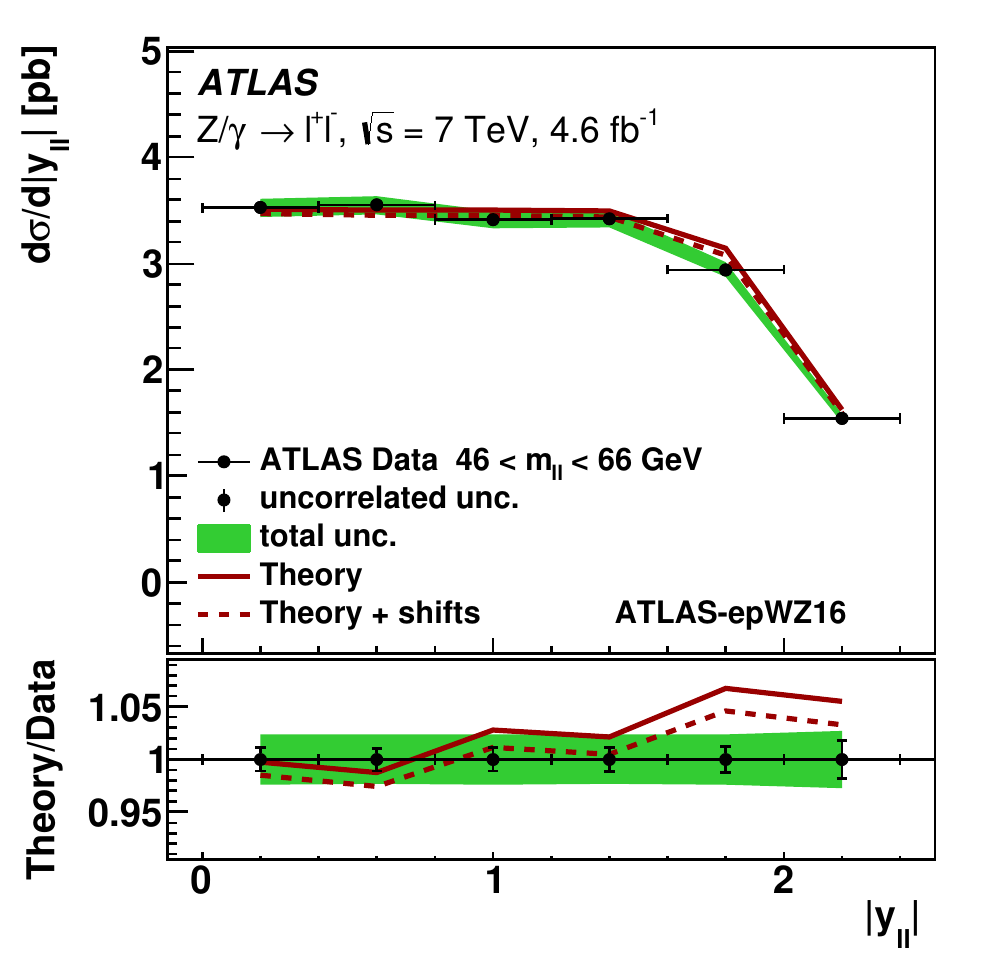}
  \end{center}
  \caption{Differential $\rd\sigma/\rd|\yll|$ cross-section measurement
    for \Zgll\ in the $Z$-peak region (upper row), as well as high
    dilepton mass $\mll=116$--$150\gev$ (middle row), and low dilepton
    mass $\mll=46$--$66\gev$ (lower row) compared to the QCD fit
    result. In the $Z$-peak region and at high dilepton mass the
    measurements are shown separately for both the central (left) and forward
    (right) regions. The predictions are shown before (solid lines)
    and after (dashed lines) the shifts due to the correlated
    uncertainties are applied. The lower box of each plot shows the
    ratio of the theoretical calculations to the data.}
  \label{fig:nnlo_resultsZ}
\end{figure}

\subsubsection{Parton distributions}
\label{sec:pdfs}

The QCD fit determines a new set of PDFs, termed \epWZ16, which has
much smaller experimental uncertainties than the previous
 \epWZ12 set. Further uncertainties in the PDFs are estimated and
classified as model uncertainties and parameterization uncertainties,
which are listed separately in \Tab~\ref{tab:modparapdf}. Model
uncertainties comprise variations of $m_c$ and $m_b$ and variations of
the starting scale value $Q_0^2$ and of the minimum $Q^2$ value
($Q^2_\mathrm{min}$) of the HERA data included in the analysis. The
variation of the heavy-quark masses follows the HERAPDF2.0
analysis~\cite{Abramowicz:2015mha}. The variation of the charm-quark mass
and the starting scale are performed simultaneously, as the constraint $Q_0^2<m_c^2$ has to be fulfilled. The parameterization uncertainties
are estimated by adding further parameters in the polynomials $P_i(x)$
and allowing $B_{\bar{s}} \neq B_{\bar{d}}$. The PDFs including all
uncertainties are shown in \Fig~\ref{fig:pdfsmodparam}. The high level
of precision of the data makes it necessary to evaluate further
uncertainties, such as those from the effect of the
renormalization and factorization scales and the limitations of the
NNLO calculations. These are detailed below in terms of their influence
on the ratio of strange quarks to the light sea.

\begin{figure}[ptb]
  \begin{center}
    \includegraphics[width=0.4\textwidth]{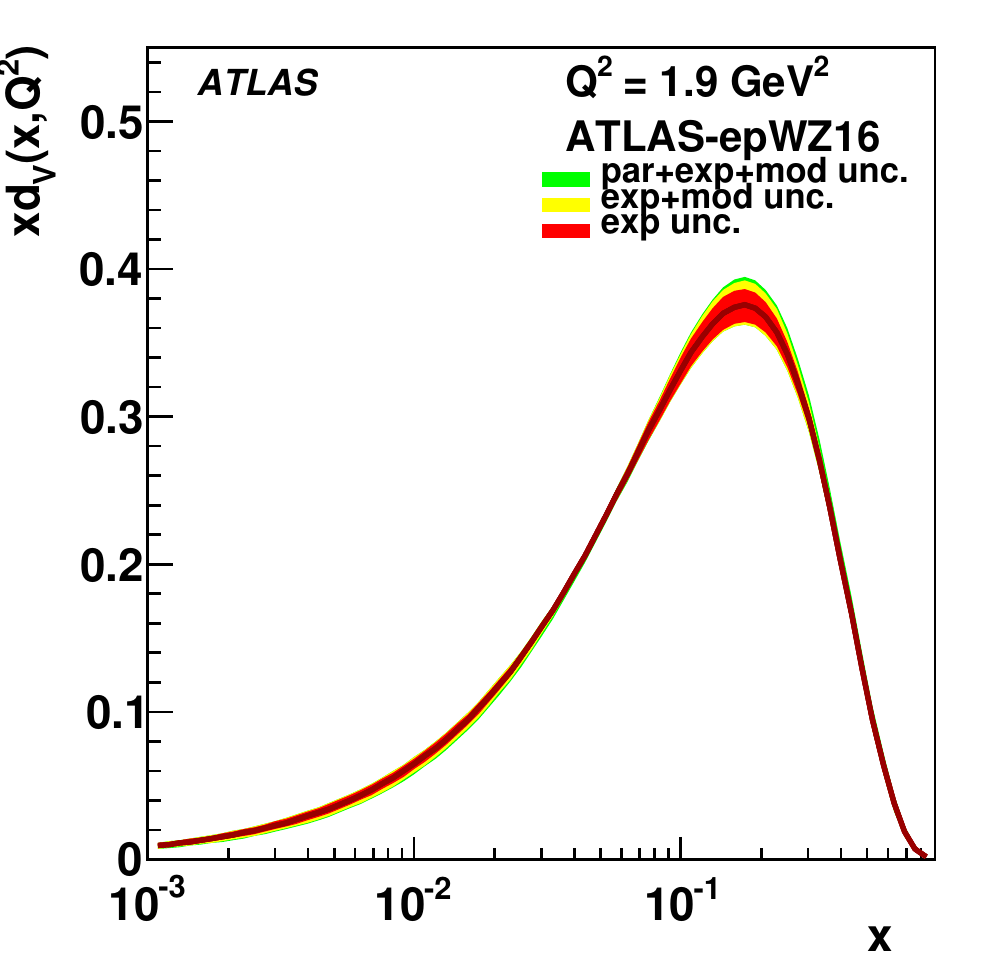}%
    \includegraphics[width=0.4\textwidth]{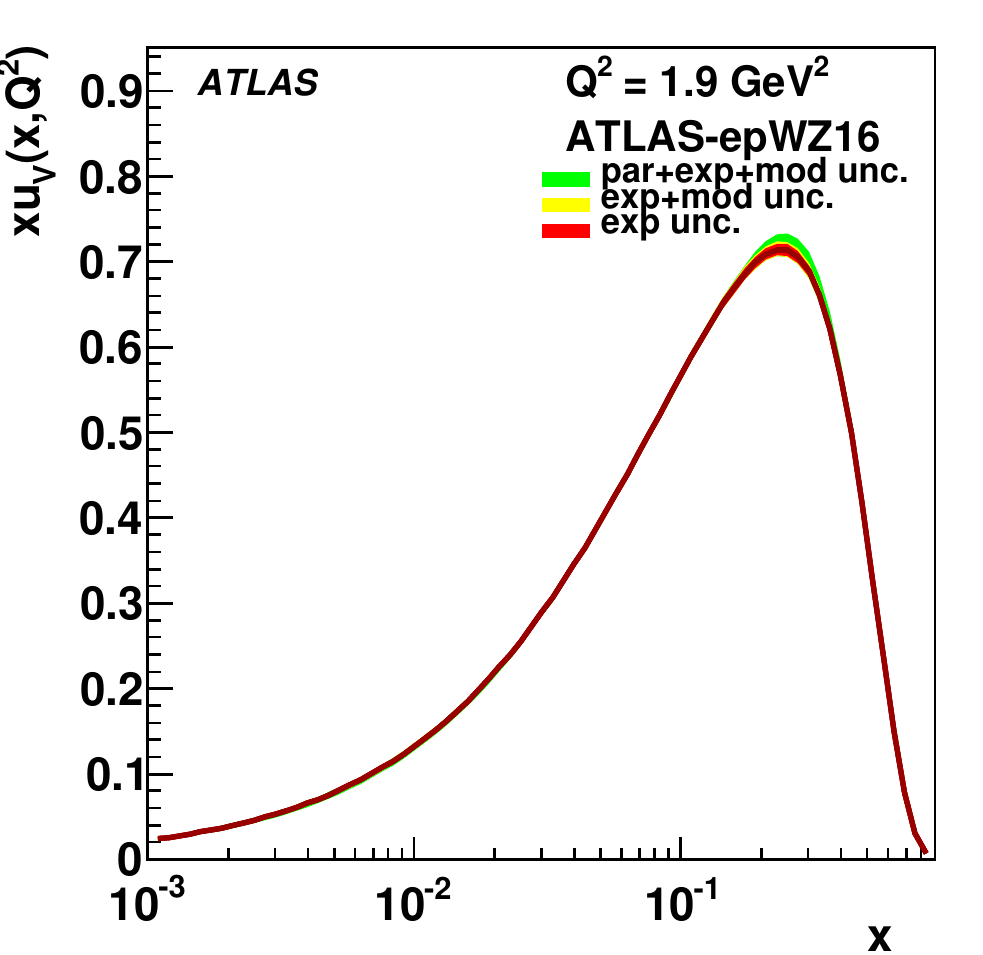}\\
    \includegraphics[width=0.4\textwidth]{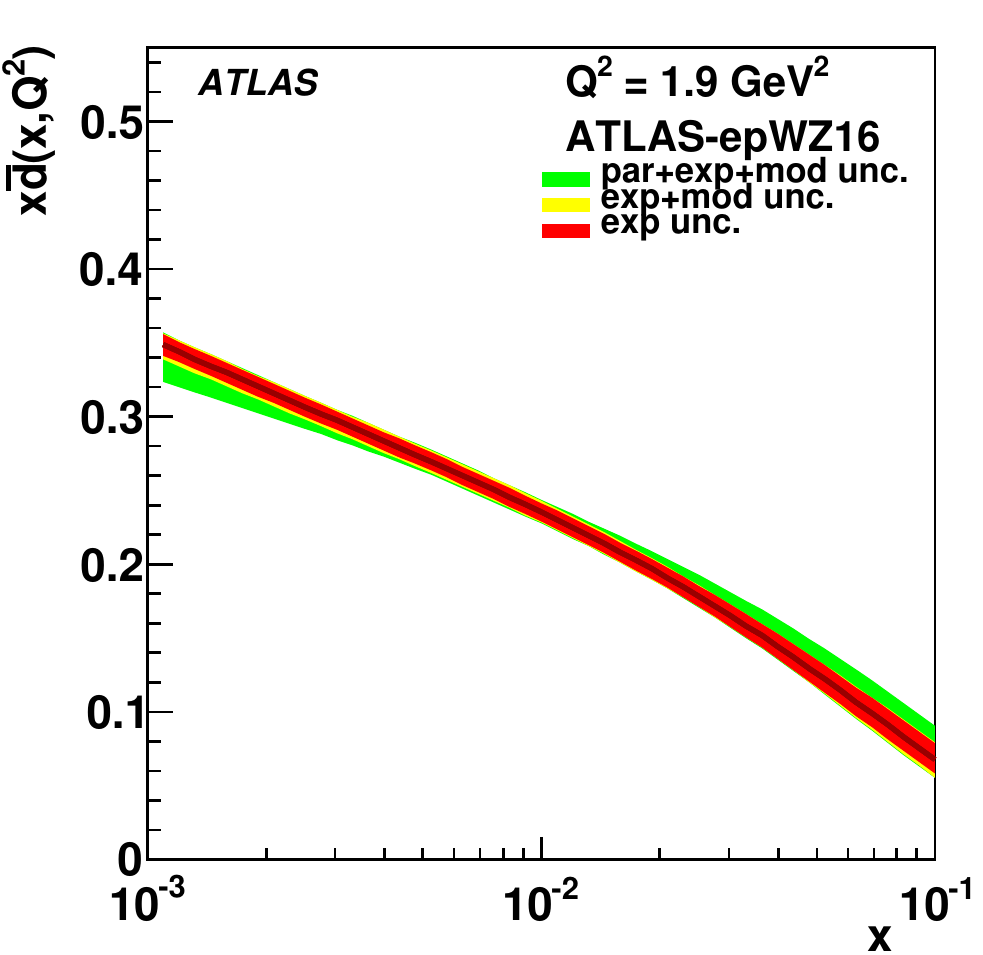}%
    \includegraphics[width=0.4\textwidth]{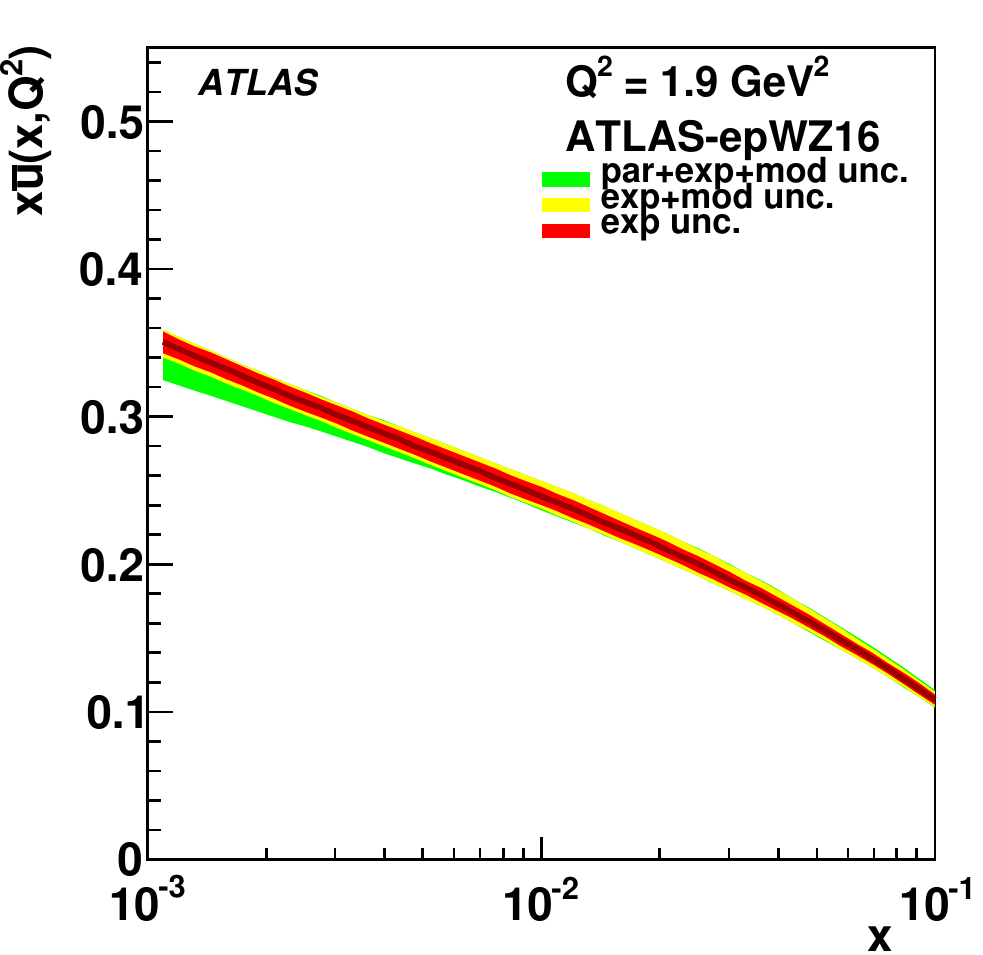}\\
    \includegraphics[width=0.4\textwidth]{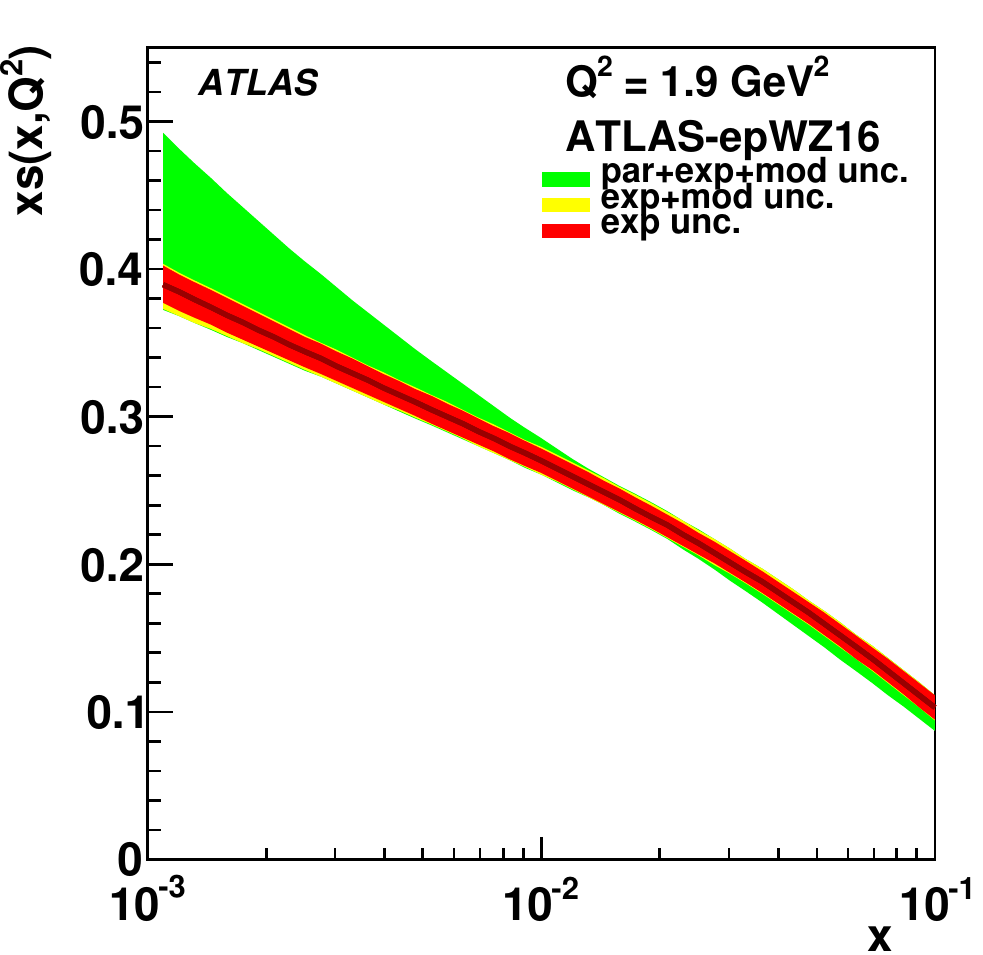}%
    \includegraphics[width=0.4\textwidth]{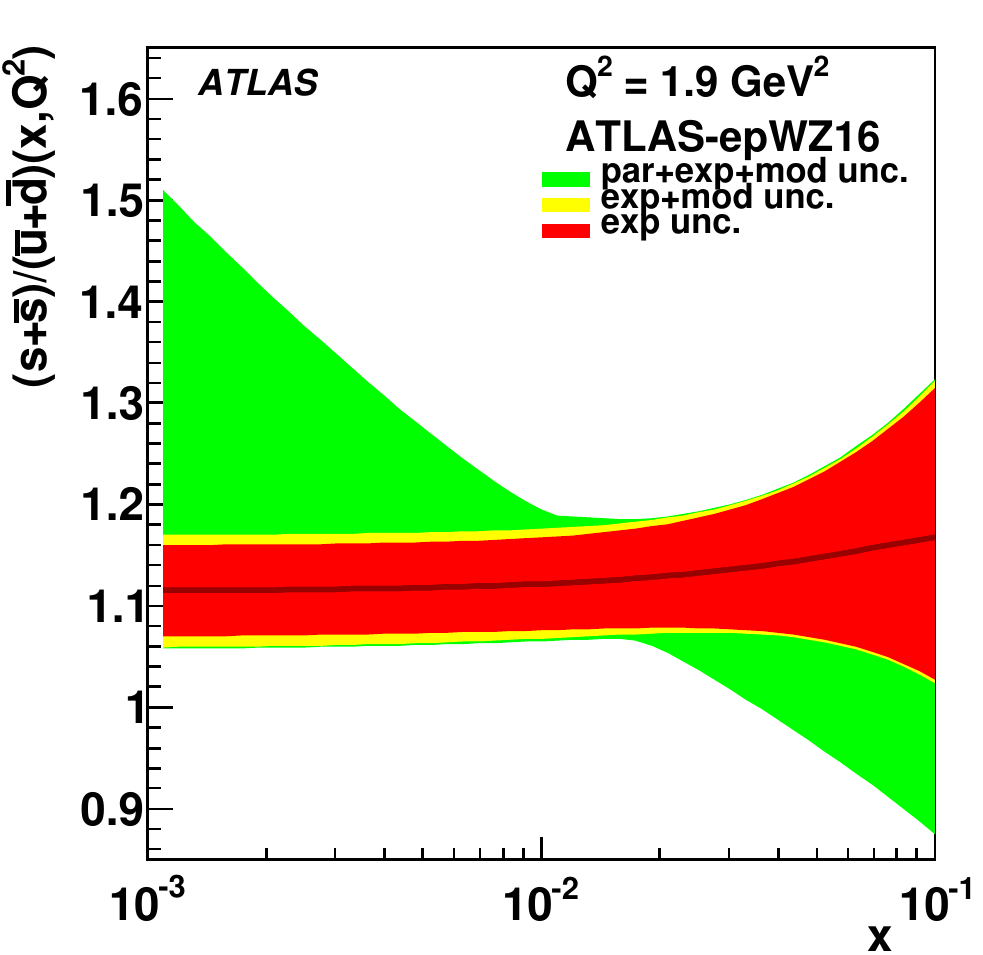}%
    \caption{PDFs from the present \epWZ{}16 determination at the
      starting scale $Q_0^2=1.9\GeV^2$. Top: valence PDFs
      $x\dv(x)$, $x\uv(x)$; middle: light sea PDFs $x\bar{d}(x)$,
      $x\bar{u}(x)$; bottom: strange-quark distribution and ratio
      $R_s(x)$. Uncertainty bands represent the experimental (exp),
      model (mod) and parameterization (par) components in red, yellow
      and green, respectively. The PDFs are shown in the region of
     maximum sensitivity of the ATLAS $W$ and \Zg\ data, $10^{-3} < x
      < 10^{-1}$, except for the valence quarks.}
    \label{fig:pdfsmodparam}
  \end{center}
\end{figure}

\begin{table}
  \begin{center}
    \begin{tabular}{lccc}
      \hline
      \hline
      Variation    & Total \chindf  & $r_s=\frac{s+\bar{s}}{2\bar{d}}$ & $R_s=\frac{{s}+\bar{s}}{\bar{u}+\bar{d}}$    \\[1ex]
      \hline
     Nominal fit  & 1321 / 1102       & 1.193  & 1.131\\
\hline
\multicolumn{4}{c}{Model variations}\\
\hline
   $m_b=4.25\GeV$ &  1319 / 1102 &1.172    & 1.111\\
   $m_b=4.75\GeV$ & 1322 / 1102  & 1.211   & 1.149\\
   $Q^2_\mathrm{min} = 5\GeV^2$&  1389 / 1149   & 1.202 & 1.128\\
   $Q^2_\mathrm{min} = 10\GeV^2$& 1263 / 1062   & 1.188 & 1.129\\
   $Q^2_0=1.6\GeV^2$ and $m_c=1.37\GeV$ & 1322 / 1101  & 1.198 & 1.148\\
   $Q^2_0=2.2\GeV^2$ and $m_c=1.49\GeV$ & 1323 / 1101  & 1.197 & 1.119\\
\hline
\multicolumn{4}{c}{Parameterization variations}\\
\hline
    $B_{\bar{s}}$  & 1319 / 1101   & 1.094 & 1.067\\    
    $D_{\bar{s}}$  & 1321 / 1101   & 1.192 & 1.130\\
    $D_{\bar{u}}$  & 1318 / 1101   & 1.184 & 1.128\\
    $D_{\bar{d}}$  & 1321 / 1101   & 1.194 & 1.132\\
    $D_{\dv}$  & 1320 / 1101   & 1.195 & 1.132\\
    $D_{\uv}$    & 1320 / 1101    & 1.161 & 1.107\\
    $D_{g}$        & 1319 / 1101   & 1.209 & 1.141\\
    $F_{\uv}$  & 1321 / 1101     & 1.206 & 1.143\\
    $F_{\dv}$  & 1323 / 1101     & 1.203 & 1.141\\
\hline
\multicolumn{4}{c}{Theoretical uncertainties}\\
\hline
   $\alphas(m_Z)=0.116$&  1320 / 1102   & 1.185 & 1.121\\
   $\alphas(m_Z)=0.120$& 1323 / 1102    &  1.194 & 1.136\\
   NLO EW down &  1323 / 1102   &  1.199 & 1.132\\
   NLO EW up &  1319 / 1102     & 1.187 & 1.130\\
   FEWZ 3.1b2 & 1314 / 1102  & 1.294 & 1.211\\
   \hline
   \hline  
 \end{tabular}
 \caption{Overview of the impact of variations in the QCD fit
   regarding the model, parameterization, and further theoretical
   choices as compared to the nominal fit. For each variation the
   total fit \chindf\ is given as well as the values of the two
   quantities $r_s$ and $R_s$ which describe the
   strange-to-light-sea-quark fraction at $Q_0^2$ and $x=0.023$. In
   the part of the table corresponding to the parameterization variations, the
   name of the additional parameter considered in addition to the
   15-parameter set given in Eq.~\eqref{eq:pdf} is listed.}
 \label{tab:modparapdf}
\end{center}
\end{table}

\subsubsection{Strange-quark density}
\label{sec:strange}

The QCD analysis of the ATLAS $2010$ $W$ and $Z$
measurements~\cite{Aad:2012sb} led to the unexpected observation that
strangeness is unsuppressed at low $x$ of $\simeq 0.023$ and low $Q^2=1.9\gev^2$, which means
that the strange, down and up sea quarks are of similar strength in
that kinematic range. This was supported by the ATLAS measurement of
associated $W$ and charm production~\cite{Aad:2014xca} and not in 
contradiction with a similar measurement performed by 
CMS~\cite{Chatrchyan:2013uja, Chatrchyan:2013mza}. But a large strange-quark density had
not been expected from previous analyses of dimuon production in
neutrino scattering~\cite{Goncharov:2001qe, Mason:2007zz,
  Samoylov:2013xoa, KayisTopaksu:2011mx} within the global PDF fit
approaches~\cite{Alekhin:2014sya, Ball:2014uwa, Harland-Lang:2014zoa,
  Dulat:2015mca}.

The fraction of the strange-quark density in the proton can be characterized by a
quantity $r_s$, defined as the ratio of the strange to the down
sea-quark distributions. When evaluated at the scale $Q^2=Q^2_0=1.9\gev^2$ and
$x=0.023$,\footnote{The value of Bjorken $x=0.023$ at $Q^2_0$
  roughly corresponds
  to the region of maximum sensitivity of a measurement at central
  rapidity at $\sqrt{s}=7\tev$ and a scale of
  $Q^2=m_Z^2$~\cite{Aad:2012sb}.}
the result is
\begin{equation}
  r_s = \frac{s+\bar{s}}{2 \bar{d}}= 1.19 \pm 0.07\,\mathrm{(exp)}\;\pm 0.02\,\mathrm{(mod)}\;^{+0.02}_{-0.10}\,\mathrm{(par)}\,.
\end{equation}
Here the uncertainties relate to those of the experimental data (exp)
determined by the Hessian method.
 The model (mod) and parameterization (par)
uncertainties are discussed in \Sec~\ref{sec:pdfs} and the
corresponding individual variations of $r_s$ are listed separately in
\Tab~\ref{tab:modparapdf}. This result represents an improvement of a
factor of three in the experimental uncertainty relative to the \epWZ12
fit~\cite{Aad:2012sb}. The improvement derives from the more precise ATLAS
data, which provide the sensitivity to the strange-quark density
through the shape of the $Z$ rapidity distribution in
combination with the common, absolute normalization of both the $W^\pm$ and $Z/\gamma^*$
cross sections.
The model uncertainties are reduced by a factor of three, mainly
because of the better control of the charm-quark mass parameter from
the HERA data~\cite{Abramowicz:1900rp}. The parameterization
uncertainty is determined to be $^{+0.02}_{-0.10}$ as compared to
$^{+0.10}_{-0.15}$ in the former analysis since the new, more precise
data leave less freedom in the parameter choice. The variation to
lower $r_s$ is dominated by the variation due to adding the $B_{\bar{s}}$
parameter which was not
accounted for in the previous analysis. The result is thus a
confirmation and improvement of the previous
observation~\cite{Aad:2012sb} of an unsuppressed strange-quark density
in the proton. As a cross-check, a re-analysis of the $2010$ data with
the present theoretical framework was performed, which yields a value
of $r_s$ consistent with both the former and the new value.

One may also express the strange-quark fraction with respect to the
total light-quark sea, which is the sum of up and down sea-quark
distributions, at the scale $Q^2=Q^2_0=1.9\gev^2$ and $x=0.023$:
\begin{equation}
  R_s = \frac{s+\bar{s}}{\bar{u}+ \bar{d}}=1.13 \pm 0.05\,\mathrm{(exp)} \pm 0.02\,\mathrm{(mod)} \;^{+0.01}_{-0.06}\,\mathrm{(par)}\,.
\end{equation}

The new determinations of $r_s$ and $R_s$ are illustrated in
\Fig~\ref{fig:rscomp}. The measurement is presented with the
experimental and the PDF-fit related uncertainties, where the latter
results from adding the model and parameterization uncertainties in
quadrature. The outer band illustrates additional, mostly theoretical
uncertainties which are presented below.  The result is
compared with recent global fit analyses, ABM12, MMHT14, CT14 and
NNPDF3.0. All of these predict $r_s$ and $R_s$ to be significantly lower than
unity, with values between about $0.4$ and $0.6$. Furthermore, these
global fit analyses are seen to exhibit substantially different
uncertainties in $r_s$ and $R_s$ due to exploiting different data and
prescriptions for fit uncertainties. The new result is in agreement
with the previous \epWZ12 analysis also shown in
\Fig~\ref{fig:rscomp}. It is also consistent with an earlier analysis
by the NNPDF group~\cite{Ball:2012cx} based on collider data only,
which obtains a value near unity, albeit with large
uncertainties.~\footnote{The CT10nnlo PDF set~\cite{Gao:2013xoa} is
observed to have a less suppressed strange-quark distribution with
$R_s = 0.80^{+0.20}_{-0.16}$ and $r_s=0.76^{+0.19}_{-0.16}$, which is
in slightly better agreement with the data than the newer CT14 PDF
set.}

\begin{figure}
  \begin{center}
    \includegraphics[width=0.5\textwidth]{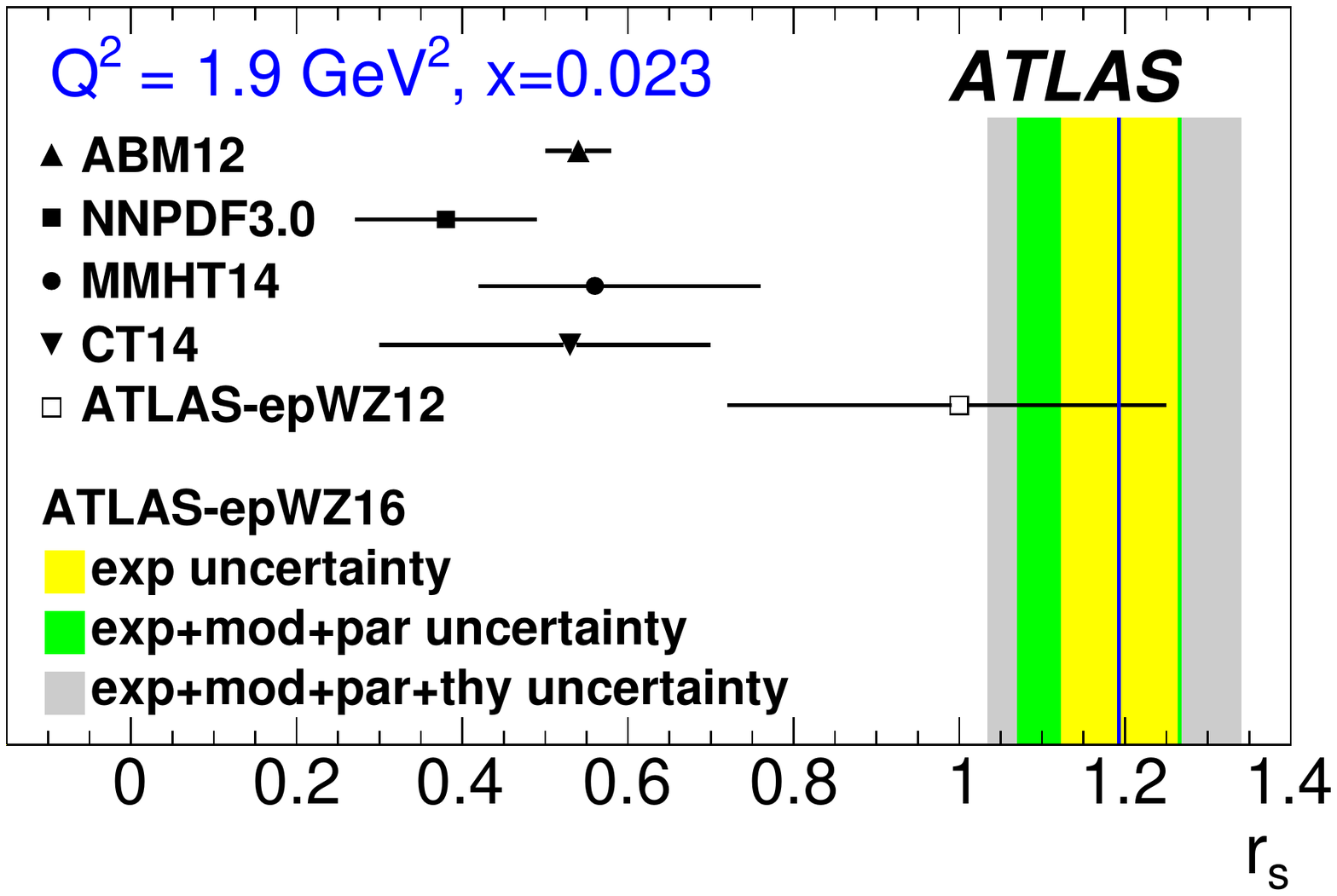}%
    \includegraphics[width=0.5\textwidth]{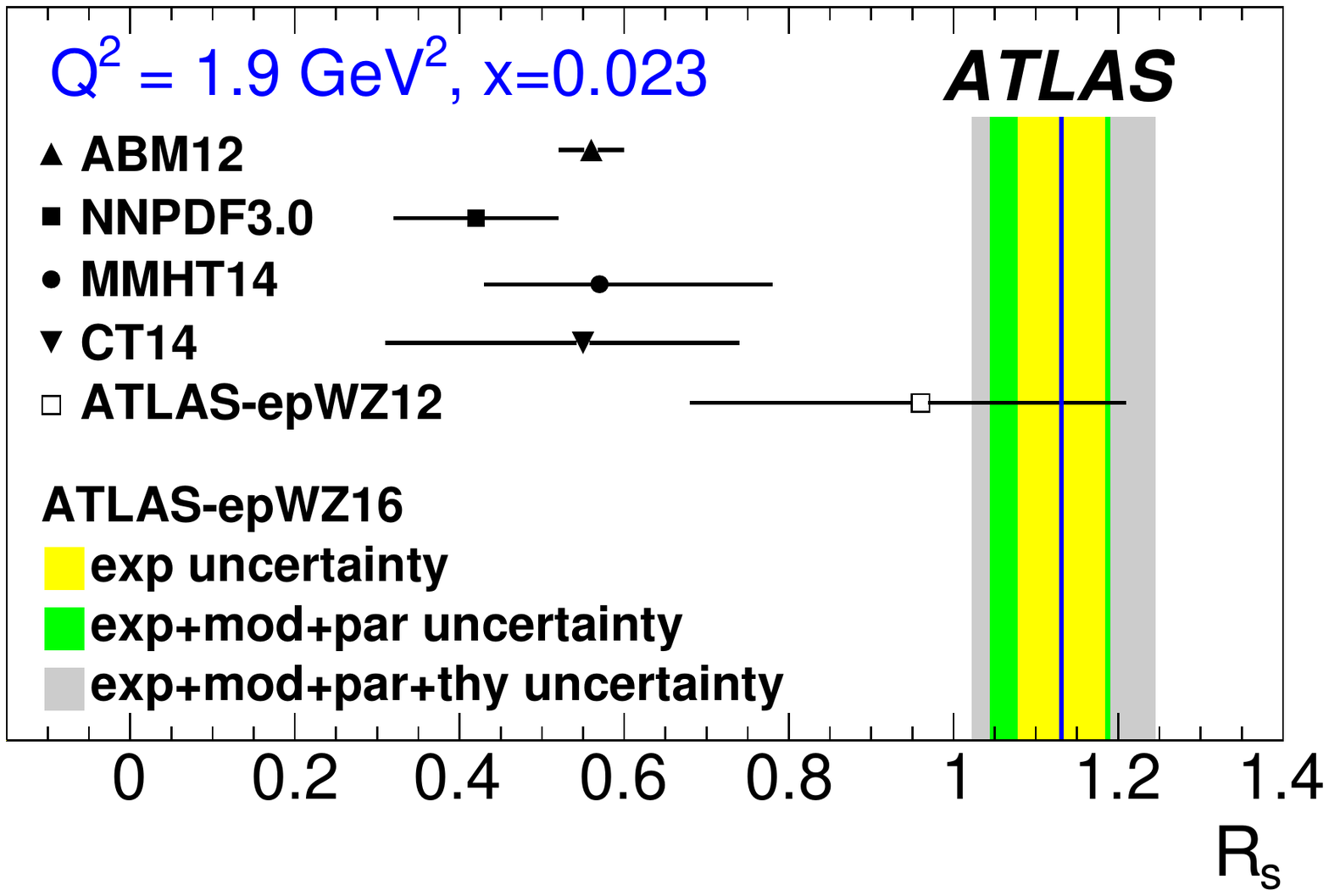}
    \caption{Determination of the relative strange-to-down sea quark
      fractions $r_s$ (left) and $R_s$ (right). Bands: Present result
      and its uncertainty contributions from experimental data, QCD
      fit, and theoretical uncertainties, see text; Closed symbols
      with horizontal error bars:  predictions from
      different NNLO PDF sets; Open square: previous ATLAS result
      ~\cite{Aad:2012sb}.  The ratios are calculated at the initial
      scale $Q_0^2=1.9\gev^2$ and at $x=0.023$ corresponding to the
      point of largest sensitivity at central rapidity of the ATLAS
      data.  }
    \label{fig:rscomp}
  \end{center}
\end{figure}

A careful evaluation of the value of $r_s$ requires the consideration
of a number of additional, mostly theoretical uncertainties. These
lead to the more complete result for $r_s$
\begin{equation}
  r_s=1.19 \pm 0.07\,\mathrm{(exp)}\;^{+0.13}_{-0.14}\,\mathrm{(mod+par+thy)}\,.
\end{equation}
Here the previously discussed model and parameterization uncertainties
are summarized and added together with further theoretical
uncertainties (thy) as follows: 
i) the uncertainty in
$\alphas(m_Z^2)$ is taken to be $\pm 0.002$ with a very small effect
on $r_s$; 
ii) the electroweak corrections and their
application, as described in \Sec\,\ref{sec:thyframe}, introduce a one
percent additional error for $r_s$; iii) the whole analysis was
repeated with predictions obtained with the FEWZ program (version
3.1b2) leading to a value of $r_s$ enlarged by $+0.10$ as compared to
the DYNNLO result; iv) finally the variation of the renormalization
($\mur$) and factorization ($\muf$) scales changes the result by
$10\%$ if one varies these by factors of $2$ up and
$1/2$ down (see below for further details). \TTab~\ref{tab:rsunc}
details all uncertainty components of $r_s$ and also $R_s$.

\begin{table}[tbh]
  \begin{center}  
    \renewcommand{\arraystretch}{1.6}
    \begin{tabular}{lcc}
      \hline
      \hline
      & $r_s=\frac{s+\bar{s}}{2\bar{d}}$ 
      & $R_s=\frac{s+\bar{s}}{\bar{u}+\bar{d}}$ \\[1ex]
      \hline                                                                    
      Central value & $1.19$ & $1.13$\\
      \hline                                                                    
      Experimental data & $\pm 0.07$ & $\pm 0.05$\\
      Model ($m_b$, $Q^2_\mathrm{min}$, $Q^2_0\,\&\,m_c$) 
      & $\pm 0.02$ & $\pm 0.02$\\
      Parameterization & {\Large $^{+0.02}_{-0.10}$} & {\Large $^{+0.01}_{-0.06}$} \\
      $\alphas$ & {\Large $^{+0.00}_{-0.01}$} & $\pm 0.01$\\
      EW corrections & $\pm 0.01$ & $\pm 0.00$\\
      QCD scales & {\Large $^{+0.08}_{-0.10}$} & {\Large $^{+0.06}_{-0.07}$}\\
      \FEWZ~3.1b2 & $+0.10$ & $+0.08$\\
      \hline
      Total uncertainty & {\Large $^{+0.15}_{-0.16}$} & $\pm 0.11$\\
      \hline
      \hline
    \end{tabular}
    \caption{Summary of the central value and all uncertainties in the
      variables $r_s$ and $R_s$ evaluated at $Q^2=1.9\gev^2$ and
      $x=0.023$ characterizing the fraction of the strange-quark
      density in the proton.}
    \label{tab:rsunc}
  \end{center}
\end{table}

Various further cross-checks are performed in order to assess the
reliability of the strange-quark density measurement.
\begin{itemize}
\item To test the sensitivity to assumptions about the low-$x$ behaviour
  of the light-quark sea, the constraint on $\bar{u}=\bar{d}$ as $x\to
  0$ is removed by allowing $A_{\bar{d}}$ and $B_{\bar{d}}$ to vary 
  independently from the respective $A_{\bar{u}}$ and
  $B_{\bar{u}}$. The resulting $\bar{u}$ is compatible with $\bar{d}$
  within uncertainties of $\simeq 8\%$ at $x \sim 0.001$ and
  $Q^2_0$, while $s+\bar{s}$ is found to be unsuppressed with $r_s
  = 1.16$.
\item The \epWZ16 PDF set results in a slightly negative central value
  of $x\bar{d}-x\bar{u}$ at $x\sim 0.1$, which with large uncertainties
  is compatible with zero. This result is about two standard deviations
  below the determination from E866 fixed-target Drell--Yan
  data~\cite{Towell:2001nh} according to which $x\bar{d}-x\bar{u} \sim
  0.04$ at $x\sim 0.1$. It has been suggested that the ATLAS
  parameterization forces a too small $x\bar{d}$ distribution if the
  strange-quark PDF is unsuppressed~\cite{Alekhin:2014sya}. However,
  the E866 observation is made at $x \sim 0.1$, while the 
  ATLAS $W,~Z$ data have the largest constraining power at $x\sim 0.023$.  For a
  cross-check, the E866 cross-section data was added to the QCD fit
  with predictions computed at NLO QCD. In this fit
  $x\bar{d}-x\bar{u}$ is enhanced and nevertheless the strange-quark
  distribution is found to be unsuppressed with $r_s$ near unity.
\item Separate analyses of the electron and muon data give results
  about one standard deviation above and below the result using their
  combination. If the $W^{\pm}$ and $Z$-peak data are used without the
  \Zg\ data at lower and higher \mll, a value of $r_s=1.23$ is found
  with a relative experimental uncertainty almost the same as in
  the nominal fit.
\item A suppressed strange-quark PDF may be enforced by fixing $r_s = 0.5$
  and setting $C_{\bar{s}} = C_{\bar{d}}$. The total $\chi^2$ obtained
  this way is $1503$, which is 182 units higher than the fit allowing
  these two parameters to be free. The ATLAS partial $\chi^2$
  increases from 108 units to 226 units for the 61 degrees of
  freedom. A particularly large increase is observed for the $Z$-peak
  data, where $\chindf = 53/12$ is found for a fit with suppressed
  strangeness.
\end{itemize}

A final estimate of uncertainties is performed with regard to choosing
the renormalization and factorization scales in the calculation of the
Drell--Yan cross sections.  The central fit is performed using the
dilepton and $W$ masses, \mll\ and $m_W$, as default scale choices.
Conventionally both scales are varied by a factor of $2$ and $0.5$ as
an estimate of missing higher-order QCD terms. 
\Tab~\ref{tab:scales} presents the results of varying the scales
separately and jointly. It is observed that a choice of half the mass
values leads to a significant improvement of the $\chi^2$ by about 24
units. All separate variations of $\mur$ and $\muf$ cause the
resulting strange fraction values to be inside the envelope obtained
from the joint variation $\mur=\muf$ up or down.
 
\begin{table}
  \begin{center}
      \begin{tabular}{cccccc}
        \hline
        \hline
        $\mur$& $\muf$ & \multicolumn{2}{c}{ \chindf}  & $r_s=\frac{s+\bar{s}}{2\bar{d}}$ &$R_s=\frac{s+\bar{s}}{\bar{u}+\bar{d}}$\\
        & & Total & ATLAS & &\\
        \hline
        1 & 1  & 1321 / 1102 & 108 / 61  &1.193   & 1.131\\
        \hline
        1/2 & 1/2 &  1297 / 1102 & 85 / 61 & 1.093  & 1.066\\
        2 &2 & 1329 / 1102  & 115 / 61  & 1.270 & 1.186\\
        1 & 1/2  &  1307 / 1102 & 94 / 61  & 1.166 & 1.115\\
        1& 2 & 1312 / 1102  & 100 / 61    &1.201 & 1.130\\
        1/2 & 1 &  1304 / 1102 & 94 / 61   & 1.128& 1.088\\
        2 & 1 & 1321 / 1102  & 107 / 61   & 1.241&1.165\\
        \hline
        \hline
      \end{tabular}
      \caption{Effect of varying the scales for the Drell--Yan data in
        the NNLO QCD fit.  The renormalization, $\mur$, and
        factorization, $\muf$, scales, are expressed relatively to the
        dilepton mass for NC and the $W$ mass for the CC cross
        section.  Changes of the total fit $\chi^2$ values are almost
        exclusively due to variations of the ATLAS values while the
        HERA $\chi^2$, given by their difference, remains nearly
        constant.  Right columns: resulting $r_s$ and $R_s$ values,
        quoted at $Q^2=Q_0^2$ and $x=0.023$. }
\label{tab:scales}
\end{center}
\end{table}

\subsubsection{Determination of \Vcs}
\label{sec:vcs}

As discussed in the preceding section, the combination of HERA DIS and
newly presented ATLAS measurements results in a precise determination of
the light-quark composition of the proton and specifically of the
strange-quark density. The most significant contributions
to $W$-boson production are from the Cabibbo-favoured initial states
$ud$ and $cs$, where the rate is also controlled by the 
magnitude of the CKM
matrix elements $|V_{ud}|$ and \Vcs. While $|V_{ud}|$ is
experimentally measured to very high precision, this is not true for
the \Vcs\ element.
The contributions from the Cabibbo-suppressed initial state $cd$,
which are sensitive to $|V_{cd}|$, are suppressed by one order of
magnitude compared to the $cs$ contribution.
Both the $W^\pm$ production rates and the lepton
pseudorapidity distributions contain information about the $cs\to W$
contribution to the CC Drell--Yan cross section. A PDF fit as described
above is performed, but in addition the \Vcs\ parameter is allowed
to vary freely while all other CKM matrix elements are fixed to the
values given in \Tab~\ref{ewqcd:tabEWpar}, which were 
obtained from a global fit imposing unitarity.
The following value and corresponding uncertainties are found
\begin{equation}
  \Vcs =  0.969 \pm 0.013\,\mathrm{(exp)}~^{+0.006}_{-0.003}\,\mathrm{(mod)}~^{+0.003}_{-0.027}\,\mathrm{(par)}~^{+0.011}_{-0.005}\,\mathrm{(thy)}\,.
\end{equation}
\TTab~\ref{tab:Vcsunc} details all the uncertainty components of
\Vcs. In this fit the value of $r_s$ is found to be $1.18$, 
compared to $1.19$ when \Vcs\ is fixed to the value assuming
unitarity of the CKM matrix. The experimental uncertainty of
\Vcs\ is $66\%$ correlated with the parameter $A_s$ controlling the
normalization of the strange-quark density, while the parameter $B_s$
is fixed to $B_{\bar{d}}$. The correlation with $C_s$ is found to be
$10\%$.

\begin{table}[tbh]
  \begin{center}  
    \renewcommand{\arraystretch}{1.6}
    \begin{tabular}{lc}
      \hline
      \hline
      & \Vcs\\
      \hline                                                                    
      Central value & $0.969$ \\
      \hline                                                                    
      Experimental data & $\pm 0.013$ \\
      Model ($m_b$, $Q^2_\mathrm{min}$, $Q^2_0\,\&\,m_c$) 
      & {\Large$^{+0.006}_{-0.003}$}\\
      Parameterization & {\Large $^{+0.003}_{-0.027}$} \\
      $\alphas$ & $\pm 0.000$ \\
      EW corrections & $\pm 0.004$ \\
      QCD scales & {\Large $^{+0.000}_{-0.003}$} \\
      \FEWZ~3.1b2 & $+0.011$ \\
      \hline
      Total uncertainty& {\Large $^{+0.018}_{-0.031}$}\\
      \hline
      \hline
    \end{tabular}
    \caption{Summary of the central value and all uncertainties in the
      CKM matrix element \Vcs.}
    \label{tab:Vcsunc}
  \end{center}
\end{table}

The dominant uncertainty of \Vcs\ arises from the parameterization
variation associated with the extra freedom given to the strange-quark
distribution by releasing the assumption $B_{\bar{d}}=B_{\bar{s}}$
that fixes the rise of $x\bar{d}(x)$ and $x\bar{s}(x)$ to be the same
at low $x$. 

This determination represents a new, competitive
measurement of \Vcs. \FFig~\ref{fig:Vcs} compares the result to
determinations of \Vcs\ extracted from leptonic $D_s$ meson decays,
$D_s \to \ell\nu$~\cite{Zupanc:2013byn, Alexander:2009ux, delAmoSanchez:2010jg,Onyisi:2009th,Naik:2009tk,Aoki:2013ldr},
and from semileptonic $D$ meson decays, $D \to
K\ell\nu$~\cite{Besson:2009uv,Widhalm:2006wz,Aubert:2007wg,Aoki:2013ldr}, from data by the CLEO-c, BABAR, and Belle experiments as
reported in Ref.~\cite{Agashe:2014kda}. In addition, an early
determination of \Vcs\ by the NNPDF Collaboration from a QCD fit is
shown~\cite{Ball:2009mk}.

\begin{figure}
  \begin{center}   
    \includegraphics[width=0.59\textwidth]{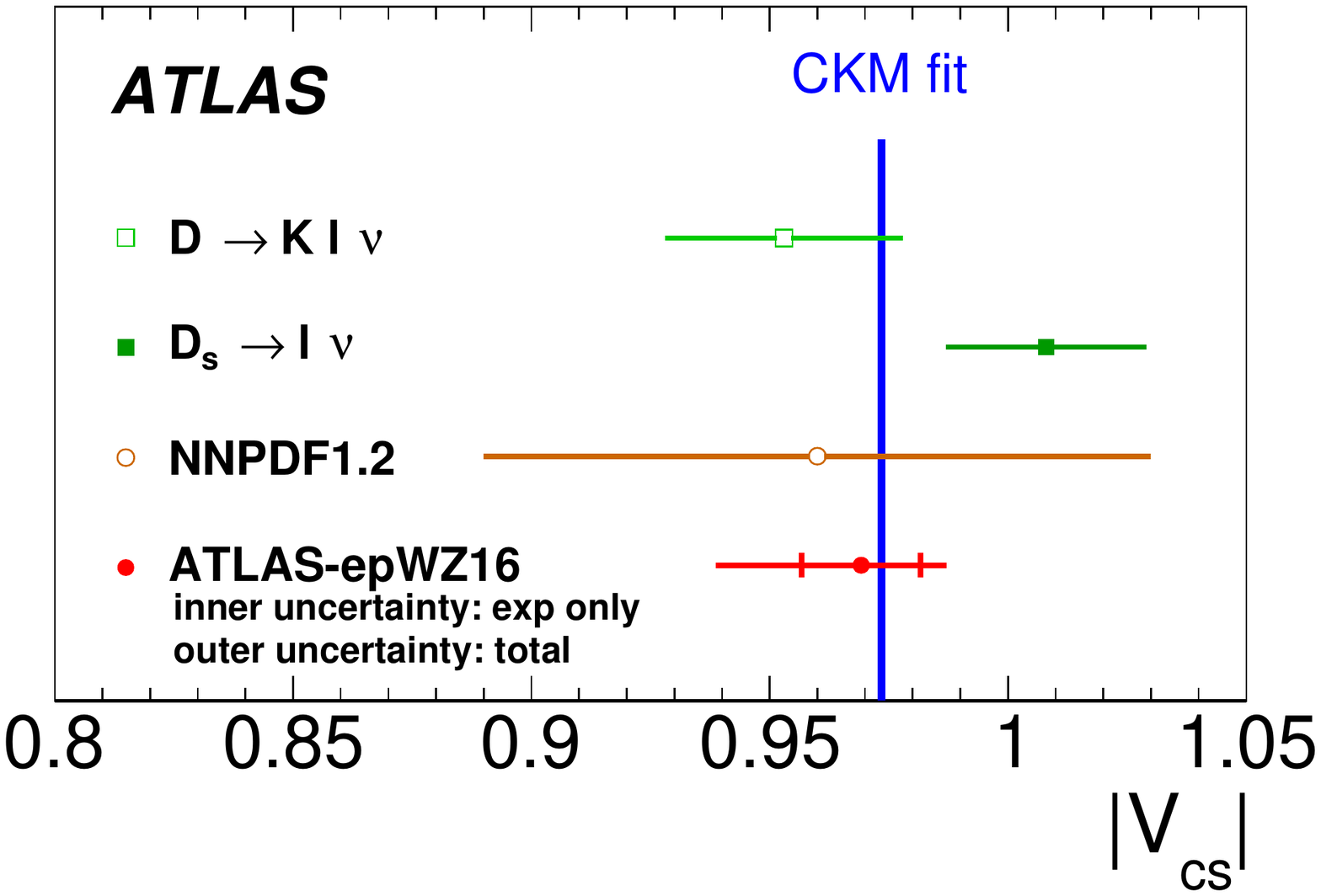}
  \end{center}
  \caption{\Vcs\ as determined in the global CKM
    fit cited by the PDG~\cite{Agashe:2014kda} (blue vertical line) compared to
    extractions from $D_s \to \ell\nu$ and $D \to K\ell\nu$
    decays~\cite{Agashe:2014kda} and the NNPDF1.2
    fit~\cite{Ball:2009mk}. The \epWZ16 fit result is shown with
    uncertainty contributions from the experimental data (inner error
    bar) and the total uncertainty including all fit and further
    theoretical uncertainties (outer error bar). The uncertainty in
    \Vcs\ from the CKM fit with unitarity constraint is smaller
    than the width of the vertical line.}
  \label{fig:Vcs}
\end{figure}


\section{Summary}
\label{sec:summary}
New cross-section measurements by the ATLAS Collaboration
are presented for inclusive Drell--Yan
production in the neutral-current channel, \Zgll, and the charged-current
channel, \Wpluslnu\ and \Wminuslnu. The measurement
is based on data taken in $pp$ collisions at the LHC at a centre-of-mass energy
of $\sqrt{s}=7\TeV$ with an integrated luminosity of \lumi.
Cross sections are provided in the electron and muon decay
channels, integrated over the fiducial regions and differentially. The
\Wpluslnu\ and \Wminuslnu\ cross sections are measured as a function of lepton
pseudorapidity \etal. The \Zgll\ cross sections are measured as a
function of the dilepton rapidity, \yll, in three dilepton mass bins
$46 < \mll <150\GeV$ in the central region and extended into
the forward region up to $|\yll|=3.6$ for $66 < \mll <150\GeV$.

The electron and muon channel results are combined considering all
sources of correlated and uncorrelated uncertainties.  A new
sensitive test of electron--muon universality in on-shell $W$ and $Z$ decays
is presented. The combined integrated fiducial $W^+,~W^-,$ and $Z$ cross
sections are measured to an experimental precision of
$0.6,~0.5,$ and $0.32\%$, respectively, apart from the common $1.8\%$
normalization uncertainty through the luminosity determination. The
differential measurements are nearly as precise as the integrated
cross-section results except at the edges of the phase space. 
With the full information about correlated uncertainties given, the data
provide correspondingly precise results of cross-section ratios and the $W^{\pm}$
lepton charge asymmetry as well.

A measurement precision at sub-percent level represents an opportunity
and challenge for the QCD interpretation. Predictions for the
Drell--Yan processes \Wpmln\ and \Zgll\ are calculated at NNLO fixed
order in QCD and including NLO electroweak corrections. A quantitative
comparison of the differential cross sections shows deviations of the
predictions obtained with many of the contemporary PDF sets, 
hinting to a special impact of the data on the determination of the 
strange-quark distribution. 

An NNLO QCD analysis is performed on the new
\Wpmln\ and \Zgll\ ATLAS data together with the final, combined data from H1 and
ZEUS on inclusive neutral-current and charged-current deep inelastic
scattering. A new set of parton distribution functions, 
termed \epWZ16, is provided. A detailed fit
analysis supports the previous observation by ATLAS of a large ratio of the
strange-quark distribution  to the lighter sea-quark
distributions at low $x$. Specifically, 
the ratio of the strange to
the down sea-quark distributions, evaluated at a scale of
$Q^2=1.9\gev^2$ at a mean $x=0.023$, is found to be $r_s=1.19$
with a total uncertainty
of $0.16$. Experimentally, $r_s$ is determined with 
an uncertainty of $0.07$ which is a threefold reduction
relative to the previous determination by the
ATLAS Collaboration.

A complete set of uncertainties in the QCD fit result is provided in addition
to the experimental uncertainties. This covers the effects of model,
parameterization, and further theoretical uncertainties. Detailed
studies are performed regarding the accuracy with which NNLO QCD
predictions for the Drell--Yan process can be computed, including the
differences in existing codes, DYNNLO and FEWZ, and the effect of the
choice of scales. The uncertainties in the strange-quark density from
the limitations of NNLO QCD calculations of the fiducial cross sections
are found to significantly exceed the
experimental errors. An interesting
observation is the significant improvement in the description of the
ATLAS data when factorization and renormalization scales are set to a
half of the canonically used dilepton mass scales. Several cross-checks
are presented to evaluate the reliability of the measured enhancement of
the strange-quark density.
The paper finally presents a determination of
the CKM matrix element \Vcs\ which has a precision comparable to
extractions from charm meson decays.

\section*{Acknowledgements}


We thank CERN for the very successful operation of the LHC, as well as the
support staff from our institutions without whom ATLAS could not be
operated efficiently.

We acknowledge the support of ANPCyT, Argentina; YerPhI, Armenia; ARC, Australia; BMWFW and FWF, Austria; ANAS, Azerbaijan; SSTC, Belarus; CNPq and FAPESP, Brazil; NSERC, NRC and CFI, Canada; CERN; CONICYT, Chile; CAS, MOST and NSFC, China; COLCIENCIAS, Colombia; MSMT CR, MPO CR and VSC CR, Czech Republic; DNRF and DNSRC, Denmark; IN2P3-CNRS, CEA-DSM/IRFU, France; SRNSF, Georgia; BMBF, HGF, and MPG, Germany; GSRT, Greece; RGC, Hong Kong SAR, China; ISF, I-CORE and Benoziyo Center, Israel; INFN, Italy; MEXT and JSPS, Japan; CNRST, Morocco; NWO, Netherlands; RCN, Norway; MNiSW and NCN, Poland; FCT, Portugal; MNE/IFA, Romania; MES of Russia and NRC KI, Russian Federation; JINR; MESTD, Serbia; MSSR, Slovakia; ARRS and MIZ\v{S}, Slovenia; DST/NRF, South Africa; MINECO, Spain; SRC and Wallenberg Foundation, Sweden; SERI, SNSF and Cantons of Bern and Geneva, Switzerland; MOST, Taiwan; TAEK, Turkey; STFC, United Kingdom; DOE and NSF, United States of America. In addition, individual groups and members have received support from BCKDF, the Canada Council, CANARIE, CRC, Compute Canada, FQRNT, and the Ontario Innovation Trust, Canada; EPLANET, ERC, ERDF, FP7, Horizon 2020 and Marie Sk{\l}odowska-Curie Actions, European Union; Investissements d'Avenir Labex and Idex, ANR, R{\'e}gion Auvergne and Fondation Partager le Savoir, France; DFG and AvH Foundation, Germany; Herakleitos, Thales and Aristeia programmes co-financed by EU-ESF and the Greek NSRF; BSF, GIF and Minerva, Israel; BRF, Norway; CERCA Programme Generalitat de Catalunya, Generalitat Valenciana, Spain; the Royal Society and Leverhulme Trust, United Kingdom.

The crucial computing support from all WLCG partners is acknowledged gratefully, in particular from CERN, the ATLAS Tier-1 facilities at TRIUMF (Canada), NDGF (Denmark, Norway, Sweden), CC-IN2P3 (France), KIT/GridKA (Germany), INFN-CNAF (Italy), NL-T1 (Netherlands), PIC (Spain), ASGC (Taiwan), RAL (UK) and BNL (USA), the Tier-2 facilities worldwide and large non-WLCG resource providers. Major contributors of computing resources are listed in Ref.~\cite{ATL-GEN-PUB-2016-002}.

\clearpage

\printbibliography

\clearpage

\appendix

\part*{Appendix}
\addcontentsline{toc}{part}{Appendix}

\section{Differential measurements in electron and muon channels}
\label{sec:simple_xsectable}

\begin{table}[bhp]
  \begin{minipage}{7.5cm}
    {
      \tiny
      \begin{tabular}{cc|c|cccc}
        \hline
        \hline
        $|\etal|^\mathrm{min}$ &
        $|\etal|^\mathrm{max}$ &
        $\mathrm{d}\sigma/\mathrm{d}|\etal|$ &
        $\delta_\mathrm{sta}$ &
        $\delta_\mathrm{unc}$ &
        $\delta_\mathrm{sys}$ &
        $\delta_\mathrm{tot}$ \\
        & & $[\mathrm{pb}]$ & $[\%]$ & $[\%]$ & $[\%]$ & $[\%]$ \\\hline
0.00 & 0.21 & 436.8 & 0.15 & 0.15 & 0.91 & 0.93\\
0.21 & 0.42 & 433.1 & 0.14 & 0.17 & 0.89 & 0.91\\
0.42 & 0.63 & 430.0 & 0.14 & 0.15 & 0.90 & 0.92\\
0.63 & 0.84 & 424.5 & 0.14 & 0.13 & 0.99 & 1.01\\
0.84 & 1.05 & 415.3 & 0.15 & 0.17 & 1.08 & 1.10\\
1.05 & 1.37 & 405.1 & 0.13 & 0.16 & 1.36 & 1.38\\
1.52 & 1.74 & 371.0 & 0.17 & 0.17 & 1.31 & 1.34\\
1.74 & 1.95 & 367.6 & 0.18 & 0.26 & 1.26 & 1.30\\
1.95 & 2.18 & 345.8 & 0.17 & 0.18 & 1.28 & 1.31\\
2.18 & 2.50 & 322.3 & 0.2 & 0.2 & 2.2 & 2.2\\
        \hline\hline
      \end{tabular}
    }
  \end{minipage}%
  \begin{minipage}{7.5cm}
    {
      \tiny
      \begin{tabular}{cc|c|cccc}
        \hline
        \hline
        $|\etal|^\mathrm{min}$ &
        $|\etal|^\mathrm{max}$ &
        $\mathrm{d}\sigma/\mathrm{d}|\etal|$ &
        $\delta_\mathrm{sta}$ &
        $\delta_\mathrm{unc}$ &
        $\delta_\mathrm{sys}$ &
        $\delta_\mathrm{tot}$ \\
        & & $[\mathrm{pb}]$ & $[\%]$ & $[\%]$ & $[\%]$ & $[\%]$ \\\hline
0.00 & 0.21 & 577.2 & 0.13 & 0.14 & 1.00 & 1.01\\
0.21 & 0.42 & 577.5 & 0.12 & 0.15 & 0.94 & 0.96\\
0.42 & 0.63 & 583.2 & 0.12 & 0.14 & 0.93 & 0.95\\
0.63 & 0.84 & 588.7 & 0.12 & 0.12 & 0.97 & 0.98\\
0.84 & 1.05 & 588.4 & 0.12 & 0.16 & 0.94 & 0.96\\
1.05 & 1.37 & 598.5 & 0.10 & 0.15 & 1.13 & 1.14\\
1.52 & 1.74 & 593.7 & 0.14 & 0.14 & 1.17 & 1.19\\
1.74 & 1.95 & 610.8 & 0.14 & 0.19 & 1.03 & 1.05\\
1.95 & 2.18 & 594.6 & 0.12 & 0.15 & 1.04 & 1.05\\
2.18 & 2.50 & 559.6 & 0.13 & 0.15 & 1.55 & 1.56\\
        \hline\hline
      \end{tabular}
    }
  \end{minipage}
  \caption{Differential cross section for the \Wenum\ (left) and \Wenup\ (right) processes,
    extrapolated to the common fiducial region. The relative statistical
    ($\delta_\mathrm{sta}$), uncorrelated systematic
    ($\delta_\mathrm{unc}$), correlated systematic
    ($\delta_\mathrm{sys}$), and total ($\delta_\mathrm{tot}$)
    uncertainties are given in percent.
    The overall \dlumi\% luminosity uncertainty is not
    included.}
  \label{tab:w_eta_e}
\end{table}

\begin{table}[bhp]
  \begin{center}  
    {\small
    \begin{tabular}{cc|c|cccc}
      \hline
      \hline
      $|\yll|^\mathrm{min}$ &
      $|\yll|^\mathrm{max}$ &
      $\mathrm{d}\sigma/\mathrm{d}|\yll|$ &
      $\delta_\mathrm{sta}$ &
      $\delta_\mathrm{unc}$ &
      $\delta_\mathrm{sys}$ &
      $\delta_\mathrm{tot}$ \\
      & & $[\mathrm{pb}]$ & $[\%]$ & $[\%]$ & $[\%]$ & $[\%]$ \\\hline
0.00 & 0.40 & 3.595 & 1.5 & 0.9 & 1.3 & 2.2\\
0.40 & 0.80 & 3.622 & 1.5 & 0.8 & 1.2 & 2.1\\
0.80 & 1.20 & 3.456 & 1.8 & 0.9 & 1.4 & 2.4\\
1.20 & 1.60 & 3.382 & 2.0 & 1.0 & 1.5 & 2.7\\
1.60 & 2.00 & 2.968 & 2.3 & 1.1 & 1.5 & 2.9\\
2.00 & 2.40 & 1.567 & 2.9 & 1.2 & 1.2 & 3.4\\
      \hline\hline
    \end{tabular}
  }
  \caption{Differential cross section for the \Zgee\ process in the
    central region with $46<\mll<66\gev$, extrapolated to the common
    fiducial region. The relative statistical ($\delta_\mathrm{sta}$),
    uncorrelated systematic ($\delta_\mathrm{unc}$), correlated
    systematic ($\delta_\mathrm{sys}$), and total
    ($\delta_\mathrm{tot}$) uncertainties are given in percent.
    The overall \dlumi\% luminosity uncertainty
    is not included.}
    \label{tab:zlow_yll_e}
  \end{center}
\end{table}

\begin{table}[bhp]
  \begin{minipage}{8.5cm}
    {
      \tiny
      \begin{tabular}{cc|c|cccc}
        \hline
        \hline
        $|\yll|^\mathrm{min}$ &
        $|\yll|^\mathrm{max}$ &
        $\mathrm{d}\sigma/\mathrm{d}|\yll|$ &
        $\delta_\mathrm{sta}$ &
        $\delta_\mathrm{unc}$ &
        $\delta_\mathrm{sys}$ &
        $\delta_\mathrm{tot}$ \\
        & & $[\mathrm{pb}]$ & $[\%]$ & $[\%]$ & $[\%]$ & $[\%]$ \\\hline
0.00 & 0.20 & 135.6 & 0.28 & 0.18 & 0.40 & 0.52\\
0.20 & 0.40 & 135.3 & 0.29 & 0.16 & 0.39 & 0.52\\
0.40 & 0.60 & 133.9 & 0.30 & 0.16 & 0.39 & 0.51\\
0.60 & 0.80 & 133.7 & 0.31 & 0.17 & 0.40 & 0.54\\
0.80 & 1.00 & 132.9 & 0.32 & 0.18 & 0.41 & 0.55\\
1.00 & 1.20 & 129.4 & 0.34 & 0.20 & 0.41 & 0.57\\
1.20 & 1.40 & 120.2 & 0.36 & 0.19 & 0.44 & 0.60\\
1.40 & 1.60 & 106.5 & 0.38 & 0.19 & 0.43 & 0.61\\
1.60 & 1.80 & 89.3 & 0.44 & 0.23 & 0.54 & 0.73\\
1.80 & 2.00 & 68.7 & 0.51 & 0.30 & 0.39 & 0.71\\
2.00 & 2.20 & 46.03 & 0.59 & 0.39 & 0.47 & 0.85\\
2.20 & 2.40 & 21.86 & 0.91 & 0.67 & 0.74 & 1.35\\
        \hline\hline
      \end{tabular}
    }
  \end{minipage}%
  \begin{minipage}{8.5cm}
    {
      \tiny
      \begin{tabular}{cc|c|cccc}
        \hline
        \hline
        $|\yll|^\mathrm{min}$ &
        $|\yll|^\mathrm{max}$ &
        $\mathrm{d}\sigma/\mathrm{d}|\yll|$ &
        $\delta_\mathrm{sta}$ &
        $\delta_\mathrm{unc}$ &
        $\delta_\mathrm{sys}$ &
        $\delta_\mathrm{tot}$ \\
        & & $[\mathrm{pb}]$ & $[\%]$ & $[\%]$ & $[\%]$ & $[\%]$ \\\hline
1.20 & 1.40 & 7.71 & 1.8 & 1.8 & 3.2 & 4.1\\
1.40 & 1.60 & 17.95 & 1.0 & 1.1 & 3.0 & 3.4\\
1.60 & 1.80 & 32.57 & 0.7 & 0.7 & 2.7 & 2.9\\
1.80 & 2.00 & 50.5 & 0.6 & 1.8 & 2.6 & 3.2\\
2.00 & 2.20 & 68.5 & 0.6 & 2.7 & 2.2 & 3.5\\
2.20 & 2.40 & 86.6 & 0.5 & 1.9 & 1.9 & 2.8\\
2.40 & 2.80 & 86.1 & 0.3 & 3.0 & 1.7 & 3.5\\
2.80 & 3.20 & 40.71 & 0.5 & 0.6 & 5.5 & 5.6\\
3.20 & 3.60 & 11.00 & 1.2 & 3.7 & 6.4 & 7.5\\
        \hline\hline
      \end{tabular}
    }
  \end{minipage}
  \caption{Differential cross section for the \Zgee\ process in the
    central (left) and forward (right) region with $66<\mll<116\gev$, extrapolated to the common
    fiducial region. The relative statistical ($\delta_\mathrm{sta}$),
    uncorrelated systematic ($\delta_\mathrm{unc}$), correlated
    systematic ($\delta_\mathrm{sys}$), and total
    ($\delta_\mathrm{tot}$) uncertainties are given in percent.
    The overall \dlumi\% luminosity uncertainty
    is not included.}
  \label{tab:zpeak_yll_e}
\end{table}

\begin{table}[bhp]
  \begin{minipage}{8.5cm}
    {
      \tiny
      \begin{tabular}{cc|c|cccc}
        \hline
        \hline
        $|\yll|^\mathrm{min}$ &
        $|\yll|^\mathrm{max}$ &
        $\mathrm{d}\sigma/\mathrm{d}|\yll|$ &
        $\delta_\mathrm{sta}$ &
        $\delta_\mathrm{unc}$ &
        $\delta_\mathrm{sys}$ &
        $\delta_\mathrm{tot}$ \\
        & & $[\mathrm{pb}]$ & $[\%]$ & $[\%]$ & $[\%]$ & $[\%]$ \\\hline
0.00 & 0.40 & 1.503 & 2.0 & 2.5 & 1.4 & 3.5\\
0.40 & 0.80 & 1.422 & 2.1 & 0.9 & 1.4 & 2.7\\
0.80 & 1.20 & 1.329 & 2.3 & 1.3 & 1.4 & 3.0\\
1.20 & 1.60 & 1.181 & 2.6 & 1.6 & 1.5 & 3.4\\
1.60 & 2.00 & 0.754 & 3.3 & 2.4 & 2.0 & 4.6\\
2.00 & 2.40 & 0.328 & 4.9 & 2.4 & 1.8 & 5.7\\
        \hline\hline
      \end{tabular}
    }
  \end{minipage}%
  \begin{minipage}{8.5cm}
    {
      \tiny
      \begin{tabular}{cc|c|cccc}
        \hline
        \hline
        $|\yll|^\mathrm{min}$ &
        $|\yll|^\mathrm{max}$ &
        $\mathrm{d}\sigma/\mathrm{d}|\yll|$ &
        $\delta_\mathrm{sta}$ &
        $\delta_\mathrm{unc}$ &
        $\delta_\mathrm{sys}$ &
        $\delta_\mathrm{tot}$ \\
        & & $[\mathrm{pb}]$ & $[\%]$ & $[\%]$ & $[\%]$ & $[\%]$ \\\hline
1.20 & 1.60 & 0.300 & 6.8 & 6.6 & 9.1 & 13.1\\
1.60 & 2.00 & 0.547 & 5.2 & 7.8 & 7.3 & 11.9\\
2.00 & 2.40 & 0.912 & 4.0 & 13.5 & 4.5 & 14.8\\
2.40 & 2.80 & 0.931 & 3.9 & 20.9 & 4.0 & 21.6\\
2.80 & 3.20 & 0.438 & 5.3 & 14.4 & 6.8 & 16.8\\
3.20 & 3.60 & 0.070 & 14.5 & 11.6 & 7.2 & 19.9\\
        \hline\hline
      \end{tabular}
    }
  \end{minipage}
  \caption{Differential cross section for the \Zgee\ process in the
    central (left) and forward (right) region with $116<\mll<150\gev$, extrapolated to the common
    fiducial region. The relative statistical ($\delta_\mathrm{sta}$),
    uncorrelated systematic ($\delta_\mathrm{unc}$), correlated
    systematic ($\delta_\mathrm{sys}$), and total
    ($\delta_\mathrm{tot}$) uncertainties are given in percent.
    The overall \dlumi\% luminosity uncertainty
    is not included.}
  \label{tab:zhigh_yll_e}
\end{table}

\begin{table}[bhp]
  \begin{minipage}{7.5cm}
    {
      \tiny
      \begin{tabular}{cc|c|cccc}
        \hline
        \hline
        $|\etal|^\mathrm{min}$ &
        $|\etal|^\mathrm{max}$ &
        $\mathrm{d}\sigma/\mathrm{d}|\etal|$ &
        $\delta_\mathrm{sta}$ &
        $\delta_\mathrm{unc}$ &
        $\delta_\mathrm{sys}$ &
        $\delta_\mathrm{tot}$ \\
        & & $[\mathrm{pb}]$ & $[\%]$ & $[\%]$ & $[\%]$ & $[\%]$ \\\hline
0.00 & 0.21 & 439.0 & 0.16 & 0.41 & 0.67 & 0.80\\
0.21 & 0.42 & 437.0 & 0.15 & 0.52 & 0.55 & 0.77\\
0.42 & 0.63 & 431.4 & 0.14 & 0.27 & 0.59 & 0.67\\
0.63 & 0.84 & 425.6 & 0.15 & 0.33 & 0.62 & 0.72\\
0.84 & 1.05 & 413.5 & 0.16 & 0.29 & 0.60 & 0.69\\
1.05 & 1.37 & 406.8 & 0.12 & 0.29 & 0.56 & 0.65\\
1.37 & 1.52 & 389.2 & 0.17 & 0.34 & 0.55 & 0.67\\
1.52 & 1.74 & 380.6 & 0.14 & 0.43 & 0.60 & 0.75\\
1.74 & 1.95 & 367.1 & 0.15 & 0.32 & 0.62 & 0.71\\
1.95 & 2.18 & 345.0 & 0.14 & 0.38 & 0.63 & 0.75\\
2.18 & 2.50 & 318.3 & 0.15 & 0.50 & 0.67 & 0.85\\
        \hline\hline
      \end{tabular}
    }
  \end{minipage}%
  \begin{minipage}{7.5cm}
    {
      \tiny
      \begin{tabular}{cc|c|cccc}
        \hline
        \hline
        $|\etal|^\mathrm{min}$ &
        $|\etal|^\mathrm{max}$ &
        $\mathrm{d}\sigma/\mathrm{d}|\etal|$ &
        $\delta_\mathrm{sta}$ &
        $\delta_\mathrm{unc}$ &
        $\delta_\mathrm{sys}$ &
        $\delta_\mathrm{tot}$ \\
        & & $[\mathrm{pb}]$ & $[\%]$ & $[\%]$ & $[\%]$ & $[\%]$ \\\hline
0.00 & 0.21 & 581.3 & 0.14 & 0.41 & 0.63 & 0.77\\
0.21 & 0.42 & 583.6 & 0.13 & 0.46 & 0.58 & 0.75\\
0.42 & 0.63 & 583.2 & 0.12 & 0.25 & 0.57 & 0.64\\
0.63 & 0.84 & 587.3 & 0.13 & 0.31 & 0.59 & 0.67\\
0.84 & 1.05 & 585.6 & 0.14 & 0.37 & 0.59 & 0.71\\
1.05 & 1.37 & 601.5 & 0.10 & 0.26 & 0.59 & 0.65\\
1.37 & 1.52 & 599.1 & 0.13 & 0.33 & 0.57 & 0.67\\
1.52 & 1.74 & 607.5 & 0.11 & 0.31 & 0.57 & 0.66\\
1.74 & 1.95 & 604.4 & 0.11 & 0.50 & 0.57 & 0.76\\
1.95 & 2.18 & 598.7 & 0.10 & 0.57 & 0.60 & 0.83\\
2.18 & 2.50 & 563.1 & 0.11 & 0.60 & 0.63 & 0.88\\
        \hline\hline
      \end{tabular}
    }
  \end{minipage}
  \caption{Differential cross section for the \Wmunum\ (left) and \Wmunup\ (right) processes,
    extrapolated to the common fiducial region. The relative statistical
    ($\delta_\mathrm{sta}$), uncorrelated systematic
    ($\delta_\mathrm{unc}$), correlated systematic
    ($\delta_\mathrm{sys}$), and total ($\delta_\mathrm{tot}$)
    uncertainties are given in.
    The overall \dlumi\% luminosity uncertainty is not
    included.}
  \label{tab:w_eta_mu}
\end{table}

\begin{table}[bhp]
  \tiny
  \begin{tabular}{cc|c|cccc}
    \hline
    \hline
    $|\yll|^\mathrm{min}$ &
    $|\yll|^\mathrm{max}$ &
    $\mathrm{d}\sigma/\mathrm{d}|\yll|$ &
    $\delta_\mathrm{sta}$ &
    $\delta_\mathrm{unc}$ &
    $\delta_\mathrm{sys}$ &
    $\delta_\mathrm{tot}$ \\
    & & $[\mathrm{pb}]$ & $[\%]$ & $[\%]$ & $[\%]$ & $[\%]$ \\\hline
0.00 & 0.20 & 134.8 & 0.25 & 0.12 & 0.41 & 0.50\\
0.20 & 0.40 & 134.2 & 0.26 & 0.12 & 0.41 & 0.50\\
0.40 & 0.60 & 134.3 & 0.26 & 0.11 & 0.41 & 0.50\\
0.60 & 0.80 & 132.5 & 0.26 & 0.11 & 0.41 & 0.50\\
0.80 & 1.00 & 132.2 & 0.25 & 0.12 & 0.40 & 0.48\\
1.00 & 1.20 & 128.8 & 0.26 & 0.13 & 0.40 & 0.49\\
1.20 & 1.40 & 119.6 & 0.26 & 0.11 & 0.42 & 0.50\\
1.40 & 1.60 & 107.6 & 0.28 & 0.16 & 0.41 & 0.52\\
1.60 & 1.80 & 89.9 & 0.30 & 0.13 & 0.46 & 0.57\\
1.80 & 2.00 & 68.7 & 0.34 & 0.17 & 0.49 & 0.62\\
2.00 & 2.20 & 45.39 & 0.45 & 0.27 & 0.44 & 0.69\\
2.20 & 2.40 & 22.43 & 0.78 & 0.43 & 0.52 & 1.03\\
    \hline\hline
  \end{tabular}
  \caption{Differential cross section for the \Zgmumu\ process in the
    region with $66<\mll<116\gev$, extrapolated to the common
    fiducial region. The relative statistical ($\delta_\mathrm{sta}$),
    uncorrelated systematic ($\delta_\mathrm{unc}$), correlated
    systematic ($\delta_\mathrm{sys}$), and total
    ($\delta_\mathrm{tot}$) uncertainties are given in percent.
    The overall \dlumi\% luminosity uncertainty
    is not included.}
  \label{tab:zpeak_yll_mu}
\end{table}

\begin{table}[bhp]
  \begin{minipage}{8.5cm}
    {
      \tiny
      \begin{tabular}{cc|c|cccc}
        \hline
        \hline
        $|\yll|^\mathrm{min}$ &
        $|\yll|^\mathrm{max}$ &
        $\mathrm{d}\sigma/\mathrm{d}|\yll|$ &
        $\delta_\mathrm{sta}$ &
        $\delta_\mathrm{unc}$ &
        $\delta_\mathrm{sys}$ &
        $\delta_\mathrm{tot}$ \\
        & & $[\mathrm{pb}]$ & $[\%]$ & $[\%]$ & $[\%]$ & $[\%]$ \\\hline
0.00 & 0.40 & 3.444 & 1.3 & 0.6 & 1.6 & 2.2\\
0.40 & 0.80 & 3.479 & 1.2 & 0.6 & 1.5 & 2.0\\
0.80 & 1.20 & 3.375 & 1.2 & 0.6 & 1.5 & 2.0\\
1.20 & 1.60 & 3.412 & 1.2 & 0.5 & 1.4 & 1.9\\
1.60 & 2.00 & 2.914 & 1.3 & 0.5 & 1.4 & 1.9\\
2.00 & 2.40 & 1.522 & 2.0 & 0.7 & 1.5 & 2.6\\
        \hline\hline
      \end{tabular}
    }
  \end{minipage}%
  \begin{minipage}{8.5cm}
    {
      \tiny
      \begin{tabular}{cc|c|cccc}
        \hline
        \hline
        $|\yll|^\mathrm{min}$ &
        $|\yll|^\mathrm{max}$ &
        $\mathrm{d}\sigma/\mathrm{d}|\yll|$ &
        $\delta_\mathrm{sta}$ &
        $\delta_\mathrm{unc}$ &
        $\delta_\mathrm{sys}$ &
        $\delta_\mathrm{tot}$ \\
        & & $[\mathrm{pb}]$ & $[\%]$ & $[\%]$ & $[\%]$ & $[\%]$ \\\hline
0.00 & 0.40 & 1.505 & 1.8 & 0.8 & 1.4 & 2.4\\
0.40 & 0.80 & 1.467 & 1.8 & 0.8 & 1.4 & 2.4\\
0.80 & 1.20 & 1.356 & 1.9 & 0.9 & 1.3 & 2.5\\
1.20 & 1.60 & 1.172 & 1.9 & 0.8 & 1.3 & 2.5\\
1.60 & 2.00 & 0.766 & 2.5 & 0.9 & 1.7 & 3.2\\
2.00 & 2.40 & 0.324 & 4.2 & 1.5 & 1.9 & 4.8\\
        \hline\hline
      \end{tabular}
    }
  \end{minipage}
  \caption{Differential cross section for the \Zgmumu\ process in the
    region $46<\mll<66\gev$ (left) and  $116<\mll<150\gev$ (right), extrapolated to the common
    fiducial region. The relative statistical ($\delta_\mathrm{sta}$),
    uncorrelated systematic ($\delta_\mathrm{unc}$), correlated
    systematic ($\delta_\mathrm{sys}$), and total
    ($\delta_\mathrm{tot}$) uncertainties are given in percent.
    The overall \dlumi\% luminosity uncertainty
    is not included.}
  \label{tab:zlowhigh_yll_mu}
\end{table}


\clearpage
\begin{flushleft}
{\Large The ATLAS Collaboration}

\bigskip

M.~Aaboud$^\textrm{\scriptsize 137d}$,
G.~Aad$^\textrm{\scriptsize 88}$,
B.~Abbott$^\textrm{\scriptsize 115}$,
J.~Abdallah$^\textrm{\scriptsize 8}$,
O.~Abdinov$^\textrm{\scriptsize 12}$,
B.~Abeloos$^\textrm{\scriptsize 119}$,
O.S.~AbouZeid$^\textrm{\scriptsize 139}$,
N.L.~Abraham$^\textrm{\scriptsize 151}$,
H.~Abramowicz$^\textrm{\scriptsize 155}$,
H.~Abreu$^\textrm{\scriptsize 154}$
% ...
\bigskip
\\
% ...
$^{8}$ Department of Physics, The University of Texas at Arlington, Arlington TX, United States of America\\
% ...
$^{12}$ Institute of Physics, Azerbaijan Academy of Sciences, Baku, Azerbaijan\\
% ...
$^{88}$ CPPM, Aix-Marseille Universit{\'e} and CNRS/IN2P3, Marseille, France\\
% ...
$^{115}$ Homer L. Dodge Department of Physics and Astronomy, University of Oklahoma, Norman OK, United States of America\\
% ...
$^{119}$ LAL, Univ. Paris-Sud, CNRS/IN2P3, Universit{\'e} Paris-Saclay, Orsay, France\\
% ...
$^{137}$ $^{(a)}$ Facult{\'e} des Sciences Ain Chock, R{\'e}seau Universitaire de Physique des Hautes Energies - Universit{\'e} Hassan II, Casablanca; $^{(b)}$ Centre National de l'Energie des Sciences Techniques Nucleaires, Rabat; $^{(c)}$ Facult{\'e} des Sciences Semlalia, Universit{\'e} Cadi Ayyad, LPHEA-Marrakech; $^{(d)}$ Facult{\'e} des Sciences, Universit{\'e} Mohamed Premier and LPTPM, Oujda; $^{(e)}$ Facult{\'e} des sciences, Universit{\'e} Mohammed V, Rabat, Morocco\\
% ...
$^{139}$ Santa Cruz Institute for Particle Physics, University of California Santa Cruz, Santa Cruz CA, United States of America\\
% ...
$^{151}$ Department of Physics and Astronomy, University of Sussex, Brighton, United Kingdom\\
% ...
$^{154}$ Department of Physics, Technion: Israel Institute of Technology, Haifa, Israel\\
$^{155}$ Raymond and Beverly Sackler School of Physics and Astronomy, Tel Aviv University, Tel Aviv, Israel\\
% ...
$^{a}$ Also at Department of Physics, King's College London, London, United Kingdom\\
$^{b}$ Also at Institute of Physics, Azerbaijan Academy of Sciences, Baku, Azerbaijan\\
$^{c}$ Also at Novosibirsk State University, Novosibirsk, Russia\\
$^{d}$ Also at TRIUMF, Vancouver BC, Canada\\
$^{e}$ Also at Department of Physics {\&} Astronomy, University of Louisville, Louisville, KY, United States of America\\
% ...
\end{flushleft}


\end{document}